\documentclass[fleqn,twocolumn,10pt]{wlscirep}
\usepackage[utf8]{inputenc}
\usepackage[T1]{fontenc}
\usepackage{longtable}
\usepackage{tabularx}
\usepackage{multirow}

\usepackage{caption}
\usepackage{subcaption}

\title{Social network analysis of manga: similarities to real-world social networks and trends over decades}

\author[1]{Kashin Sugishita}
\author[2,3,4, *]{Naoki Masuda}

\affil[1]{Department of Transdisciplinary Science and Engineering, Tokyo Institute of Technology, Tokyo, 152-8550, Japan}
\affil[2]{Department of Mathematics, State University of New York at Buffalo, Buffalo, 14260-2900, NY, USA}
\affil[3]{Computational and Data-Enabled Science and Engineering Program, State University of New York at Buffalo, Buffalo, 14260-5030, NY, USA}
\affil[4]{Center for Computational Social Science, Kobe University, Kobe, 657-8501, Japan}

\affil[*]{Corresponding author}

\begin{abstract}
Manga, Japanese comics, has been popular on a global scale.
Social networks among characters, which are often called character networks, may be a significant contributor to their popularity. 
We collected data from 162 popular manga that span over 70 years and analyzed their character networks.
First, we found that many of static and temporal properties of the character networks are similar to those of real human social networks.
Second, the character networks of most manga are protagonist-centered such that a single protagonist interacts with the majority of other characters.
Third, the character networks for manga mainly targeting boys have shifted to denser and less protagonist-centered networks and with fewer characters over decades.
Manga mainly targeting girls showed the opposite trend except for the downward trend in the number of characters. 
The present study, which relies on manga data sampled on an unprecedented scale, paves the way for further population studies of character networks and other aspects of comics.
\end{abstract}


\begin{document}

\flushbottom
\maketitle
%
%
\thispagestyle{empty}


\section{Introduction}
The global comic market has been rapidly growing. 
The market size was valued at USD 14.7 billion in 2021 and is expected to expand at a compound annual growth rate of 4.8\% from 2022 to 2029 \cite{MangaGlobalMarketProjection2022}.
A driving force underlying the rapid expansion of the market is the widespread availability of e-books, which enables us to easily access comics around the world.
The COVID-19 pandemic also helped the growth of the comic market because the lockdown in many countries increased the demand for comics. 
Manga, Japanese comics, has been especially popular on a global scale.
Many manga have been translated into different languages.
For example, over 510 million copies of {\it One Piece} have circulated worldwide; \textit{One Piece} is recorded in the Guinness Book of Records as the best-selling comic in the history \cite{OnePiece2022Mainichi}.
The history of the development of manga culture in Japan, its social background, and its global prevalence have been studied \cite{ito2005history, brenner2007understanding, schodt2013dreamland}.

Our social networks are complex but characterized by common properties such as heterogeneity and community structure \cite{barabasi2016network, newman2018networks}. 
Because literally all storylines of manga rely on interactions among characters,  social networks among characters, which are often called character networks, may provide a backbone of the story of manga and influence its popularity. 
More generally, character networks have been analyzed for quantitatively characterizing fiction works \cite{labatut2019extraction, perc2020beauty}.
A number of problems can be addressed through analyses of character networks, such as summarization \cite{tran2015movie, bost2019remembering}, classification \cite{holanda2019character, ardanuy2014structure}, and role detection \cite{jung2013emotion, weng2007movie}.
Character networks have been studied for a variety of fiction such as novels \cite{elson2010extracting,chaturvedi2016modeling, ardanuy2014structure, bonato2016mining, min2019modeling, gessey2020narrative}, plays \cite{moretti2011network, stiller2003small, rieck2016shall, mutton2004inferring},  movies \cite{park2012social, weng2007movie, agarwal2014parsing, yeh2014clustering}, and TV series \cite{park2012social, chen2016character, bost2019remembering, nan2015social, tan2014character}.
However, character networks for comics have been rarely explored except for Marvel comics \cite{alberich2002marvel, gleiser2007become}, a graphic novel \cite{labatut2022complex}, 
and a few manga titles \cite{murakami2020creating}. 
One possible reason for this is the difficulty of automating data collection for comics \cite{labatut2019extraction}.

In the present study, we analyze character networks of 162 popular Japanese manga that span decades and test the following two hypotheses. 
First, we expect that popular manga tend to produce realistic social networks.
Therefore, we hypothesize that structural and temporal properties of the character networks of manga are similar to those of human social networks. 
If this hypothesis is supported, it could shed light on how to compose popular manga.
Second, our data enables us to investigate how trends of manga have changed over approximately 70 years. 
Therefore, we hypothesize that character networks of modern manga and old ones are systematically different. 
We also examine differences in the network structure between manga mainly targeting boys versus girls. 
Through these analyses, we provide a new understanding of manga culture and its historical development.

\section{Methods}

\subsection{Data collection}
We focus on manga of which more than 10 million copies have been published in Japan as of January 2021 \cite{MangaRanking2021}.
Since the physical size of the manga may affect the number of panels per page and we adopted the page as the unit of analysis, we excluded the manga that had not been published in the paperback pocket edition ($112 {\, \rm mm} \times 174 {\, \rm mm}$).
Note that the paperback pocket edition is the most common for the boys' and girls' manga in Japan.
As a result, we included 162 manga (see SI file for the list of the 162 manga).

We prepared a data table of time-stamped copresence of characters for each manga as follows. 
For each manga, we manually examined volumes one through three.
Examining only three volumes is due to a logistic limitation.
All the selected 162 manga had at least three volumes.
Then, we first extracted all the characters whose name, blood relation to a named character, or job title that uniquely identifies the character, is known. 
To ensure the reproducibility of this work, we avoided to use any other information sources  (e.g., other volumes of the same manga, the Internet, or anime) to attempt to identify more characters. 
In practice, such additional information sources would not contribute to identifying many more characters. 
Second, we recorded the copresence of characters on each page as interaction between the characters.
Note that the copresence is the most common definition of interaction in constructing character networks in fictional works \cite{labatut2019extraction}.
We used copresence on a page rather than in a single panel within a page because it is common that characters appearing in different panels on the same page have some interactions (e.g., two characters talk to each other by alternately occupying successive panels).

\subsection{Character networks}
\begin{figure*}[h!]
\centering
\includegraphics[width=0.7\linewidth]{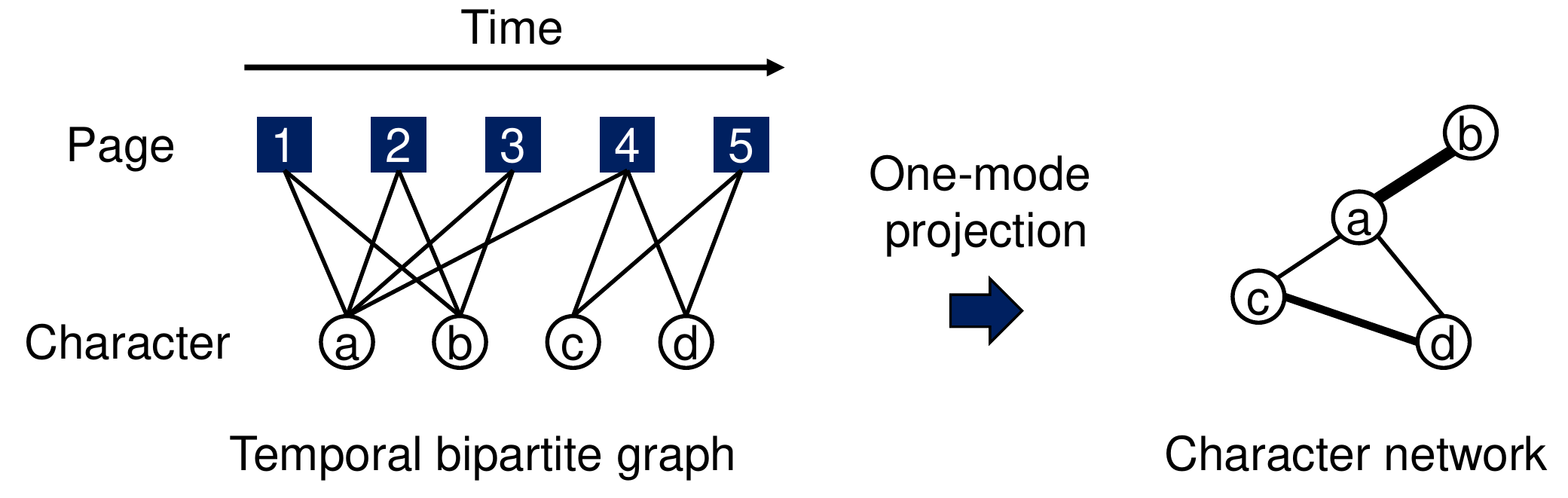}
\caption{Schematic illustration of one-mode projection, with which we construct a character network from a temporal bipartite graph of a manga.}
\label{fig:projected_network}
\end{figure*}

The original data from which we construct the temporal and static character networks are equivalent to a temporal bipartite graph in which the two types of nodes are characters and pages, and an edge connects a character and a page in which the character appears.
We construct a character network by projecting the bipartite graph onto the space of character nodes. 
In other words, in the character network, we connect two characters by an edge if and only if they appear on the same page at least once (see Fig.~\ref{fig:projected_network}).
The weight of an edge is the number of pages on which the two characters are copresent.
Because we are interested in interaction between the characters, we exclude the isolated nodes in the character networks from the analysis.

\subsection{Bipartite configuration model}
To generate a randomized bipartite network, we choose a pair of edges in the original bipartite graph uniformly at random, denoted by $(c, p)$ and $(c', p')$, where $c$ and $c'$ are characters and $p$ and $p'$ are pages.
If $c$ and $c'$ are the same or $p$ and $p'$ are the same, we discard the two edges and redraw them. 
Then, we rewire the two original edges to $(c, p')$ and $(c', p)$. 
We repeat this procedure $10{,}000$ times excluding the discarded edge pairs to generate a randomized temporal bipartite network. 
We construct a static randomized character network from the generated randomized bipartite graph by one-mode projection. 

\subsection{Coefficient of variation}
To quantify the heterogeneity of a variable, we measure the coefficient of variation (CV), which is defined as the standard deviation divided by the average.
If a variable takes only one value, the CV is $0$. 
If a continuous-valued variable obeys an exponential distribution, the CV is $1$. 
A power-law distribution has a CV value substantially larger than $1$.

\subsection{Degree assortativity coefficient}
The degree assortativity coefficient measures the extent to which nodes with similar degrees tend to be adjacent to each other in a network \cite{newman2002assortative, newman2003mixing}. 
It is defined by 
    \begin{equation}
        r=\frac{\sum_{i=1}^{N}\sum_{j=1}^{N}(A_{ij} -k_i k_j/2M)k_i k_j}{\sum_{i=1}^{N}\sum_{j=1}^{N}(k_i \delta_{ij}-k_i k_j/2M)k_i k_j}\,,
    \end{equation}
where $N$ is the number of nodes, $M$ is the number of edges, $k_i$ is the degree of node $i$, and $\delta_{ij}$ is the Kronecker delta.
Tha value of $r$ ranges between $-1$ and $1$.
A positive value indicates an assortative network, in which nodes with similar degrees tend to be adjacent to each other.
A negative value suggests a disassortative network, in which nodes with different degrees tend to be adjacent to each other.

\subsection{Clustering coefficient}
The clustering coefficient quantifies the amount of triangles in a network \cite{watts1998collective}.
The local clustering coefficient for node $i$ is defined by
\begin{equation}
    C_i=\frac{\text{(number of triangles including node $i$})}{k_i (k_i-1)/2}\,.
\end{equation}
The denominator gives the normalization such that $0 \leq C_i \leq 1$.
The clustering coefficient, denoted by $C$, is defined by the average of $C_i$ over all the nodes in the network, i.e.,
\begin{equation}
    C=\frac{1}{N}\sum_{i=1}^{N}C_i\,.
\end{equation}

\subsection{Interevent time}
An interevent time (IET) refers to the time between two consecutive events.
It should be noted that we regard a page as the time unit to simplify the analysis although manga story does not always follow the chronological order.
Copresence of two characters on a page defines a time-stamped event on the edge of the character network, where we identify the page number as discrete time for simplicity. 
Therefore, an IET on edge $(i, j)$ is the time between two consecutive copresence events of characters $i$ and $j$. 
A time-stamped event for a given node is the presence of the character on a page.
Therefore, an IET for node $i$ is the time between two consecutive appearances of character $i$.

\subsection{SI model}
We use the susceptible-infectious (SI) model \cite{kermack1927contribution} to numerically investigate contagion.
Because infection does not spread from one connected component to another, we run the SI model on the largest connected component of the character network (see SI file for the number of nodes in the largest connected component, denoted by $\overline{N}_{(L)}$, for the 162 manga).
We assume that just one character is initially infectious and that the other $\overline{N}_{(L)}-1$ characters are initially susceptible. 
If a susceptible character co-appears with an infectious character on a page, the susceptible character becomes infectious with probability $\beta$. 
Different infectious characters appearing on the same page independently attempt to infect each susceptible character on the page.
Once a character contracts infection, it stays infectious forever. 
We run the SI dynamics with each of the $\overline{N}_{(L)}$ characters as the sole character that is initially infectious.

\subsection{Temporal correlation coefficient}
We quantify the persistence of edges over time by the temporal correlation coefficient \cite{nicosia2013graph, thompson2017static}. 
First, we calculate the topological overlap for node $i$ at time $t$ by
\begin{equation}
    O_{it} = \frac{\sum_{j=1}^{N} A_{ij}^{t} A_{ij}^{t+1}}{\sqrt{\sum_{j=1}^{N} A_{ij}^{t} \sum_{j=1}^{N} A_{ij}^{t+1}}}\,,
    \label{eq:topological_overlap}
\end{equation}
where $A_{ij}^t$ is the adjacency matrix of an unweighted network at time step $t$. 
In other words, $A_{ij}^t = 1$ if characters $i$ and $j$ are copresent on page $t$, and $A_{ij}^t = 0$ otherwise.
We then define the average topological overlap for node $i$ by
\begin{equation}
    O_{i} = \frac{1}{T-1} \sum_{t=1}^{T-1} O_{it}\,,
    \label{eq:average_topological_overlap}
\end{equation}
where $T$ is the number of pages.
The temporal correlation coefficient for the entire temporal network is given by
\begin{equation}
O = \frac{1}{N} \sum_{i=1}^{N} O_{i}\,.
\label{eq:temporal_correlation_coefficient}
\end{equation}

\subsection{Partial correlation coefficient}
The partial correlation coefficient measures the extent of association between two variables while controlling for the influence of one or more additional variables.
The three-way partial correlation coefficient between $x_i$ and $x_j$ conditioned on $x_k$ is given by
\begin{equation}
    r_{ij|k}=\frac{r_{ij}-r_{ik}r_{jk}}{\sqrt{\smash[b]{1-r_{ik}^2}}\sqrt{\smash[b]{1-r_{jk}^2}}}\,,
\end{equation}
where $r_{ij}$ is the Pearson correlation coefficient between $x_i$ and $x_j$.

\section{Results}
\subsection{Similarities to empirical social networks}\label{subsec2}

\begin{figure*}[h!]
  \begin{minipage}[b]{\columnwidth}
    \centering
    \includegraphics[width=0.8\columnwidth]{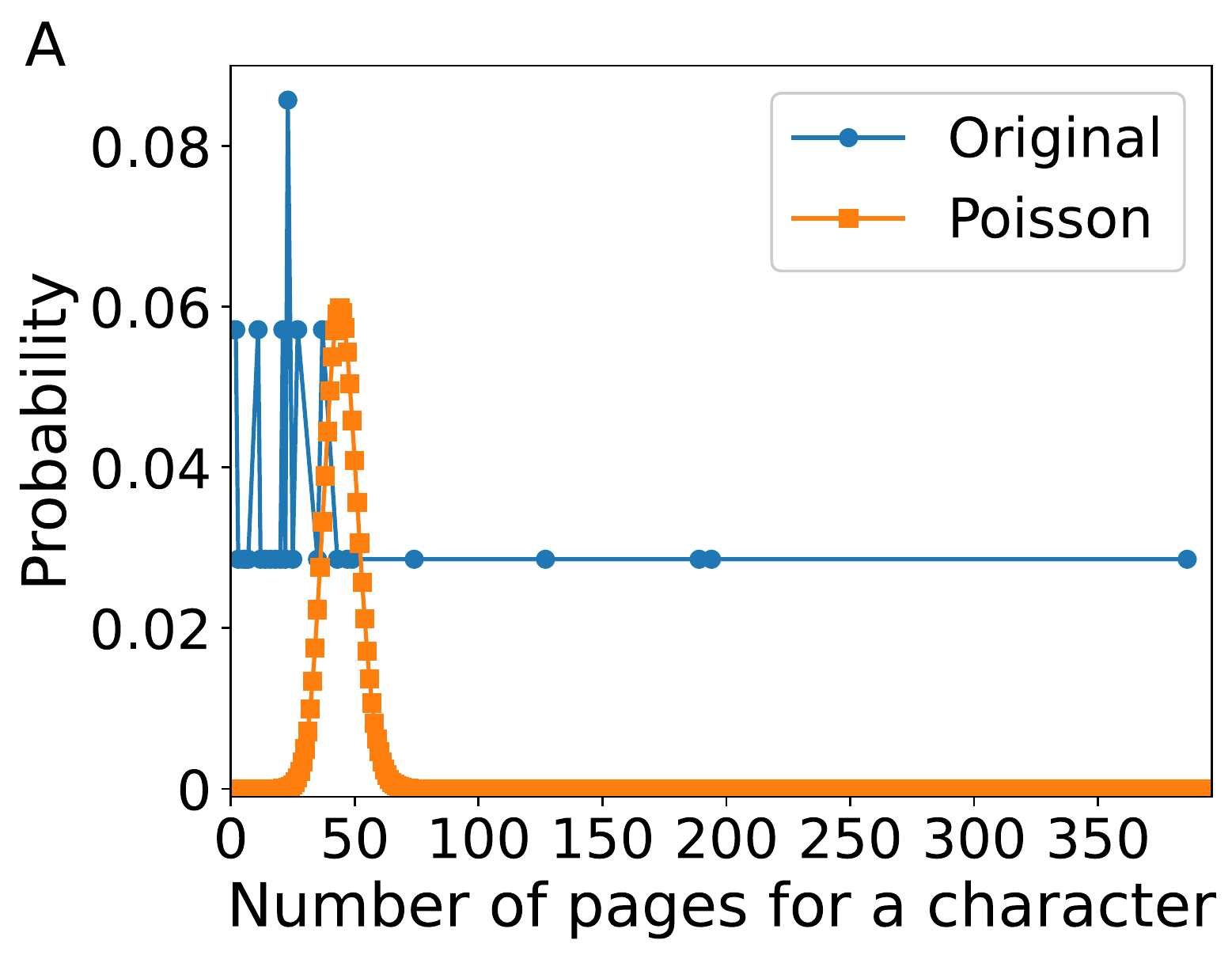}
  \end{minipage}
  \hspace{0.0\columnwidth} 
  \begin{minipage}[b]{\columnwidth}
    \centering
    \includegraphics[width=0.8\columnwidth]{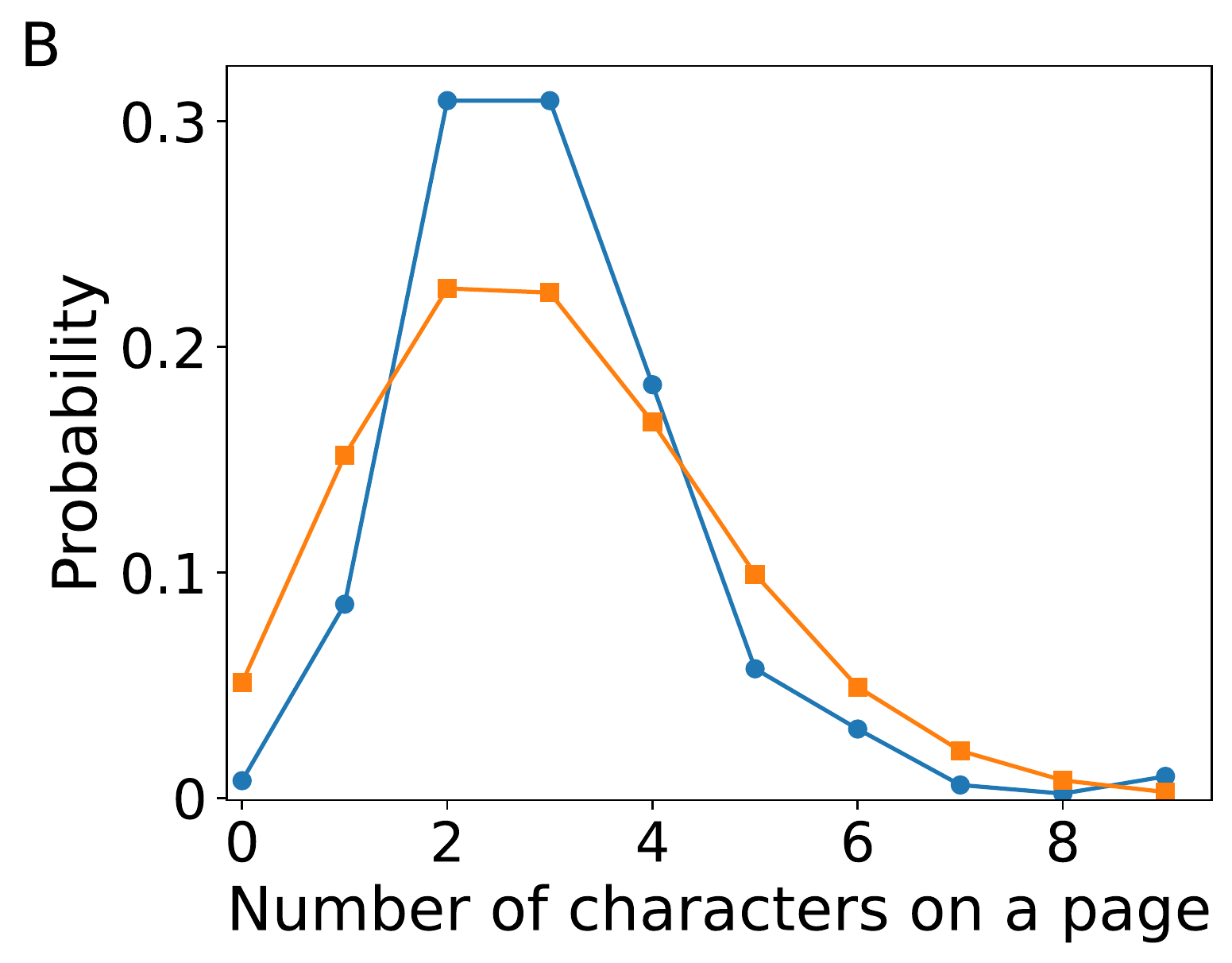}
  \end{minipage}
  \caption{Degree distribution for the (A) character and (B) page nodes in the bipartite graph for {\it One Piece}. 
  We also show the Poisson distributions with the same mean.}
\label{fig:page_distribution_onepiece}
\end{figure*}

Our original data are equivalent to a temporal bipartite graph in which the two types of nodes are characters and pages, and the edges connect characters to the pages in which they appear. 
The pages are ordered in time.
We show the descriptive statistics of the bipartite networks for 162 manga in SI file.
We show the degree distribution for the character nodes for {\it One Piece} and the Poisson distribution with the same mean in Fig.~\ref{fig:page_distribution_onepiece}A.
The CV of the original degree distribution of the characters in \textit{One Piece} is $1.66$. 
This value is approximately $11.1$ times larger than that of the Poisson distribution. 
Therefore, similar to human social networks \cite{barabasi2016network, newman2018networks}, the characters in manga have heterogeneous numbers of connections, and there are a small number of characters that appear disproportionately frequently on various pages. 
We obtained similar results for the other manga (see SI file).

In contrast, the CV of the degree distribution for the page nodes is $0.478$ for {\it One Piece}. 
The CV for the Poisson distribution with the same mean is $0.592$.
Therefore, we argue that the characters do not appear uniformly randomly over the pages. 
As we show in Fig.~\ref{fig:page_distribution_onepiece}B, there are typically two or three characters on a page in {\it One Piece}, and pages containing none or just one character are relatively rare.
We obtained similar results for the other manga (see SI file).
These results suggest that the high heterogeneity of the characters in terms of the frequency of appearance on pages and the tendency of a page typically containing two or three characters are two common properties of manga. 
Therefore, in the following analyses, we consider the bipartite configuration model, in which the degrees of all character and page nodes are preserved and the edges are otherwise randomly placed, as a null model. 
Then, we examine properties of character networks that we can explain by the null model versus those we cannot.

\begin{figure}[b!]
\centering
\includegraphics[width=\linewidth]{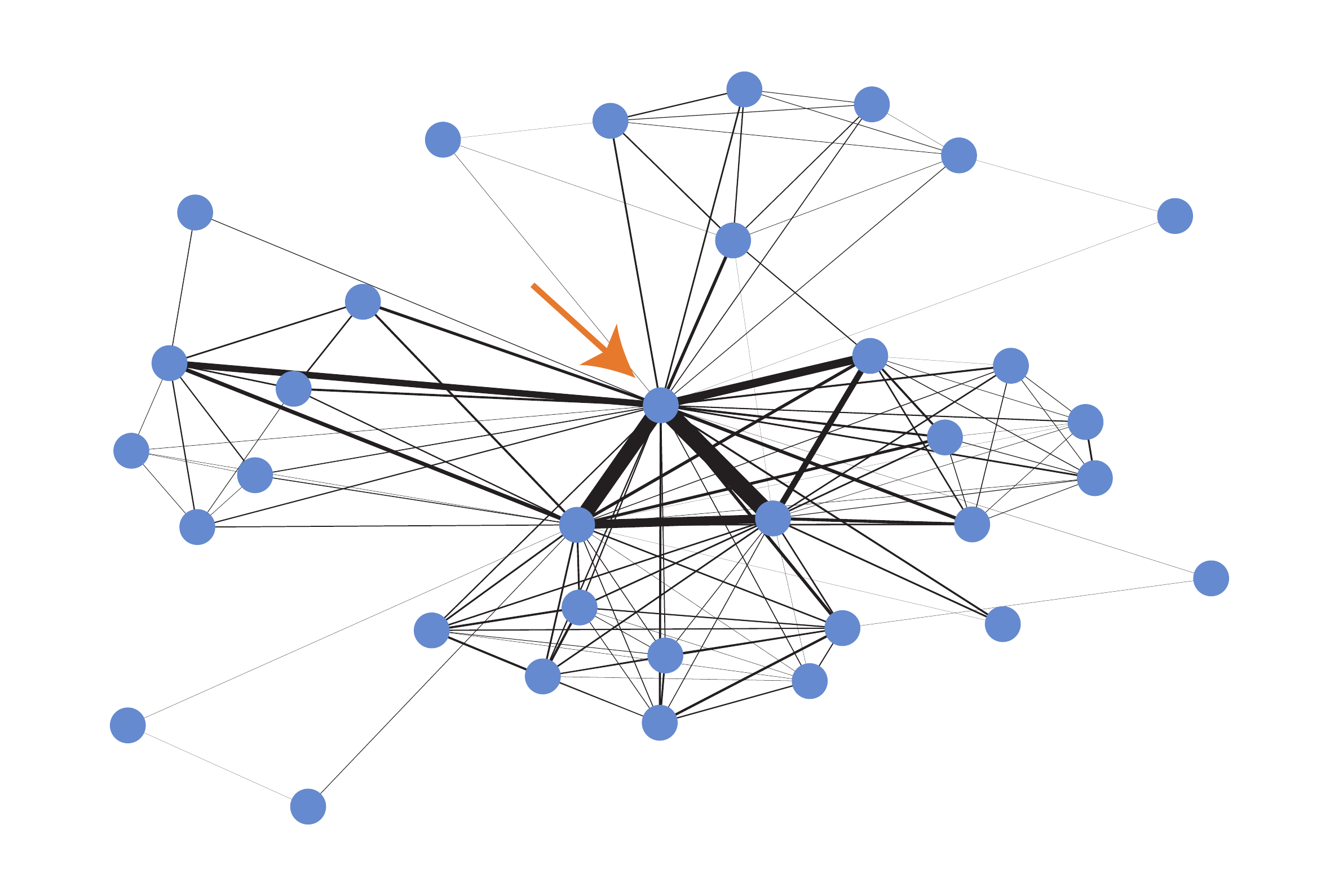}
\caption{Character network for {\it One Piece}. The thickness of an edge is proportional to the weight of the edge. The node with an arrow represents the protagonist, {\it Monkey D. Luffy}. 
}
\label{fig:network_visualization_onepiece}
\end{figure}

We show the weighted character network, which is the projection of the bipartite graph onto the space of character nodes, for {\it One Piece} in Fig.~\ref{fig:network_visualization_onepiece} (see SI Fig.~S1 for the networks for all 162 manga). 
By definition, the edge weight is equal to the number of pages in which the two characters simultaneously appear.
We show a summary of descriptive statistics of the character networks over 162 manga in Table~\ref{table:summary} (see SI file for the statistics for each manga).
The character networks vary widely in size from manga to manga.
Although multiple protagonists may exist in a manga, for simplicity, we define the protagonist as the node with the largest node strength (i.e., weighted degree) in the character network.
The protagonist is the character that appears in the largest number of pages for most manga ($97.5\%$).
In fact, the protagonist for {\it One Piece}, indicated by an arrow in Fig.~\ref{fig:network_visualization_onepiece}, is {\it Monkey D. Luffy}, who is generally known as the most central character in {\it One Piece}.

\begin{table*}[t!]
\centering
\caption{Structural properties of character networks for 162 manga. $N$: number of nodes, $M$: number of edges, $\langle s \rangle$: average strength, $s_{\rm CV}$: CV of the node strength, $s_{\rm p}$: strength of the protagonist, $\langle k \rangle$: average degree, $k_{\rm CV}$: CV of the degree, $k_{\rm p}/(N-1)$: normalized degree of the protagonist, $r$: degree assortativity coefficient, and $C$: clustering coefficient.
}
\label{table:summary}
\begin{tabular}{ccccccccccc}
\hline
& $N$  & $M$ & $\langle s \rangle$ & $s_{\rm CV}$ & $s_{\rm P}$ & $\langle k \rangle$ & $k_{\rm CV}$ & $k_{\rm P}/(N-1)$ & $r$    & $C$    \\ \hline
Mean        & 32.6 & 156 & 142                 & 1.55         & 831         & 9.08                & 0.717         & 0.899            & -0.345 & 0.790  \\
Std         & 16.6 & 129 & 95.6                & 0.400        & 325         & 3.64                & 0.250         & 0.103            & 0.123  & 0.0558 \\
Min         & 8    & 19  & 22.7                & 0.723        & 282         & 3.02                & 0.299         & 0.494            & -0.624 & 0.598  \\
Max         & 124  & 950 & 919                 & 3.22         & 2380        & 28.0                  & 1.91          & 1                & 0.192  & 0.914  \\ \hline
\end{tabular}
\end{table*}

\begin{figure*}[h!]
\centering
\includegraphics[width=0.85\linewidth]{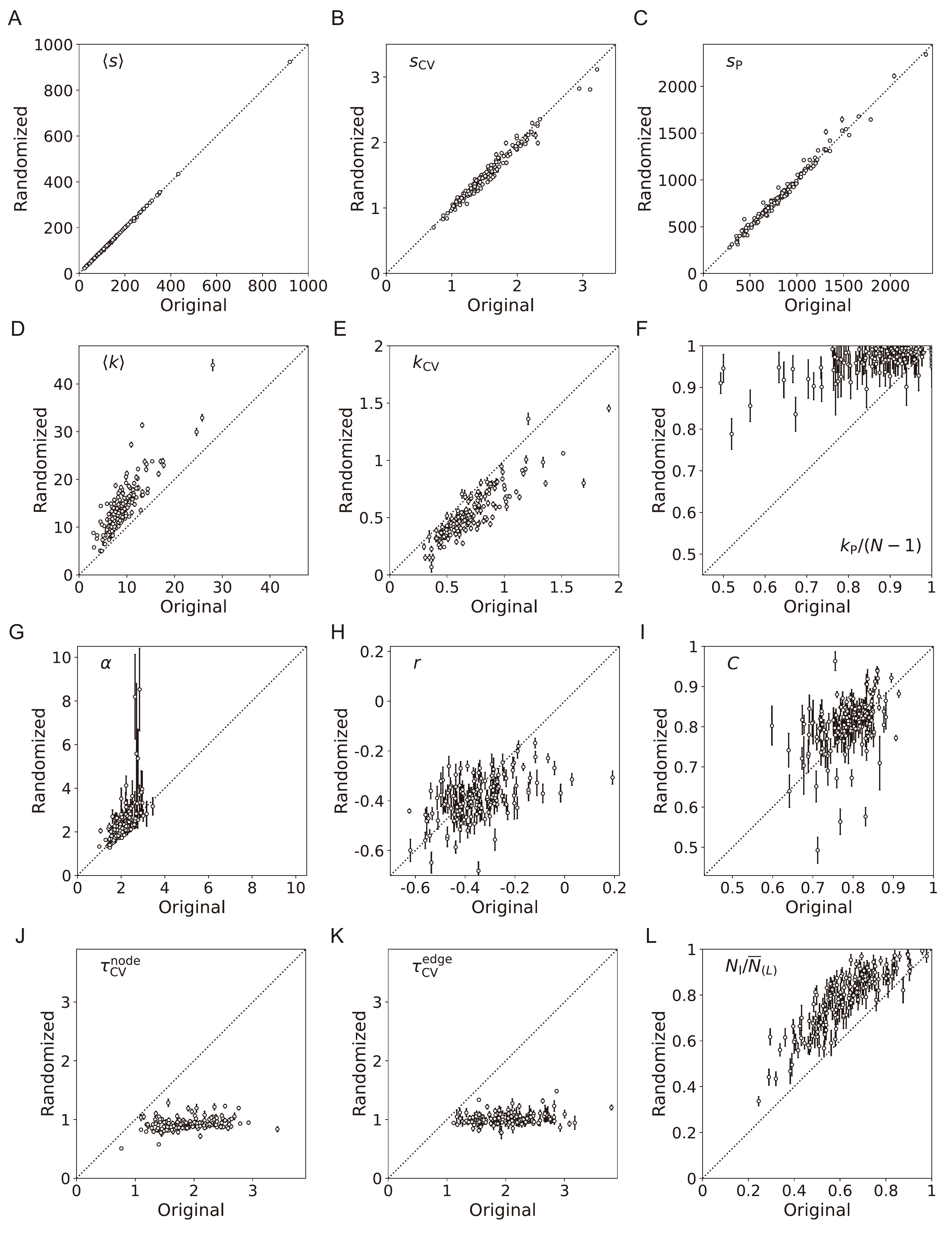}
\caption{Comparison between the original and randomized character networks for 162 manga. For the randomized networks, we show the mean and standard deviation on the basis of 1,000 realizations by the circle and error bar, respectively.
(A) $\langle s \rangle$: average node strength. 
(B) $s_{\rm CV}$: CV of the node strength. 
(C) $s_{\rm p}$: strength of the protagonist. 
(D) $\langle k \rangle$: average degree.
(E) $k_{\rm CV}$: CV of the degree.
(F) $k_{\rm p}/(N-1)$: normalized degree of the protagonist.
(G) $\alpha$: coefficient in the relationships between the degree and strength, i.e., $s_i \propto k_i^{\alpha}$.
(H) $r$: degree assortativity coefficient.
(I) $C$: clustering coefficient.
(J) $\tau_{\rm CV}^{\rm node}$: CV of IETs on nodes.
(K) $\tau_{\rm CV}^{\rm edge}$: CV of IETs on edges. 
(L) $N_{\rm I}/\overline{N}_{(L)}$: final epidemic size (i.e., fraction of infected nodes at the last time step).  
}
\label{fig:comparison_original_null_model}
\end{figure*}

We compared various properties of character networks between the empirical character networks and their randomization obtained by the one-mode projection of the bipartite network generated by the bipartite configuration model.
We first compare three strength-related indices, i.e., the average strength, $\langle s \rangle$, the CV of the strength, $s_{\rm CV}$, and the strength of the protagonist, $s_{\text{P}}$, of the character networks for all 162 manga between the original and randomized networks in Figs.~\ref{fig:comparison_original_null_model}A-C.
These figures suggest that randomized character networks well explain the strength-related indices of the original character networks.

\begin{figure*}[t!]
\centering
\includegraphics[width=\linewidth]{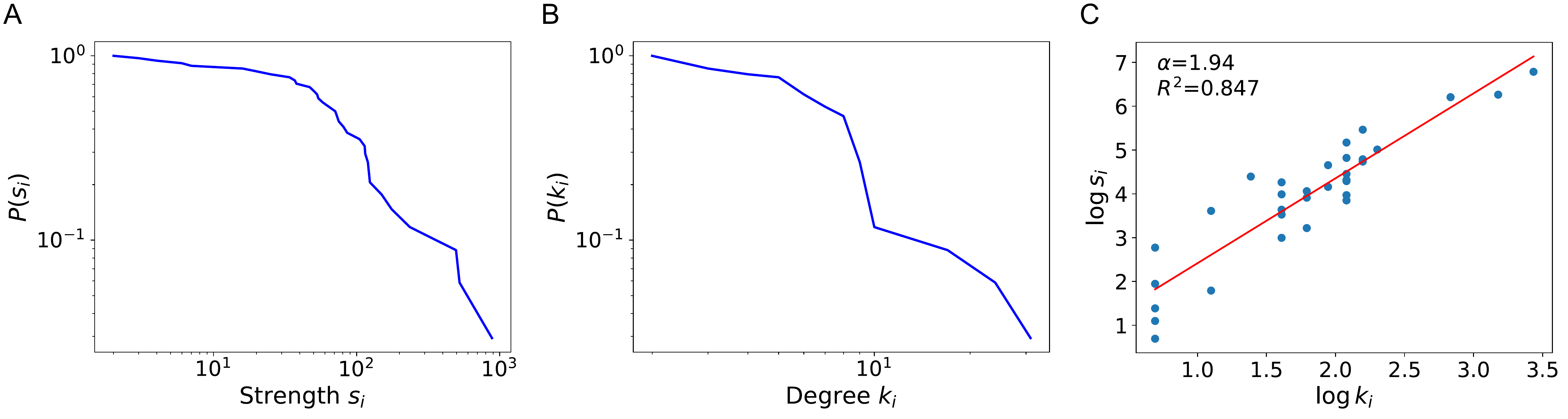}
\caption{Strength and degree distributions for the characters in {\it One Piece}. (A) Strength distribution. (B) Degree distribution. (C) The relationship between the degree and strength. Each circle represents a character. The coefficient $\alpha$ is determined by the linear regression $\log s_i = \alpha \log k_i +b$, where $b$ is an intercept; the solid line shows the linear regression. Variable $R^2$ represents the coefficient of determination. }
\label{fig:degree_strength_onepiece}
\end{figure*}

Figure~\ref{fig:degree_strength_onepiece}A shows the strength distribution for \textit{One Piece}. 
We find that the strength obeys a heavy-tailed distribution over some scale.
This result is consistent with Fig.~\ref{fig:comparison_original_null_model}B, which shows that the CV of the strength for most manga is substantially larger than 1; the CV for \textit{One Piece} is 1.48.
The heavy-tailed strength distributions shown in Fig.~\ref{fig:degree_strength_onepiece}A for \textit{One Piece} and implied in Fig.~\ref{fig:comparison_original_null_model}B for a majority of manga are consistent with those for empirical social networks \cite{barrat2004architecture, wang2006structure}.

In contrast to the strength-related indices, the randomized character networks do not explain the degree-related indices of the original character networks. 
The randomized networks tend to overestimate the average degree, $\langle k \rangle$, of the original character networks (see Fig.~\ref{fig:comparison_original_null_model}D).
This result suggests that characters tend to repeat interacting with relatively few characters. 
The CV of the degree, $k_{\rm CV}$, for the original networks tends to be larger than that for the randomized networks (see Fig.~\ref{fig:comparison_original_null_model}E).
Let $k_{\rm p}$ denote the degree of the protagonist.
The fraction of the other characters that the protagonist is adjacent to, $k_{\rm P}/(N-1)$, is $0.899 \pm 0.103 
 \, ({\rm mean} \pm {\rm standard \, deviation} $  based on the 162 manga) and $0.971 \pm 0.0413$ for the original and randomized character networks, respectively  (see Table~\ref{table:summary} and Fig.~\ref{fig:comparison_original_null_model}F).
Therefore, the protagonist is adjacent to most of the other characters in both original and randomized networks.
Based on these results, we conclude that the character networks are strongly protagonist-centered, in which the protagonist interacts with most of the other characters, while other characters tend to interact only with fewer characters than expected for the randomized networks, but including the protagonist. 
This interpretation is consistent with the aforementioned observation that $\langle k\rangle$ and $k_{\rm CV}$ are smaller and larger for the original than randomized networks, respectively.

We show the degree distribution for {\it One Piece} in Fig.~\ref{fig:degree_strength_onepiece}B.
The CV of the degree, $k_{\rm CV}$, for {\it One Piece} is $0.77$.
Figure~\ref{fig:comparison_original_null_model}E shows that the CV of the degree for most of the manga is smaller than 1.
In fact, this result is consistent with that for empirical social networks with similar number of nodes \cite{read1954cultures, zachary1977information}, while large empirical social networks tend to have heavy-tailed degree distributions implying a large CV \cite{barabasi2002evolution, mislove2007measurement}.

We found that the strength is super-linearly scaled with the degree, i.e., $s_i \propto k_i^{\alpha}$ with $\alpha>1$, in the character networks, where $\propto$ indicates ``proportional to''.
Note that the absence of correlation between the strength and degree would yield $\alpha=1$ \cite{barrat2004architecture}.
Specifically, in Fig.~\ref{fig:degree_strength_onepiece}C, we show the relationship between $k_i$ and $s_i$ for {\it One Piece}.
We obtained $\alpha=1.94$ with the coefficient of determination $R^2=0.847$.
We obtained similar results for the other manga (see Fig.~\ref{fig:comparison_original_null_model}G). 
These results are consistent with power-law relationships between the strength and degree with $\alpha > 1$, which has been observed in empirical social networks \cite{barrat2004architecture, wang2005mutual}. 
Note that the randomized character networks also yield $\alpha > 1$ and that $\alpha$ for the original and randomized networks are highly correlated with a small number of exceptions.

The character networks are disassortative. 
In fact, the degree assortativity coefficient, denoted by $r$, of the character network is $-0.345$ on average (see Table~\ref{table:summary}).
The values of $r$ for all but two manga are negative (see Fig.~\ref{fig:comparison_original_null_model}H).
This result is in stark contrast with various observations that social networks are more often than not assortative with positive $r$ \cite{newman2002assortative, newman2003social}.
We argue that the character networks are disassortative because they are strongly protagonist-centered.
In fact, star graphs are disassortative with large negative values of $r$ \cite{piraveenan2008local, estrada2011combinatorial}. 
The degree disassortativity is also observed for empirical egocentric social networks \cite{batagelj2000some, gupta2015structural}.
Figure~\ref{fig:comparison_original_null_model}H indicates that $r$ is also negative for the randomized character networks although the correlation between $r$ for the original and randomized networks is low.
Therefore, we conclude that the degree disassortativity in our character networks is a consequence of the protagonist-centered nature of the original character-page bipartite network.

The character networks are highly clustered. 
Specifically, the clustering coefficient, denoted by $C$, over the different manga is $0.790$ with the minimum value of $0.598$ (see Table~\ref{table:summary}).
This result is consistent with the observations that empirical social networks have high clustering coefficients \cite{watts1998collective, saramaki2007generalizations}. 
In fact, randomized character networks also have similarly large $C$ although the spread is large between the empirical and randomized networks.
Therefore, we conclude that a high clustering coefficient is a consequence of one-mode projection of the bipartite graph, which is known \cite{newman2001scientific, ramasco2004self}.

\begin{figure*}[t!]
  \begin{minipage}[b]{\columnwidth}
    \centering
    \includegraphics[width=0.8\columnwidth]{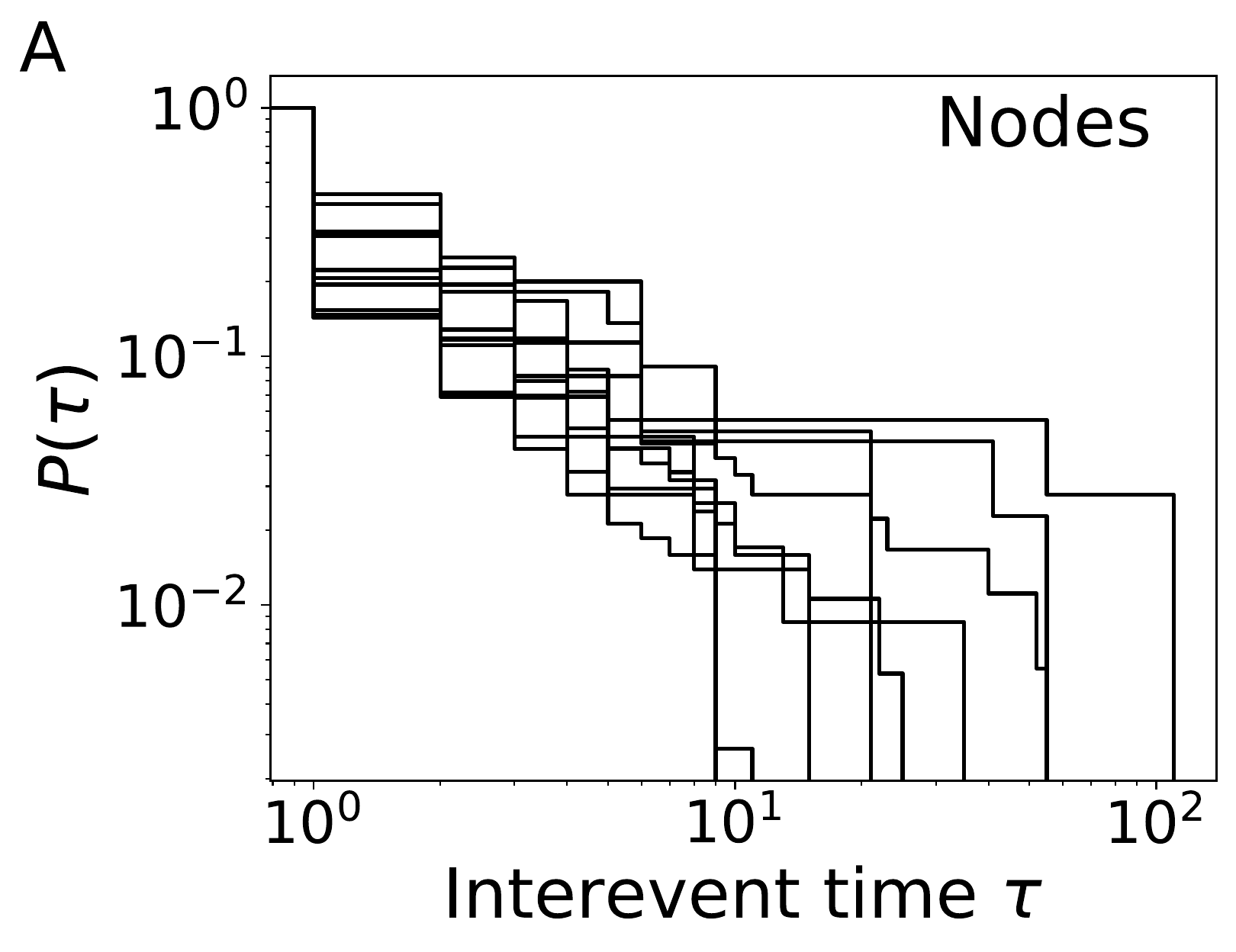}
  \end{minipage}
  \begin{minipage}[b]{\columnwidth}
    \centering
    \includegraphics[width=0.8\columnwidth]{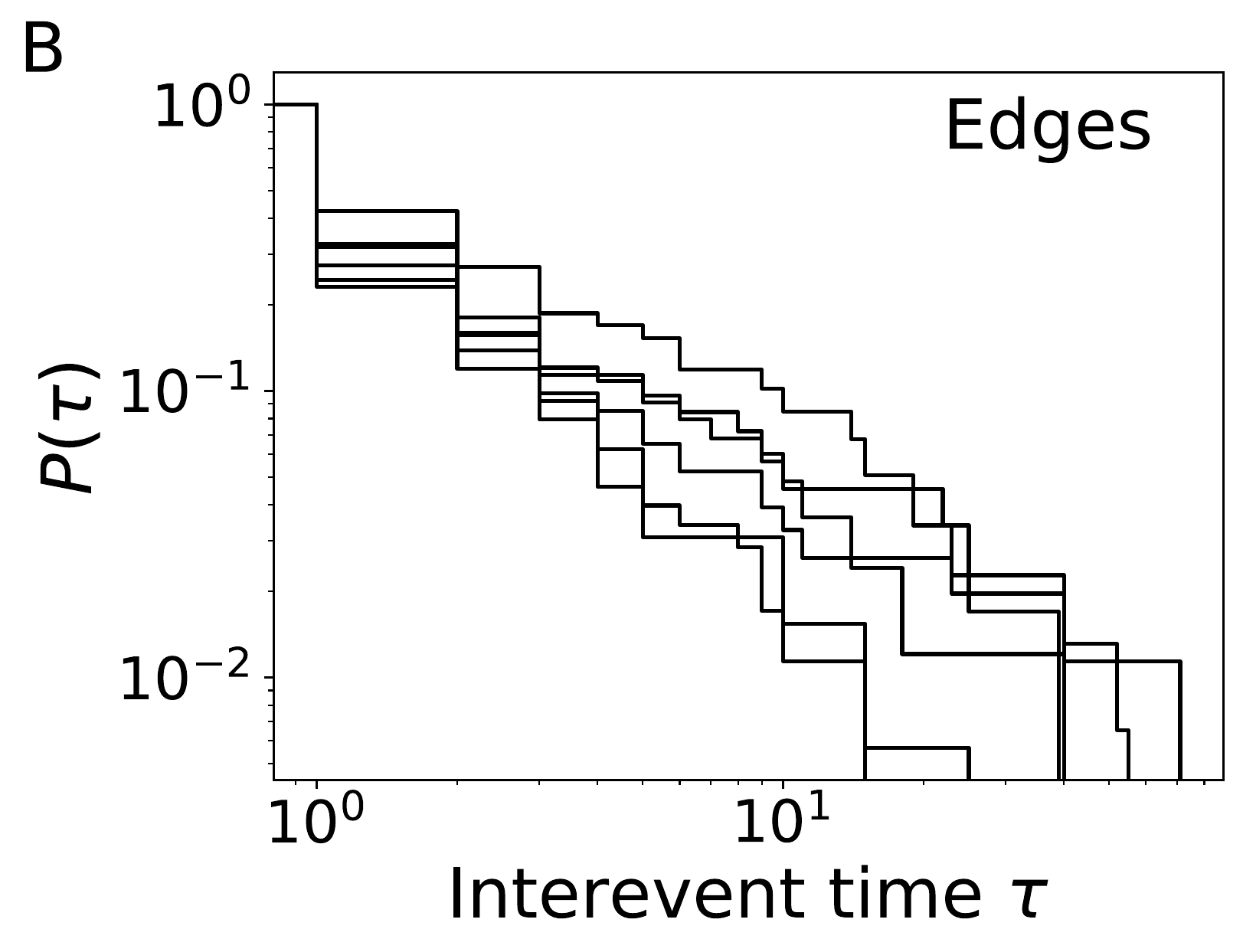}
  \end{minipage}
\caption{Survival function, $P(\tau)$, of IETs on (A) nodes and (B) edges for {\it One Piece}. Each line in (A) and (B) corresponds to a node and an edge, respectively. We only considered nodes that had at least 100 events and edges that had at least 50 events for (A) and (B), respectively.}
\label{fig:IET_onepiece}
\end{figure*}

We found that heavy-tailed distributions
of 
IETs are simultaneously present for nodes and edges in the character networks.
We show the survival functions of the IETs (i.e., probability that the IET, denoted by $\tau$, is larger than the specified value) for individual nodes and edges in \textit{One Piece} in Figs.~\ref{fig:IET_onepiece}A and \ref{fig:IET_onepiece}B, respectively. 
The relatively slow decay in Fig.~\ref{fig:IET_onepiece} suggests heavy-tailed distributions for both nodes and edges across some scales of $\tau$.  
The CV values for IETs on nodes and edges for {\it One Piece} are $1.72 \pm 0.697$ and $1.92 \pm 0.535$, respectively. 
We obtained similar results for the other manga, as we show in Figs.~\ref{fig:comparison_original_null_model}J and \ref{fig:comparison_original_null_model}K.
These figures also indicate that the randomization of the bipartite network does not preserve this feature, yielding CV values close to 1 regardless of the CV value for the original temporal character networks.
It should be noted that a Poisson process produces an exponential IET distribution, which yields CV = 1.
These results support that IETs for nodes and edges in the original character networks are non-Poissonian and heterogeneously distributed, which is consistent with properties of empirical social networks \cite{dos2020generative}.

\begin{figure}[h!]
\centering
\includegraphics[width=\linewidth]{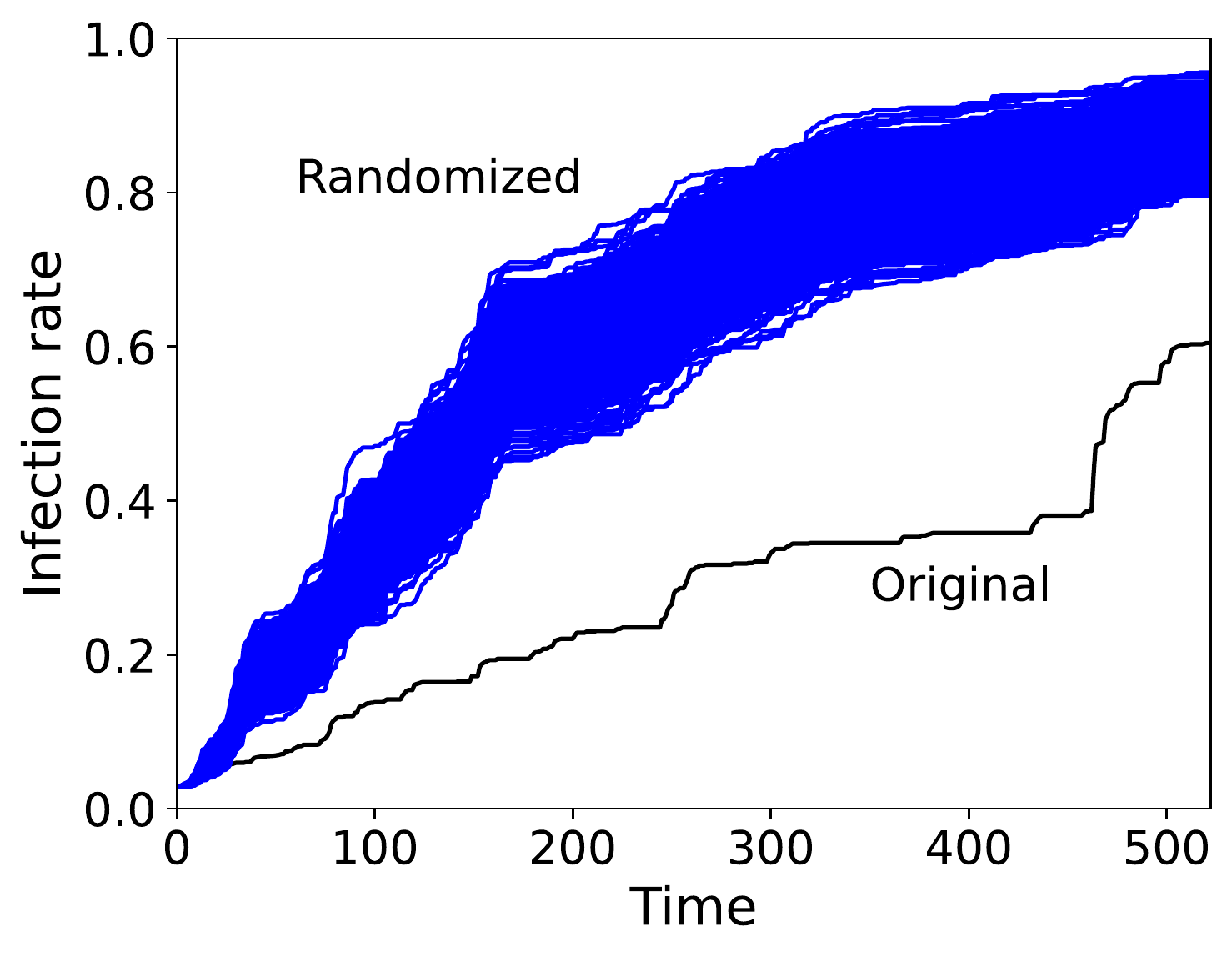}
\label{fig:Temporal_SI_model_onepiece}
\caption{Time courses of the fraction of infectious characters for the original temporal network and 1,000 randomized temporal networks for {\it One Piece}. 
We used the SI model starting from one initially infectious character.
Each curve represents the average time course of the fraction of infectious characters over $\overline{N}_{(L)}$ runs.
We set the infection probability per contact event to $\beta=0.2$.
}
\end{figure}

We simulated stochastic contagion processes on the character network, in which one character can infect another if and only if they appear on the same page. The purpose of this analysis is to examine whether contagion on the character network occurs in a manner similar to that on empirical temporal contact networks, which show different contagion patterns from the case of static contact networks.
Related to heavy-tailed IET distributions, we found that epidemic spreading occurs more slowly in the temporal character networks than in randomized counterparts. 
We ran the SI model.
By assumption, an infectious character independently infects each susceptible character coappearing on the same page with probability $\beta=0.2$.
For \textit{One Piece}, we show in Fig.~\ref{fig:Temporal_SI_model_onepiece} the time course of the fraction of the infectious characters averaged over all the runs. 
We also show the corresponding averaged time courses for each of the {1{,}000} randomized temporal networks by the blue lines.
Figure~\ref{fig:Temporal_SI_model_onepiece} indicates that the infection occurs more slowly in the original temporal network than in the randomized temporal networks. 
We obtained similar results for the other manga (see Fig.~\ref{fig:comparison_original_null_model}L). 
These results are qualitatively the same as those observed for empirical social temporal networks \cite{masuda2013predicting, karsai2011small}.

\subsection{Trends of network structure over decades}\label{subsubsec2}

The 162 manga span nearly 70 years, from {\it Astro Boy} with the first volume being published in 1952 to {\it Jujutsu Kaisen} in 2018 (see SI file for detailed information on the 162 manga).
The trend of the structure of manga character networks may have changed over the $\approx 70$ years, reflecting the transitions of Japanese society including in economics and fashion.
In this section, we explore possible existence of such trends.

\begin{figure*}[h!]
  \begin{minipage}[b]{0.65\columnwidth}
    \centering
    \includegraphics[width=\columnwidth]{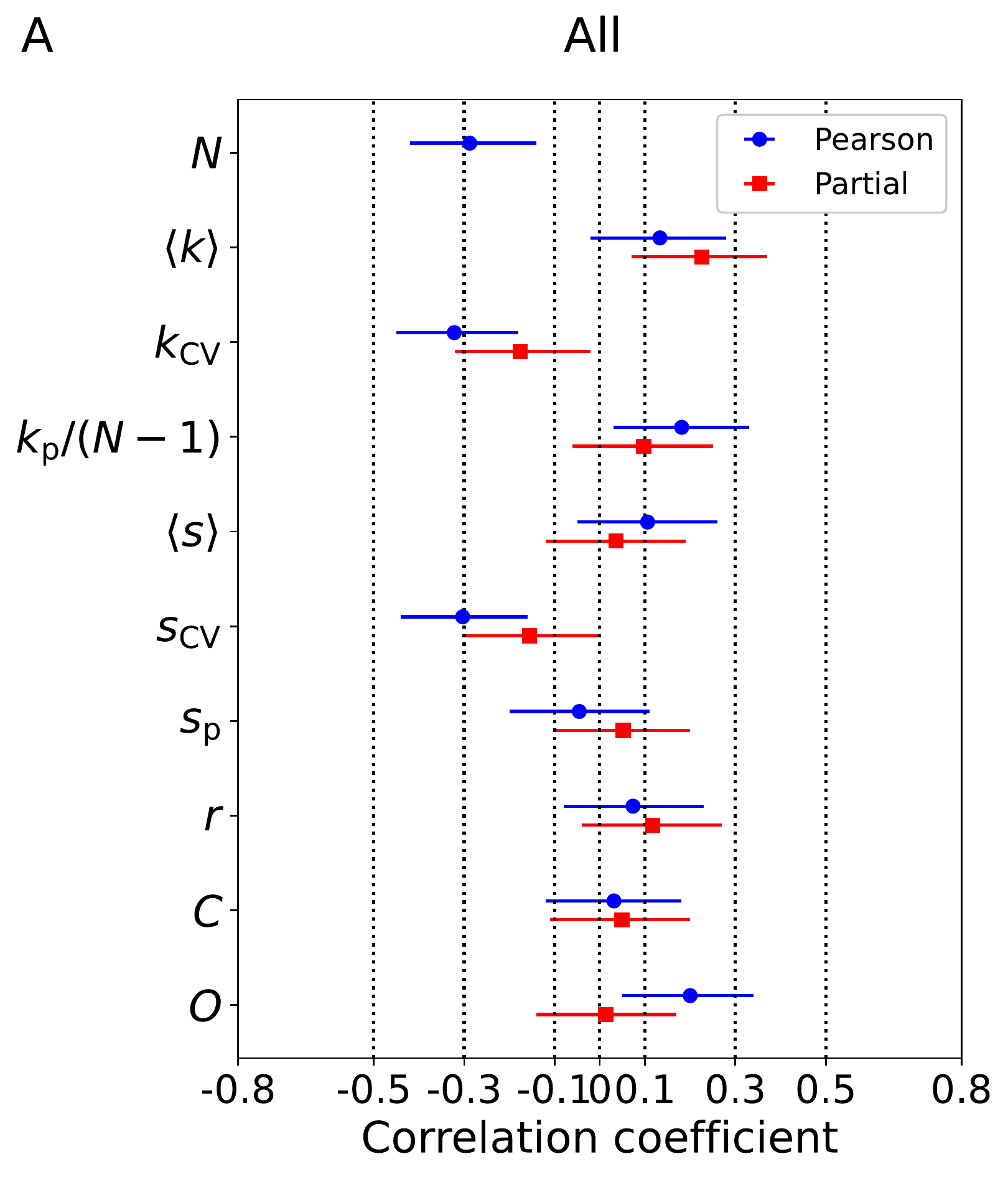}
  \end{minipage}
  \hspace{0\columnwidth} 
  \begin{minipage}[b]{0.65\columnwidth}
    \centering
    \includegraphics[width=\columnwidth]{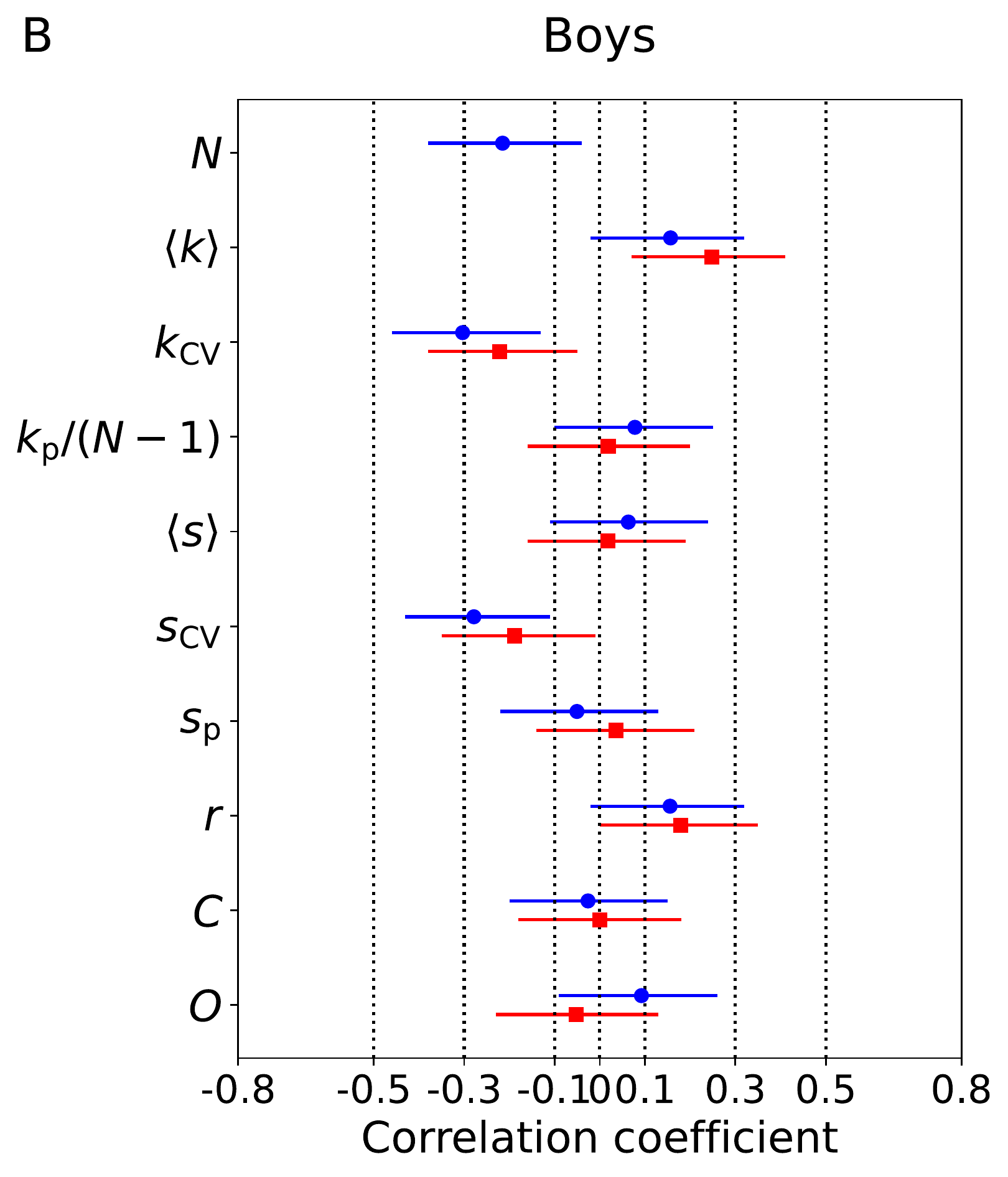}
  \end{minipage}
  \hspace{0\columnwidth} 
  \begin{minipage}[b]{0.65\columnwidth}
    \centering
    \includegraphics[width=\columnwidth]{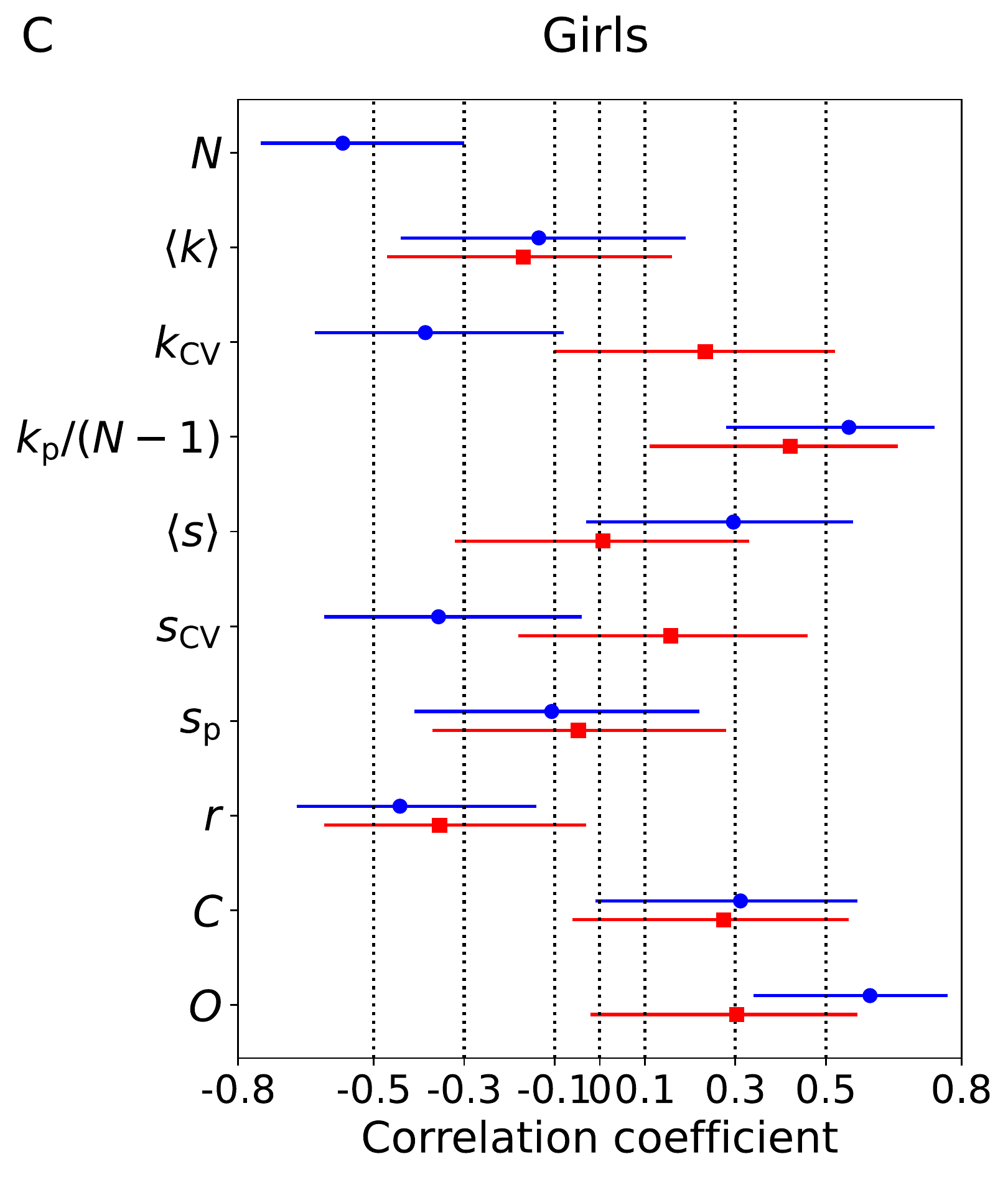}
  \end{minipage}
\caption{Correlation between the year of publication of the first volume, $y$, and indices of network structure. 
For the partial correlation coefficient, we partial out the influence of the number of nodes. 
According to a standard, the effect size is said to be large, moderate, or small when the correlation coefficient is $> 0.5$, $> 0.3$, or $> 0.1$, respectively \cite{cohen2013statistical}.
The horizontal lines indicate 95\% confidence intervals.
$N$: number of nodes, 
$\langle k \rangle$: average degree, 
$k_{\rm CV}$: CV of the degree, 
$k_{\rm p}/(N-1)$: normalized degree of the protagonist,
$\langle s \rangle$: average strength, 
$s_{\rm CV}$: CV of  strength,
$s_{\rm p}$: strength of the protagonist,
$r$: degree assortativity coefficient, 
$C$: clustering coefficient, 
and $O$: temporal correlation coefficient. 
}
\label{fig:Correlation_partial}
\end{figure*}

We first calculated the Pearson correlation coefficient between the year of publication of the first volume, denoted by $y$, and various indices of network structure on the basis of all 162 manga.
We show the correlation coefficients and their 95\% confidence intervals (CIs) in Fig.~\ref{fig:Correlation_partial}A.
We find that the number of characters, $N$, tends to decrease over decades (i.e., negative correlation between $N$ and $y$). 
The CV of the node degree, $k_{\rm CV}$, and that of the node strength, $s_{\rm CV}$, are negatively correlated with $y$ with moderate effect sizes.
In general, these and other network indices may be affected by the number of nodes, $N$ \cite{van2010comparing}. 
Therefore, we also examined the partial correlation coefficient between $y$ and each index by partialing out the effect of $N$.
The results of the partial correlation coefficient for $k_{\rm CV}$ and $s_{\rm CV}$, shown by the lines with squares in Fig.~\ref{fig:Correlation_partial}A, are consistent with those of the Pearson correlation coefficient, although the partial correlation is closer to 0 than the Pearson correlation.
We also find that the average degree, $\langle k \rangle$, is positively correlated with $y$ in terms of the partial correlation. 
These results suggest that more recent manga tend to be denser and more homogeneous in the node's connectivity, such that various characters, not just the protagonist, tend to have more connections in more recent manga.

How the trend of the character network changes over time may depend on genres of manga.
A most major categorization of Japanese manga is the one based on the gender of their main readership \cite{toku2007shojo, prough2010shojo}.
Therefore, we classify the 162 manga into 124 boys' and 38 girls' manga based on the genre section on their Wikipedia pages (see SI file for the genre of each manga), while their classification is necessarily subjective. 
We show the Pearson and partial correlation coefficients between each index and $y$, separately for the boys' and girls' manga in Figs.~\ref{fig:Correlation_partial}B and \ref{fig:Correlation_partial}C, respectively.

We find that there tend to be less characters for more recent manga for both boys' and girls' manga.
The boy's and girls' manga are different in the following aspects in terms of the partial correlation coefficients, i.e., when we control for the number of characters. 
The trends that we identified for all the 162 manga are even more pronounced in the boys' manga.
The effect sizes of the partial correlation for $\langle k \rangle$, $k_{\rm CV}$, and $s_{\rm CV}$ for the boys' manga are larger than those for the 162 manga.
In addition, the degree assortativity coefficient, $r$, is positively correlated with $y$. 
Because $r$ is negative for most manga, this result implies that $r$ tends to be closer to 0 for more recent boys' manga.
On the other hand, the two protagonist-related indices, i.e., $k_{\rm P}/(N-1)$ and $s_{\rm P}$, are not correlated with $y$. 
These results suggest that the character networks in more recent boys' manga tend to be less protagonist-centered, in which non-protagonist characters have more  connections among them.

The trend for the girls' manga is opposite to that for the boys' manga except that the downward trend in the number of characters is common.
The two indices for which the partial correlation was negative for the boys' manga, i.e., $k_{\rm CV}$ and $s_{\rm CV}$, are positively correlated with $y$ for the girls' manga in terms of the partial correlation.
On the other hand, the two indices for which the partial correlation was positive for the boys' manga, i.e., $\langle k \rangle$ and $r$, are negatively correlated with $y$ for the girls' manga.
Furthermore, the normalized degree of the protagonist, $k_{\rm P}/(N-1)$, for which we did not confirm a correlation for the boys' manga, is positively correlated with $y$.  
Overall, these results indicate that the character networks in the girls' manga have shifted toward more protagonist-centered networks.

\section{Discussion}
We have examined character networks of 162 Japanese manga that span over 70 years.
Our main findings are as follows.
First, the structural and temporal properties of the character networks of manga are similar to those of human social networks. 
Second, the trend of the manga character networks has shifted over decades. 
Third, this trend shift is substantially different between boys' and girls' manga.

Similarities between character networks and human social networks have been investigated for some fictions such as Marvel comics \cite{alberich2002marvel, gleiser2007become}, a graphic novel \cite{labatut2022complex},  Shakespeare's plays \cite{stiller2003small}, and myths \cite{mac2012universal}.
Our results of the high clustering coefficient and the heavy-tailed distribution of the strength in the character networks are consistent with these previous studies.
However, we ascribe the high clustering coefficient to the effect of one-mode projection rather than to similarity to empirical social networks because the randomized character networks also have high clustering.
Our results of the disassortativity is also consistent with the results for Marvel comics \cite{gleiser2007become}, a graphic novel \cite{labatut2022complex}, and myths \cite{mac2012universal}.
Therefore, we infer that the heavy-tailed distribution of the strength and disassortativity in the character networks are common properties for various types of fictions.
Although the aforementioned previous studies investigated only static properties of the character networks, we further revealed temporal properties of the character networks, such as the long-tailed IET distributions, which are in fact consistent with empirical social networks \cite{dos2020generative, masuda2013predicting, karsai2011small}.

The bipartite configuration model, which is a standard random bipartite graph, has been shown to be reasonably accurate at explaining some features of the empirical character networks including statistics of the node's strength, $\alpha$ values, disassortativity, and the high clustering coefficients.
In contrast, there are other features of the empirical character networks that deviate from the expectation from the bipartite configuration graph.
Such features include statistics of the degree, the IET distributions, and speeds of epidemic spreading.
We suggested that part of these differences originates from the protagonist-centered nature of the character networks. 
In other words, the protagonist interacts with most characters, while non-protagonist characters tend to interact only with the protagonist and a smaller number of other characters than expected by the configuration model.    
There are positive support of protagonist-centered social networks in the real world when the network is egocentric
 \cite{batagelj2000some, gupta2015structural}.

We also found that 
more modern manga tend to have fewer characters, be denser, and be less protagonist-centered for the boys’ manga, which may reflect a modern change in the society that places more emphasis on diversity and teamwork \cite{curcseu2013student, dell2010collaborative}.
In contrast, the character networks in the girls' manga have shifted toward more protagonist-centered, although the downward trend in the number of characters is common.
In general, girls' manga in Japan tend to revolve around issues of love and friendship with a focus on inner feelings of the protagonist \cite{prough2010shojo, takahashi2014opening}.
Our results suggest that more recent girls' manga may describe the relationships between the protagonist and a few other characters in depth.

There are many future directions of investigation.
First, we focused on 162 manga with high circulations. 
By analyzing manga with low circulation as well, we may be able to discover static and temporal properties of character networks that readers favor, contributing to understanding why some manga are more popular than others.
Second, we analyzed only the first three volumes of each manga because of a logistic limitation.
Examining all volumes would allow us to understand the narrative structure, which have been analyzed for novels \cite{gessey2020narrative}, movies \cite{weng2007movie}, and TV series \cite{park2012social}.
For this purpose, a wide variety of time series and temporal network analysis tools, such as change-point detection, temporal centrality, and temporal community structure, may be useful \cite{holme2012temporal, holme2019temporal, masuda2020guidance}.
Third, character networks are probably signed in most cases, connecting characters by positive or negative ties. 
It is worth to deploy sentiment analysis based on text \cite{min2019modeling} and facial expressions \cite{soleymani2015analysis} to construct and analyze signed networks of characters in manga.
Temporal network analysis of signed character networks may also reveal complicated dynamics of relationships among characters (e.g., an enemy 
 later becomes an ally).
Fourth, manga in different genres such as action, adventure, sports, and comedy, may  have different structures of character networks.
In addition, we have only analyzed the manga that have been published in the paperback pocket edition, which is the most common for the boys' and girls' manga in Japan. 
Analyzing manga of other sizes may help us understand variation of character networks over a wider range of genres.
Fifth, studying character networks in comics in countries other than Japan warrants future work. 
For example, character networks may reflect societal differences from country to country.

In conclusion, to the best of our knowledge, the present study of character networks for 162 manga titles is unprecedented in scale in quantitative studies of comics and has enabled us to discover their general features.
We also introduced new tools and views to the analysis of fiction character network data, such as the bipartite configuration model, the concept of protagonist-centeredness, and temporal network analysis.
We hope that this study triggers further quantitative studies of character networks in manga and other types of fictions.

\section*{Acknowledgments}
We thank Juyong Park for valuable discussion. 

\section*{Funding}
K.S. thanks the financial support by the Japan Society for the Promotion of Science (under Grant No.~19K23531). 
N.M. thanks the financial support by AFOSR European Office (under Grant No.~FA9550-19-1-7024), the Japan Science and Technology Agency (JST) Moonshot R\&D (under Grant No.~JPMJMS2021), the Japan Society for the Promotion of Science (under Grant No.~21H04595 and 23H03414), and the National Science Foundation (under Grant No.~2052720 and 2204936).

\section*{Data availability}
The datasets generated during and/or analysed during the current study are available in the GitHub repository, \url{https://github.com/KS-92/Manga}.

\section*{Authors' contributions}
N.M. conceived the research; K.S. and N.M. designed the research. K.S. collected and analyzed the data; K.S. and N.M. discussed the results and wrote the paper.

\section*{Authors' contributions}
N.M. conceived the research; K.S. and N.M. designed the research. K.S. collected and analyzed the data; K.S. and N.M. discussed the results and wrote the paper.

\section*{Competing interests}
The authors declare no competing financial or non-financial interests.

\newpage

\onecolumn
\begin{center}
\vspace*{5pt}
{\Large Supplementary Information for:\\
\vspace{5pt}
Social network analysis of manga: similarities to real-world social networks and trends over decades}
\vspace{5pt} \\
\end{center}

\setcounter{figure}{0}
\setcounter{table}{0}
\setcounter{section}{0}

\renewcommand{\thesection}{S\arabic{section}}
\renewcommand{\thefigure}{S\arabic{figure}}
\renewcommand{\thetable}{S\arabic{table}}

\begin{center}
\author{Kashin Sugishita and Naoki Masuda}
\end{center}
\begin{figure*}[b!]
     \centering
     \begin{subfigure}[b]{0.195\textwidth}
         \centering
         \includegraphics[width=\textwidth]{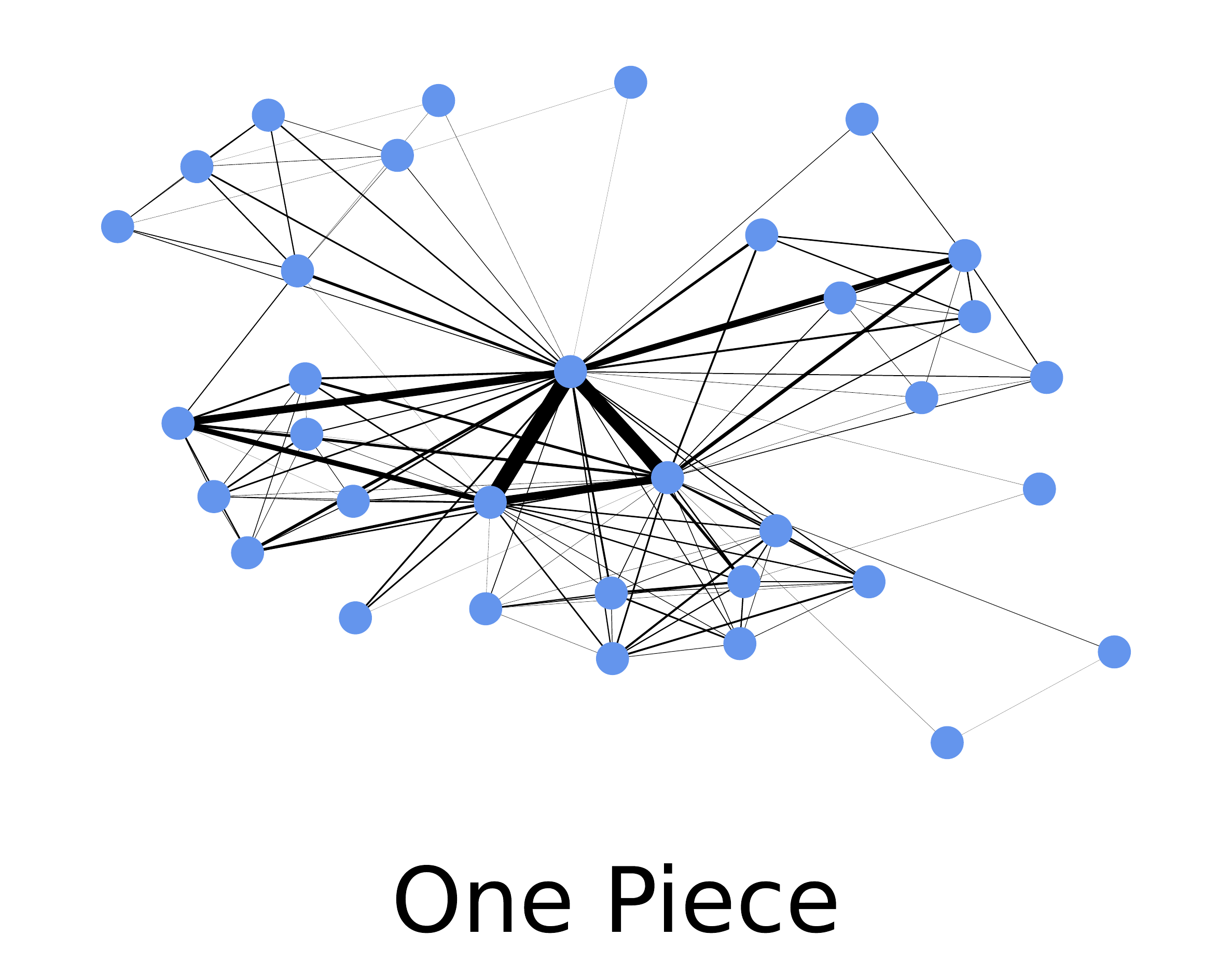}
         \label{fig:y equals x}
     \end{subfigure}
     \begin{subfigure}[b]{0.195\textwidth}
         \centering
         \includegraphics[width=\textwidth]{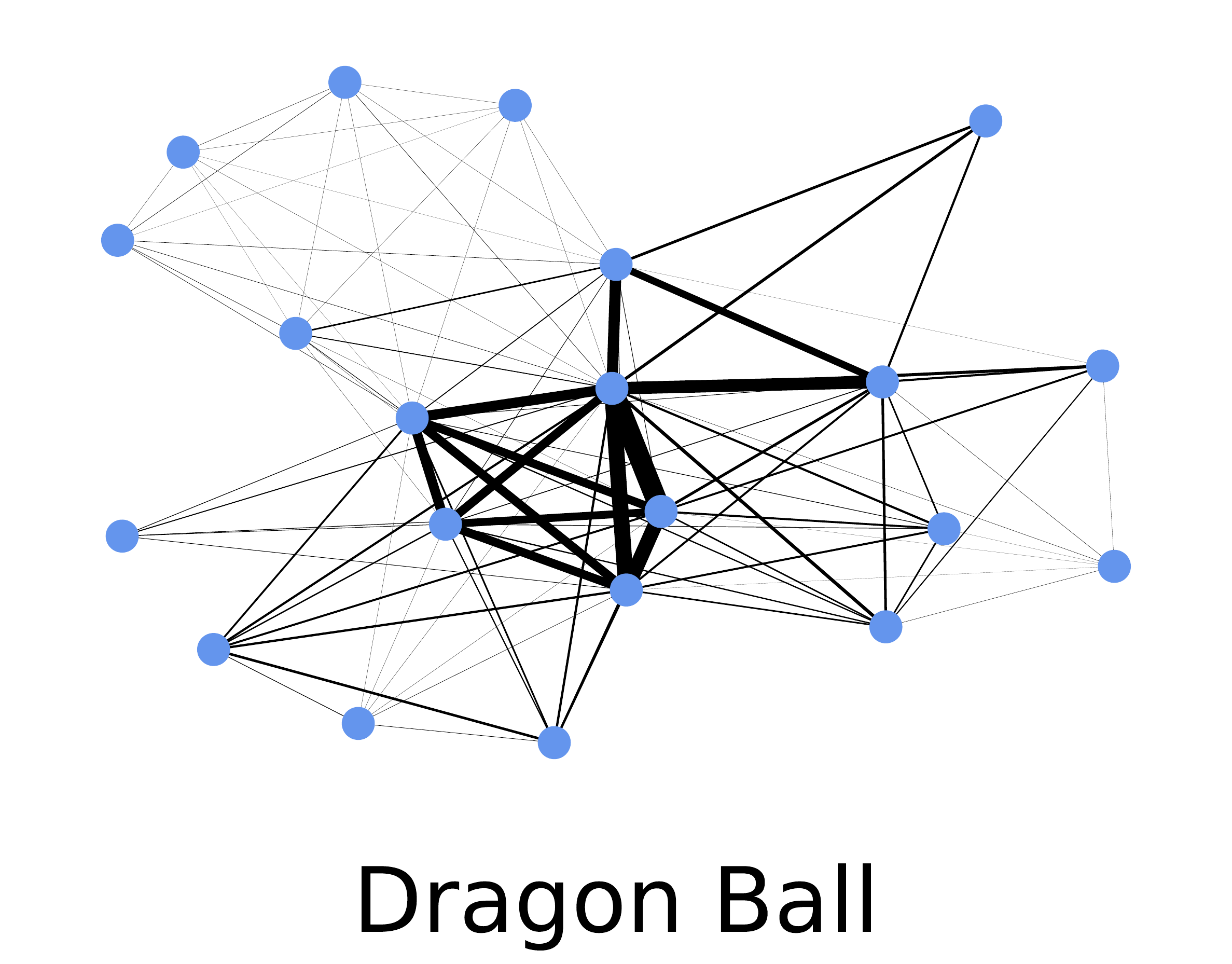}
         \label{fig:three sin x}
     \end{subfigure}
     \begin{subfigure}[b]{0.195\textwidth}
         \centering
         \includegraphics[width=\textwidth]{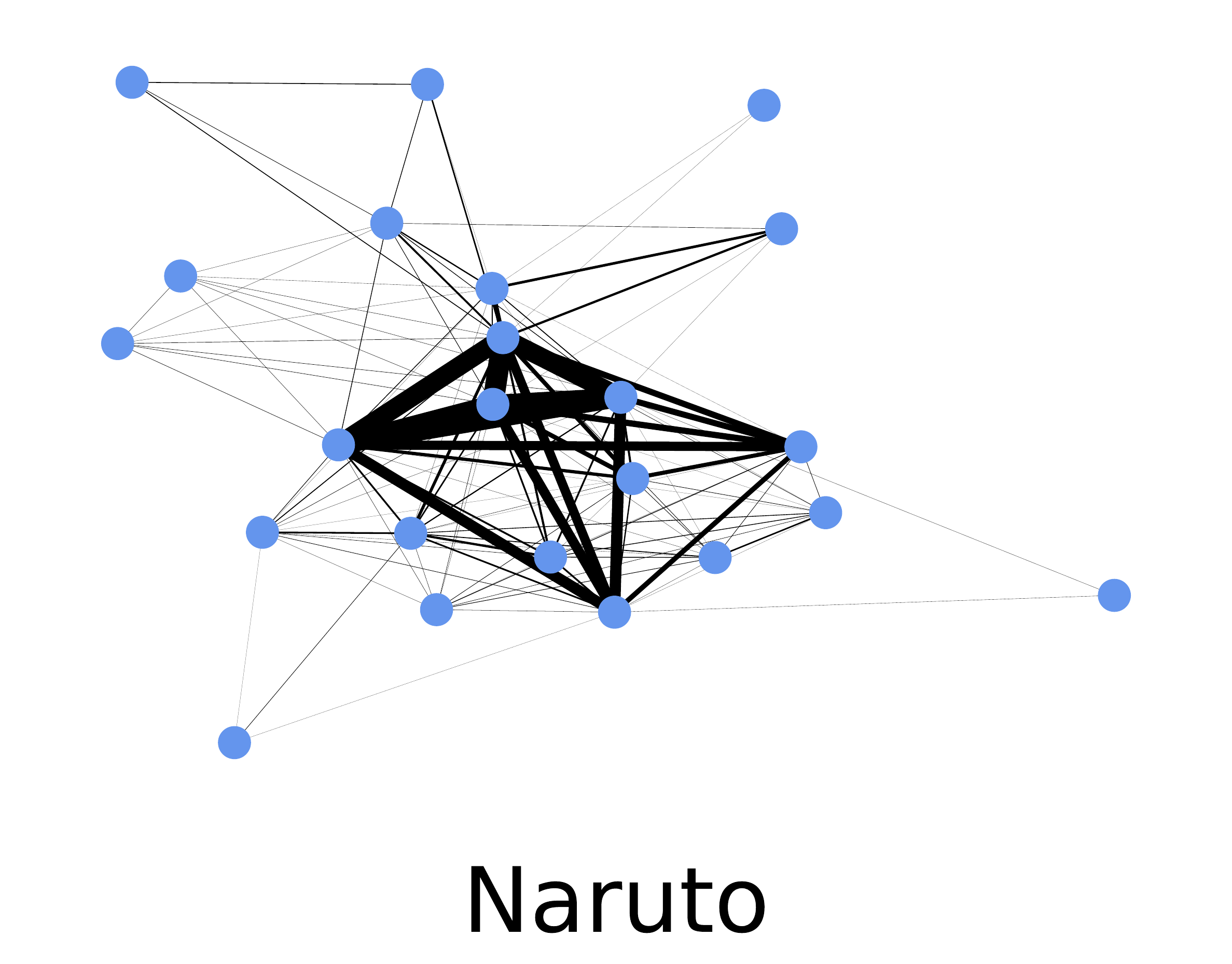}
         \label{fig:five over x}
     \end{subfigure}
     \begin{subfigure}[b]{0.195\textwidth}
         \centering
         \includegraphics[width=\textwidth]{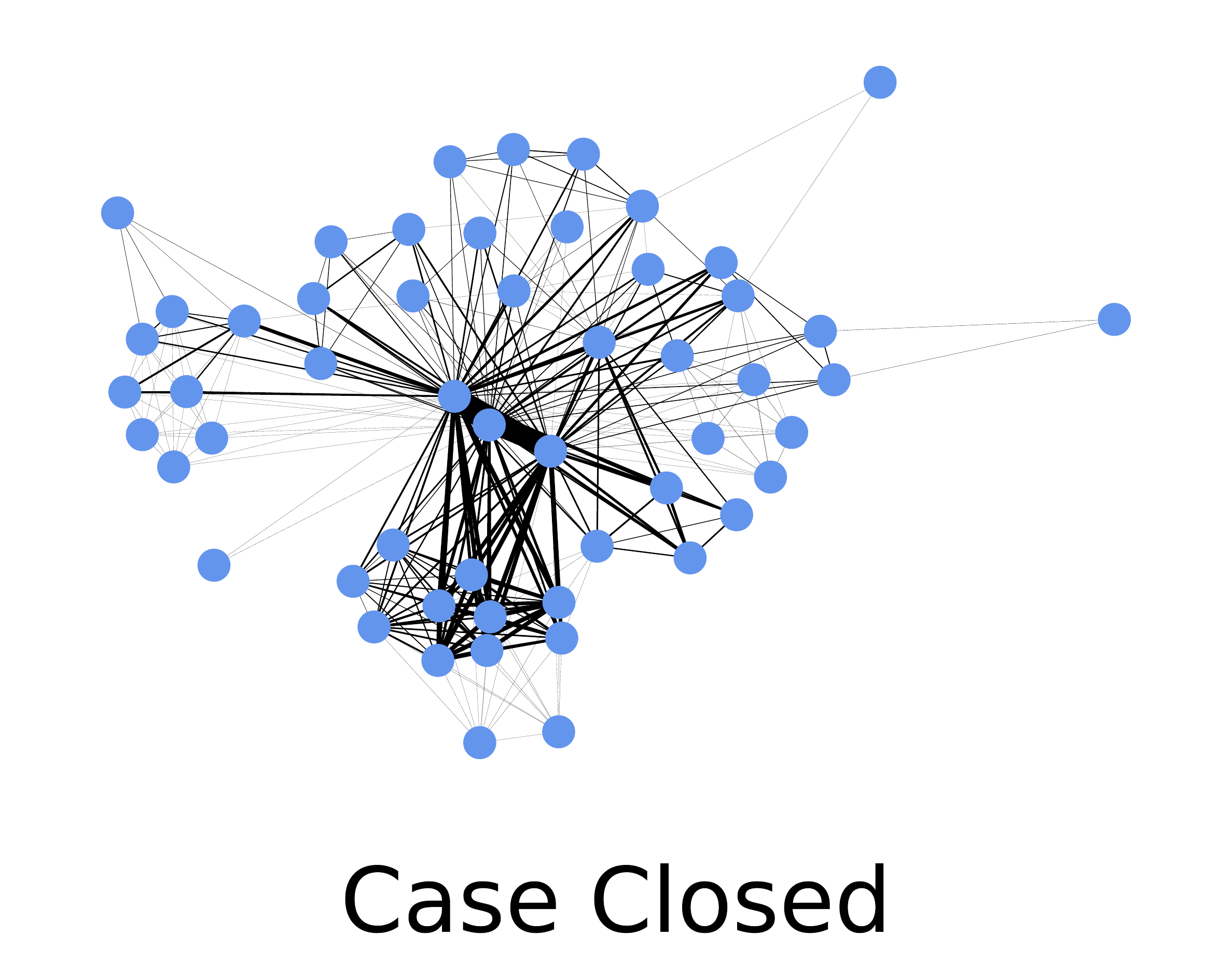}
         \label{fig:five over x}
     \end{subfigure}
     \begin{subfigure}[b]{0.195\textwidth}
         \centering
         \includegraphics[width=\textwidth]{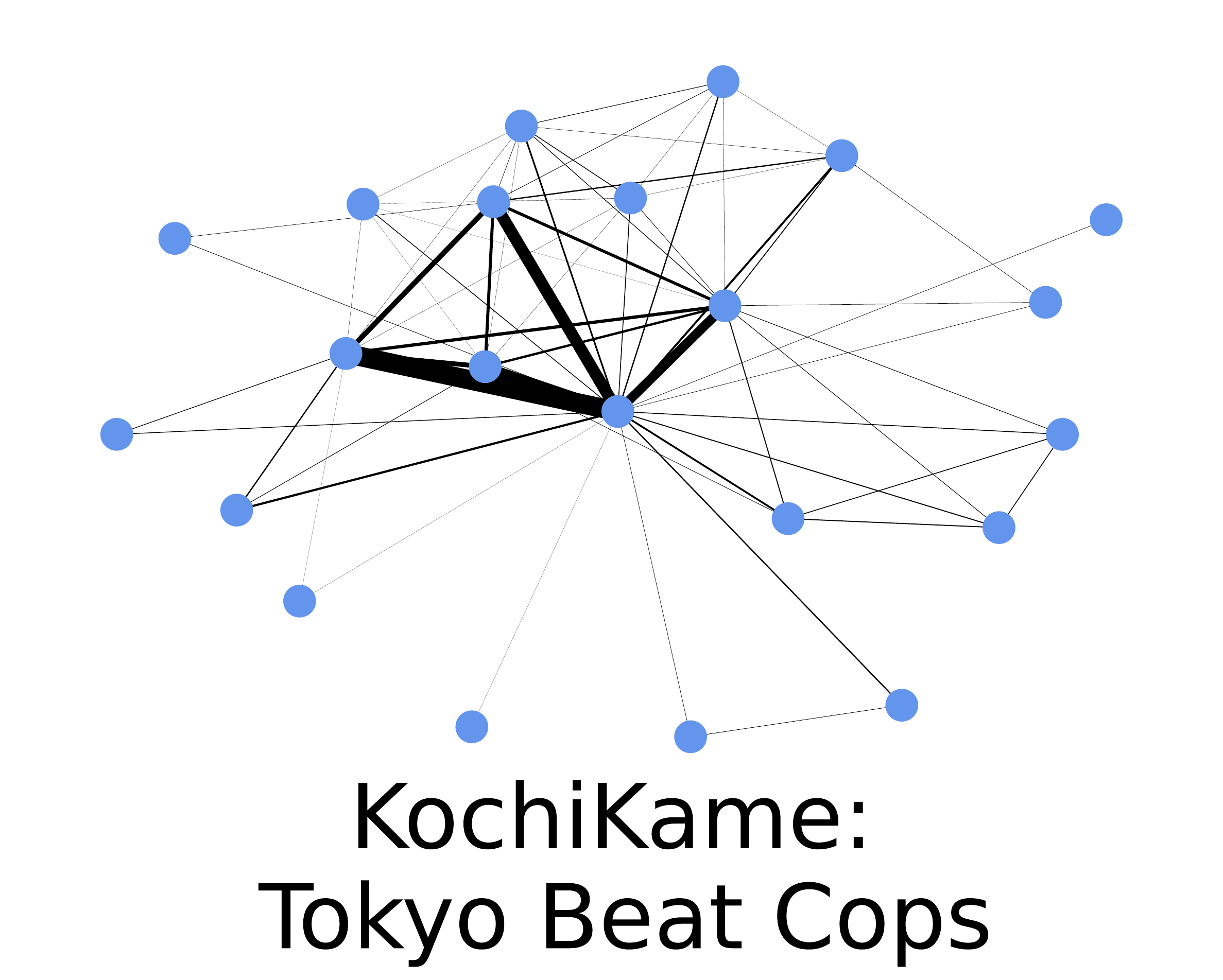}
         \label{fig:five over x}
     \end{subfigure}
     \\
      \begin{subfigure}[b]{0.195\textwidth}
         \centering
         \includegraphics[width=\textwidth]{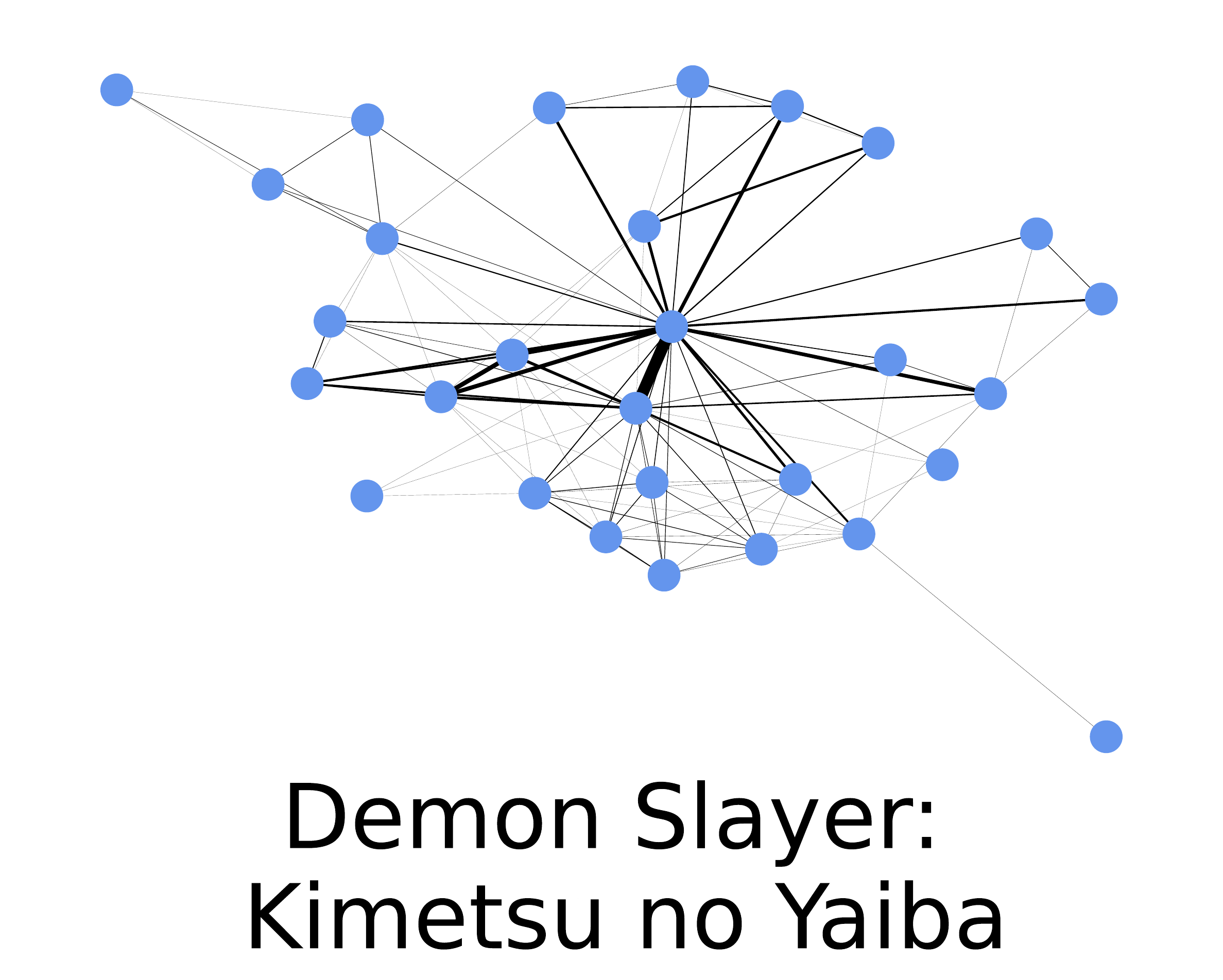}
         \label{fig:y equals x}
     \end{subfigure}
     \begin{subfigure}[b]{0.195\textwidth}
         \centering
         \includegraphics[width=\textwidth]{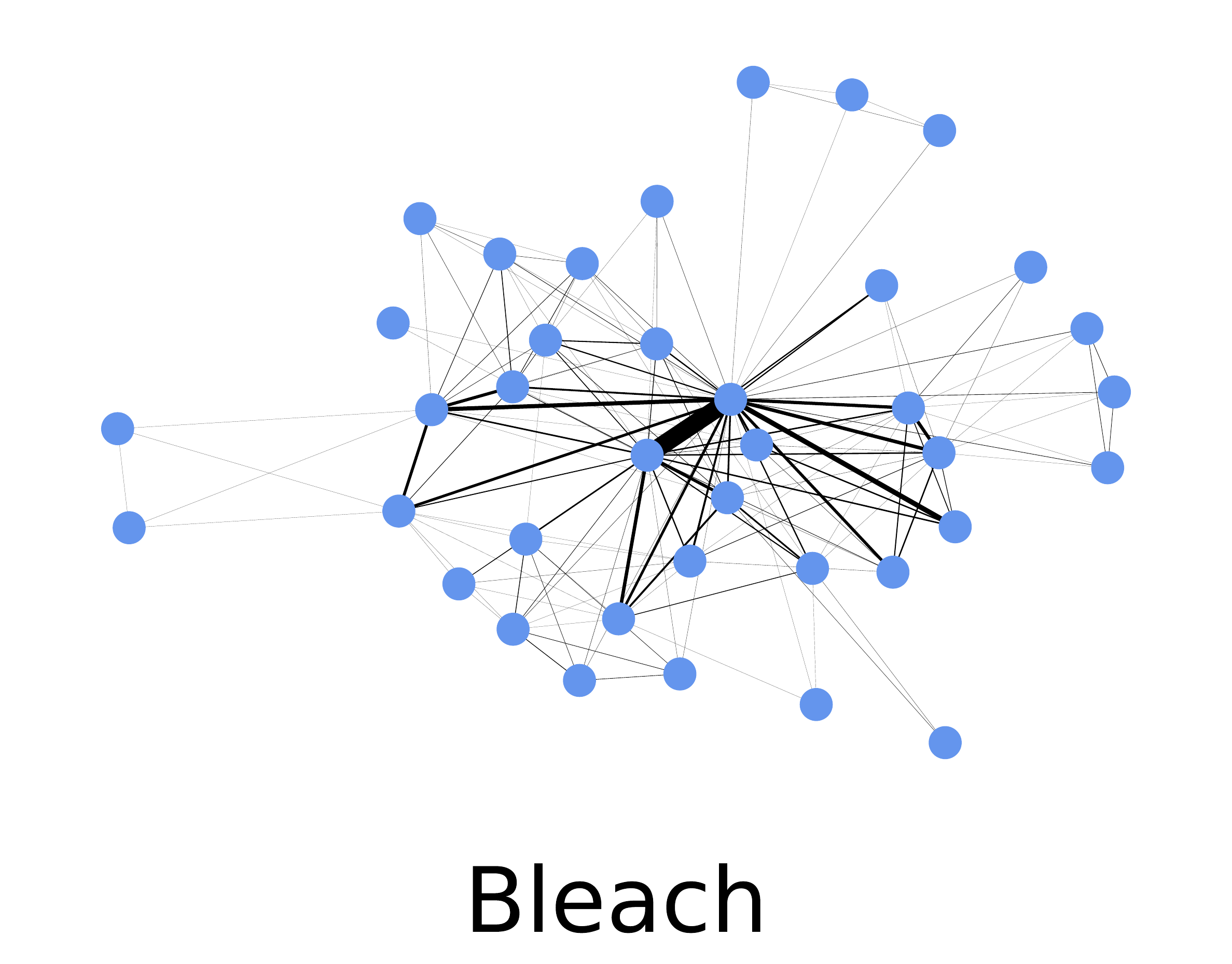}
         \label{fig:three sin x}
     \end{subfigure}
     \begin{subfigure}[b]{0.195\textwidth}
         \centering
         \includegraphics[width=\textwidth]{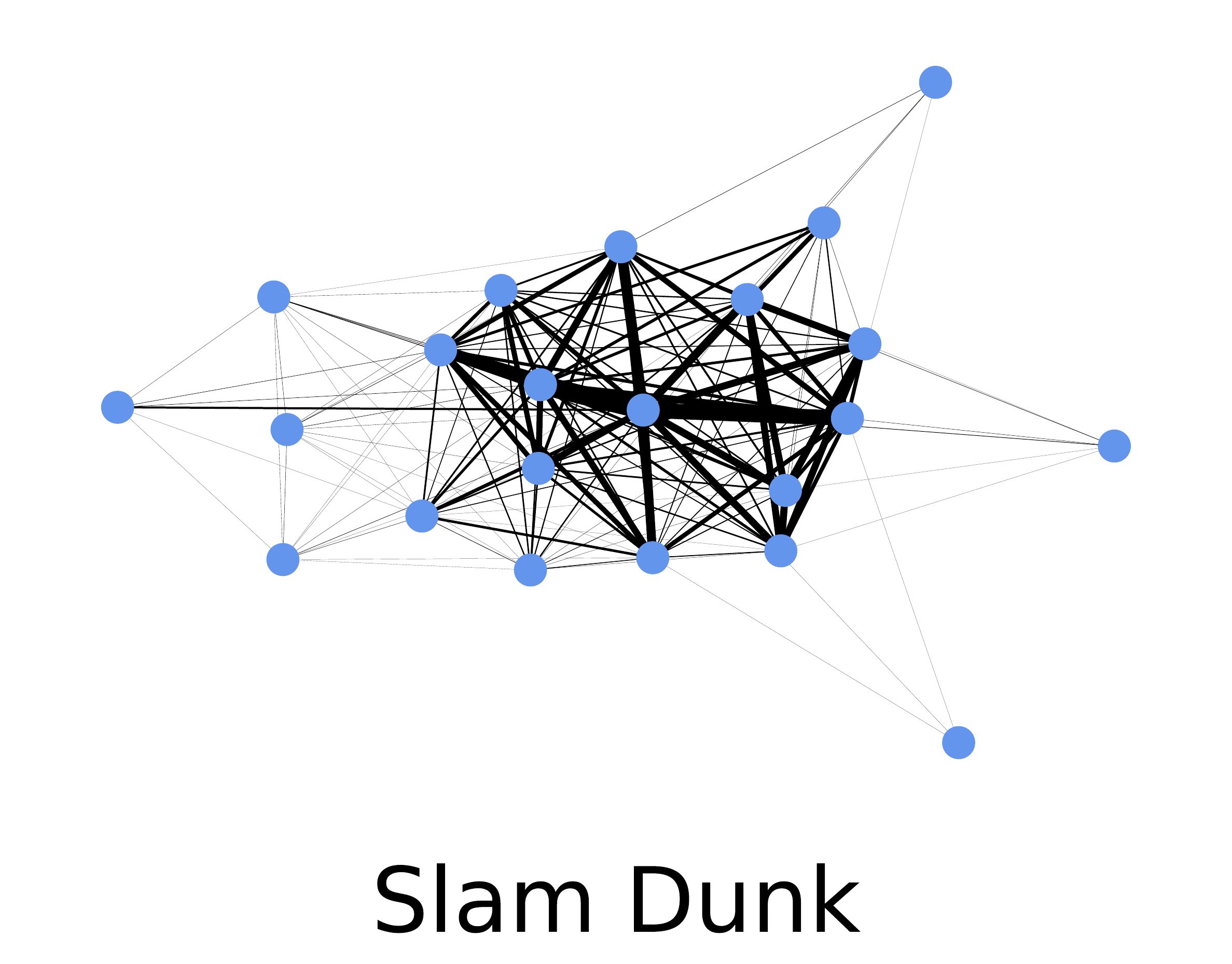}
         \label{fig:five over x}
     \end{subfigure}
     \begin{subfigure}[b]{0.195\textwidth}
         \centering
         \includegraphics[width=\textwidth]{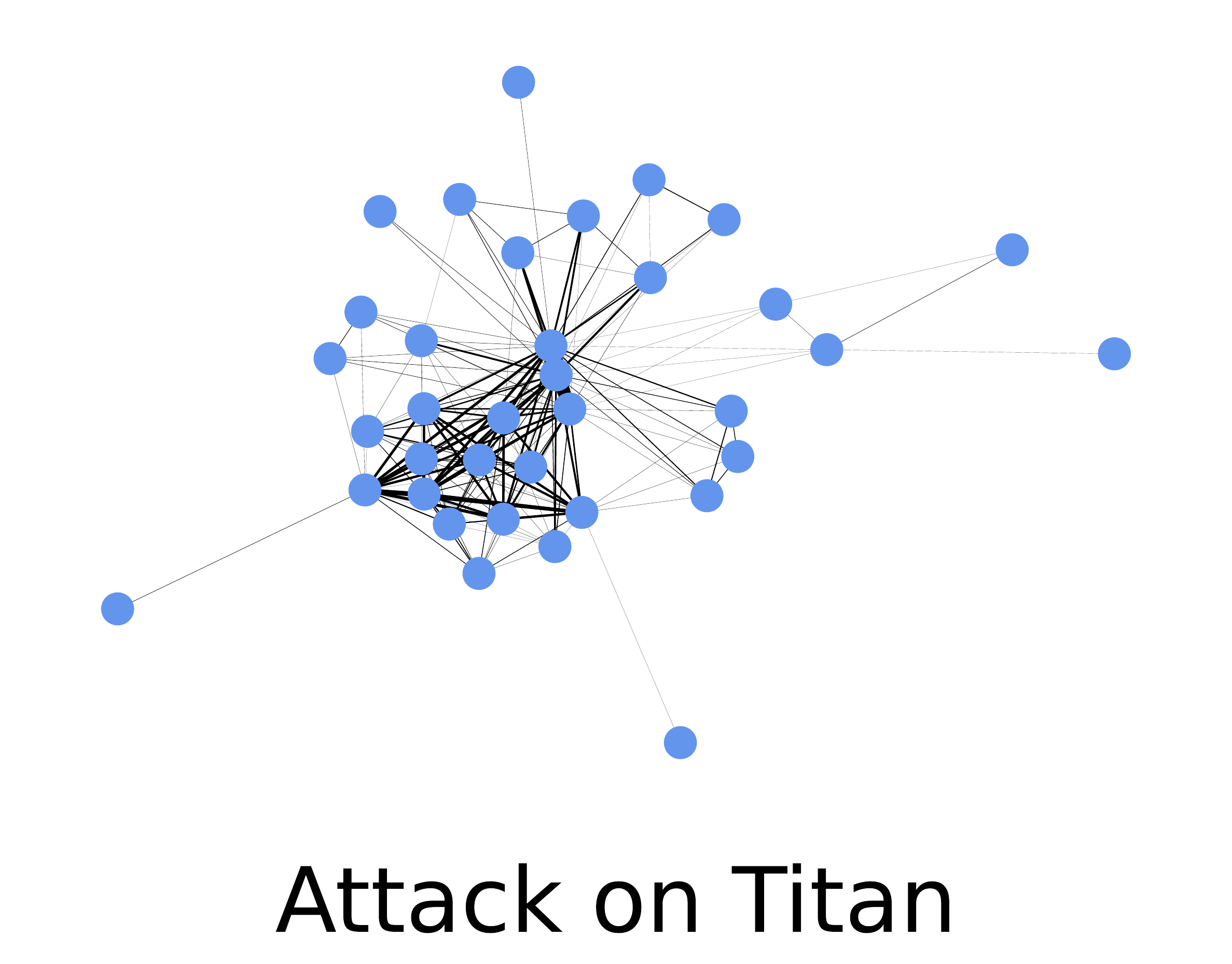}
         \label{fig:five over x}
     \end{subfigure}
     \begin{subfigure}[b]{0.195\textwidth}
         \centering
         \includegraphics[width=\textwidth]{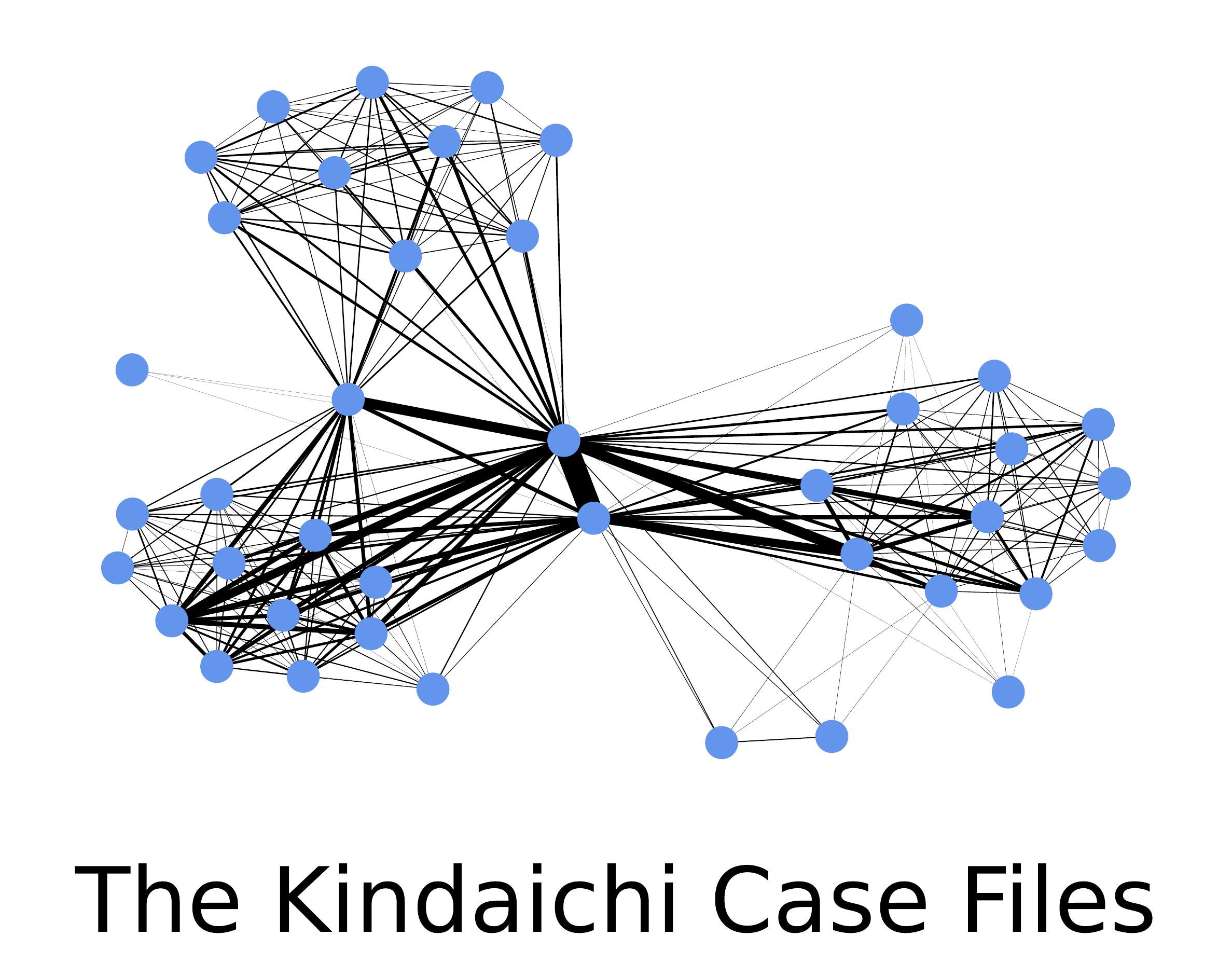}
         \label{fig:five over x}
     \end{subfigure}
     \\
      \begin{subfigure}[b]{0.195\textwidth}
         \centering
         \includegraphics[width=\textwidth]{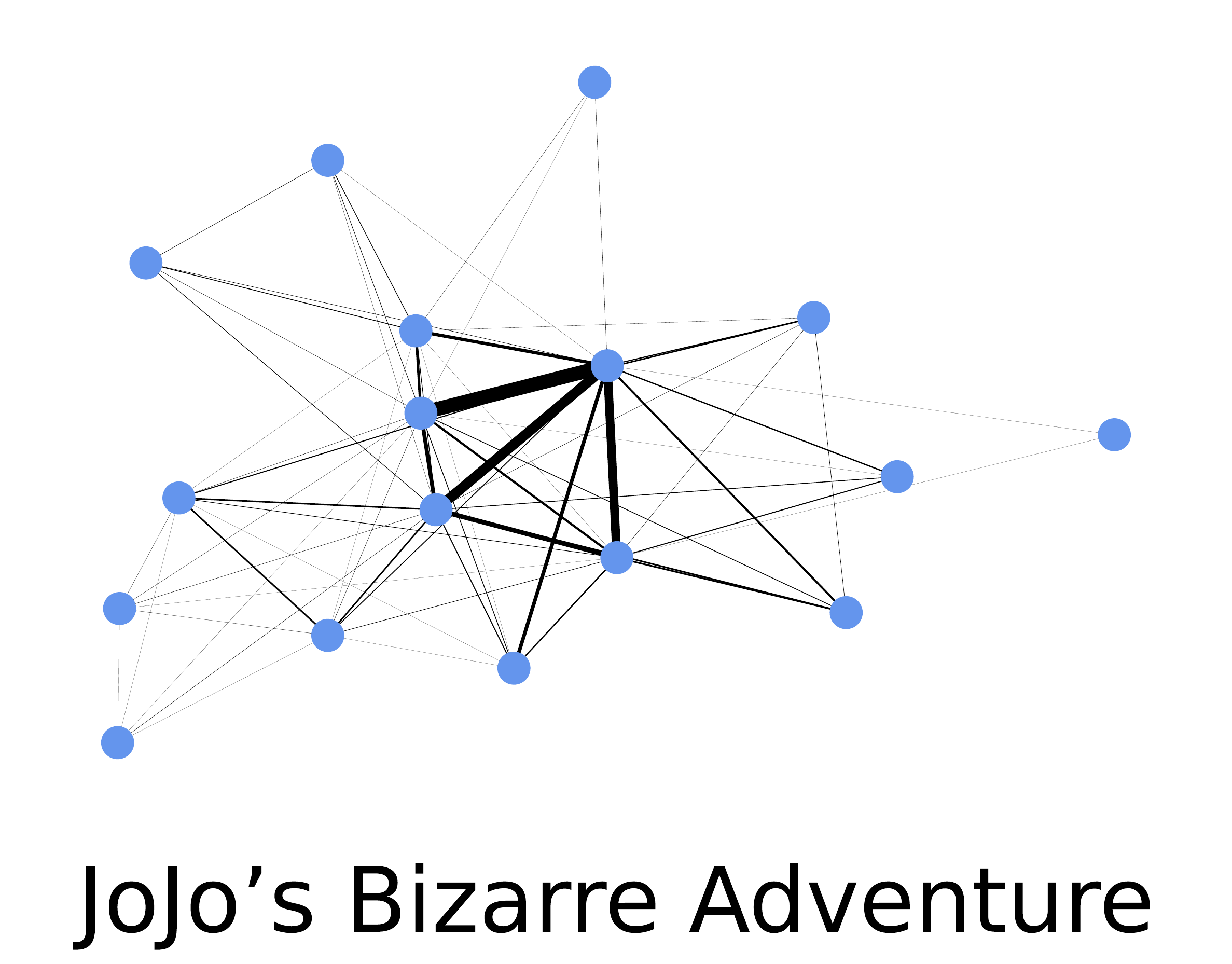}
         \label{fig:y equals x}
     \end{subfigure}
     \begin{subfigure}[b]{0.195\textwidth}
         \centering
         \includegraphics[width=\textwidth]{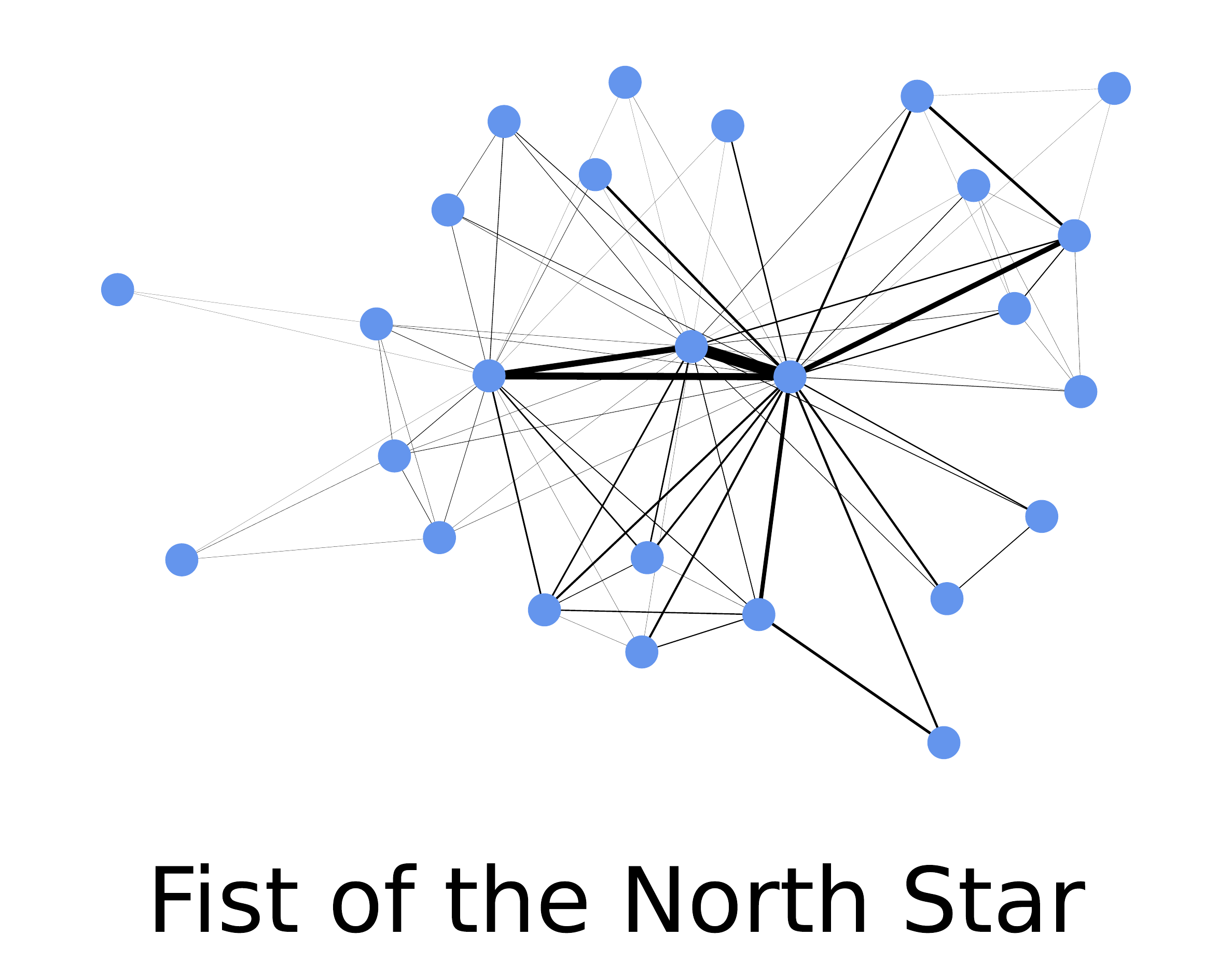}
         \label{fig:three sin x}
     \end{subfigure}
     \begin{subfigure}[b]{0.195\textwidth}
         \centering
         \includegraphics[width=\textwidth]{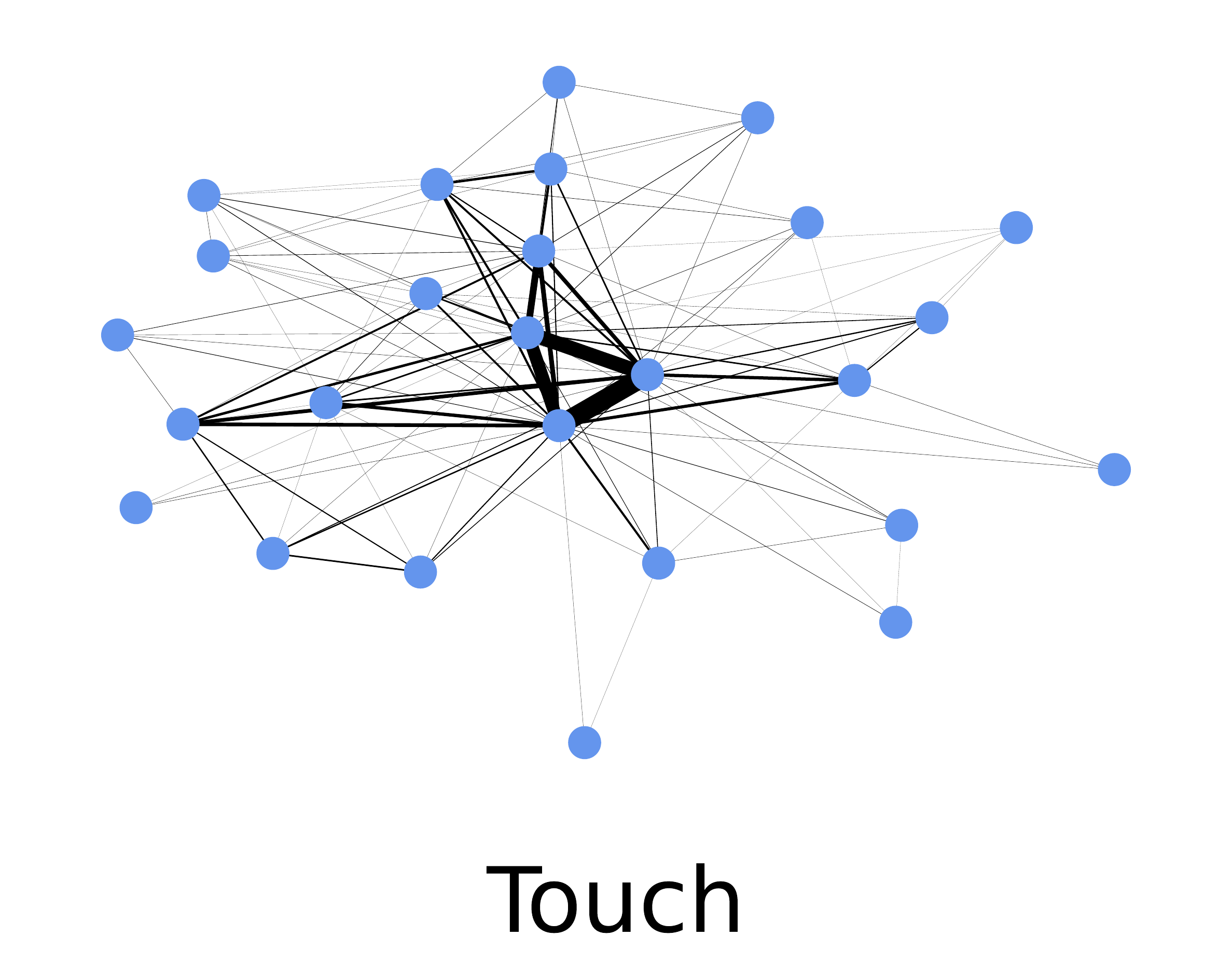}
         \label{fig:five over x}
     \end{subfigure}
     \begin{subfigure}[b]{0.195\textwidth}
         \centering
         \includegraphics[width=\textwidth]{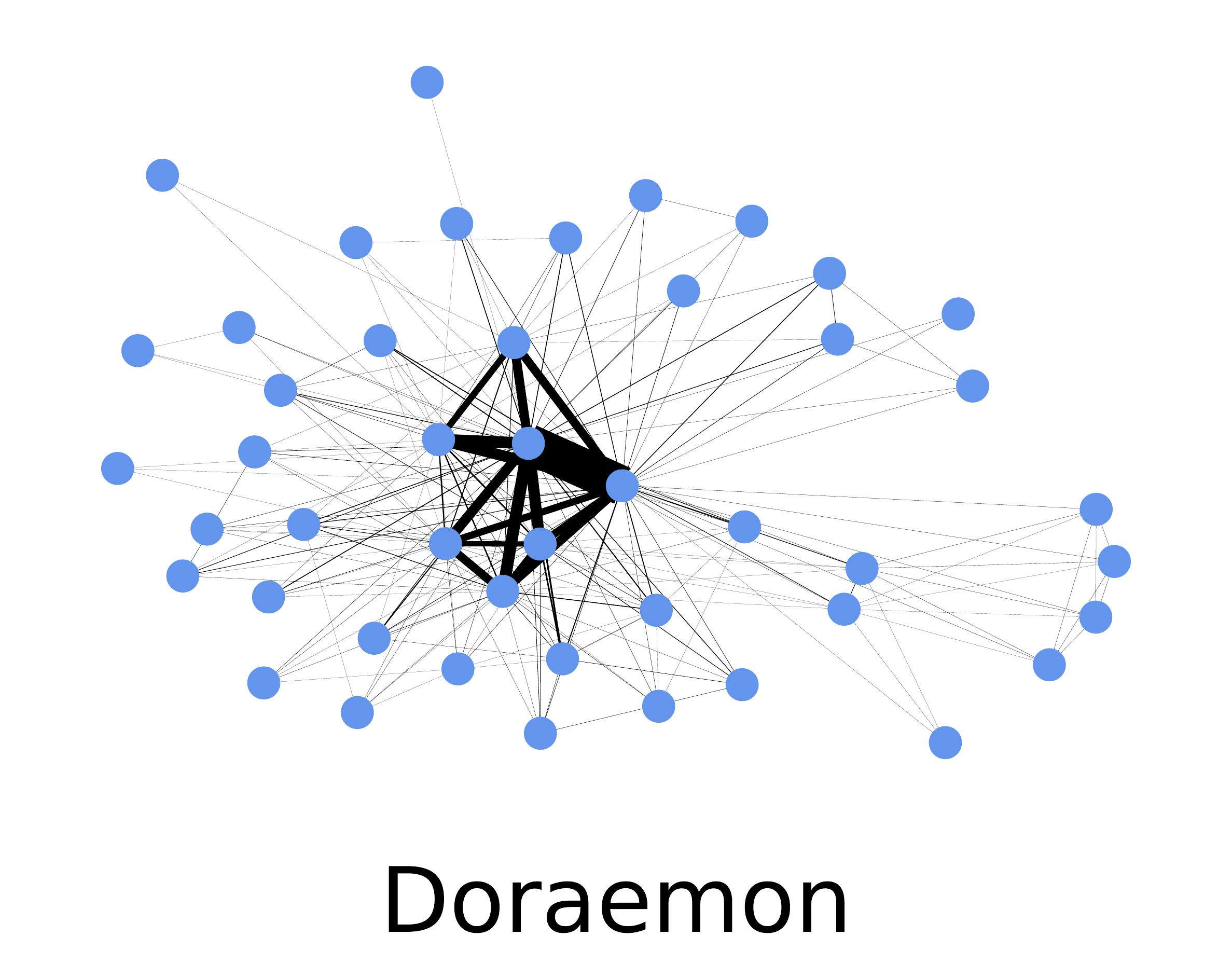}
         \label{fig:five over x}
     \end{subfigure}
     \begin{subfigure}[b]{0.195\textwidth}
         \centering
         \includegraphics[width=\textwidth]{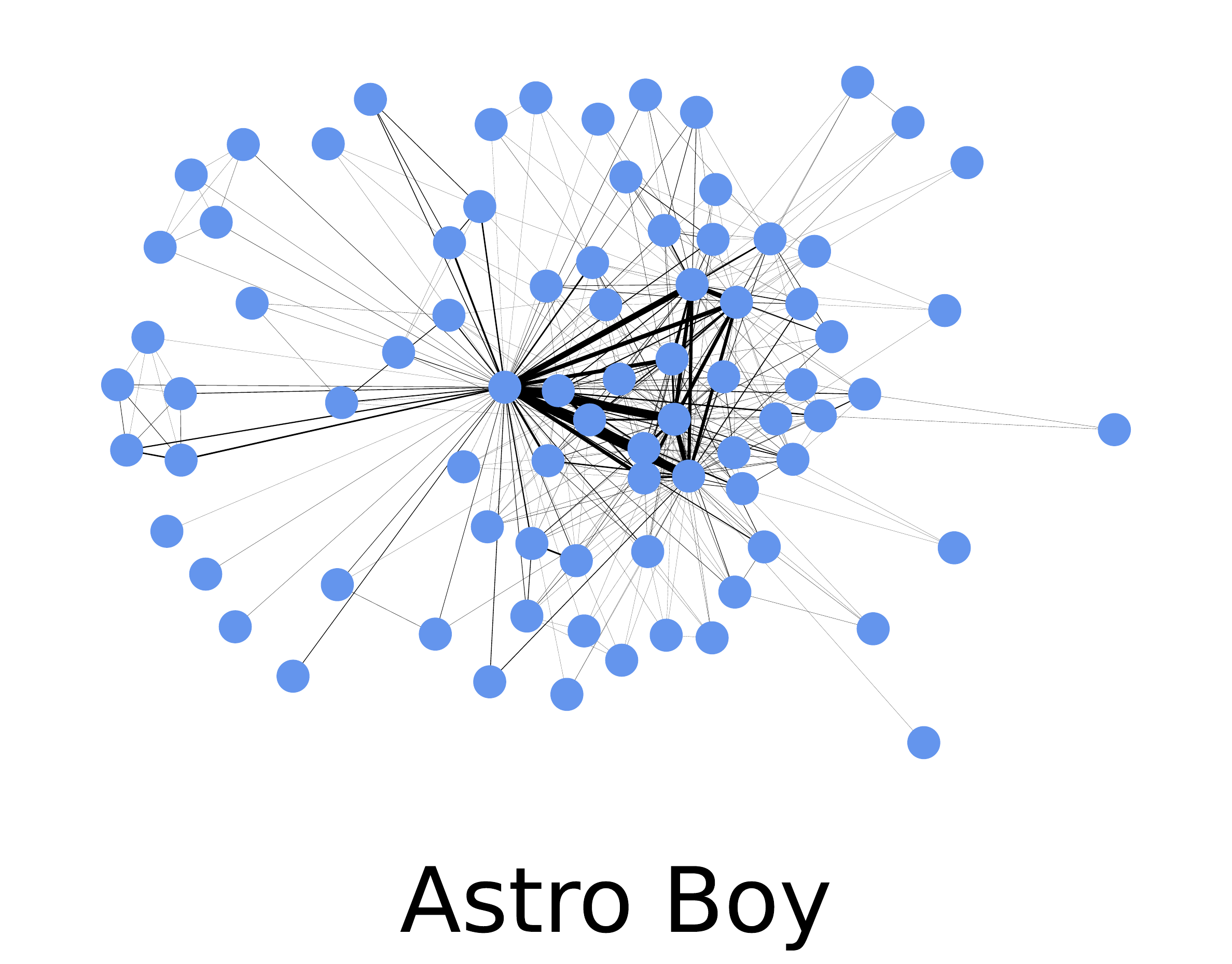}
         \label{fig:five over x}
     \end{subfigure}
     \\
      \begin{subfigure}[b]{0.195\textwidth}
         \centering
         \includegraphics[width=\textwidth]{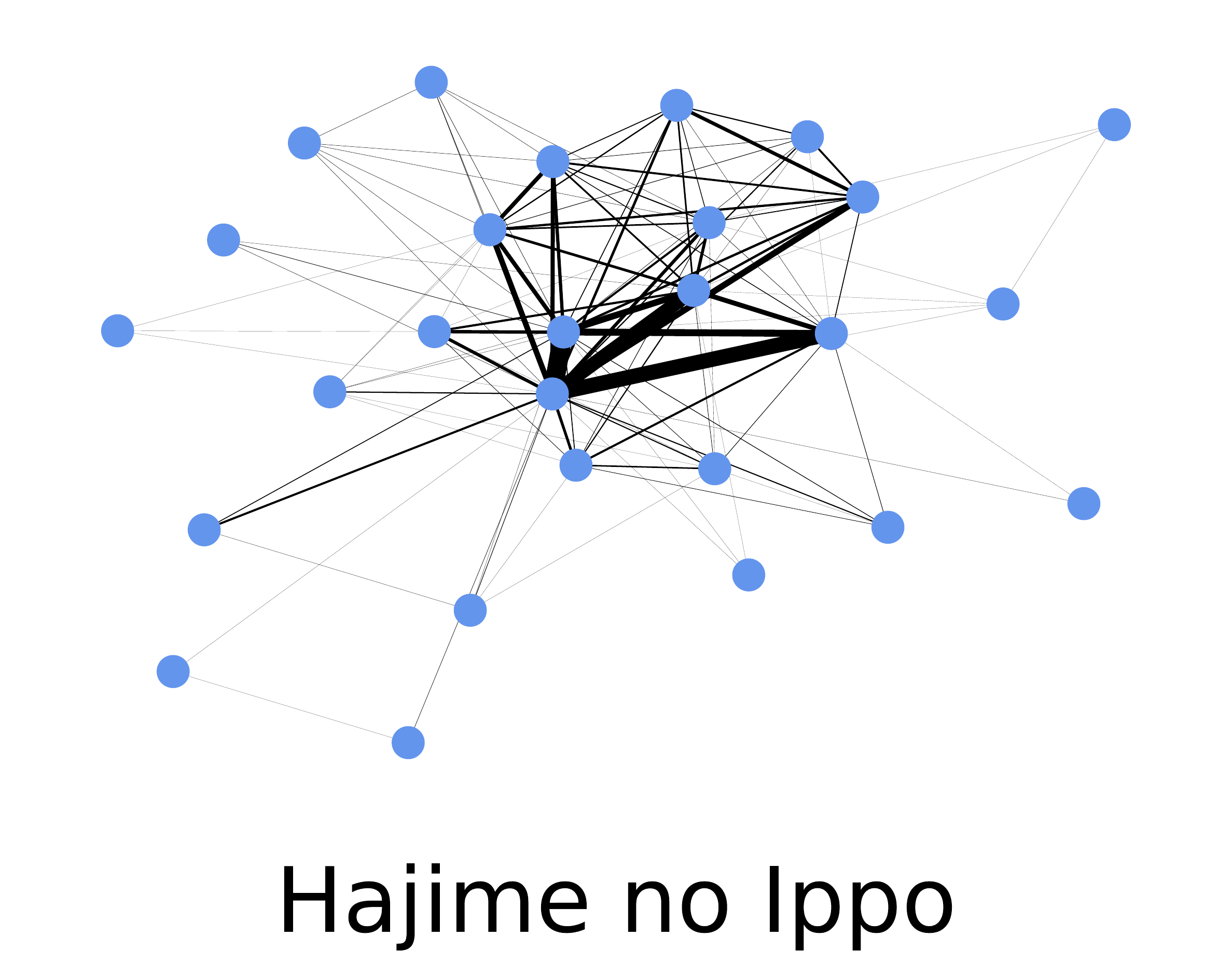}
         \label{fig:y equals x}
     \end{subfigure}
     \begin{subfigure}[b]{0.195\textwidth}
         \centering
         \includegraphics[width=\textwidth]{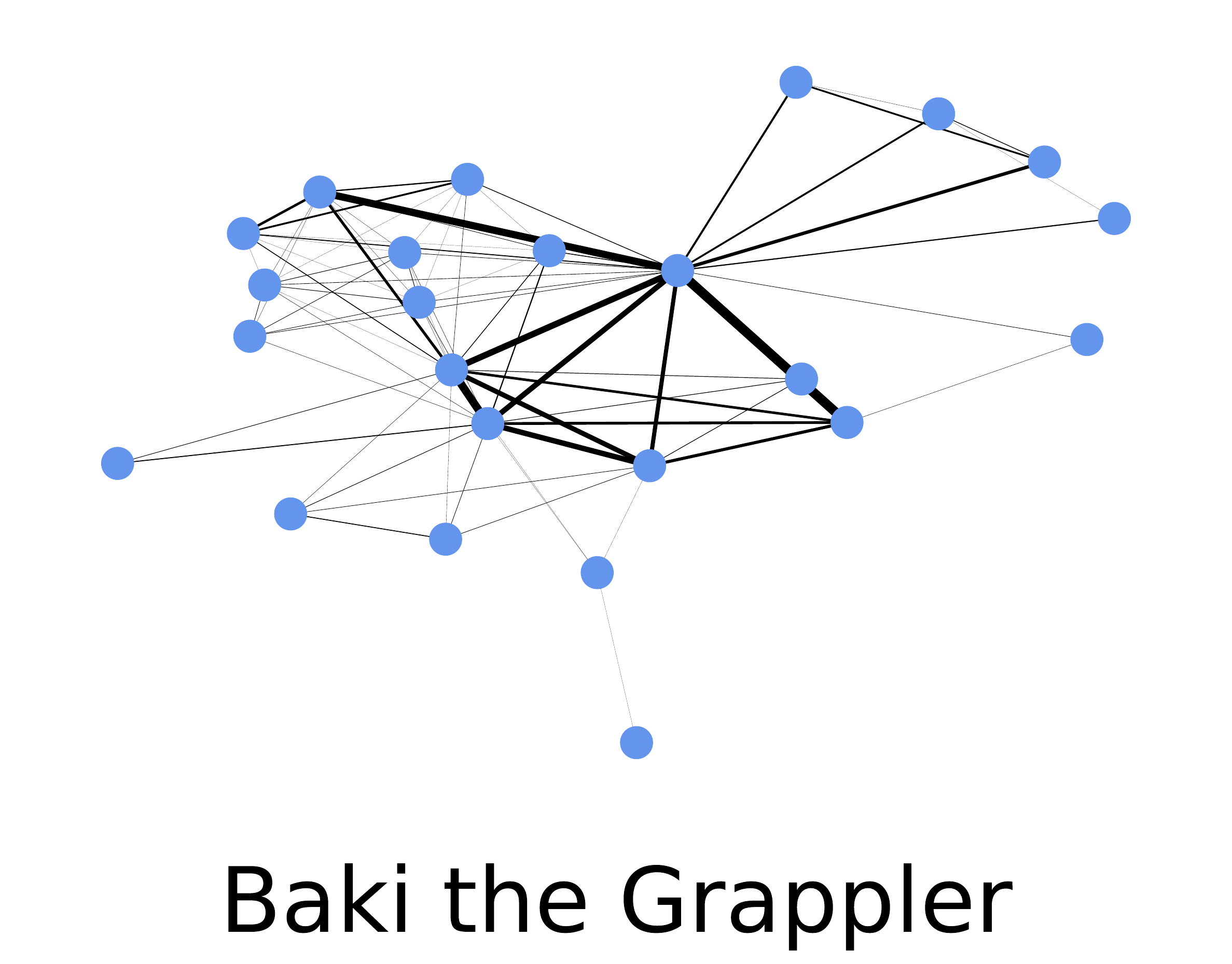}
         \label{fig:three sin x}
     \end{subfigure}
     \begin{subfigure}[b]{0.195\textwidth}
         \centering
         \includegraphics[width=\textwidth]{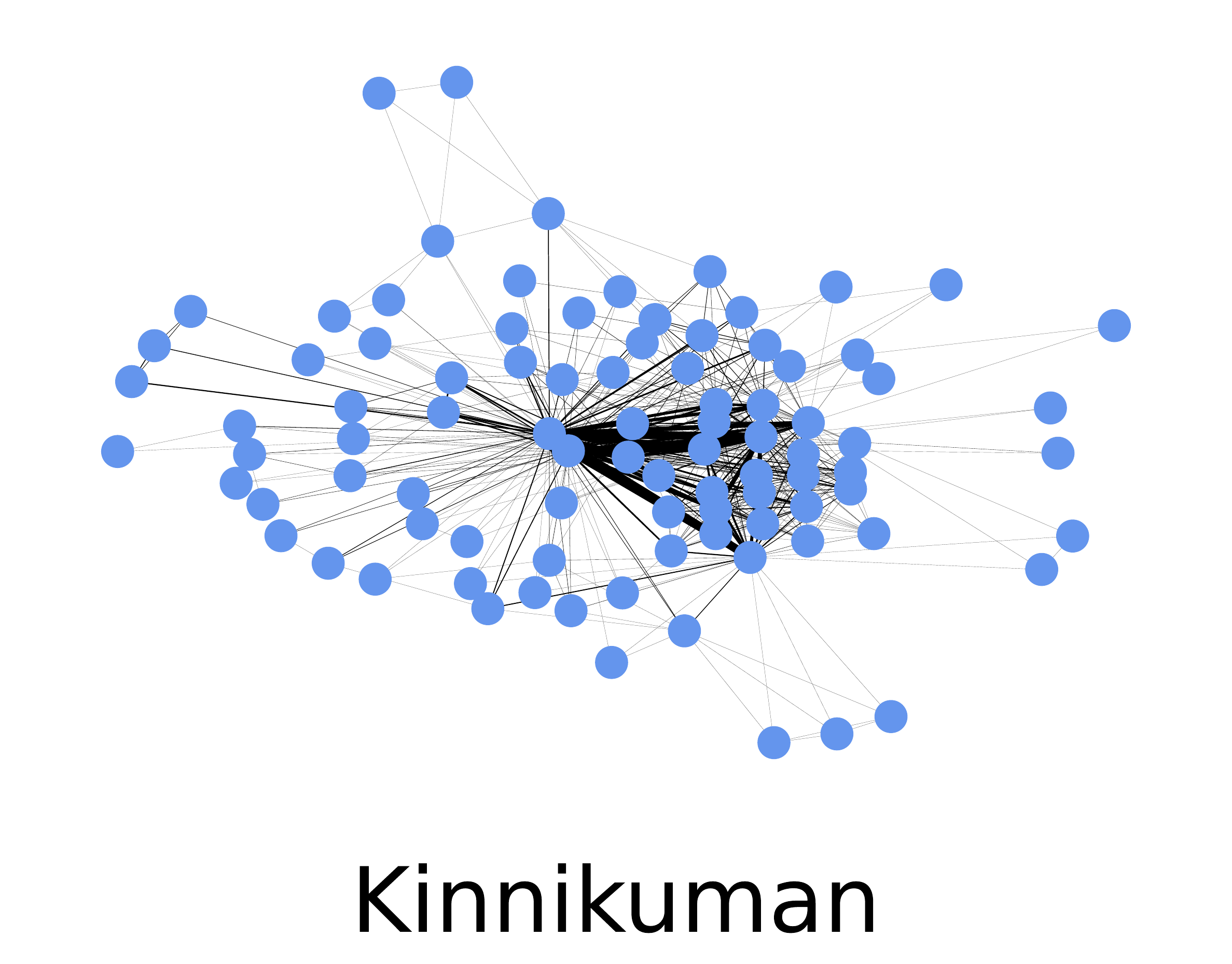}
         \label{fig:five over x}
     \end{subfigure}
     \begin{subfigure}[b]{0.195\textwidth}
         \centering
         \includegraphics[width=\textwidth]{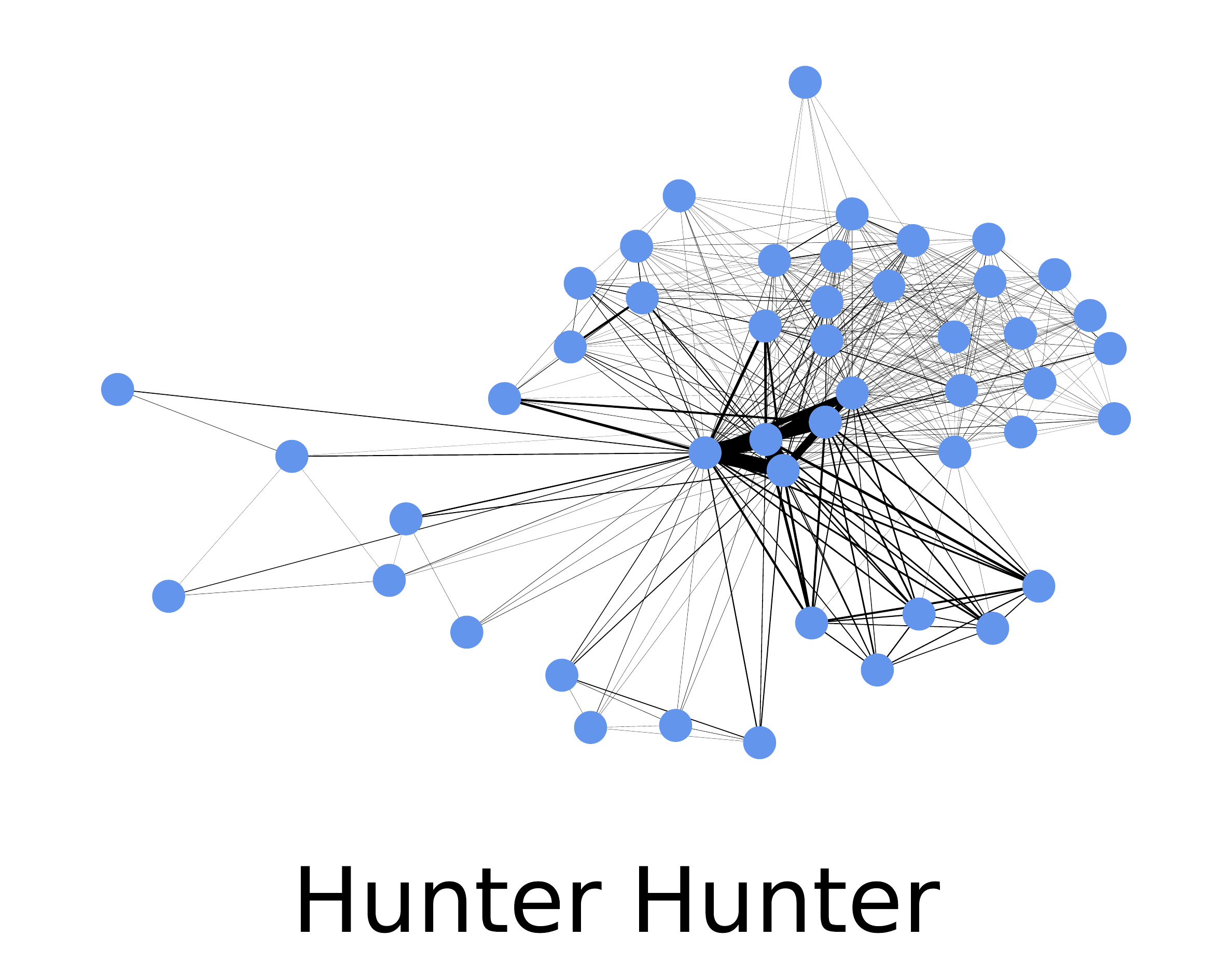}
         \label{fig:five over x}
     \end{subfigure}
     \begin{subfigure}[b]{0.195\textwidth}
         \centering
         \includegraphics[width=\textwidth]{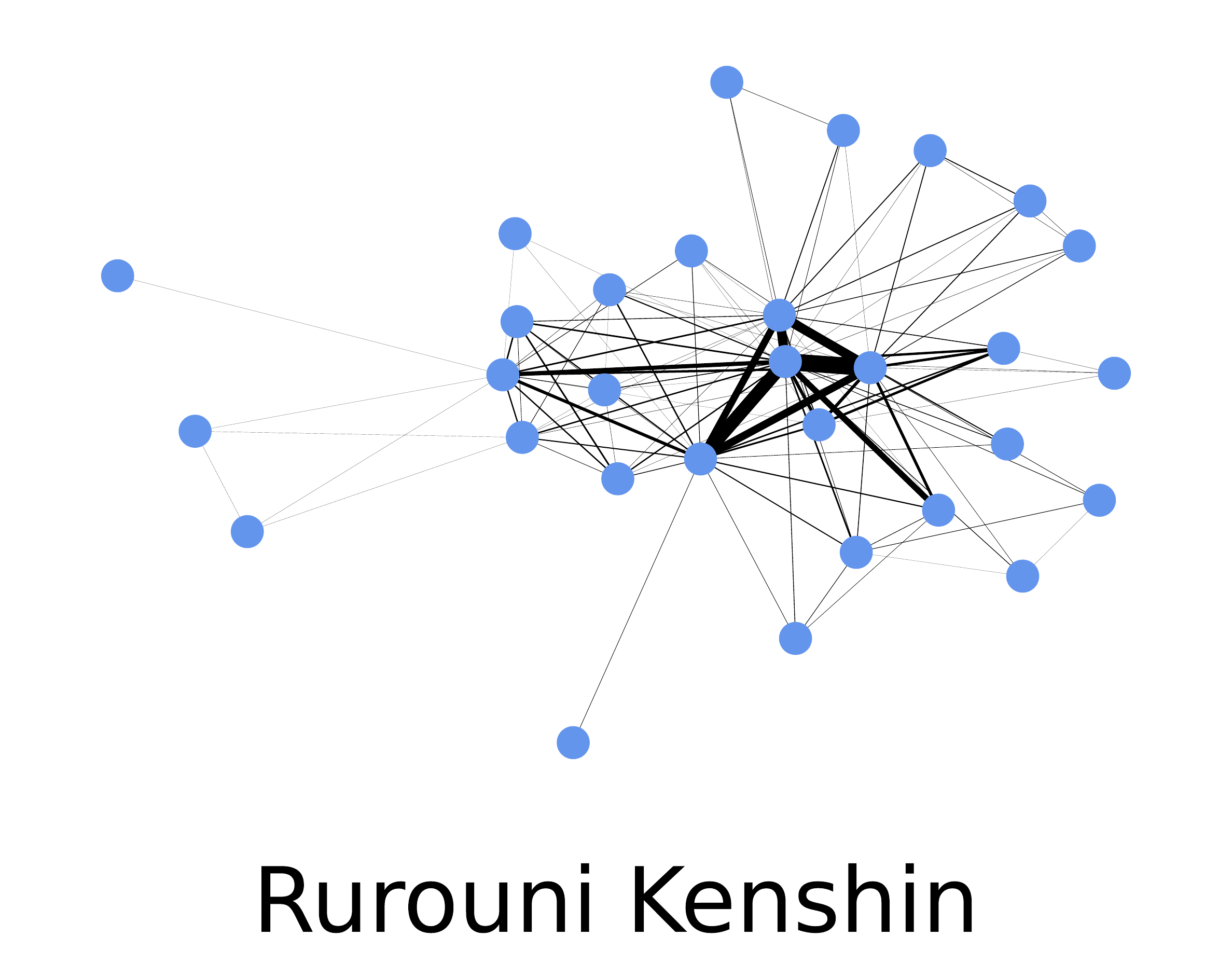}
         \label{fig:five over x}
     \end{subfigure}
          \\
      \begin{subfigure}[b]{0.195\textwidth}
         \centering
         \includegraphics[width=\textwidth]{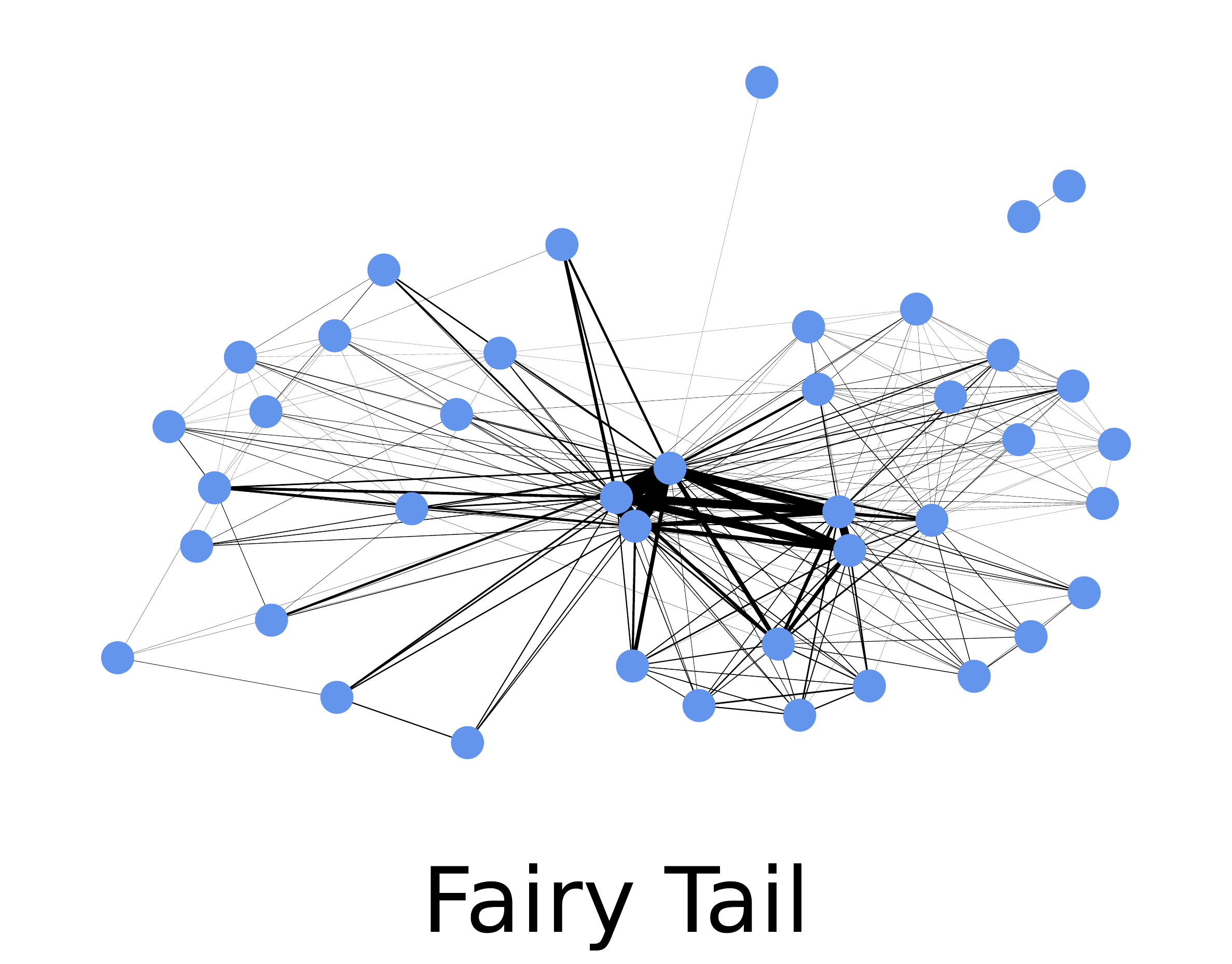}
         \label{fig:y equals x}
     \end{subfigure}
     \begin{subfigure}[b]{0.195\textwidth}
         \centering
         \includegraphics[width=\textwidth]{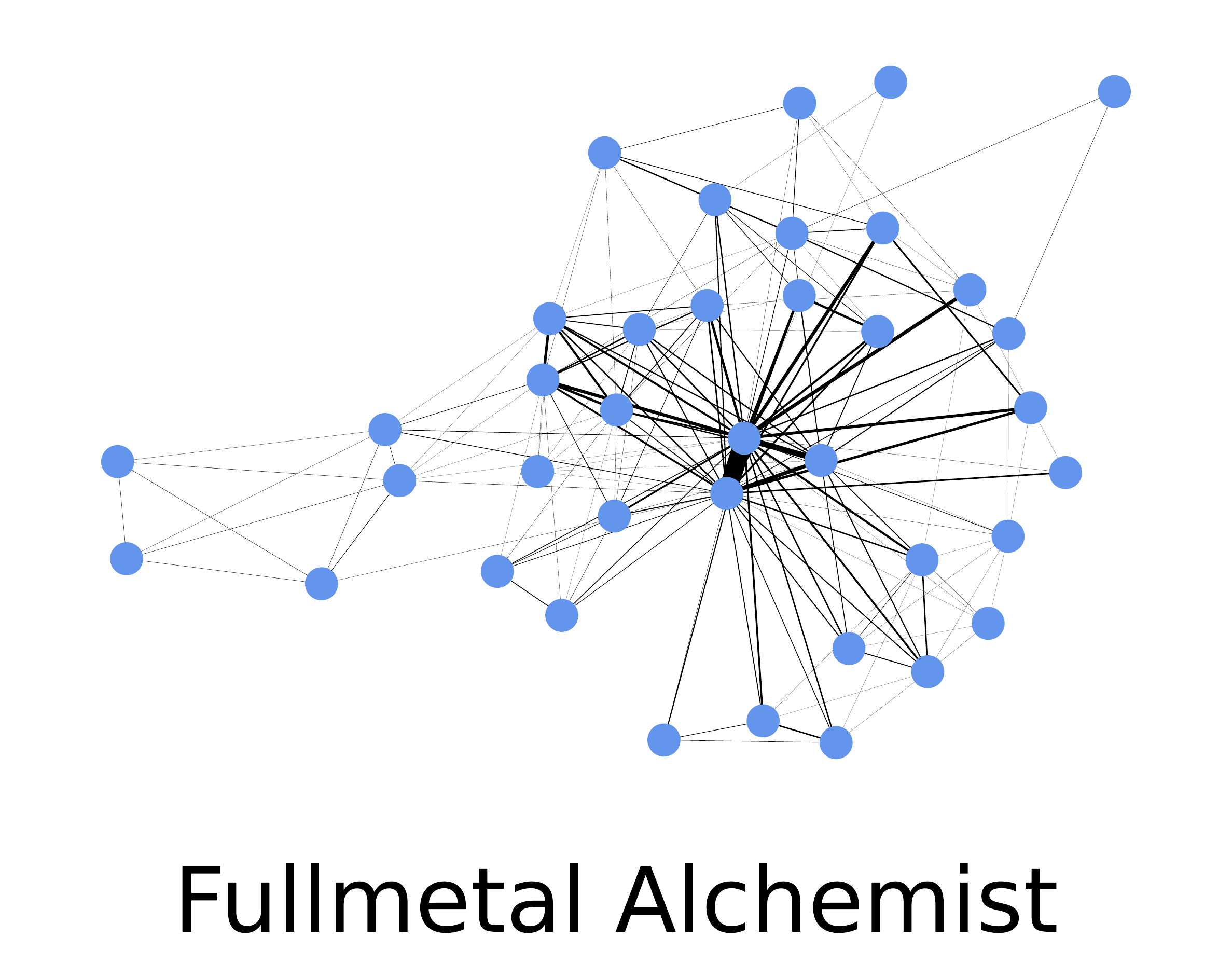}
         \label{fig:three sin x}
     \end{subfigure}
     \begin{subfigure}[b]{0.195\textwidth}
         \centering
         \includegraphics[width=\textwidth]{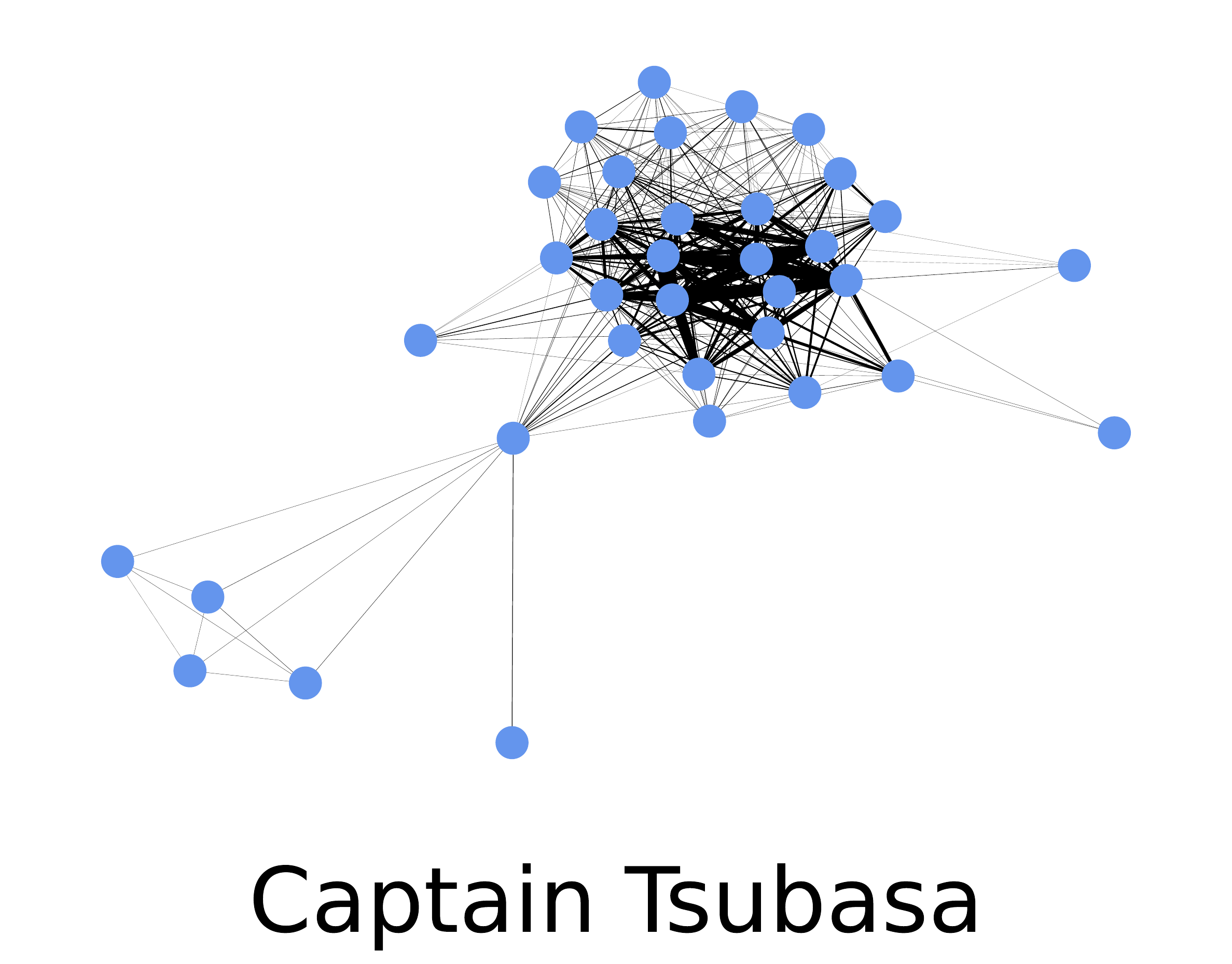}
         \label{fig:five over x}
     \end{subfigure}
     \begin{subfigure}[b]{0.195\textwidth}
         \centering
         \includegraphics[width=\textwidth]{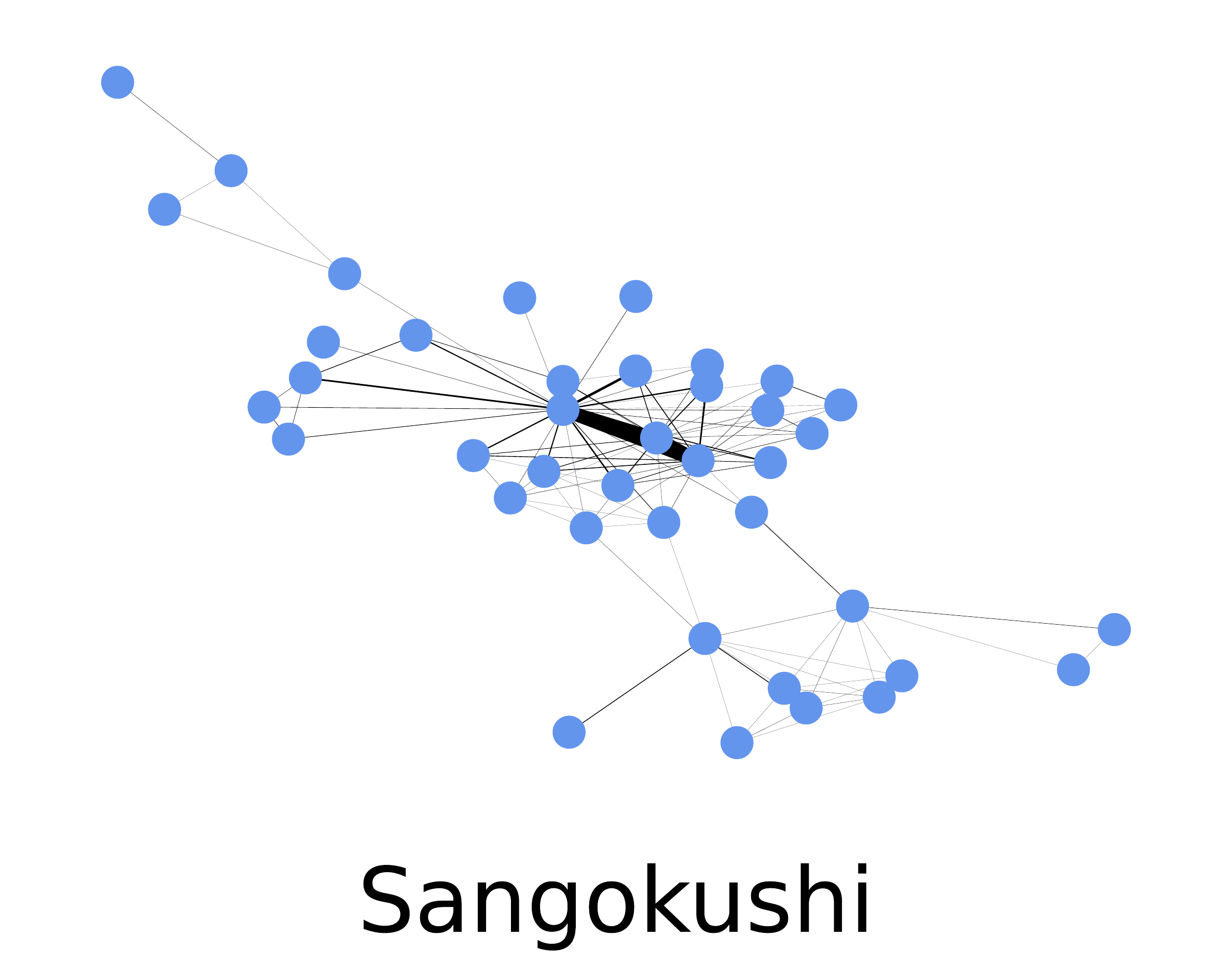}
         \label{fig:five over x}
     \end{subfigure}
     \begin{subfigure}[b]{0.195\textwidth}
         \centering
         \includegraphics[width=\textwidth]{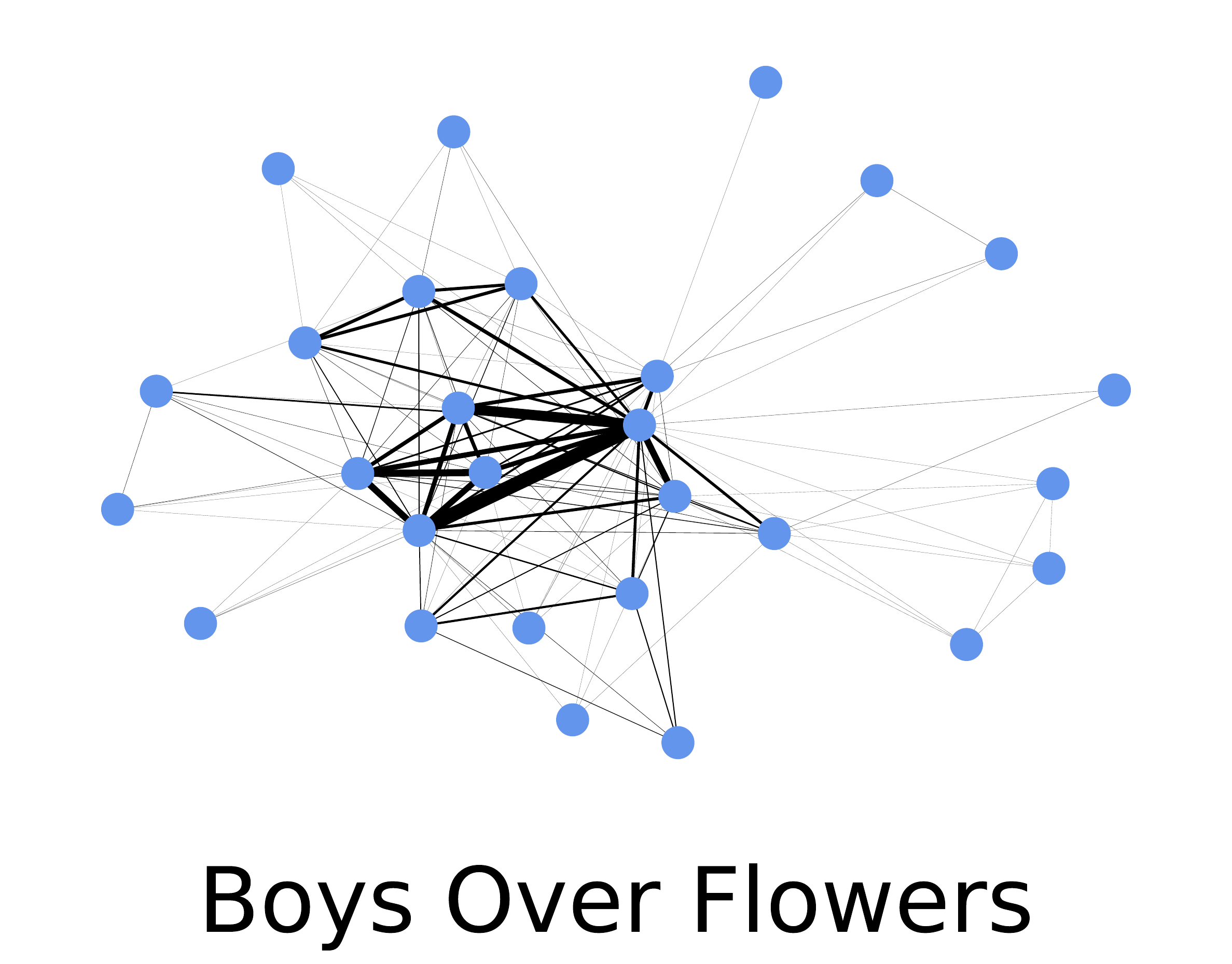}
         \label{fig:five over x}
     \end{subfigure}

    \caption{Character networks for 162 manga}
    
    \label{fig:three graphs}
\end{figure*}

\begin{figure*}\ContinuedFloat
     \centering
           \begin{subfigure}[b]{0.195\textwidth}
         \centering
         \includegraphics[width=\textwidth]{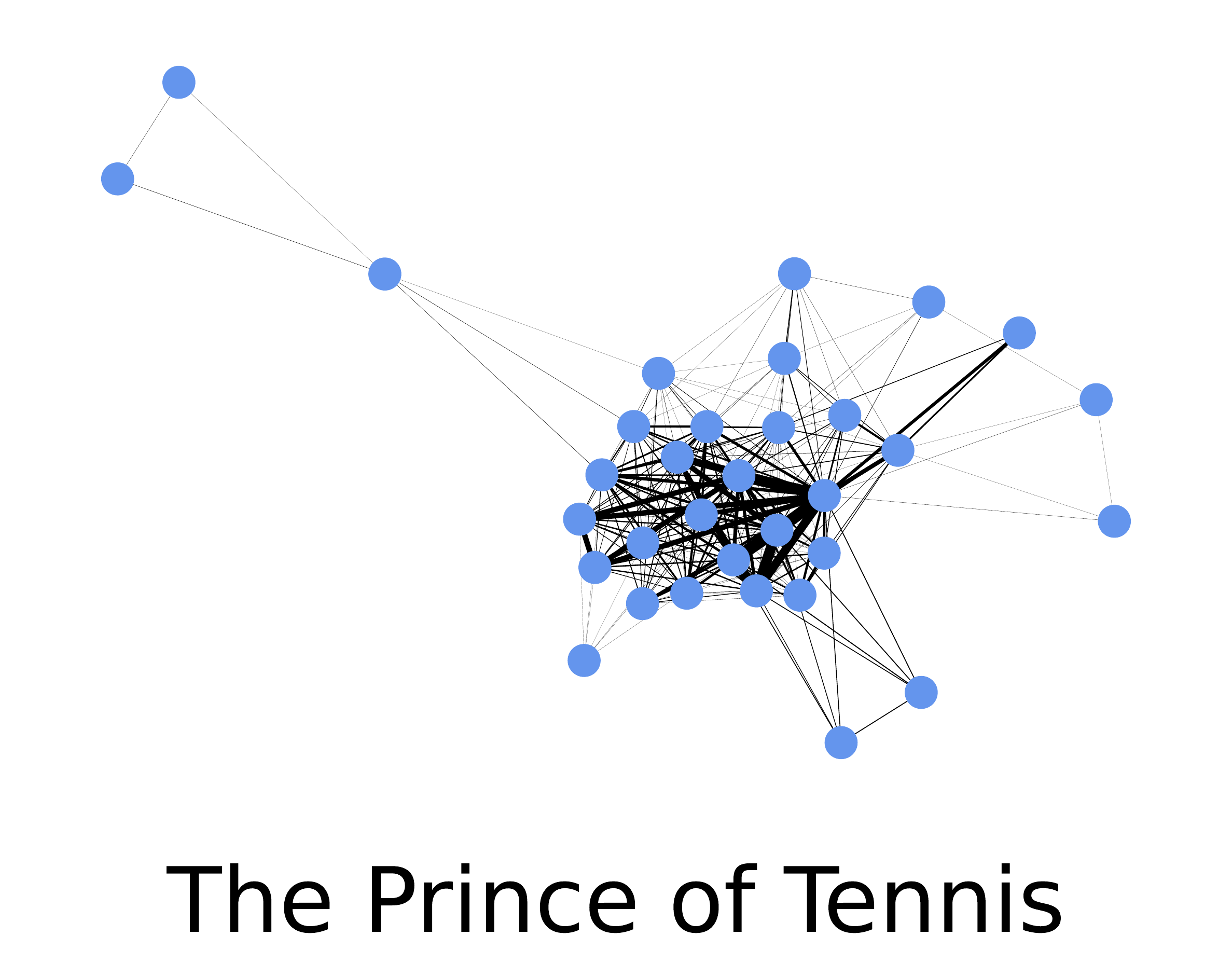}
         \label{fig:y equals x}
     \end{subfigure}
     \begin{subfigure}[b]{0.195\textwidth}
         \centering
         \includegraphics[width=\textwidth]{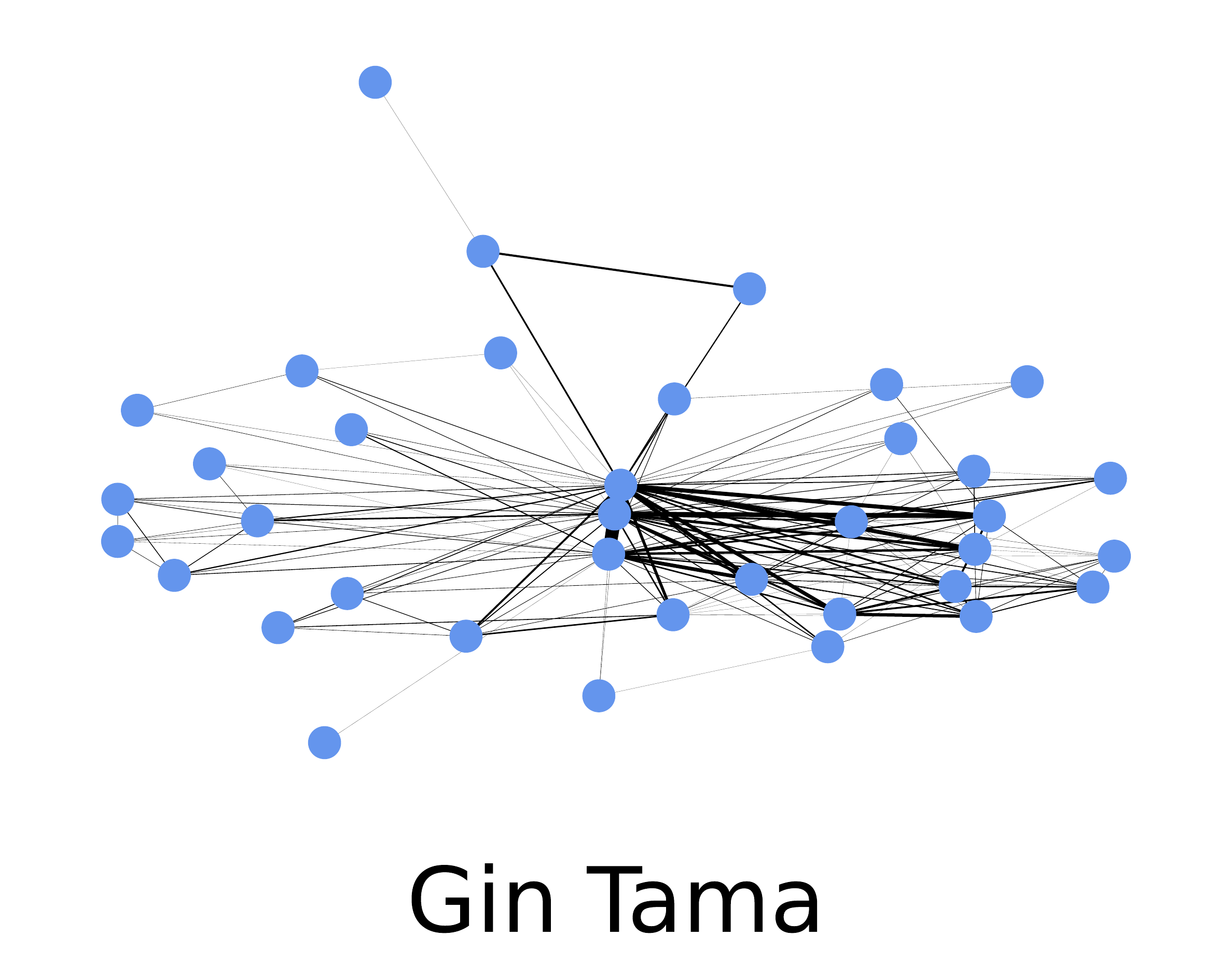}
         \label{fig:three sin x}
     \end{subfigure}
     \begin{subfigure}[b]{0.195\textwidth}
         \centering
         \includegraphics[width=\textwidth]{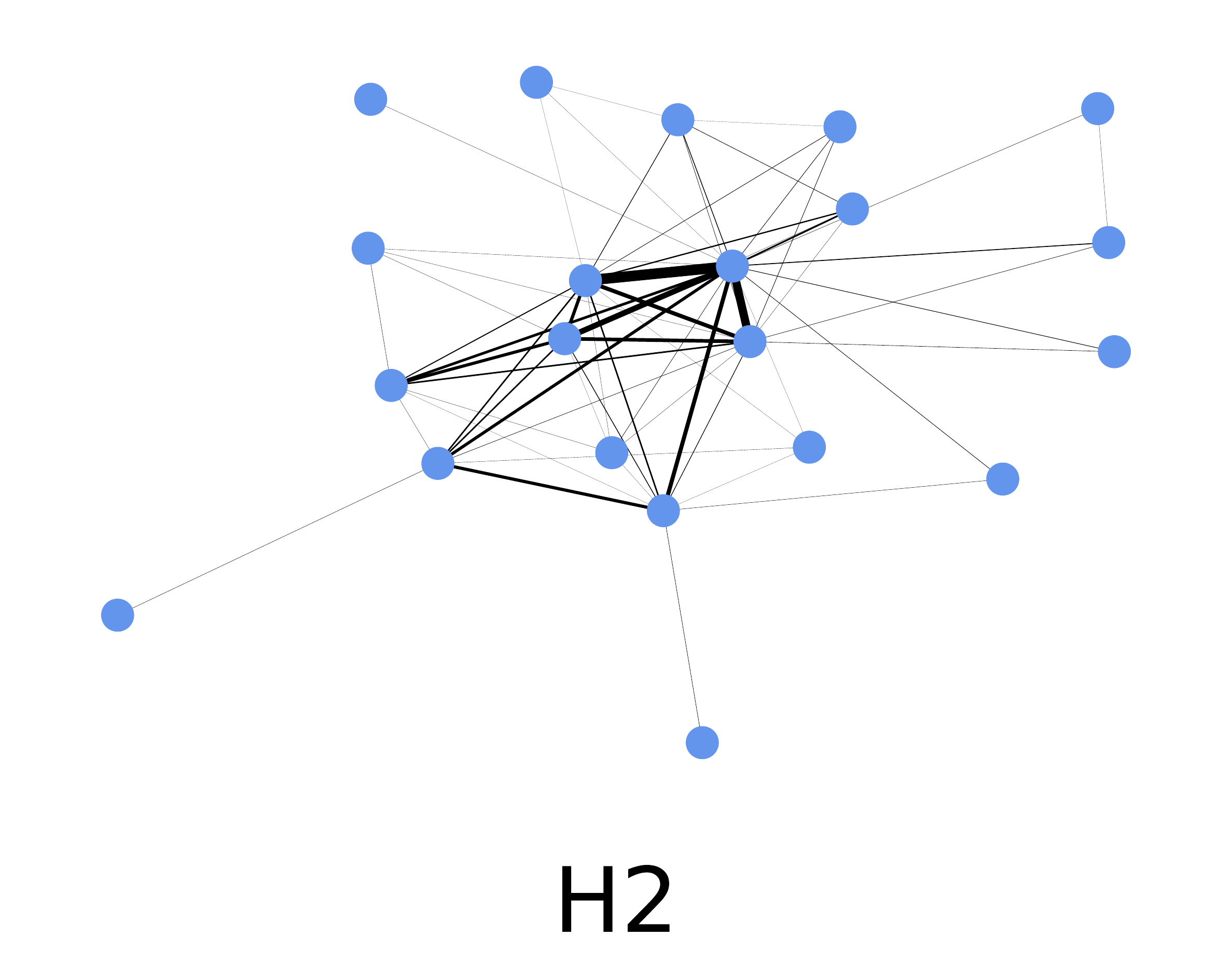}
         \label{fig:five over x}
     \end{subfigure}
     \begin{subfigure}[b]{0.195\textwidth}
         \centering
         \includegraphics[width=\textwidth]{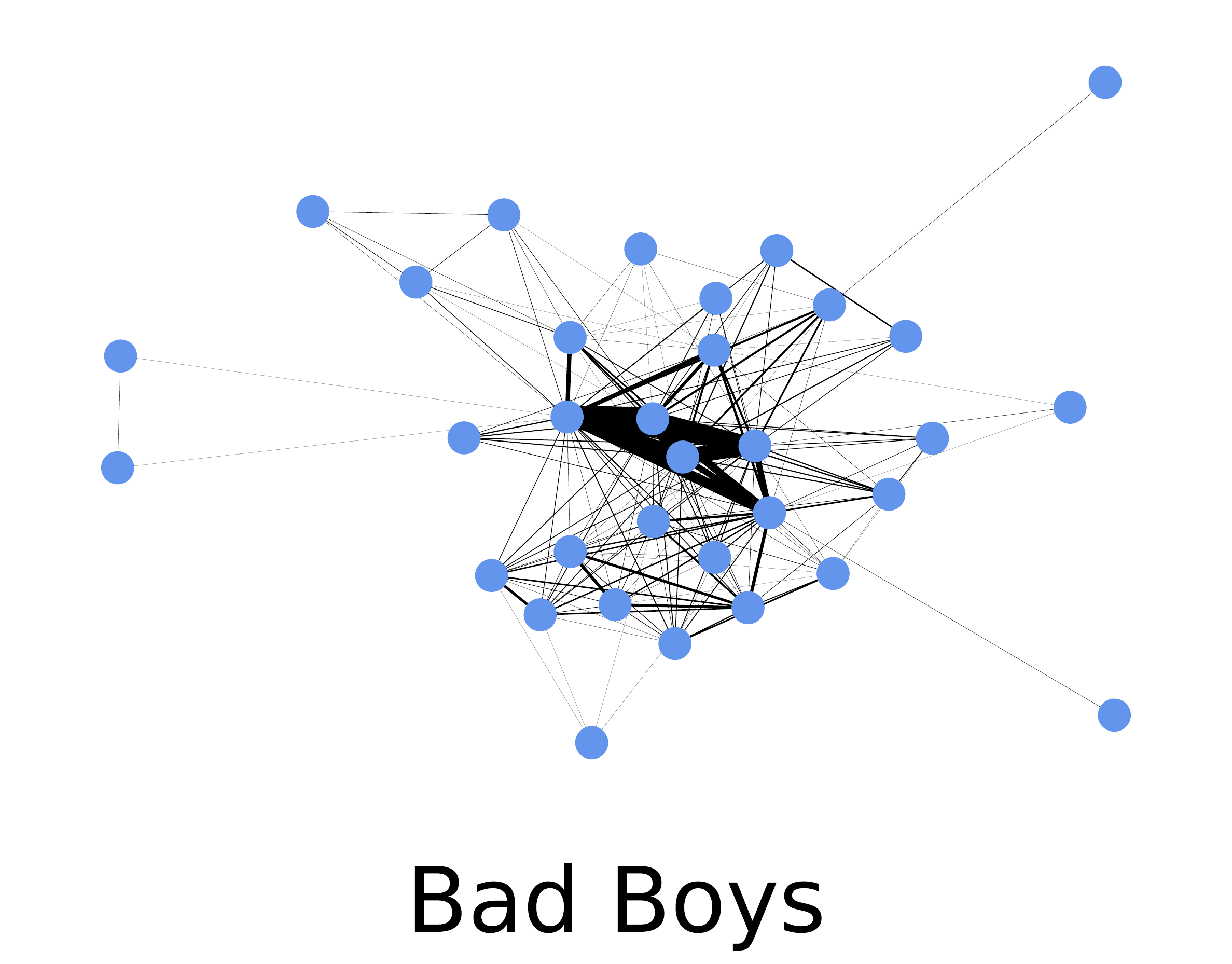}
         \label{fig:five over x}
     \end{subfigure}
     \begin{subfigure}[b]{0.195\textwidth}
         \centering
         \includegraphics[width=\textwidth]{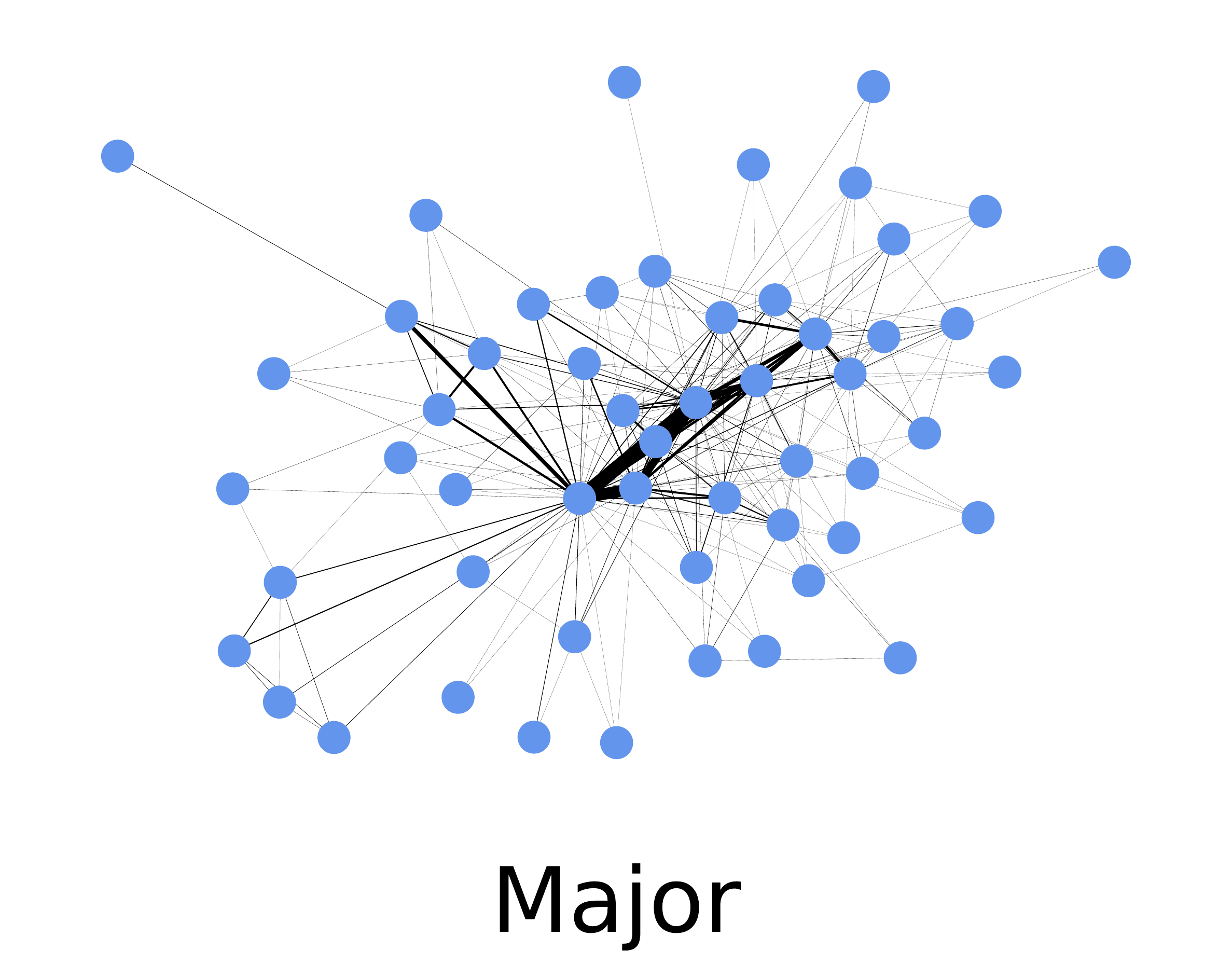}
         \label{fig:five over x}
     \end{subfigure}
     \\
      \begin{subfigure}[b]{0.195\textwidth}
         \centering
         \includegraphics[width=\textwidth]{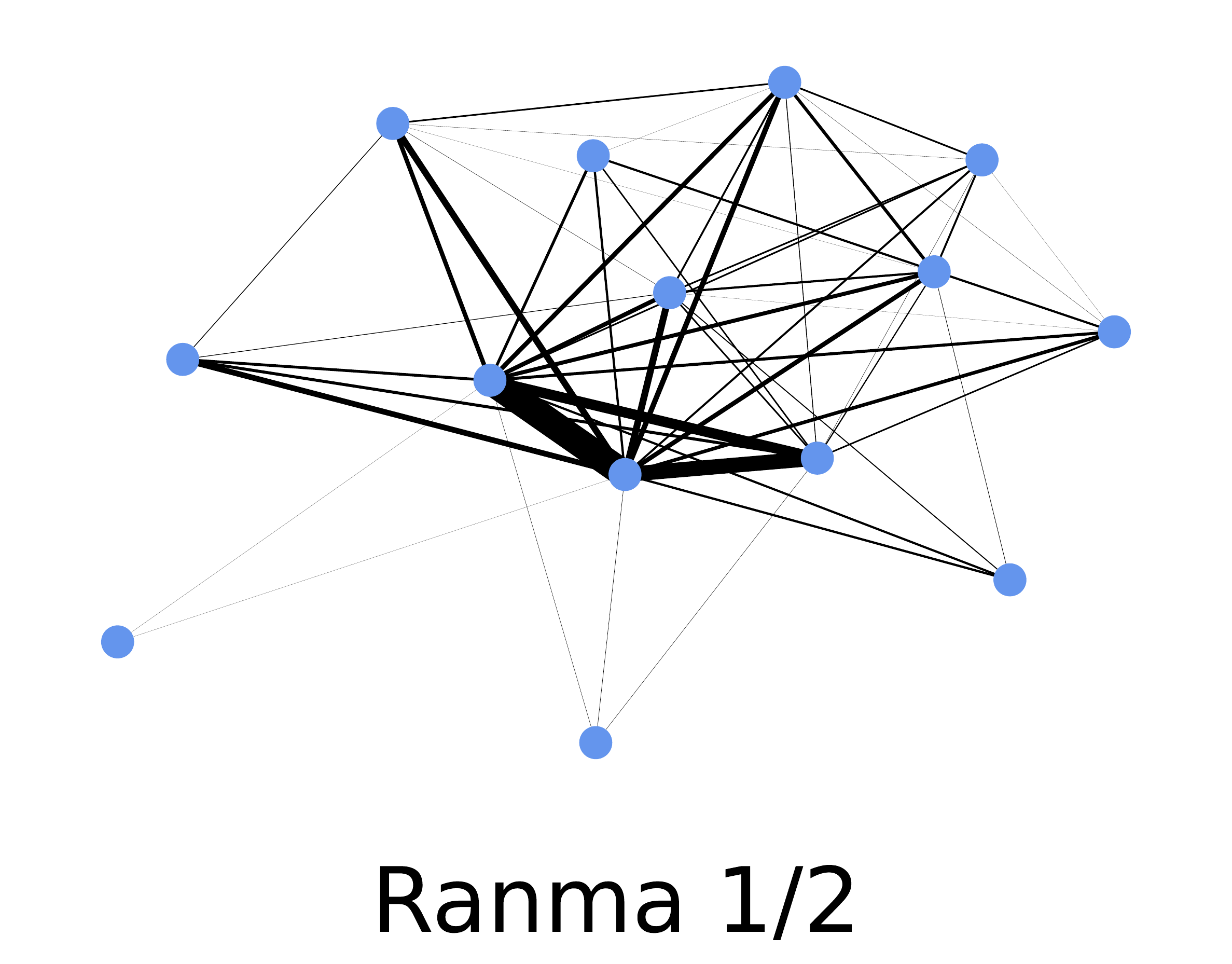}
         \label{fig:y equals x}
     \end{subfigure}
     \begin{subfigure}[b]{0.195\textwidth}
         \centering
         \includegraphics[width=\textwidth]{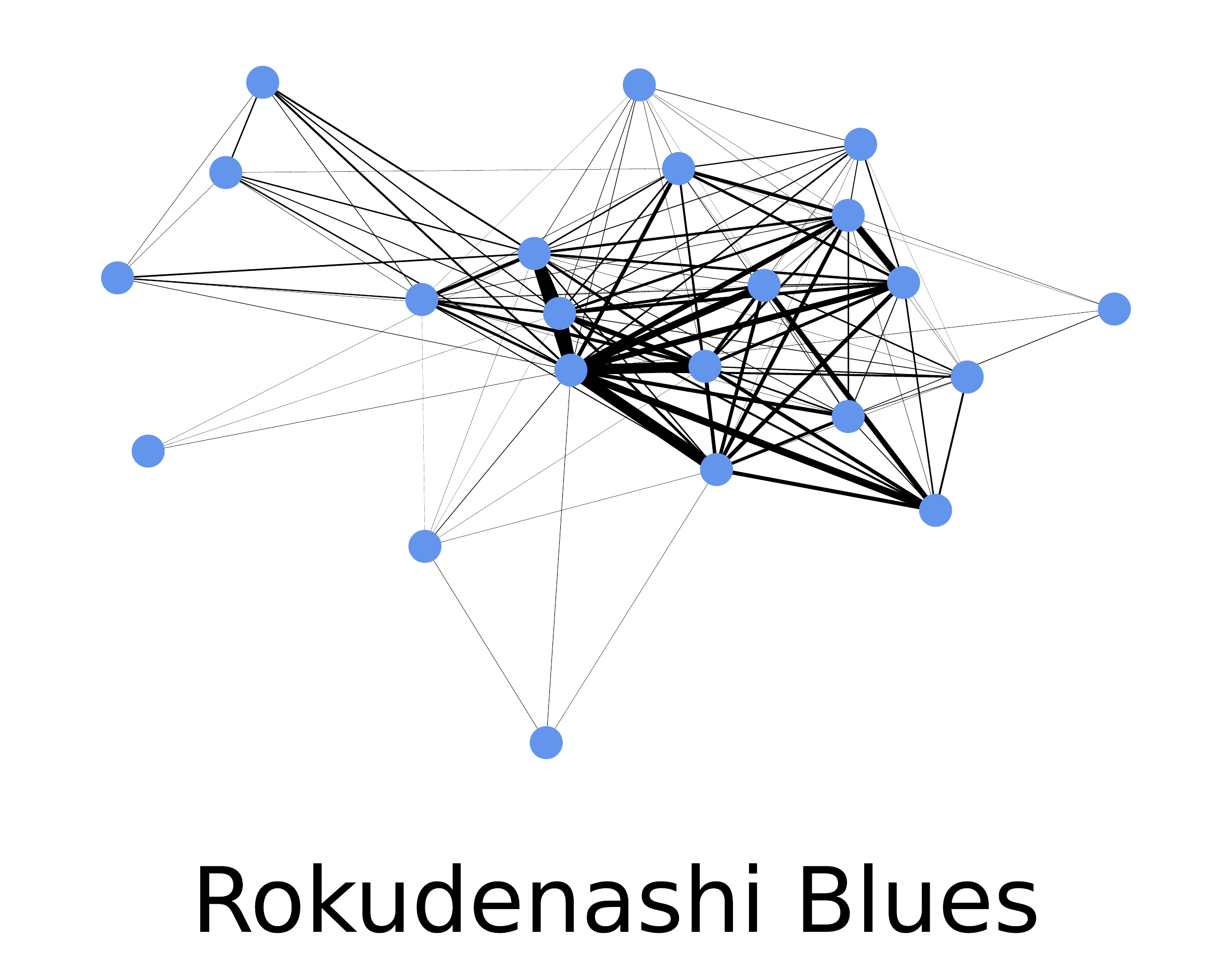}
         \label{fig:three sin x}
     \end{subfigure}
     \begin{subfigure}[b]{0.195\textwidth}
         \centering
         \includegraphics[width=\textwidth]{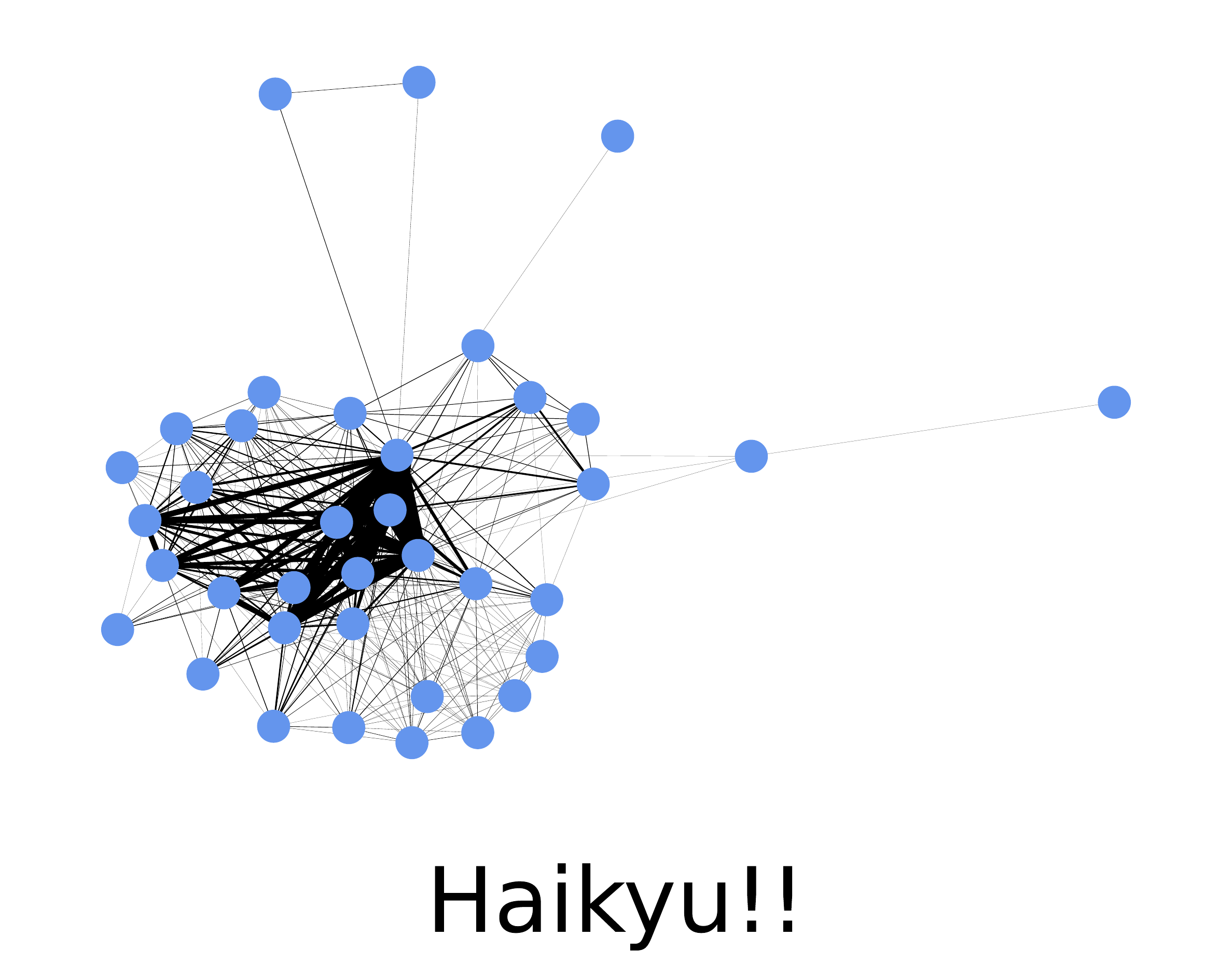}
         \label{fig:five over x}
     \end{subfigure}
     \begin{subfigure}[b]{0.195\textwidth}
         \centering
         \includegraphics[width=\textwidth]{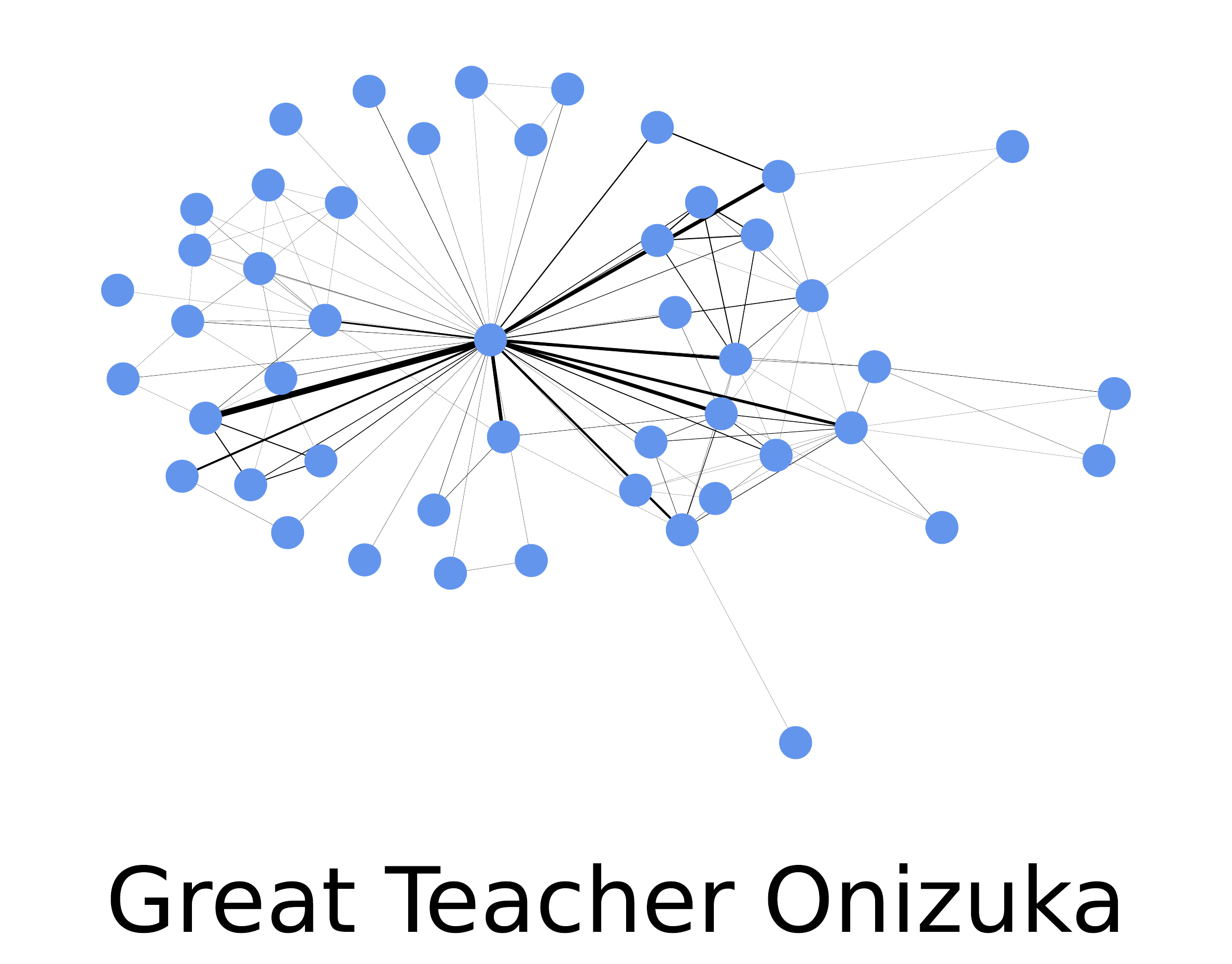}
         \label{fig:five over x}
     \end{subfigure}
     \begin{subfigure}[b]{0.195\textwidth}
         \centering
         \includegraphics[width=\textwidth]{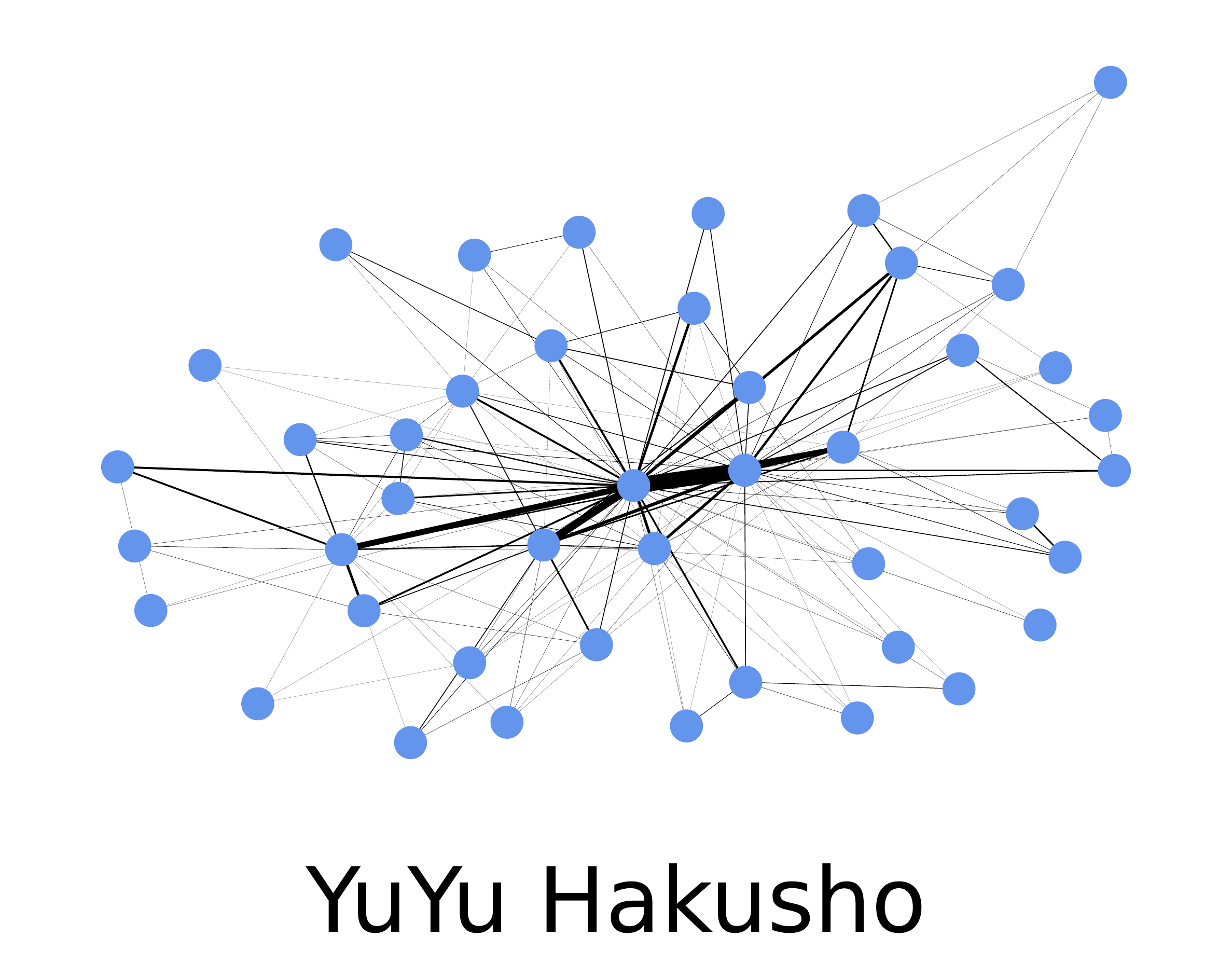}
         \label{fig:five over x}
     \end{subfigure}
    \\
     \begin{subfigure}[b]{0.195\textwidth}
         \centering
         \includegraphics[width=\textwidth]{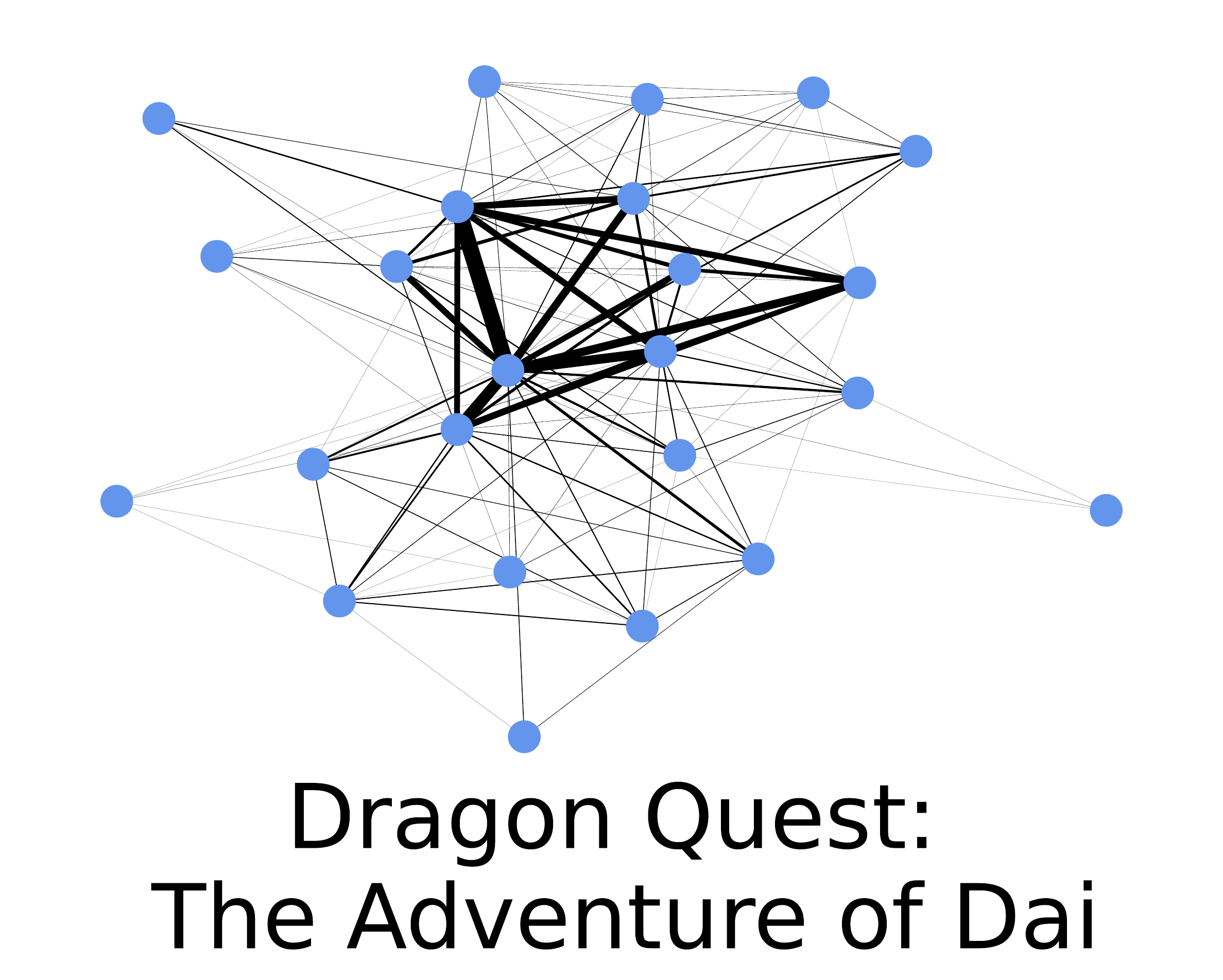}
         \label{fig:y equals x}
     \end{subfigure}
     \begin{subfigure}[b]{0.195\textwidth}
         \centering
         \includegraphics[width=\textwidth]{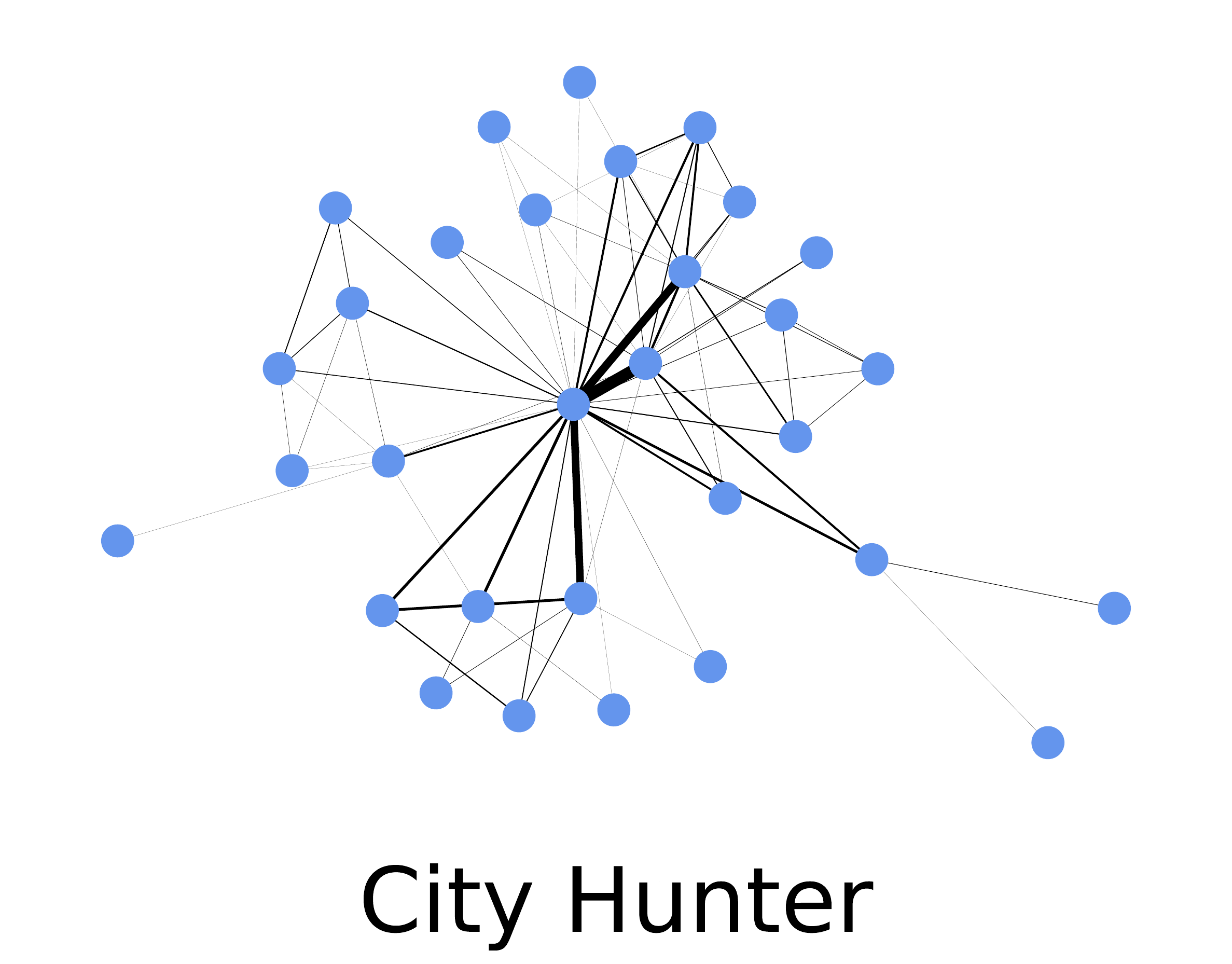}
         \label{fig:three sin x}
     \end{subfigure}
     \begin{subfigure}[b]{0.195\textwidth}
         \centering
         \includegraphics[width=\textwidth]{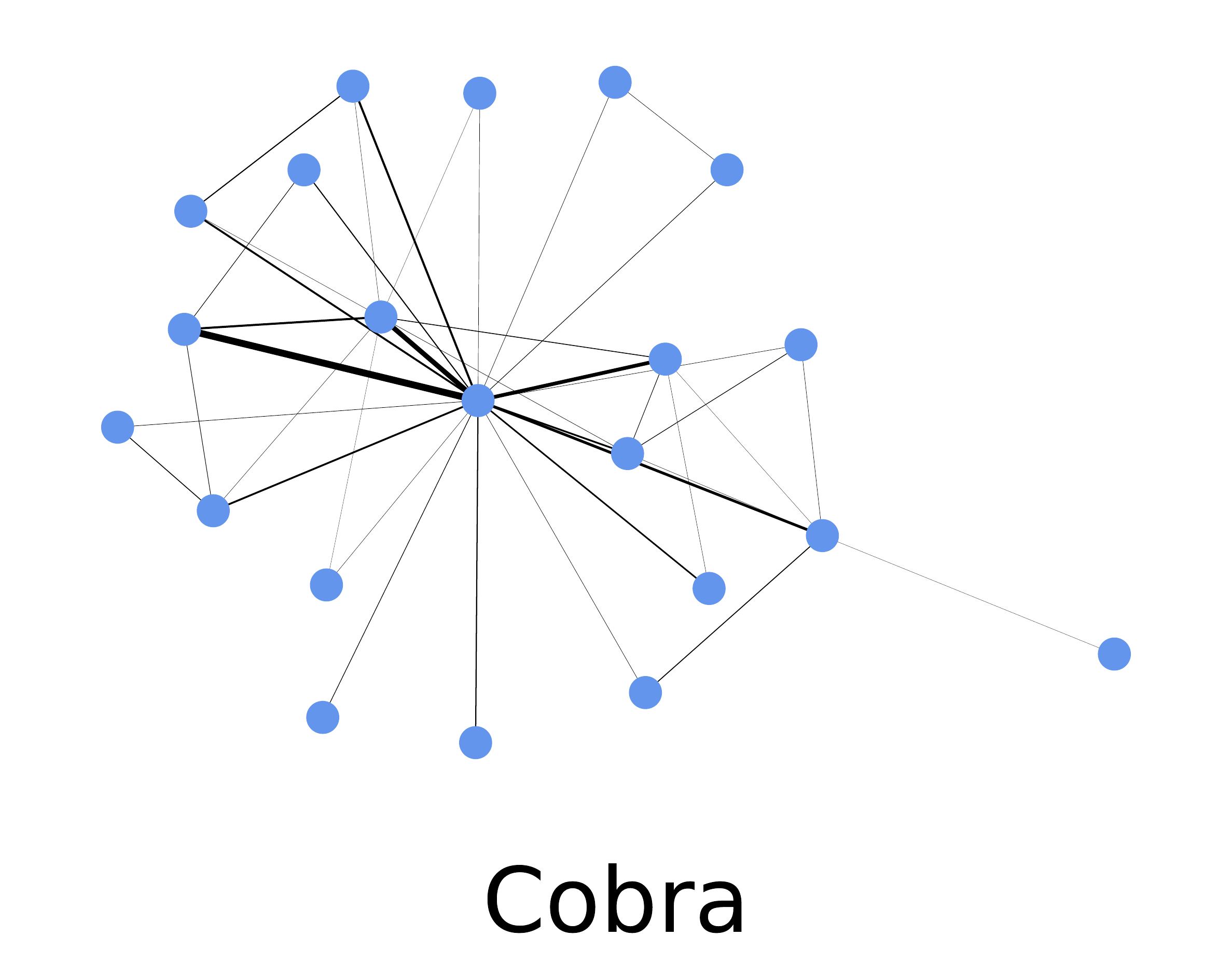}
         \label{fig:five over x}
     \end{subfigure}
     \begin{subfigure}[b]{0.195\textwidth}
         \centering
         \includegraphics[width=\textwidth]{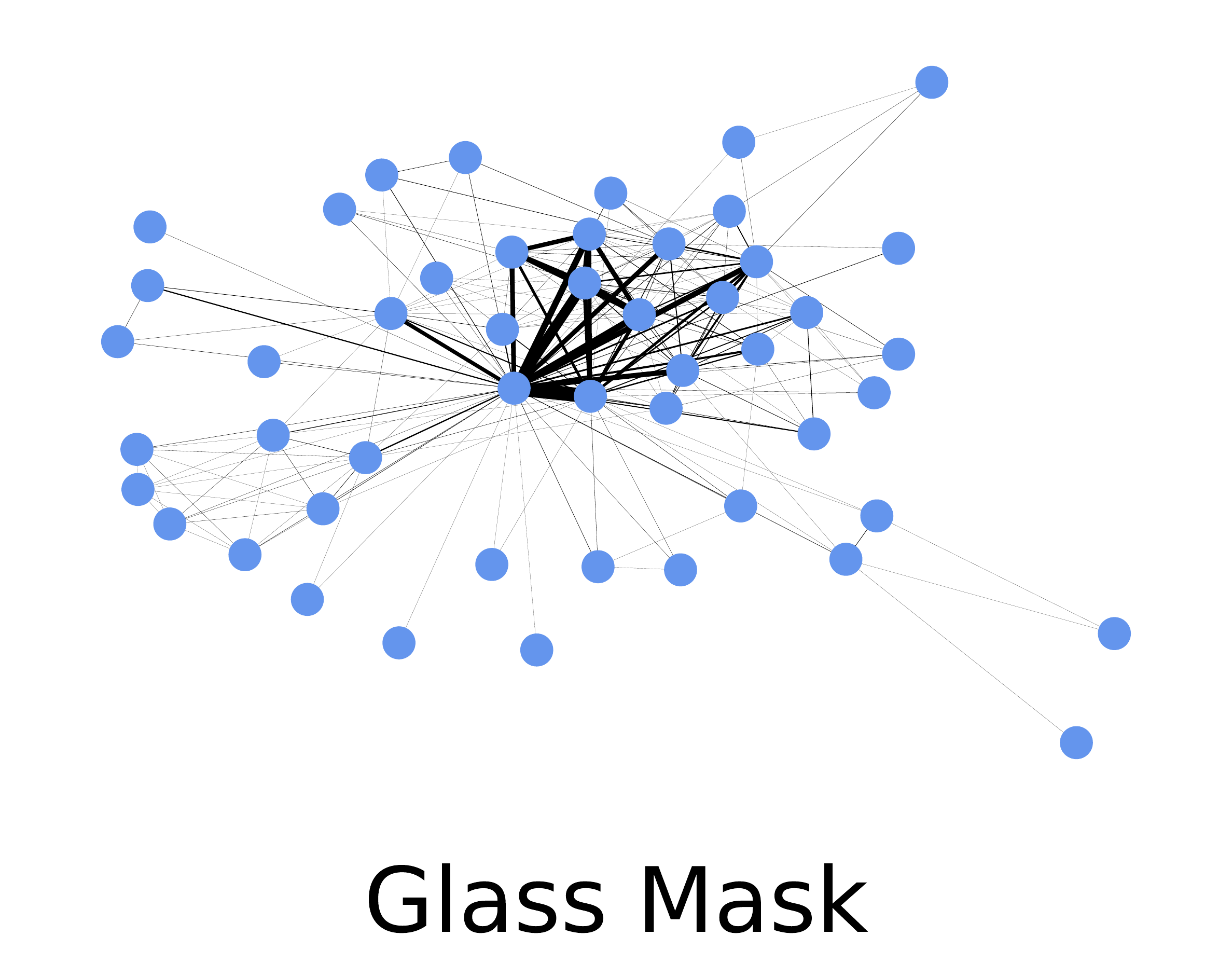}
         \label{fig:five over x}
     \end{subfigure}
     \begin{subfigure}[b]{0.195\textwidth}
         \centering
         \includegraphics[width=\textwidth]{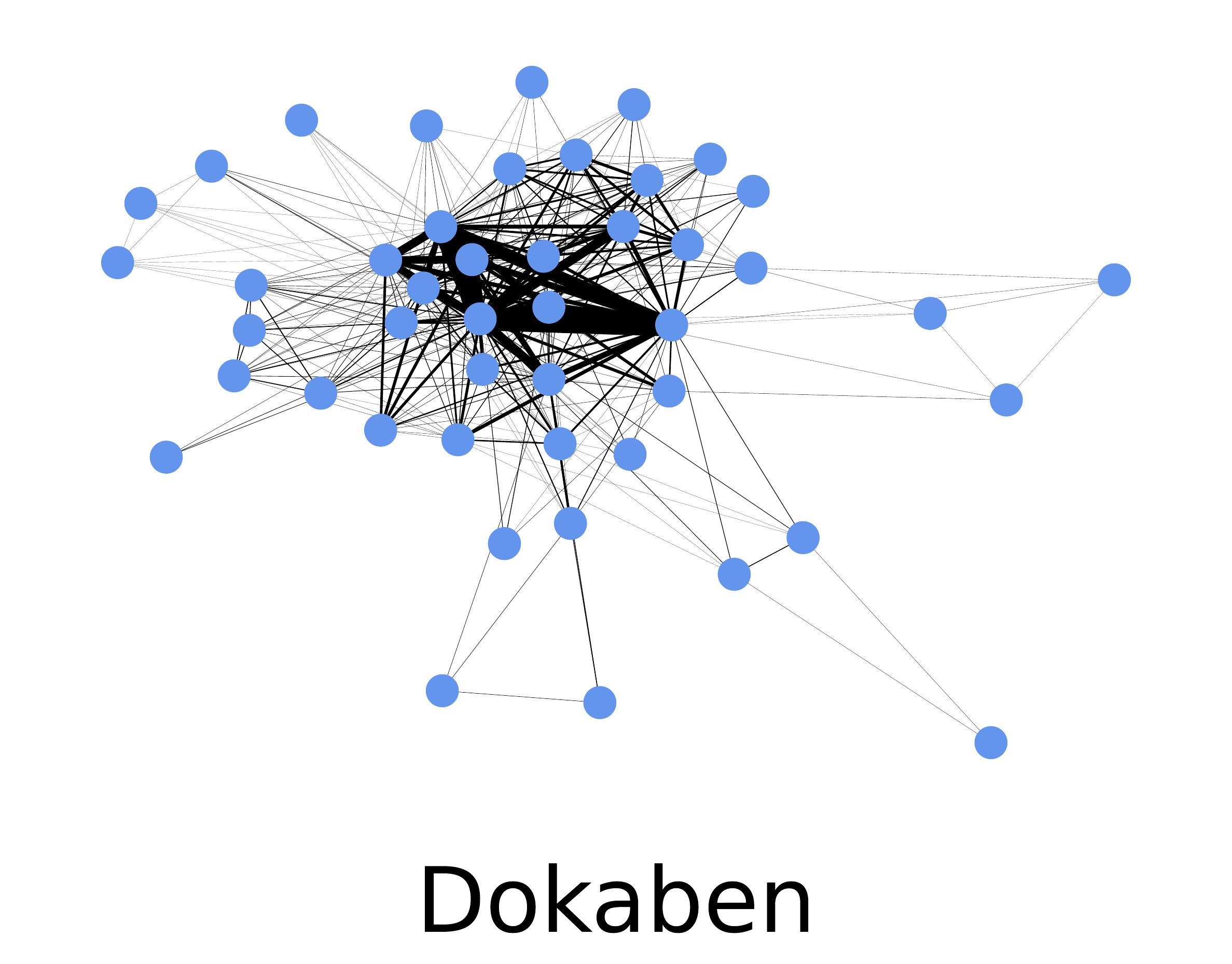}
         \label{fig:five over x}
     \end{subfigure}
     \\
      \begin{subfigure}[b]{0.195\textwidth}
         \centering
         \includegraphics[width=\textwidth]{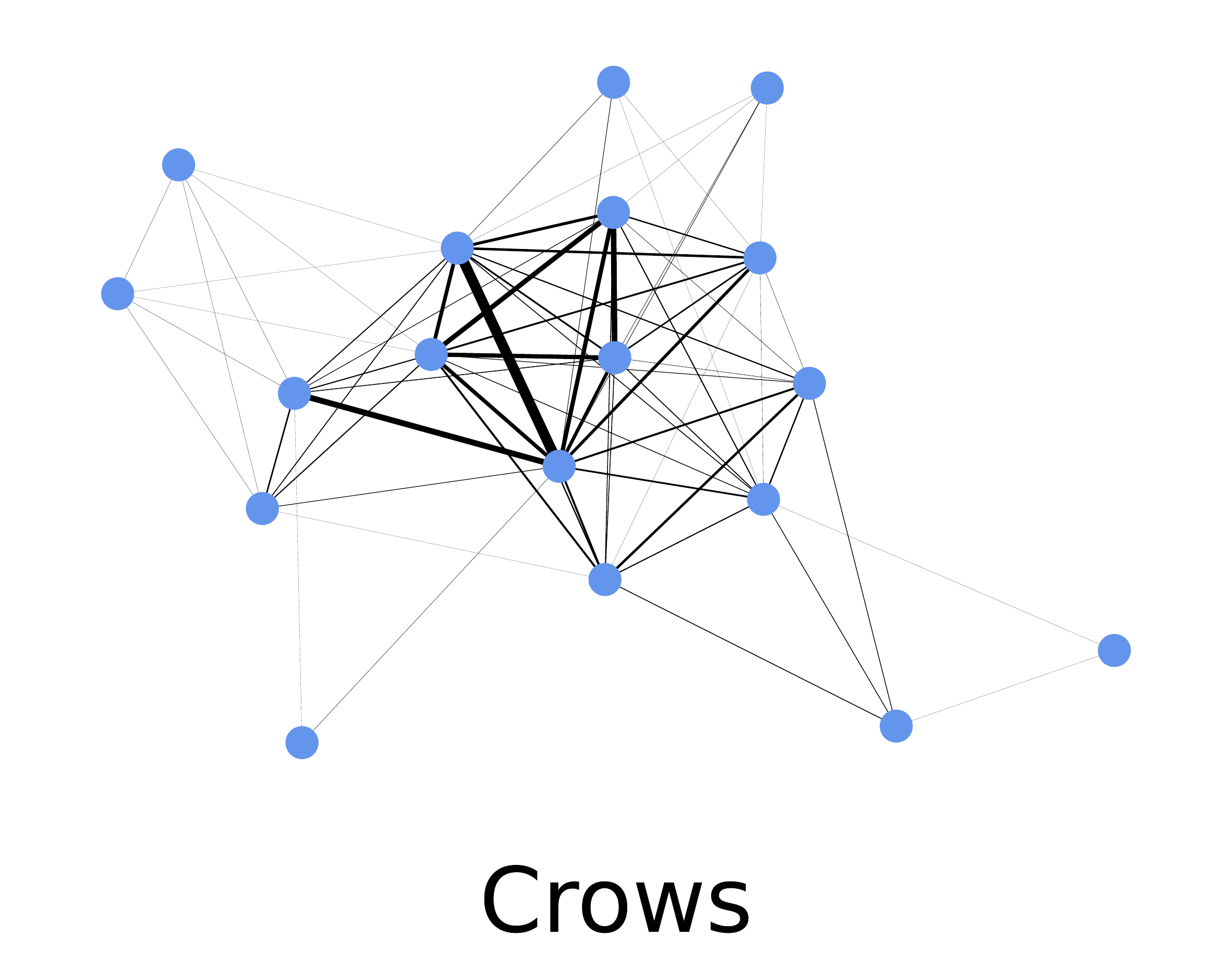}
         \label{fig:y equals x}
     \end{subfigure}
     \begin{subfigure}[b]{0.195\textwidth}
         \centering
         \includegraphics[width=\textwidth]{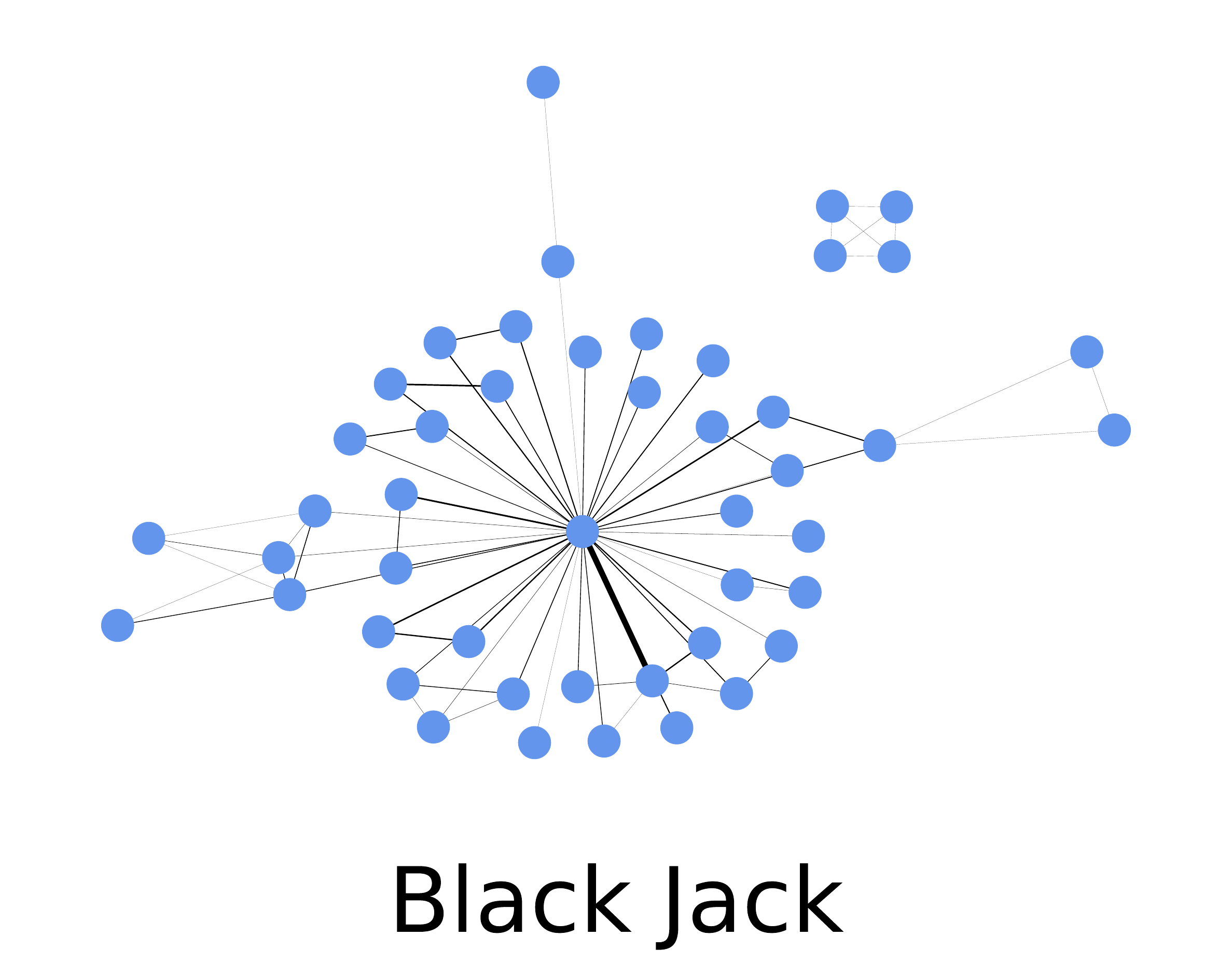}
         \label{fig:three sin x}
     \end{subfigure}
     \begin{subfigure}[b]{0.195\textwidth}
         \centering
         \includegraphics[width=\textwidth]{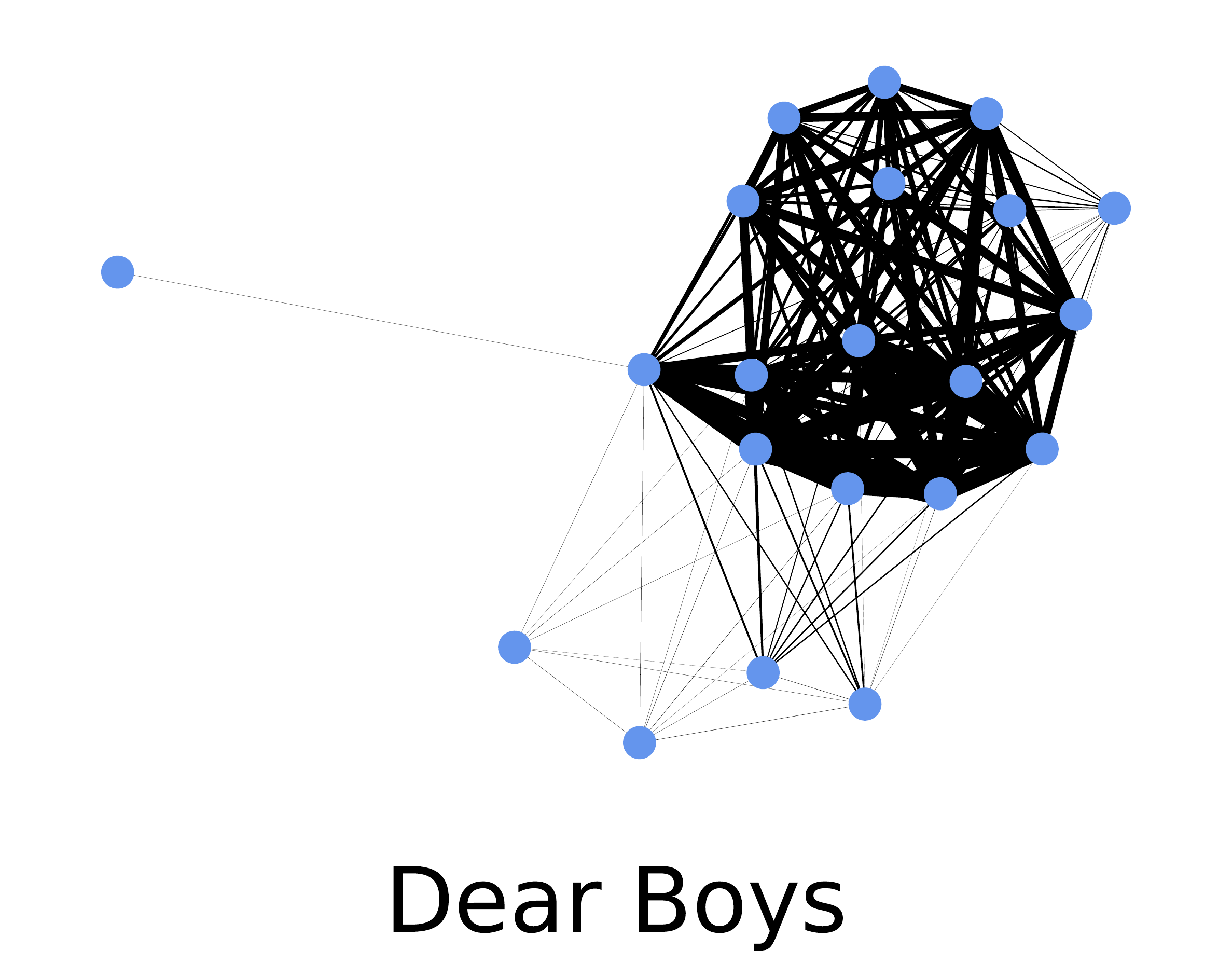}
         \label{fig:five over x}
     \end{subfigure}
     \begin{subfigure}[b]{0.195\textwidth}
         \centering
         \includegraphics[width=\textwidth]{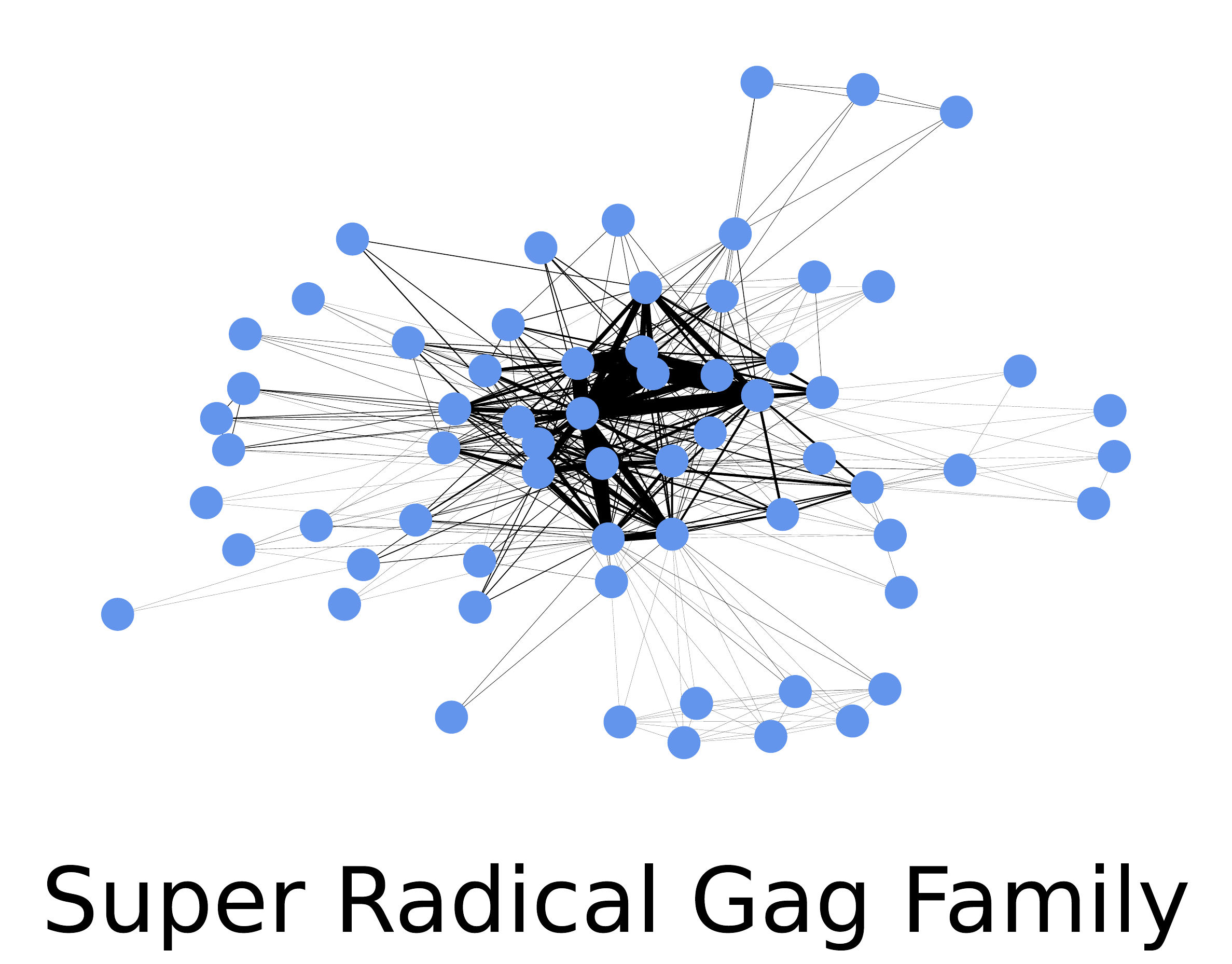}
         \label{fig:five over x}
     \end{subfigure}
     \begin{subfigure}[b]{0.195\textwidth}
         \centering
         \includegraphics[width=\textwidth]{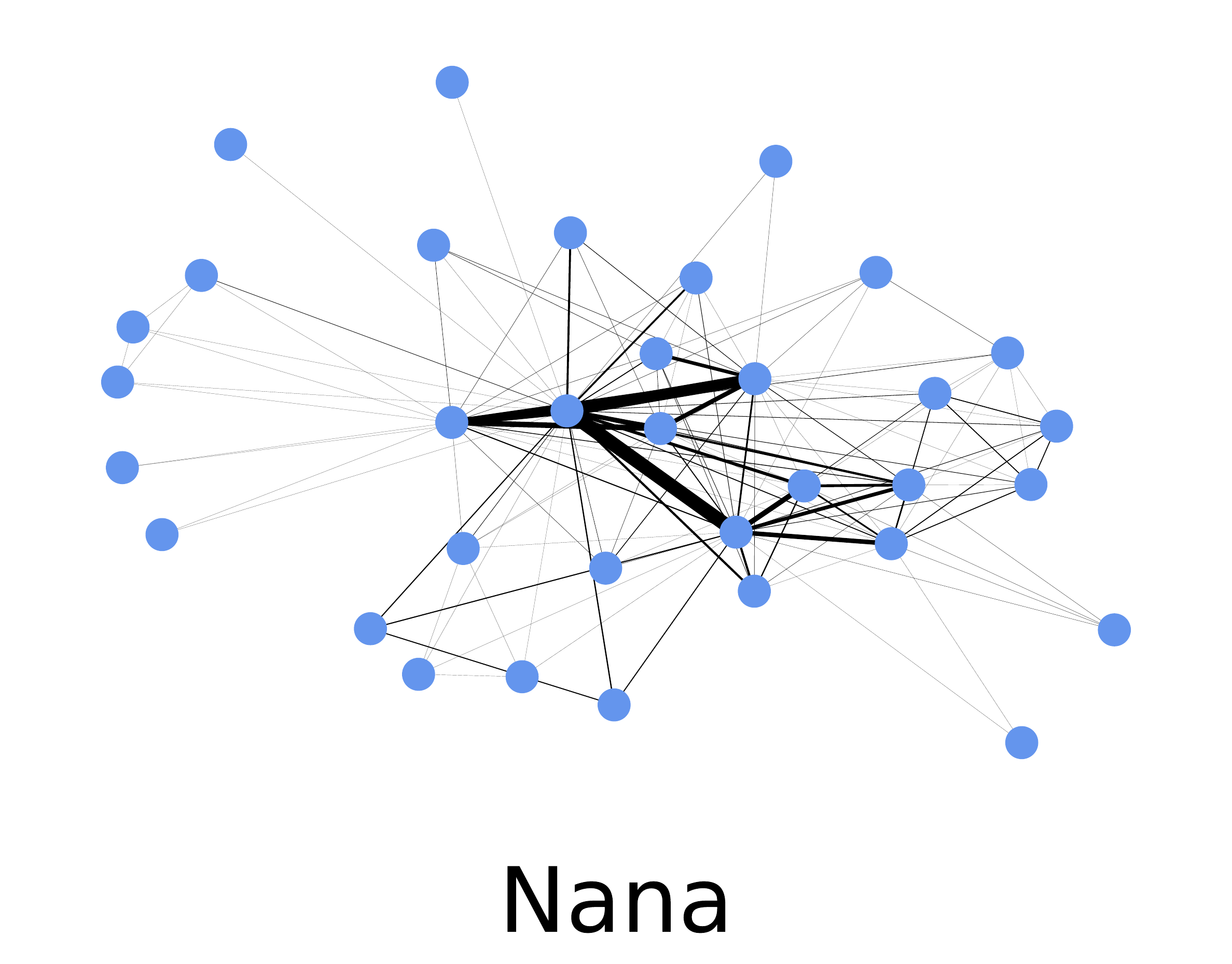}
         \label{fig:five over x}
     \end{subfigure}
     \\
      \begin{subfigure}[b]{0.195\textwidth}
         \centering
         \includegraphics[width=\textwidth]{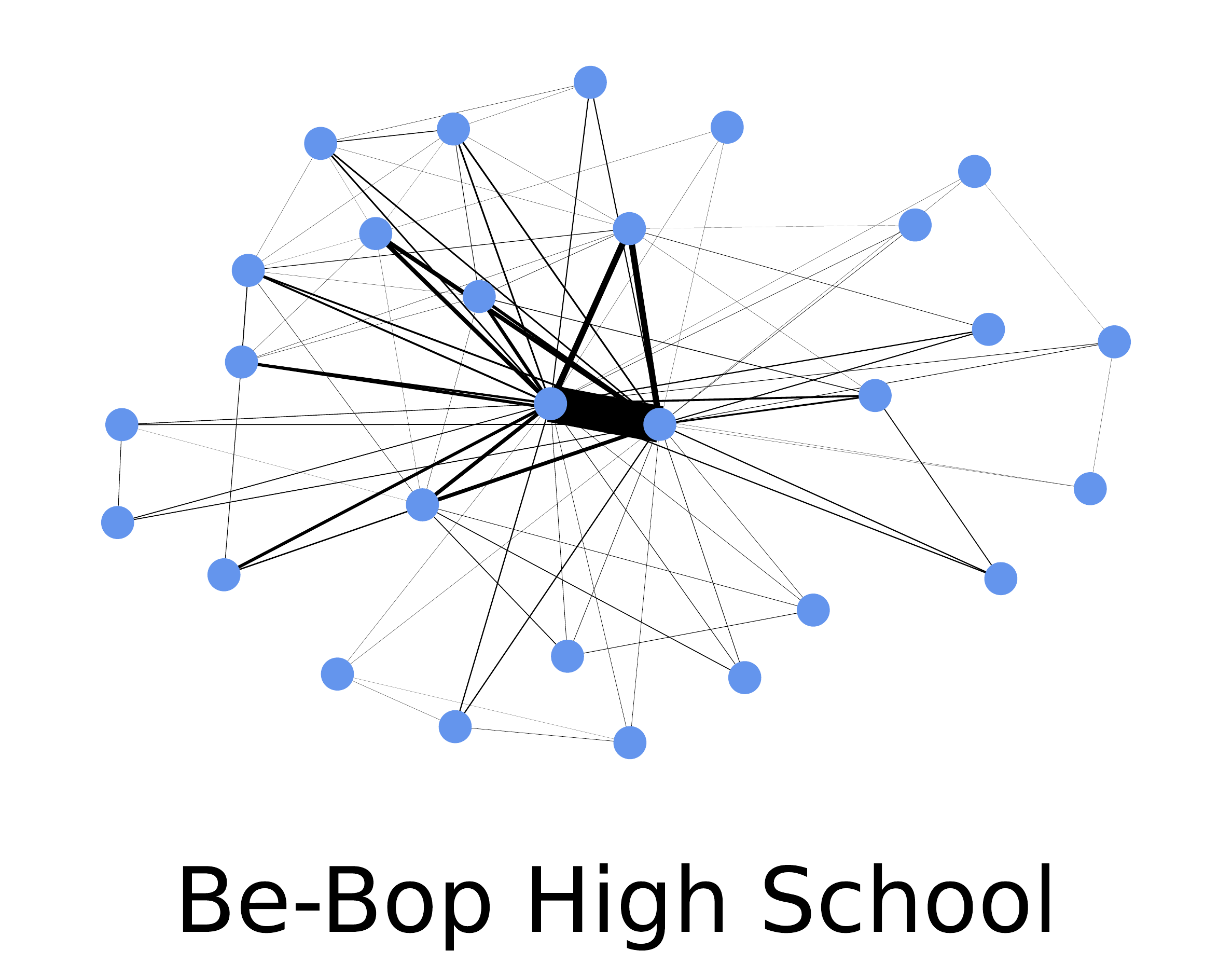}
         \label{fig:y equals x}
     \end{subfigure}
     \begin{subfigure}[b]{0.195\textwidth}
         \centering
         \includegraphics[width=\textwidth]{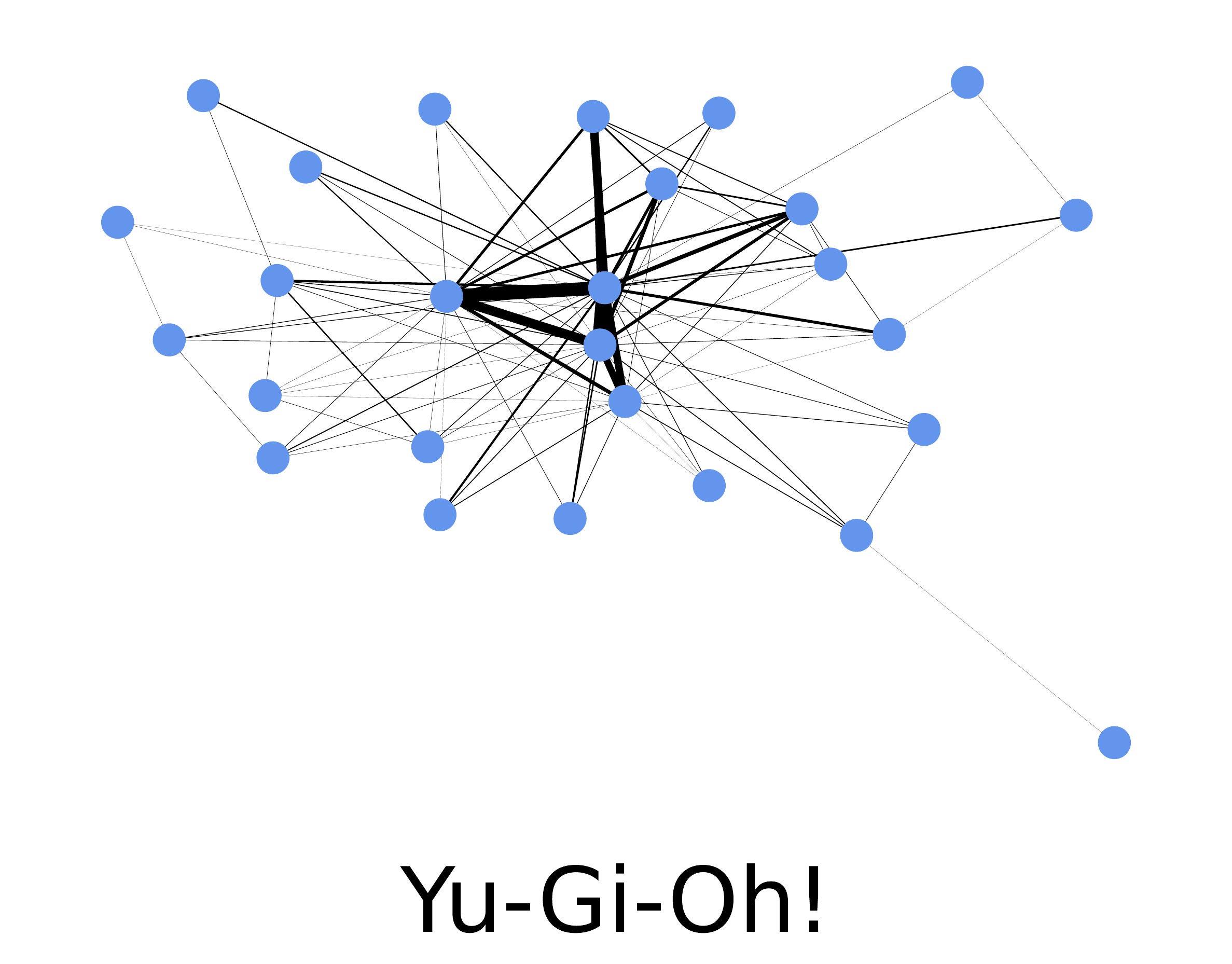}
         \label{fig:three sin x}
     \end{subfigure}
     \begin{subfigure}[b]{0.195\textwidth}
         \centering
         \includegraphics[width=\textwidth]{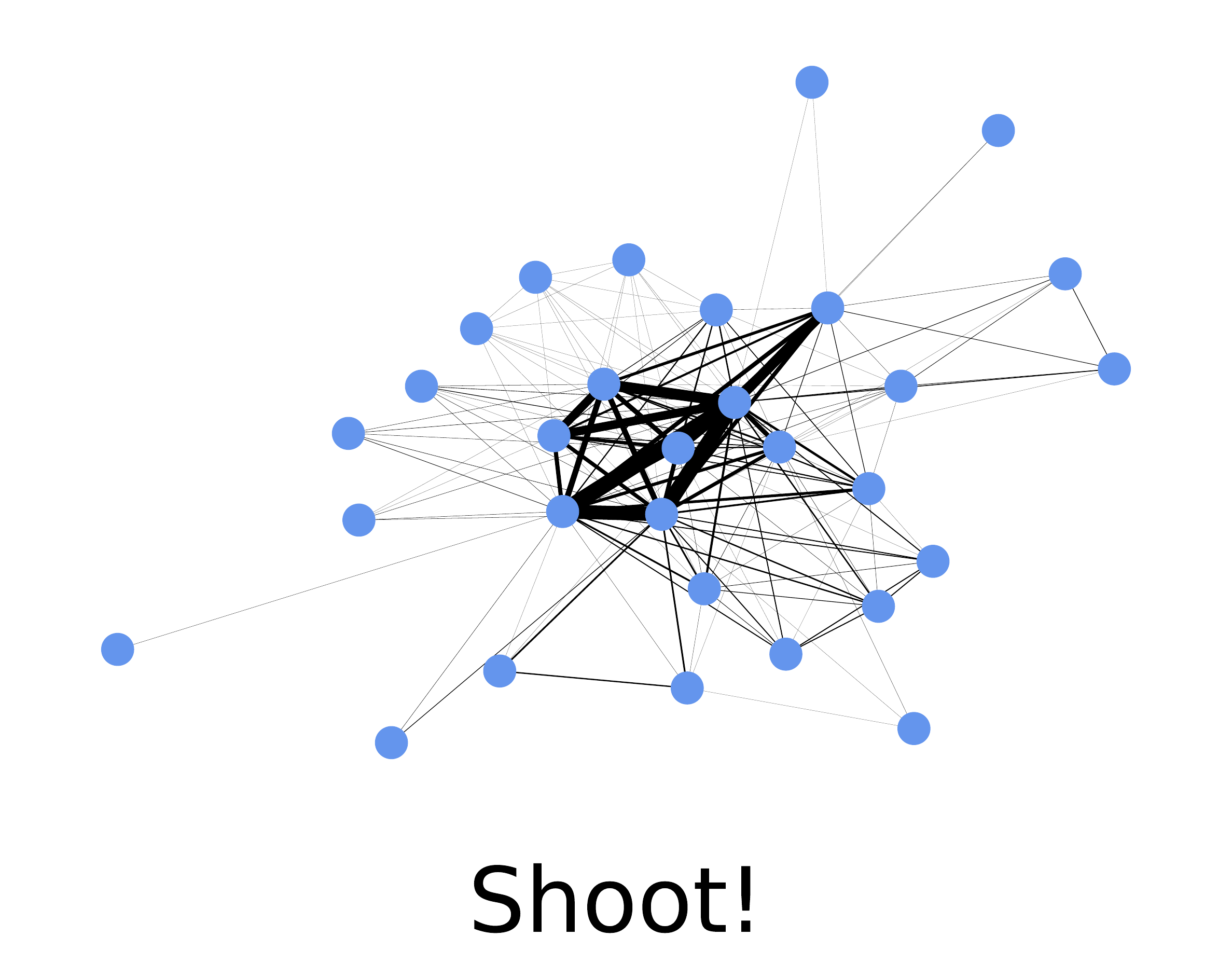}
         \label{fig:five over x}
     \end{subfigure}
     \begin{subfigure}[b]{0.195\textwidth}
         \centering
         \includegraphics[width=\textwidth]{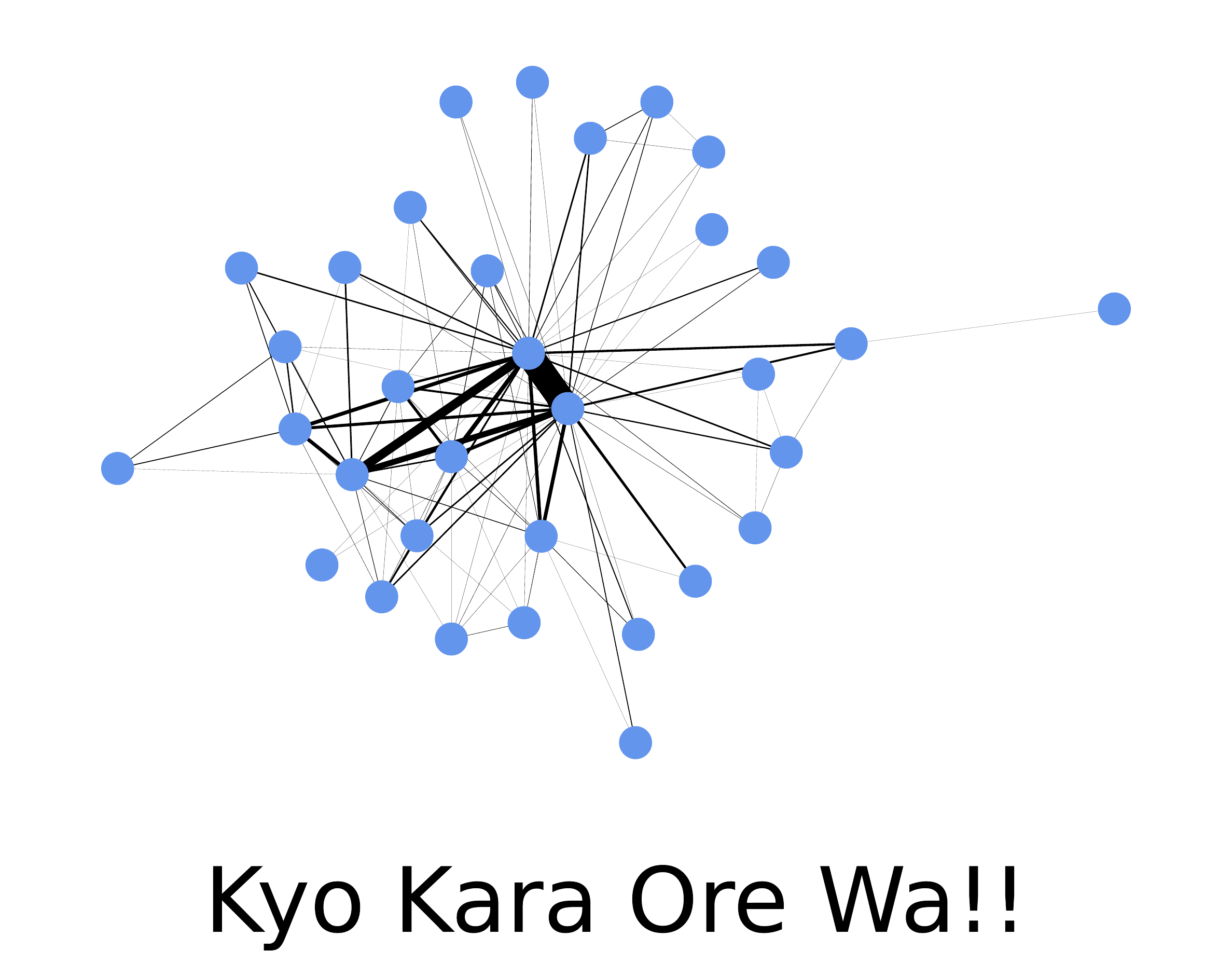}
         \label{fig:five over x}
     \end{subfigure}
     \begin{subfigure}[b]{0.195\textwidth}
         \centering
         \includegraphics[width=\textwidth]{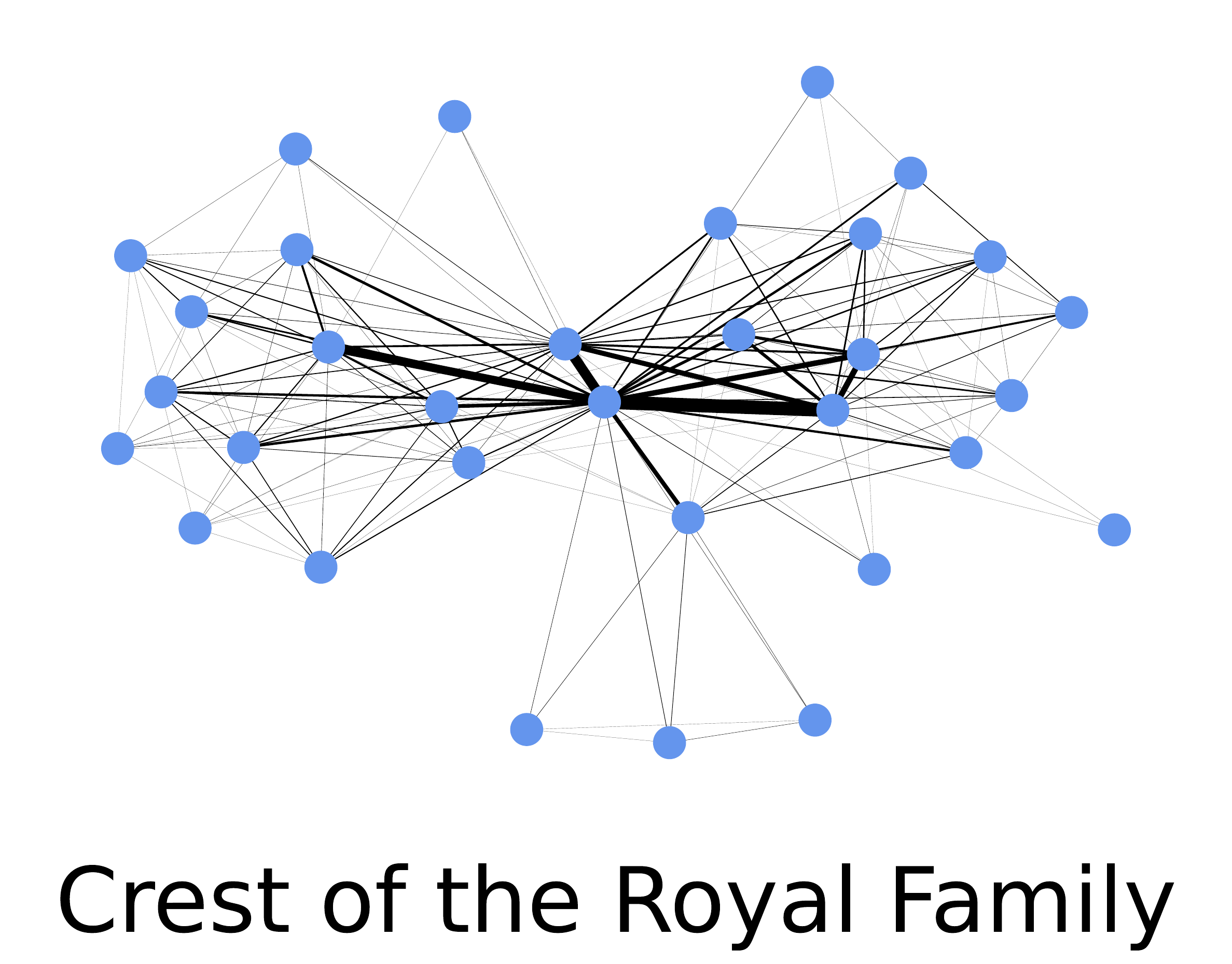}
         \label{fig:five over x}
     \end{subfigure}
     \\
      \begin{subfigure}[b]{0.195\textwidth}
         \centering
         \includegraphics[width=\textwidth]{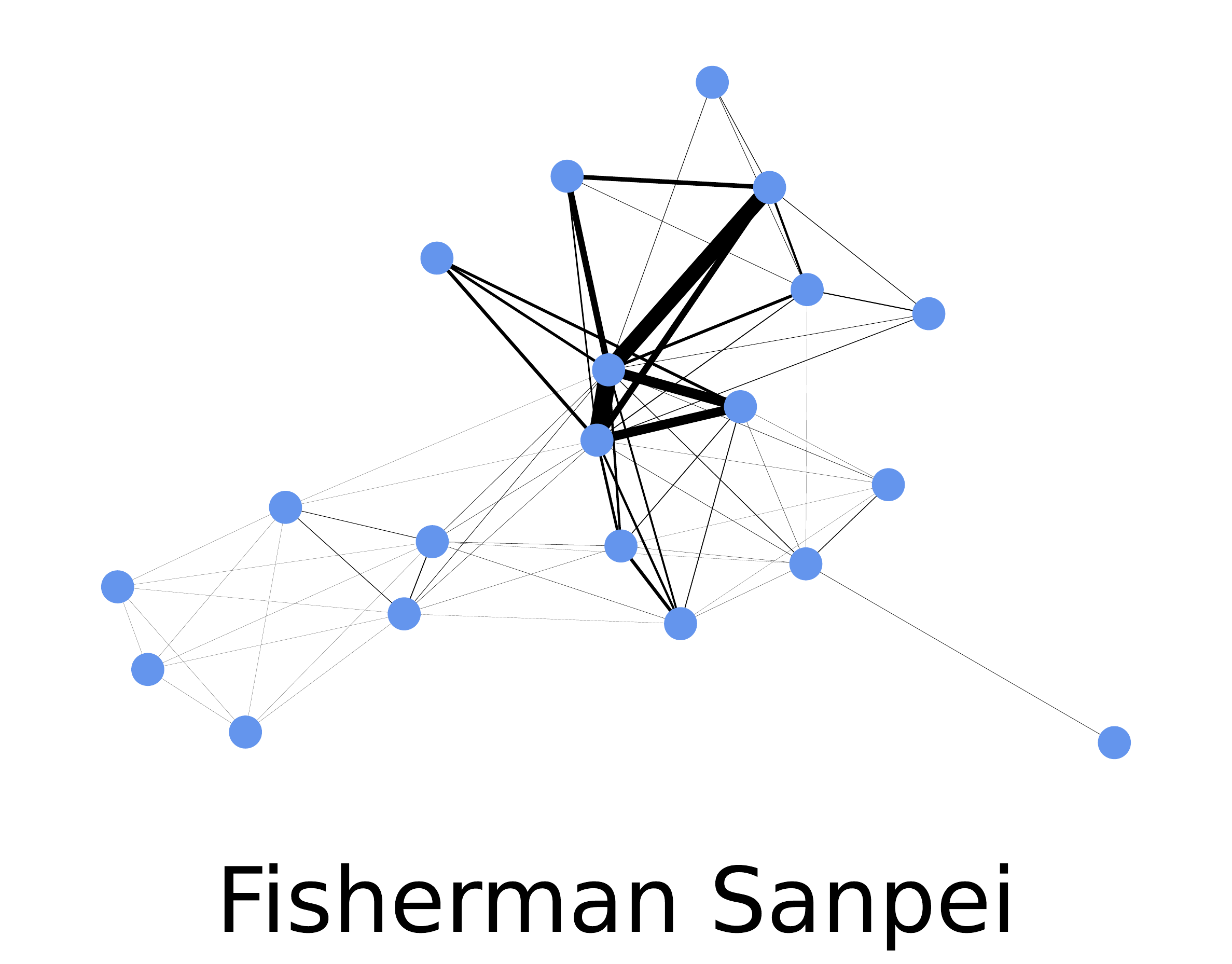}
         \label{fig:y equals x}
     \end{subfigure}
     \begin{subfigure}[b]{0.195\textwidth}
         \centering
         \includegraphics[width=\textwidth]{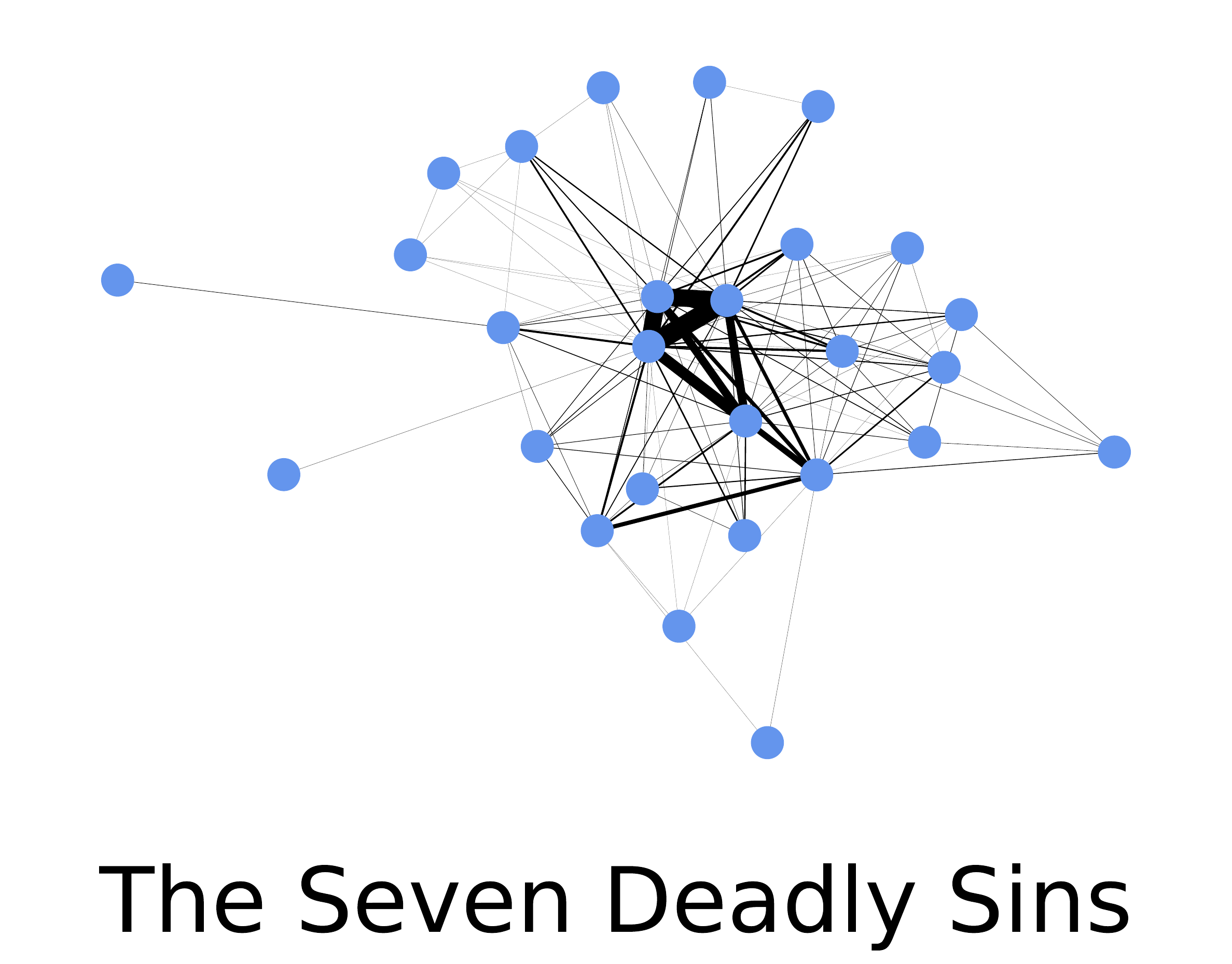}
         \label{fig:three sin x}
     \end{subfigure}
     \begin{subfigure}[b]{0.195\textwidth}
         \centering
         \includegraphics[width=\textwidth]{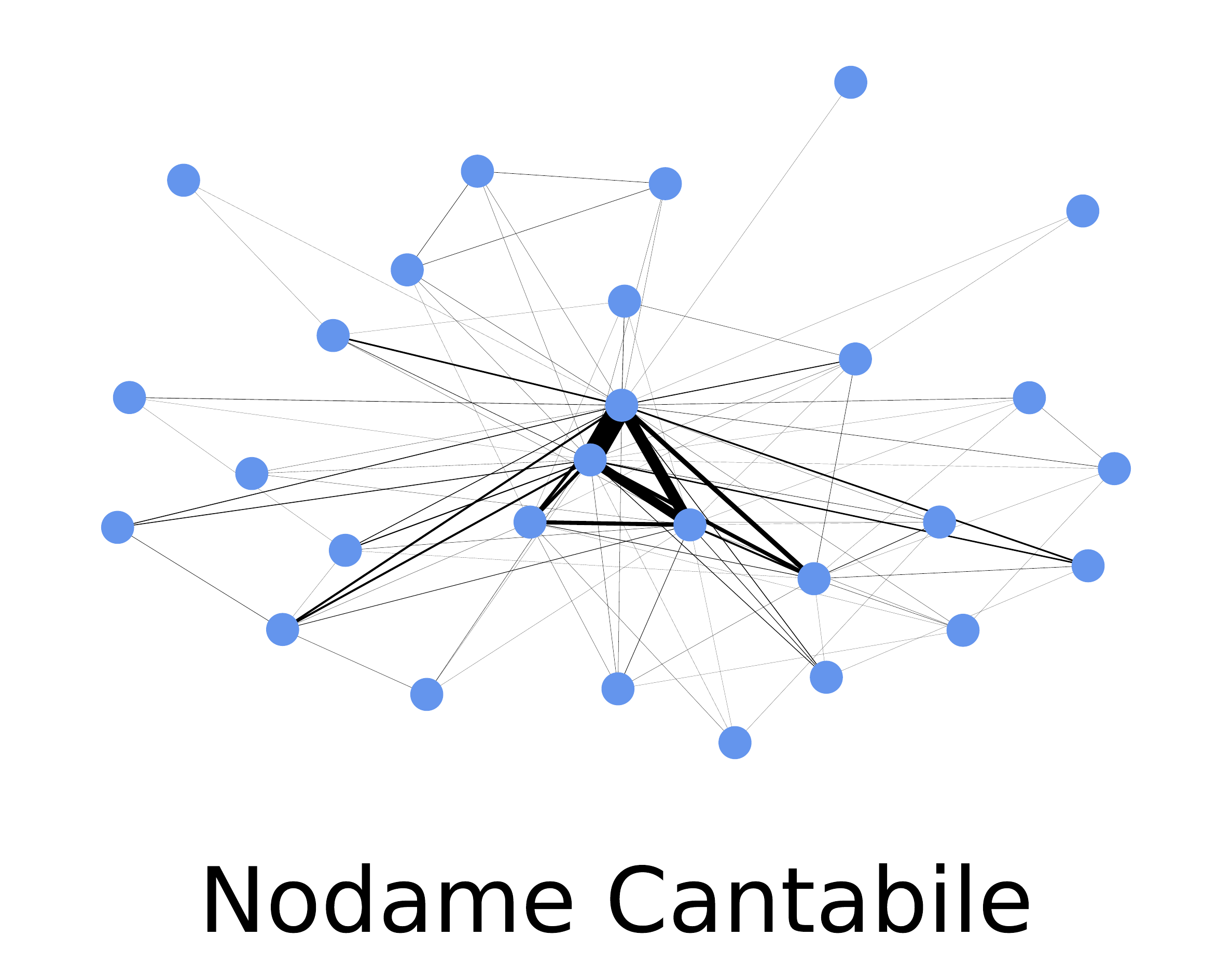}
         \label{fig:five over x}
     \end{subfigure}
     \begin{subfigure}[b]{0.195\textwidth}
         \centering
         \includegraphics[width=\textwidth]{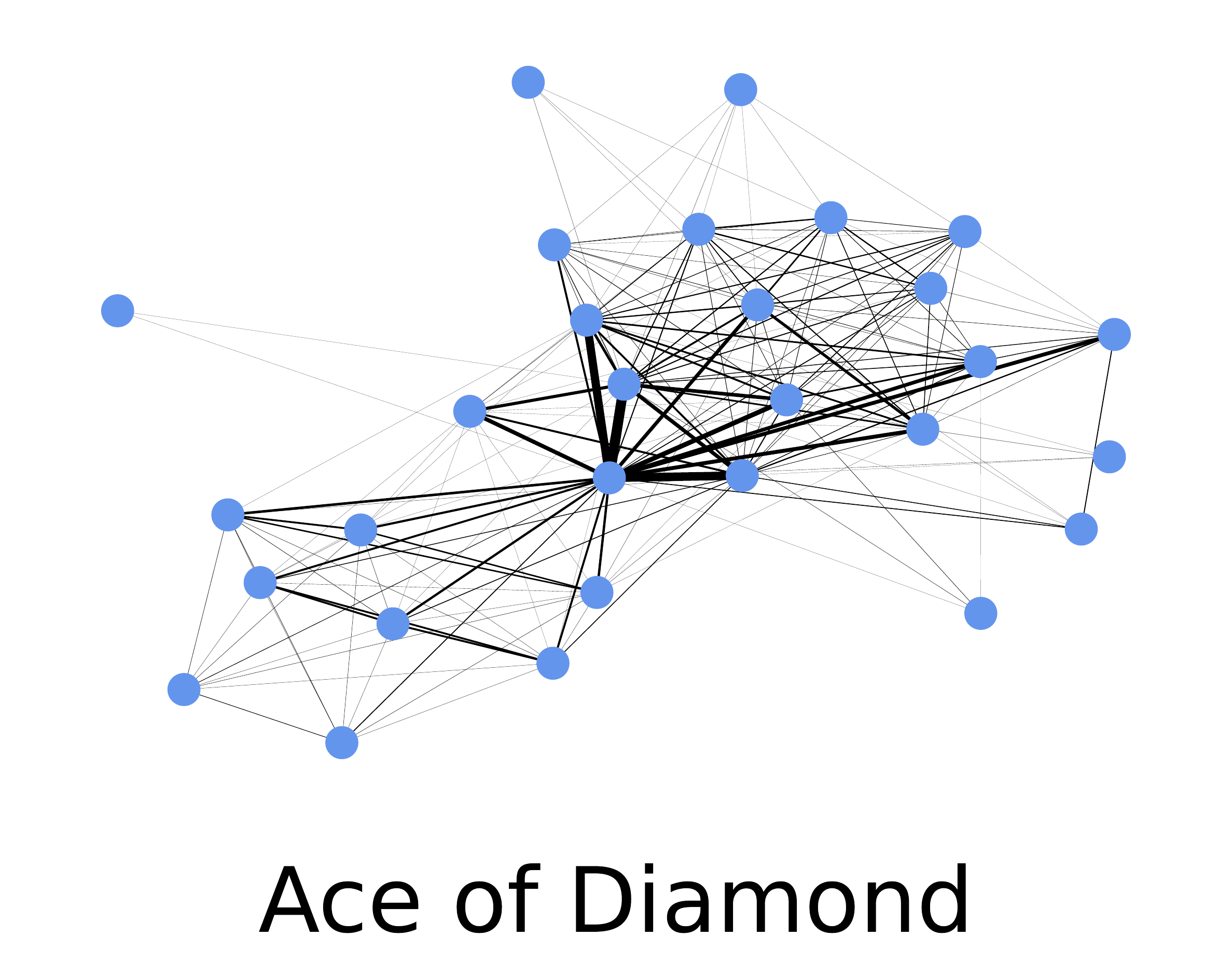}
         \label{fig:five over x}
     \end{subfigure}
     \begin{subfigure}[b]{0.195\textwidth}
         \centering
         \includegraphics[width=\textwidth]{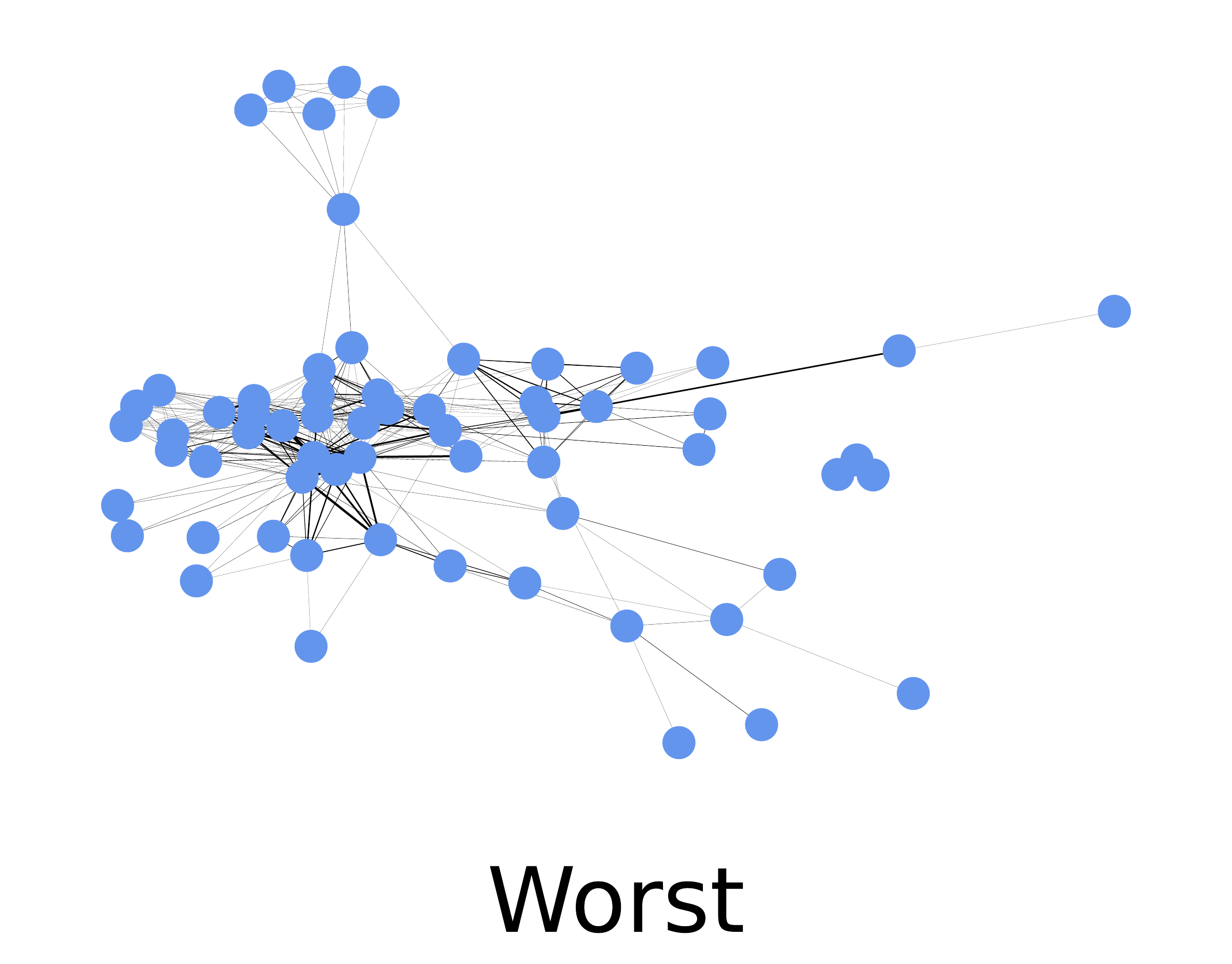}
         \label{fig:five over x}
     \end{subfigure}
     \\
      \begin{subfigure}[b]{0.195\textwidth}
         \centering
         \includegraphics[width=\textwidth]{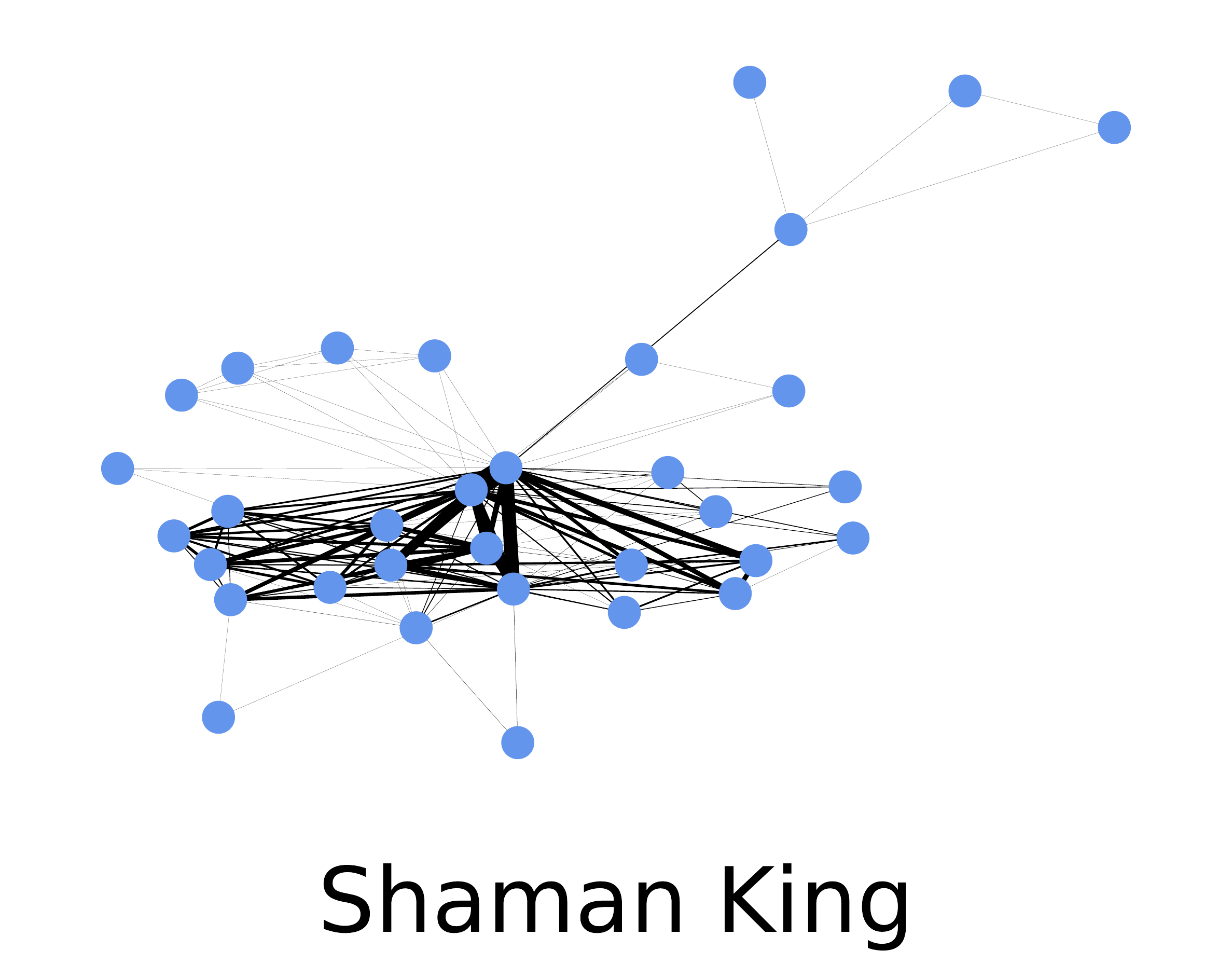}
         \label{fig:y equals x}
     \end{subfigure}
     \begin{subfigure}[b]{0.195\textwidth}
         \centering
         \includegraphics[width=\textwidth]{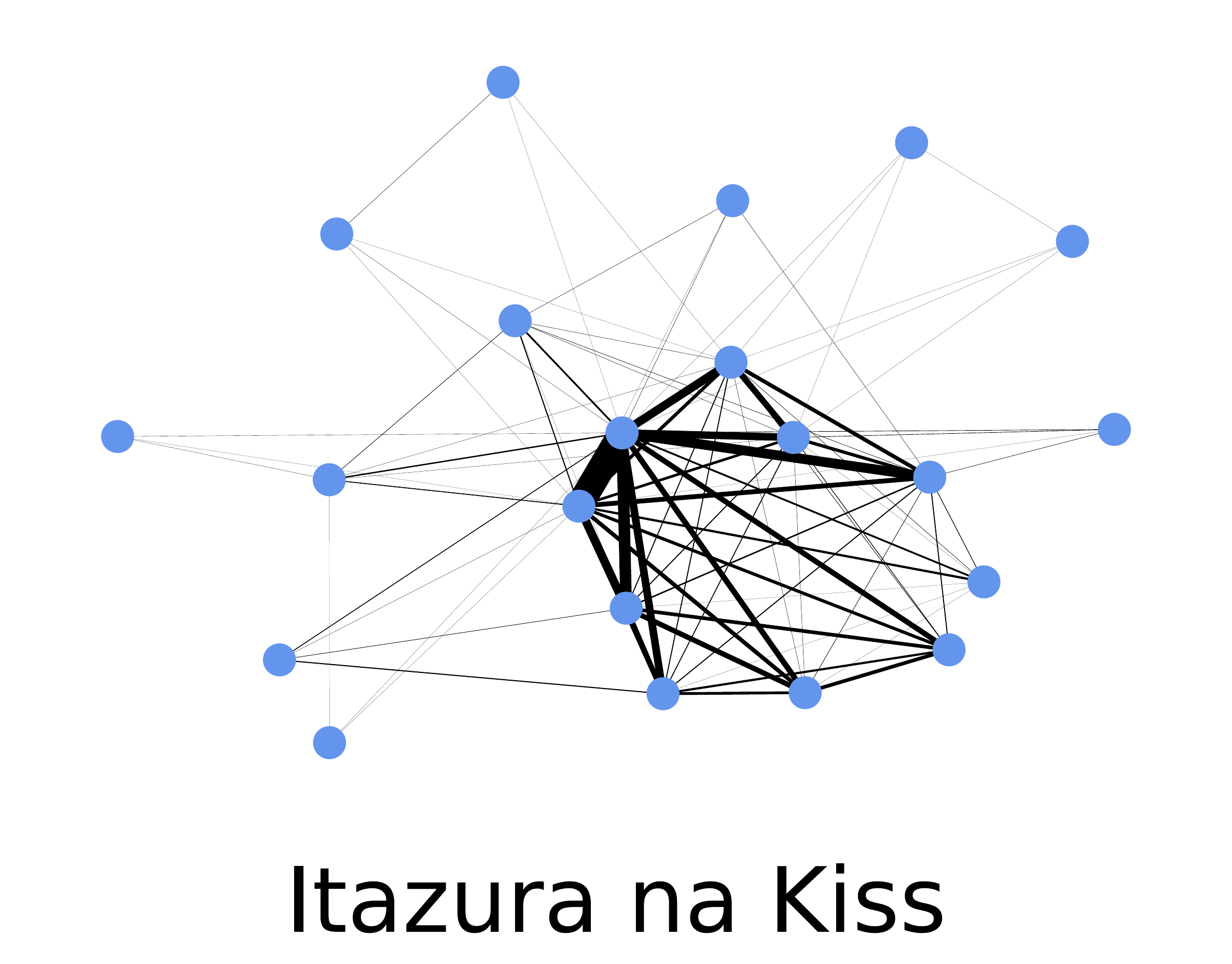}
         \label{fig:three sin x}
     \end{subfigure}
     \begin{subfigure}[b]{0.195\textwidth}
         \centering
         \includegraphics[width=\textwidth]{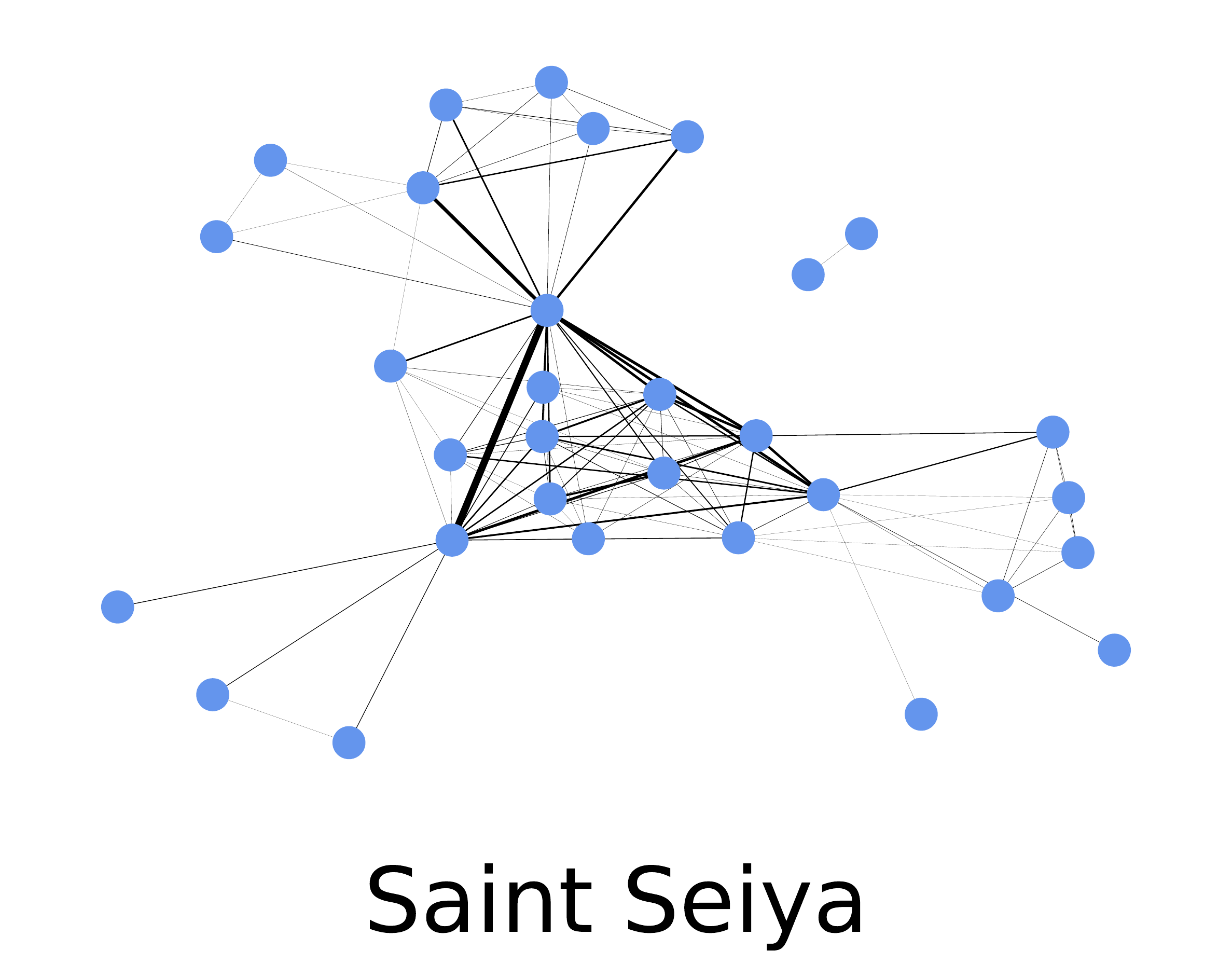}
         \label{fig:five over x}
     \end{subfigure}
     \begin{subfigure}[b]{0.195\textwidth}
         \centering
         \includegraphics[width=\textwidth]{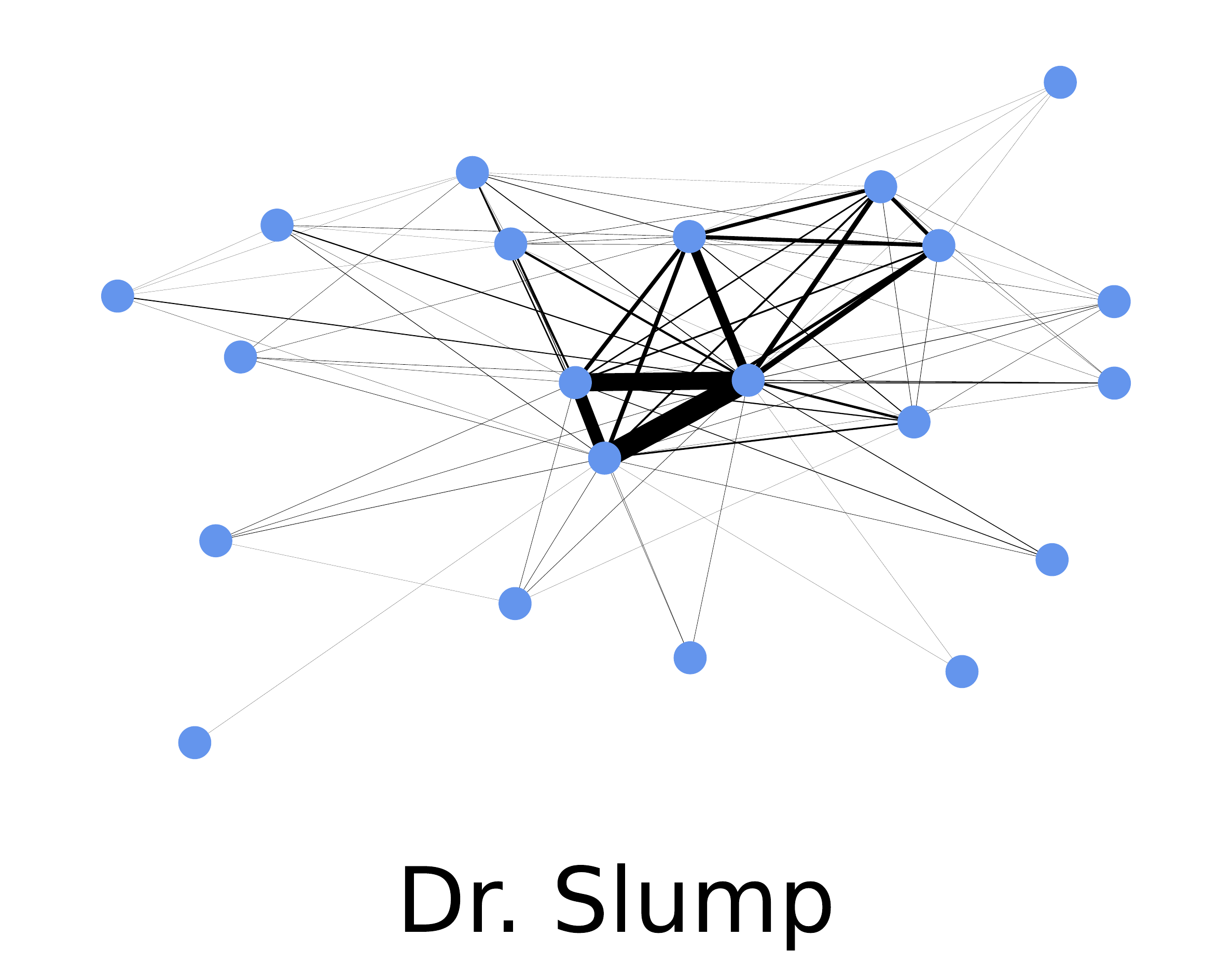}
         \label{fig:five over x}
     \end{subfigure}
     \begin{subfigure}[b]{0.195\textwidth}
         \centering
         \includegraphics[width=\textwidth]{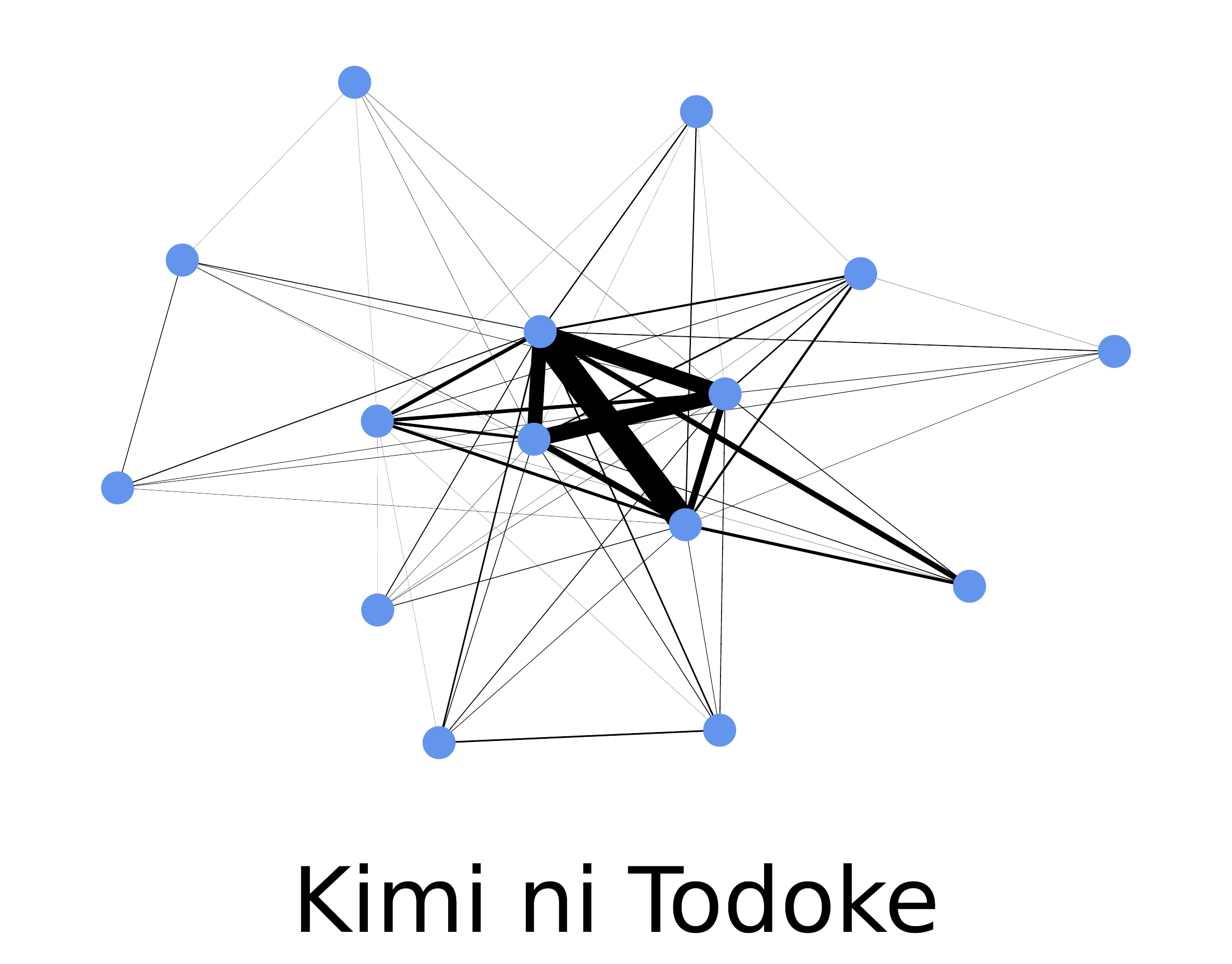}
         \label{fig:five over x}
     \end{subfigure}
     
    \caption{Character networks for 162 manga (continued)}
    
    \label{fig:three graphs}
\end{figure*}

\begin{figure*}\ContinuedFloat
     \centering
      \begin{subfigure}[b]{0.195\textwidth}
         \centering
         \includegraphics[width=\textwidth]{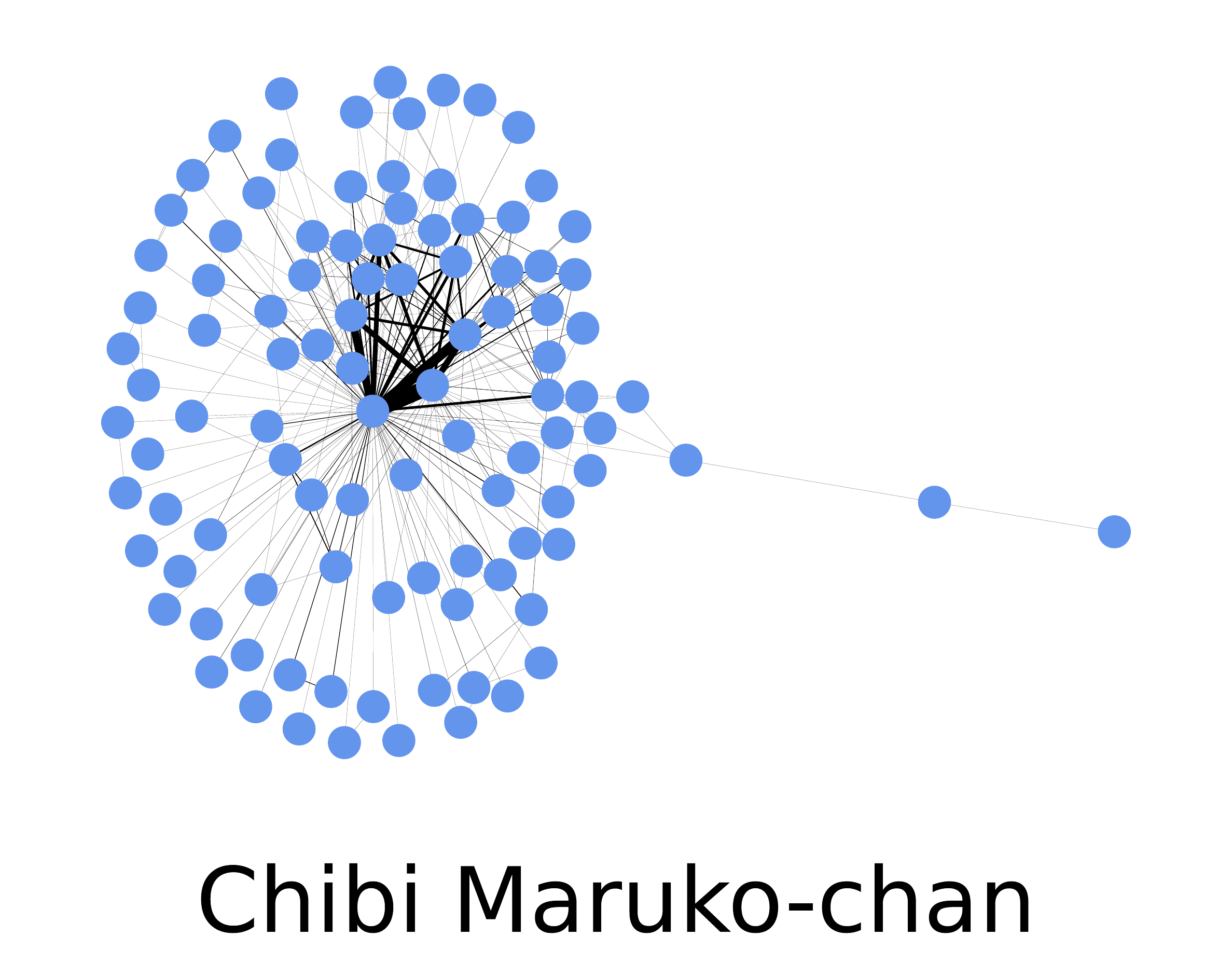}
         \label{fig:y equals x}
     \end{subfigure}
     \begin{subfigure}[b]{0.195\textwidth}
         \centering
         \includegraphics[width=\textwidth]{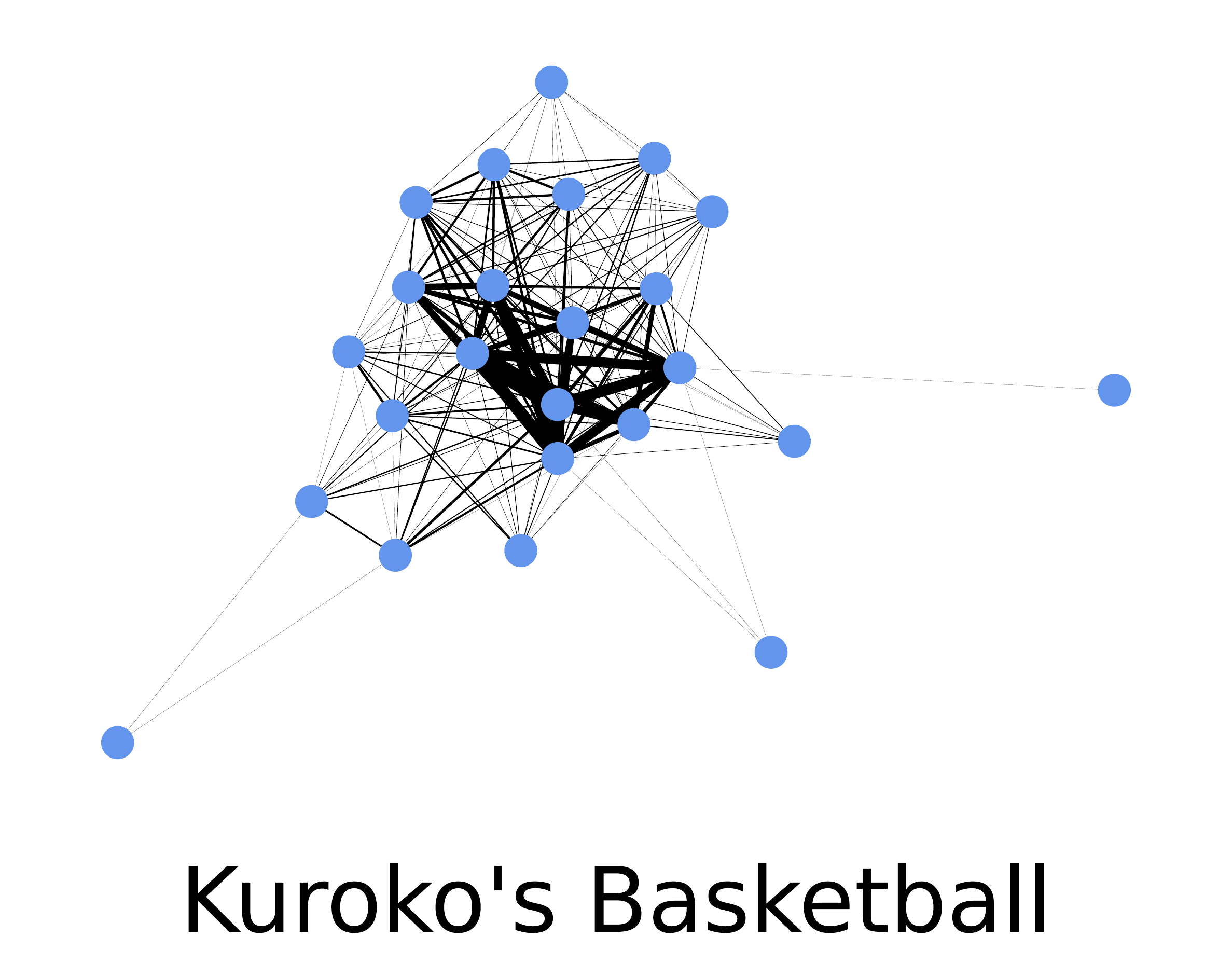}
         \label{fig:three sin x}
     \end{subfigure}
     \begin{subfigure}[b]{0.195\textwidth}
         \centering
         \includegraphics[width=\textwidth]{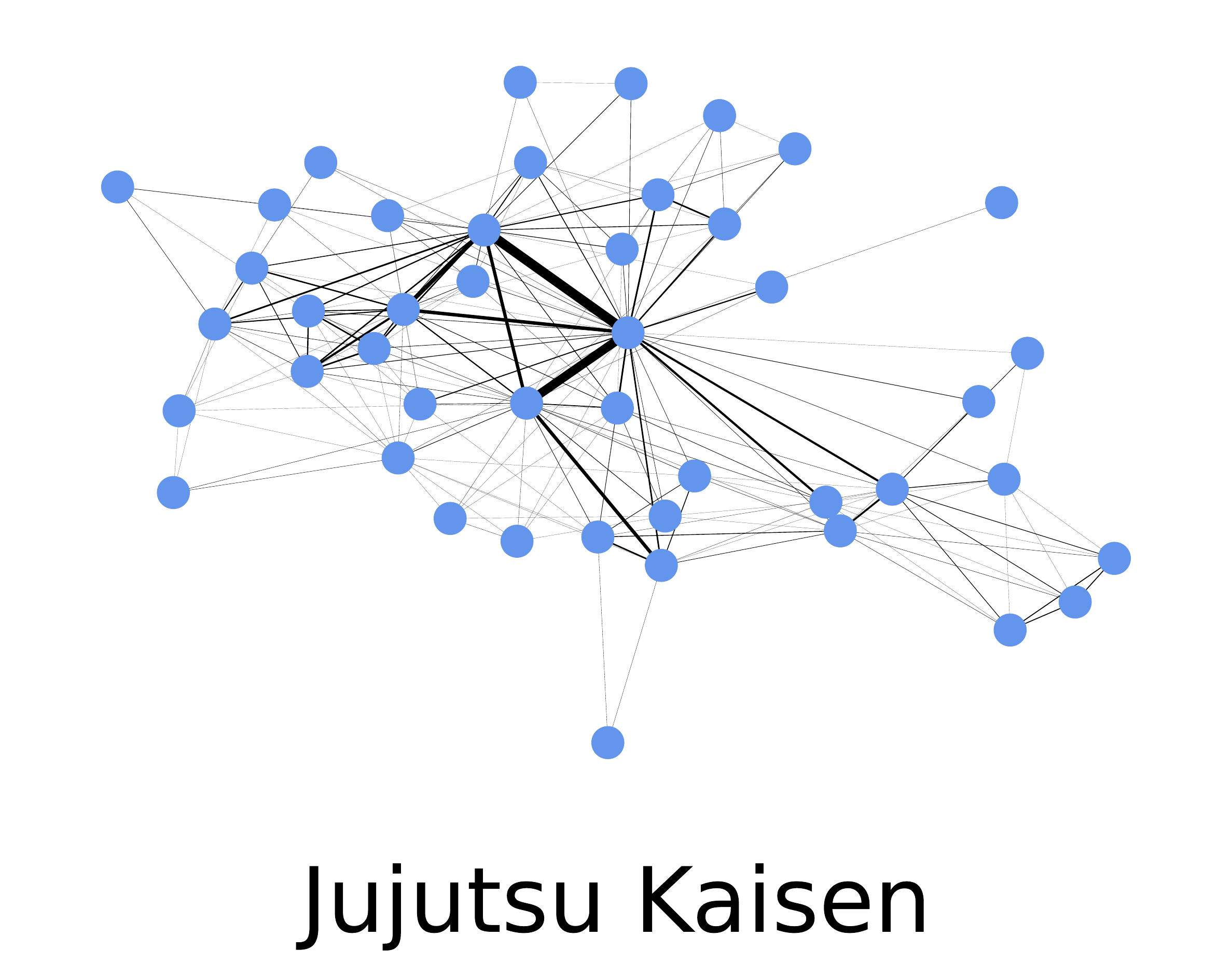}
         \label{fig:five over x}
     \end{subfigure}
     \begin{subfigure}[b]{0.195\textwidth}
         \centering
         \includegraphics[width=\textwidth]{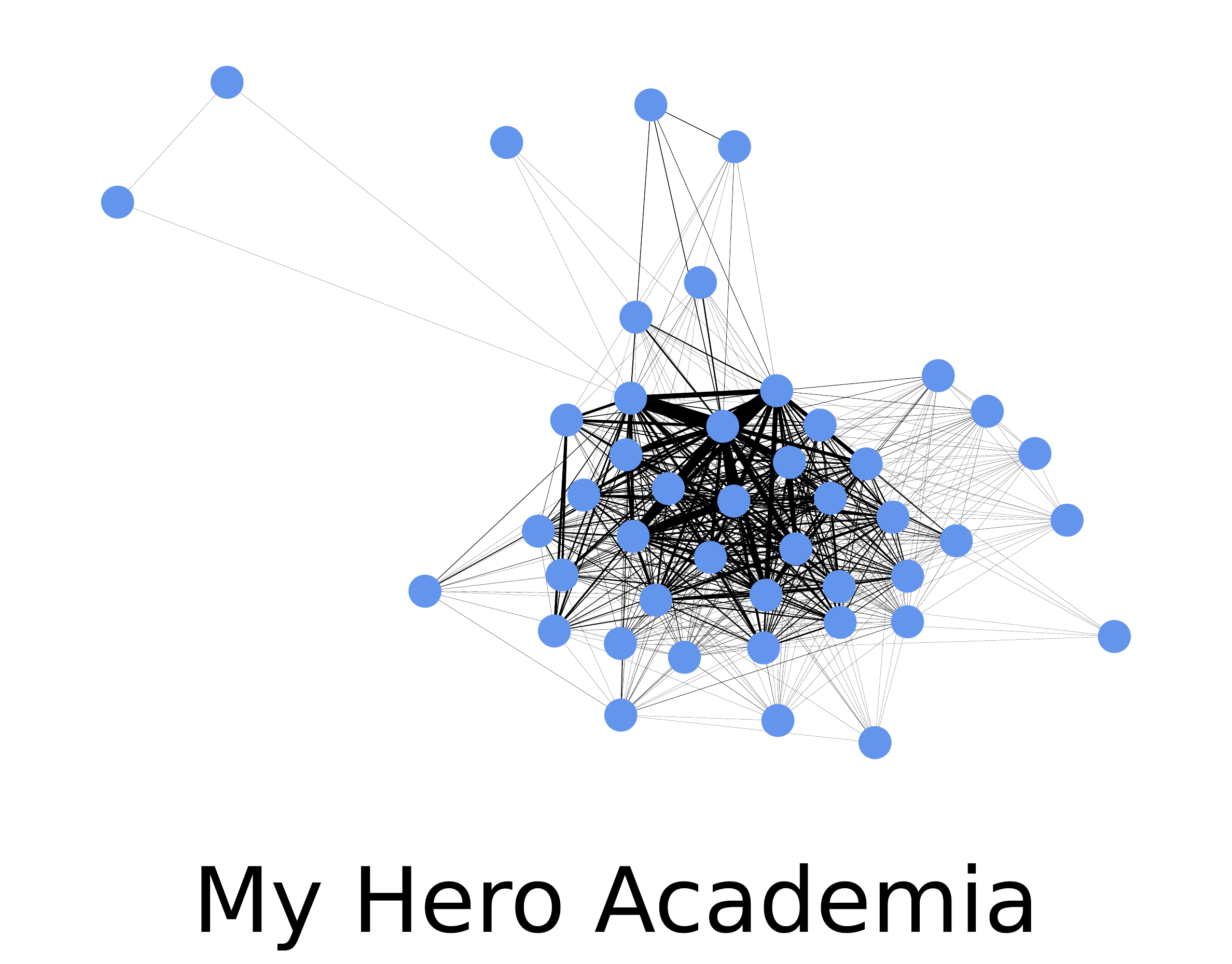}
         \label{fig:five over x}
     \end{subfigure}
     \begin{subfigure}[b]{0.195\textwidth}
         \centering
         \includegraphics[width=\textwidth]{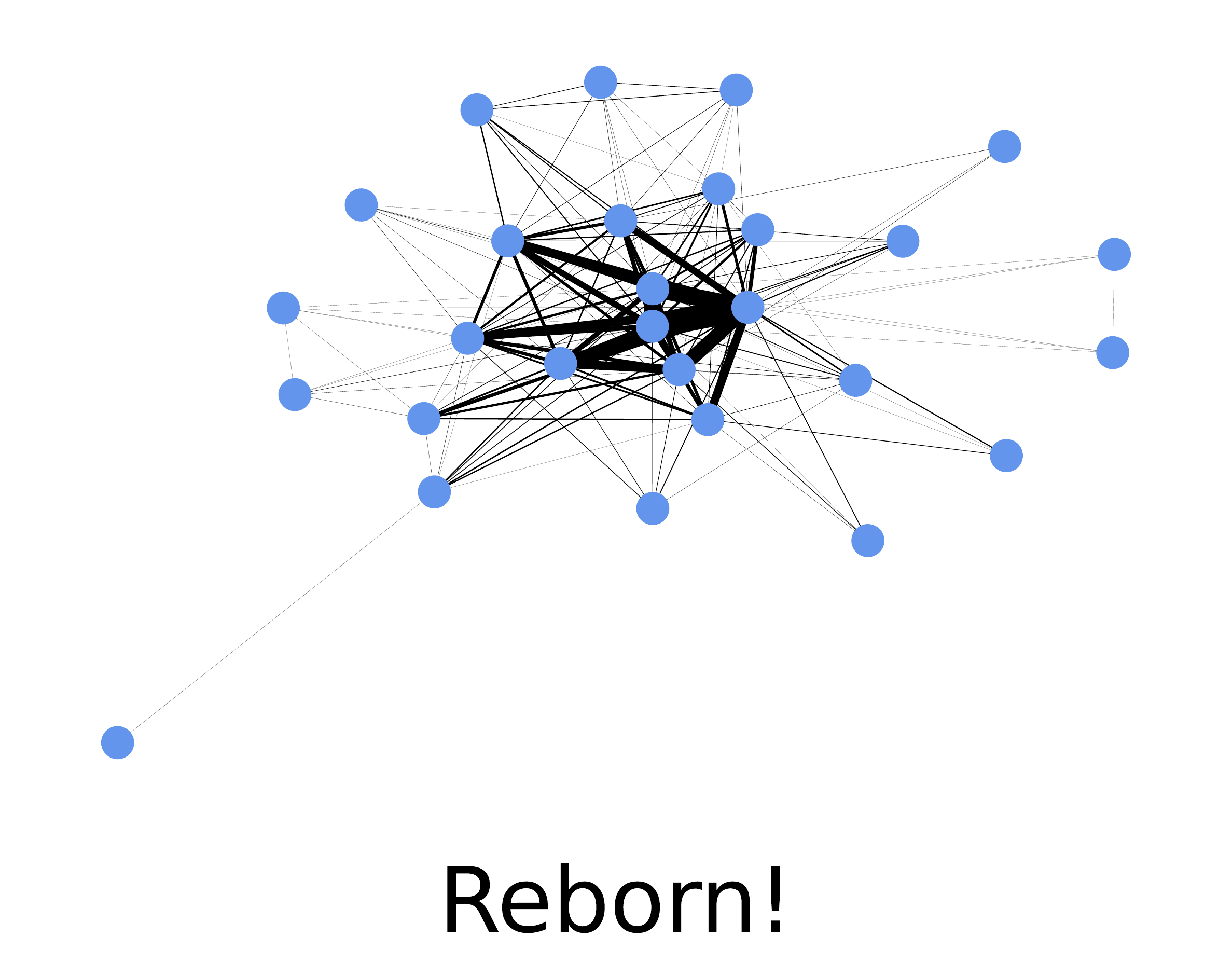}
         \label{fig:five over x}
     \end{subfigure}
     \\
      \begin{subfigure}[b]{0.195\textwidth}
         \centering
         \includegraphics[width=\textwidth]{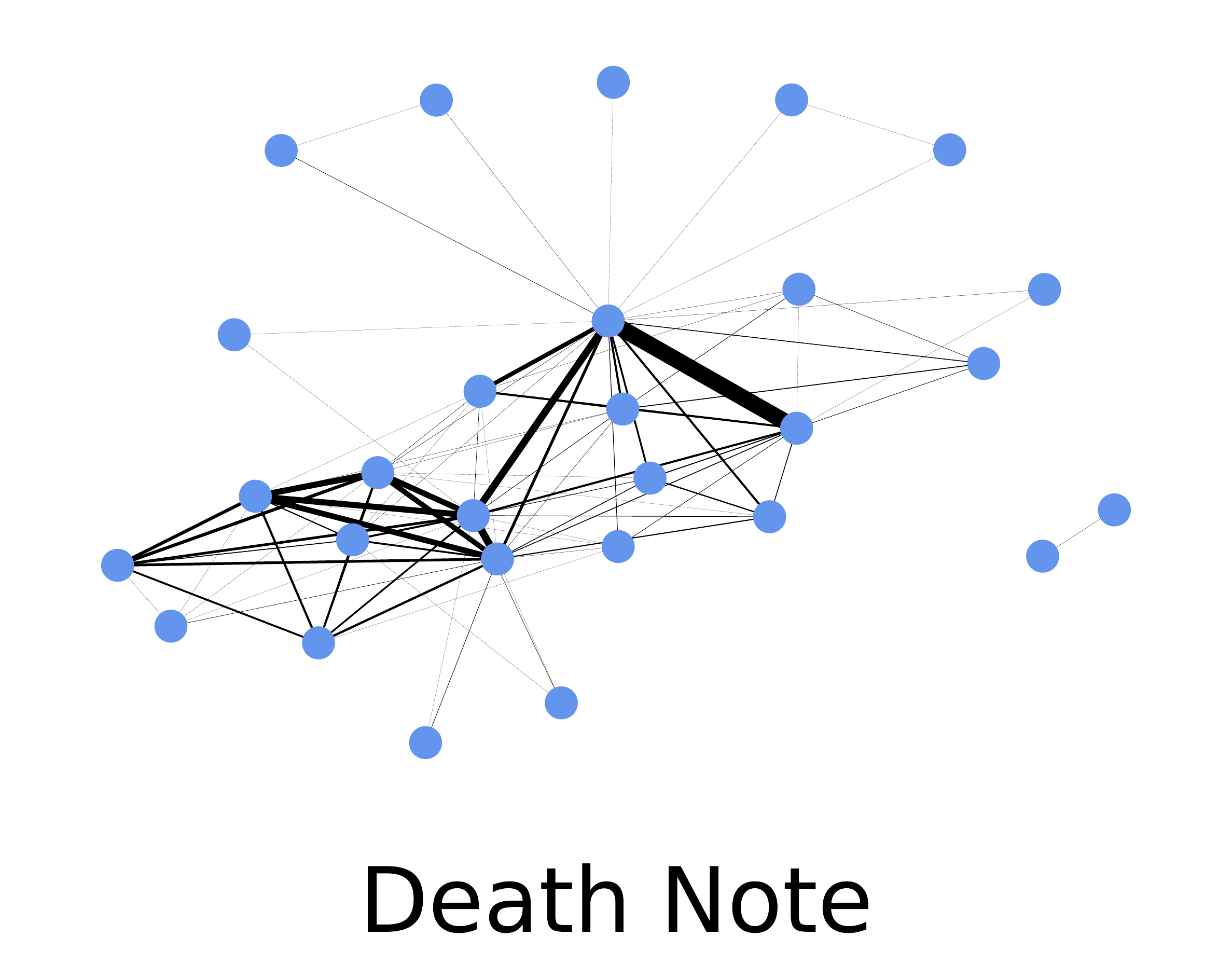}
         \label{fig:y equals x}
     \end{subfigure}
     \begin{subfigure}[b]{0.195\textwidth}
         \centering
         \includegraphics[width=\textwidth]{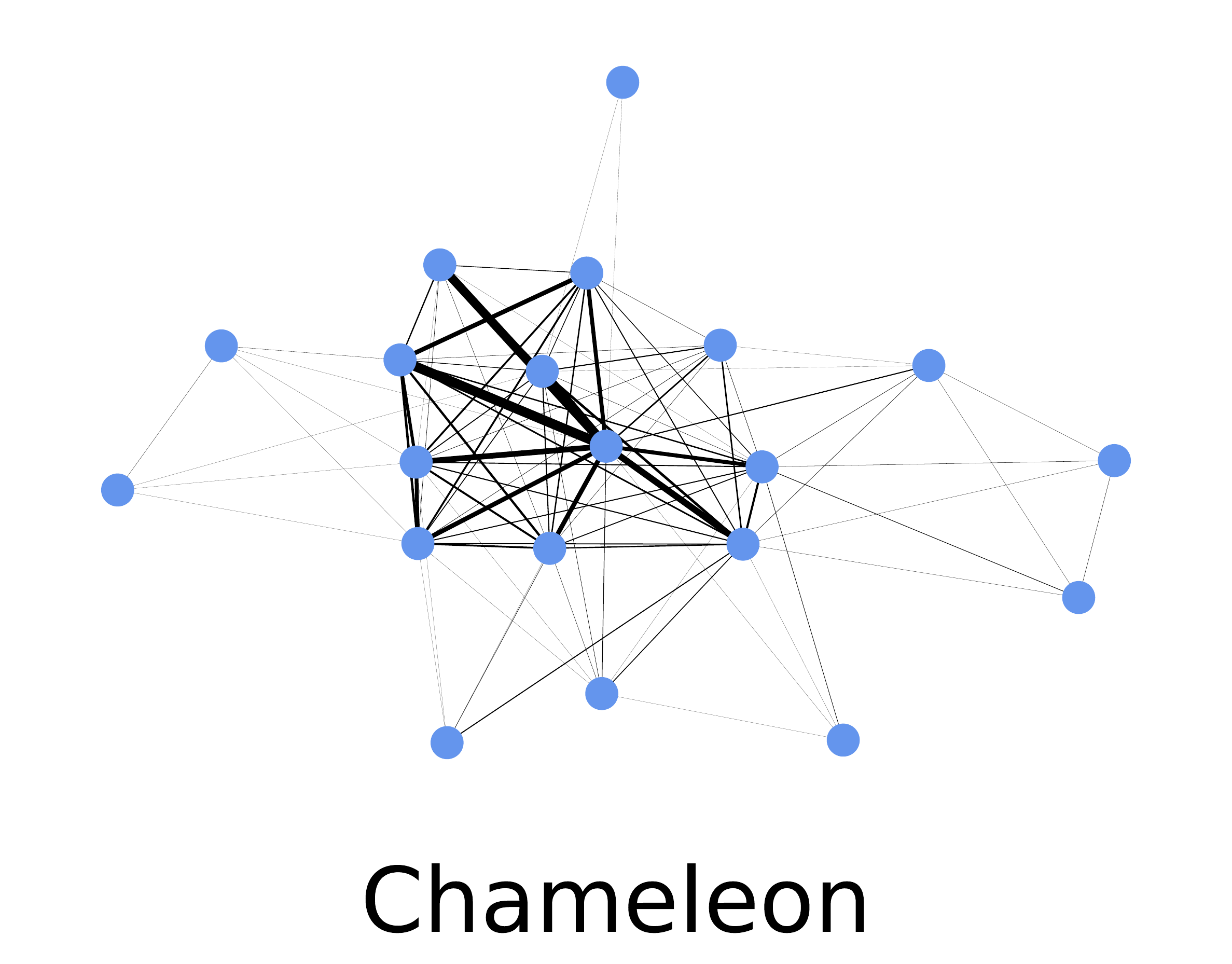}
         \label{fig:three sin x}
     \end{subfigure}
     \begin{subfigure}[b]{0.195\textwidth}
         \centering
         \includegraphics[width=\textwidth]{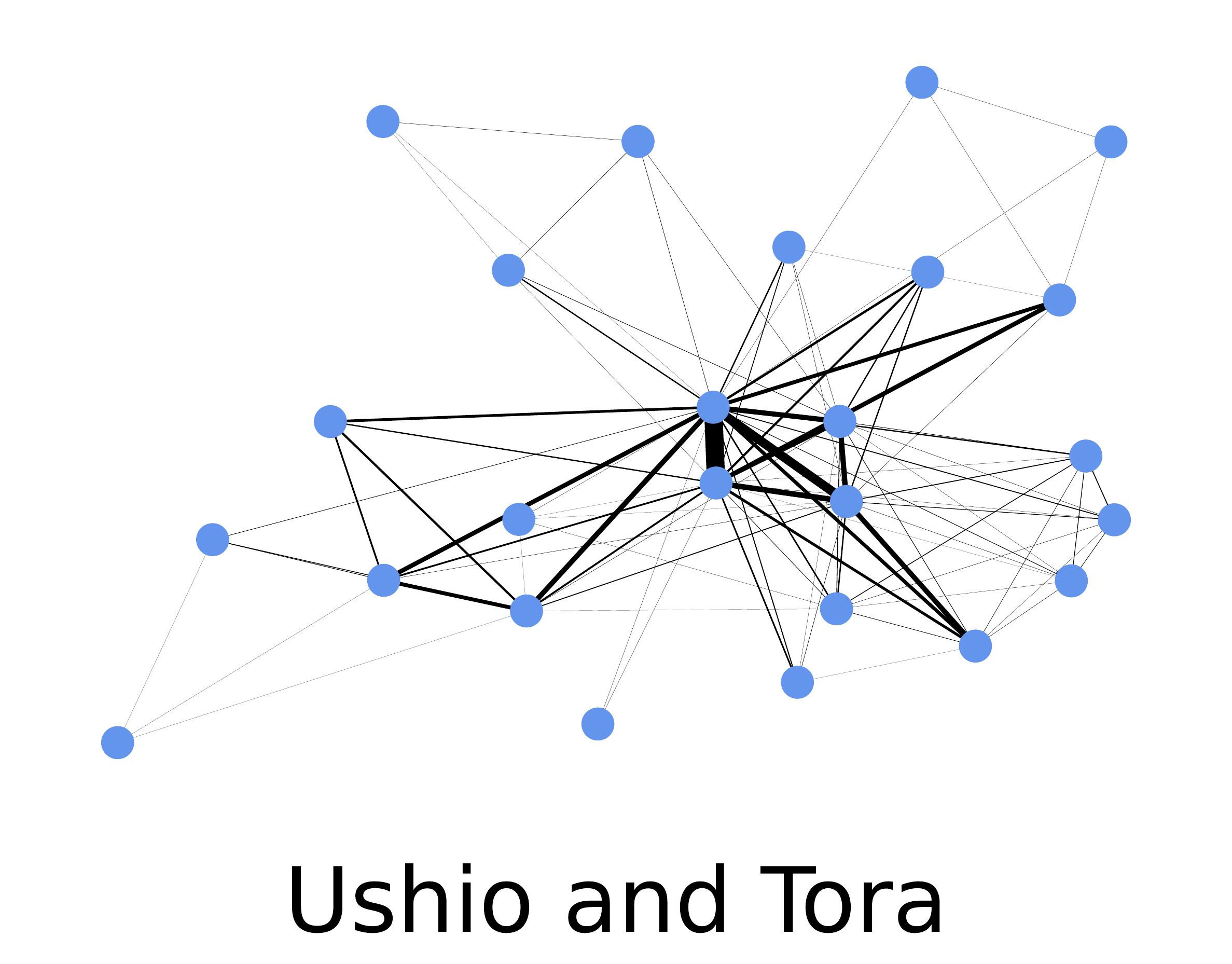}
         \label{fig:five over x}
     \end{subfigure}
     \begin{subfigure}[b]{0.195\textwidth}
         \centering
         \includegraphics[width=\textwidth]{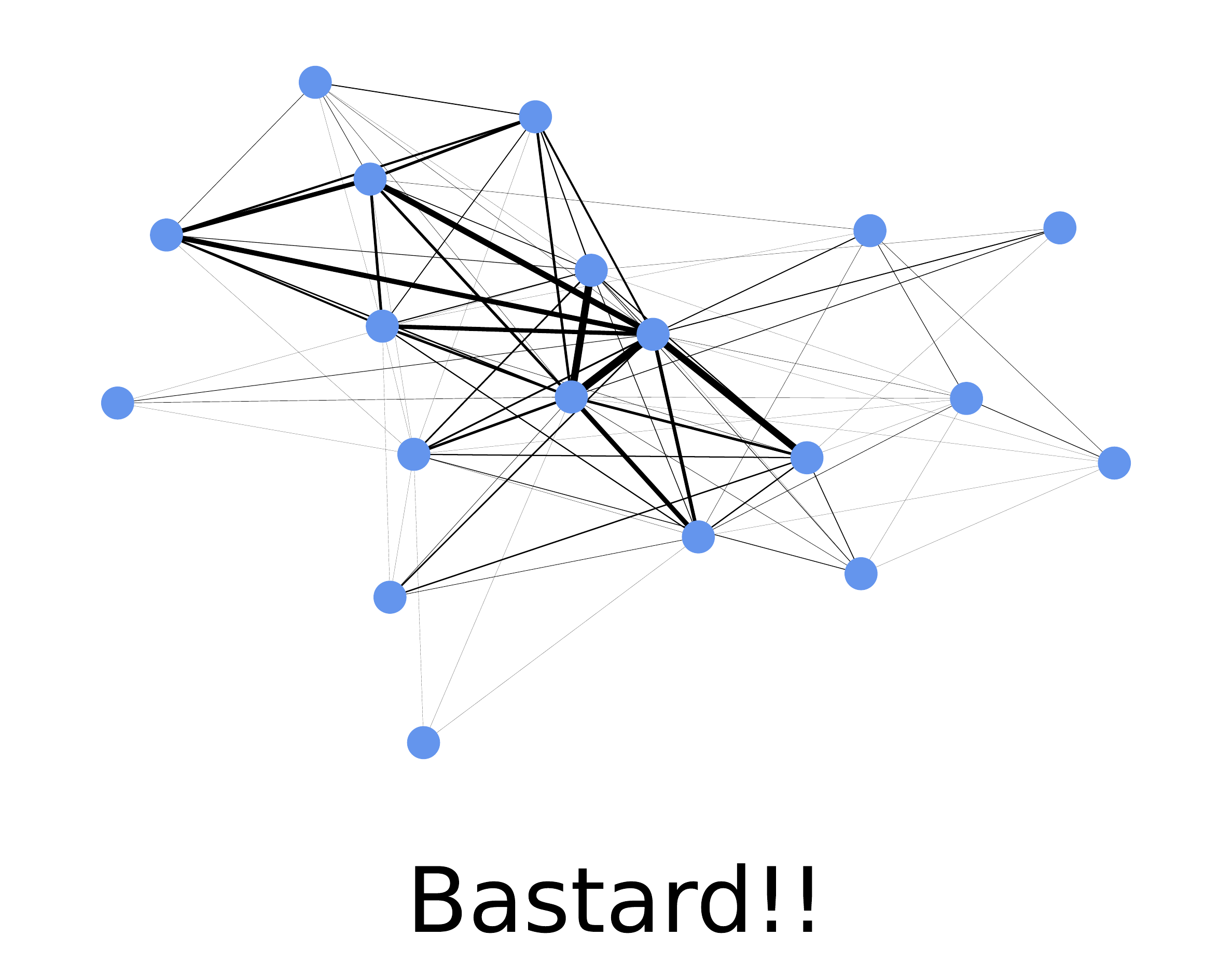}
         \label{fig:five over x}
     \end{subfigure}
     \begin{subfigure}[b]{0.195\textwidth}
         \centering
         \includegraphics[width=\textwidth]{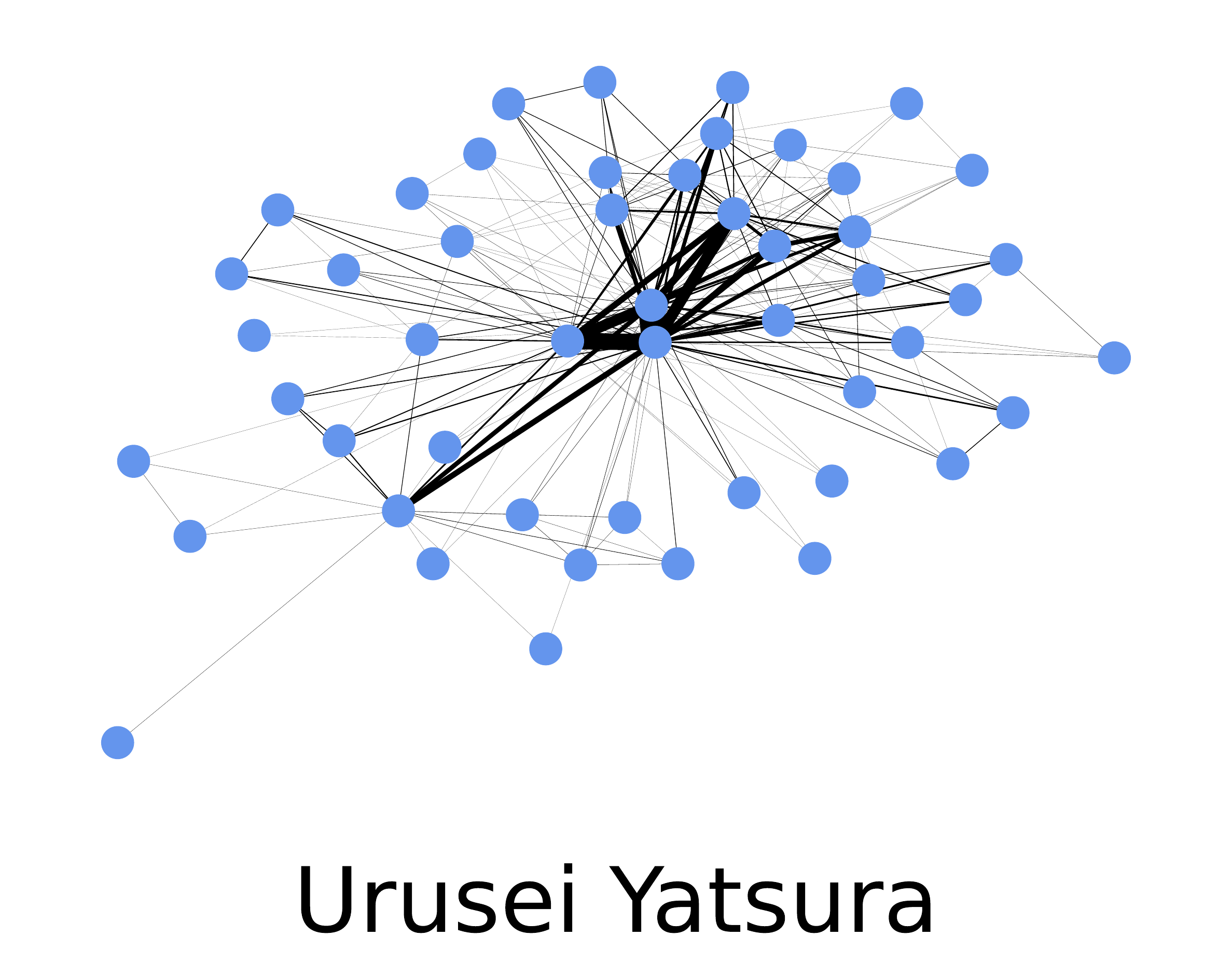}
         \label{fig:five over x}
     \end{subfigure}
     \\
     \begin{subfigure}[b]{0.195\textwidth}
         \centering
         \includegraphics[width=\textwidth]{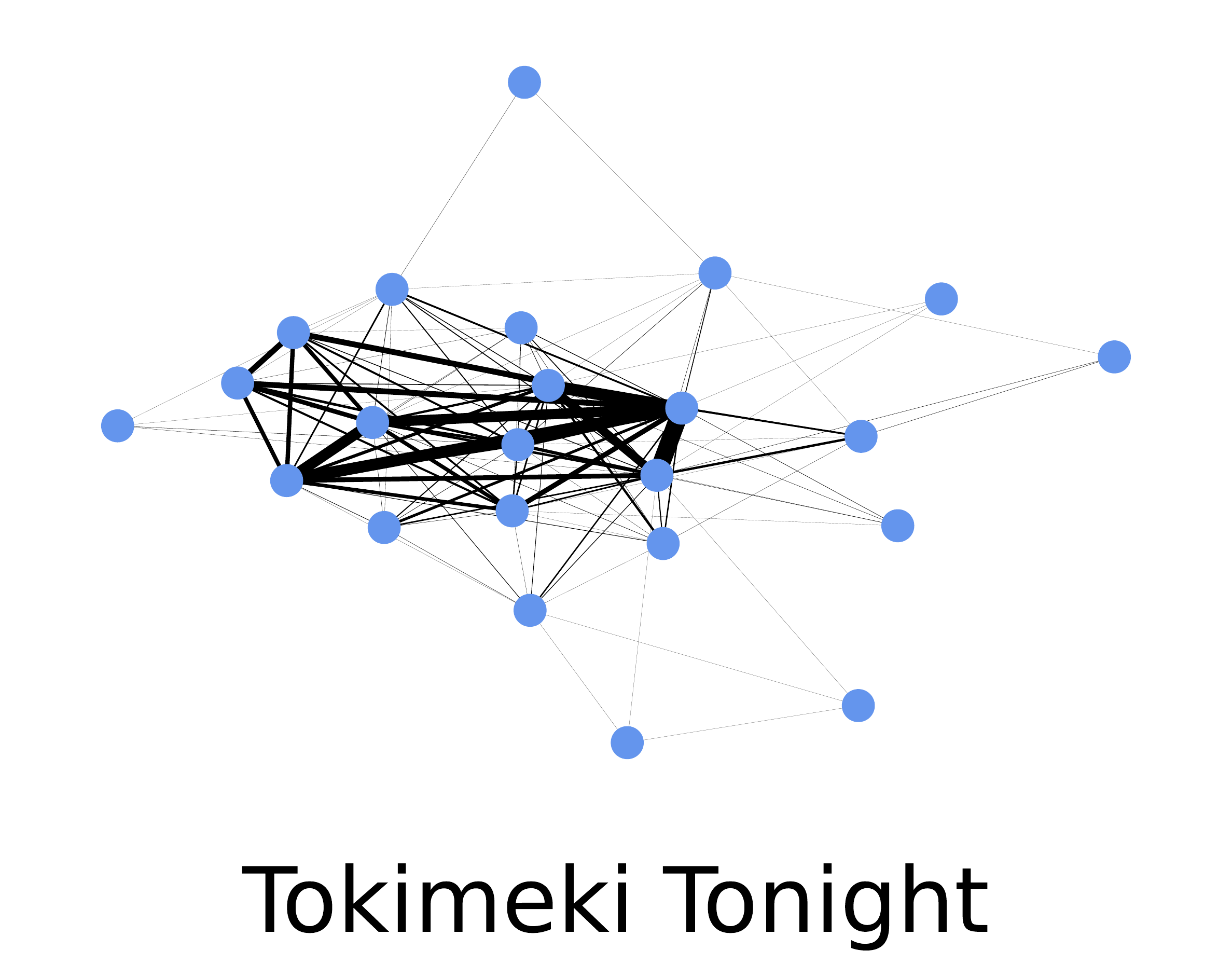}
     \end{subfigure}
     \begin{subfigure}[b]{0.195\textwidth}
         \centering
         \includegraphics[width=\textwidth]{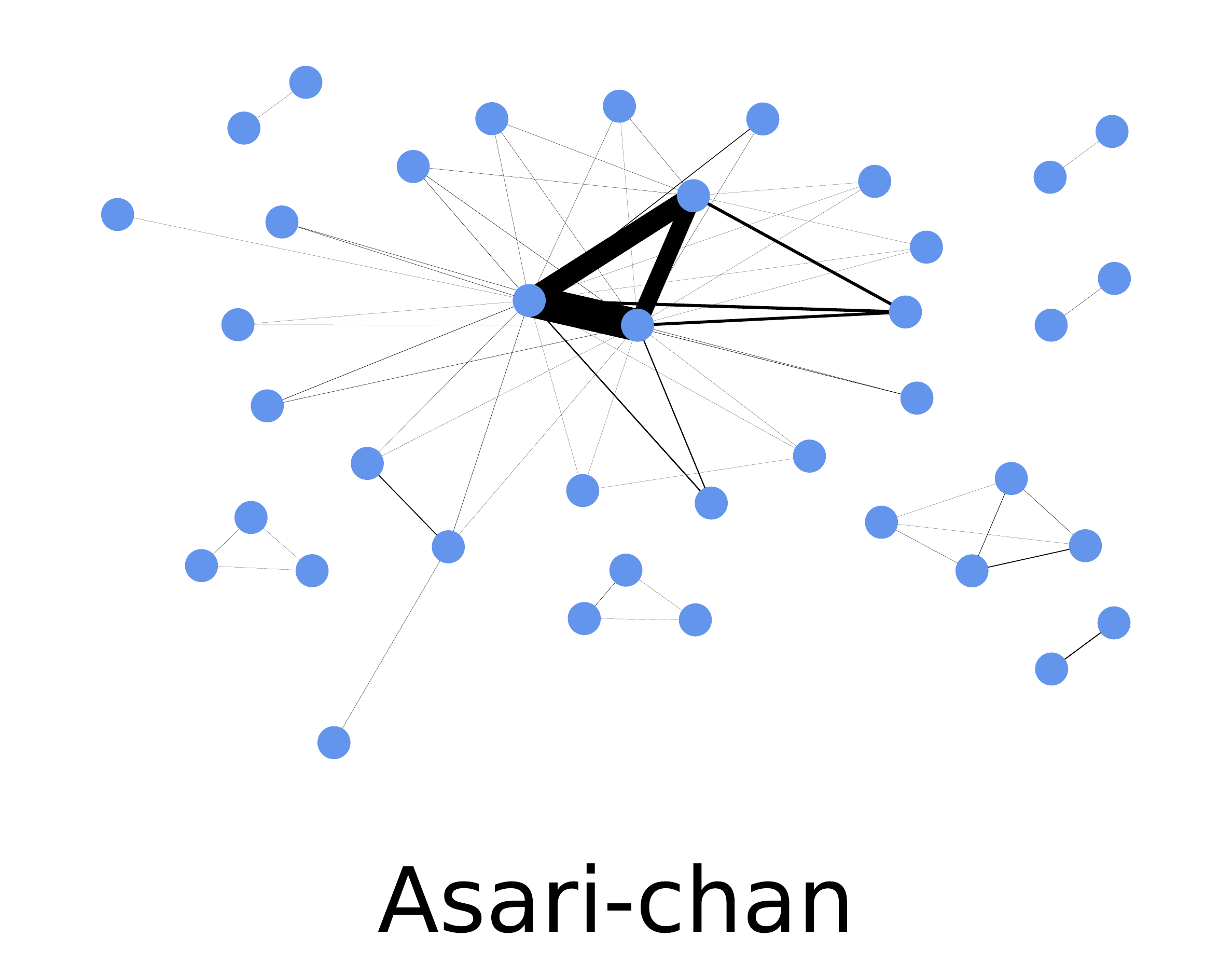}
     \end{subfigure}
     \begin{subfigure}[b]{0.195\textwidth}
         \centering
         \includegraphics[width=\textwidth]{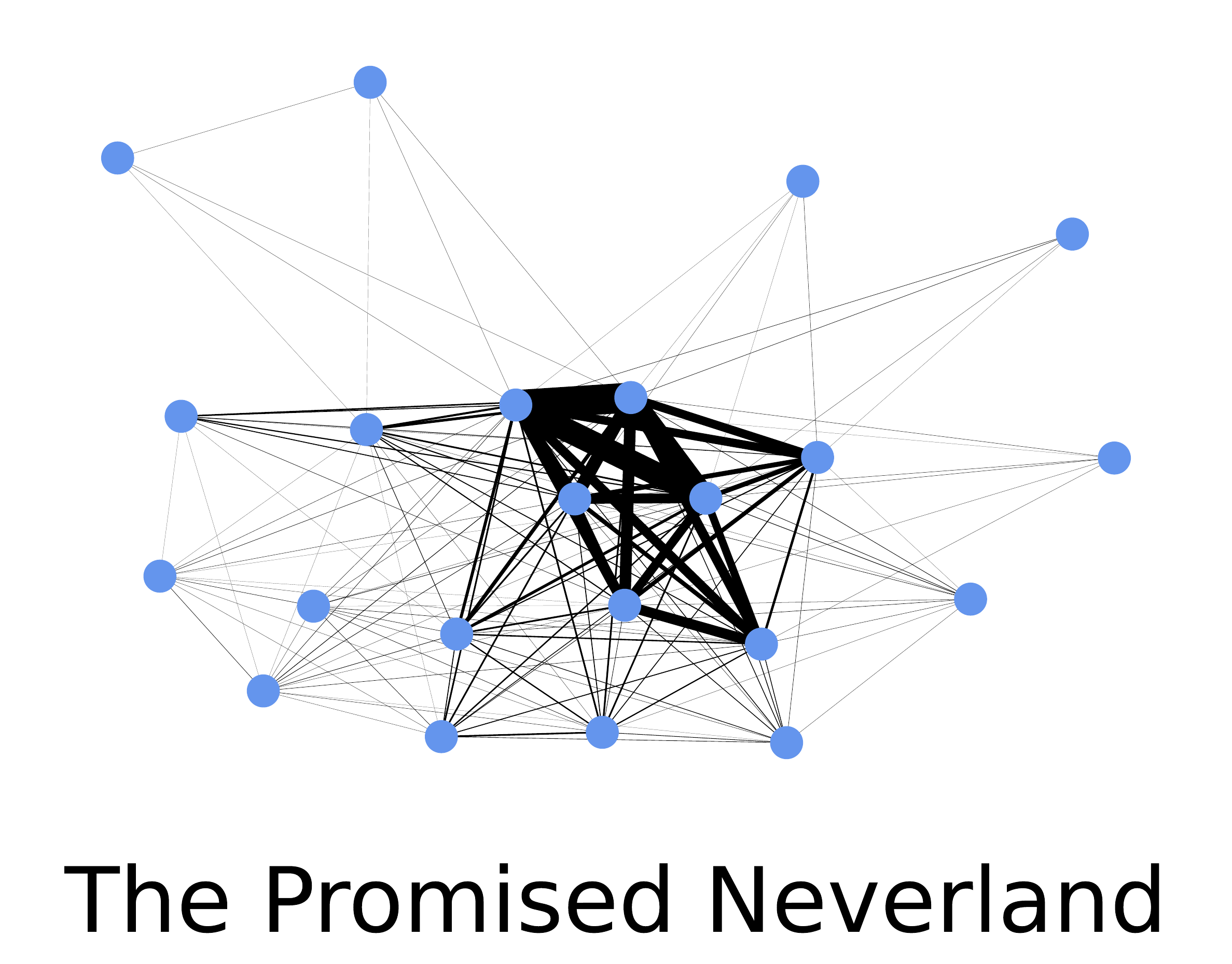}
     \end{subfigure}
     \begin{subfigure}[b]{0.195\textwidth}
         \centering
         \includegraphics[width=\textwidth]{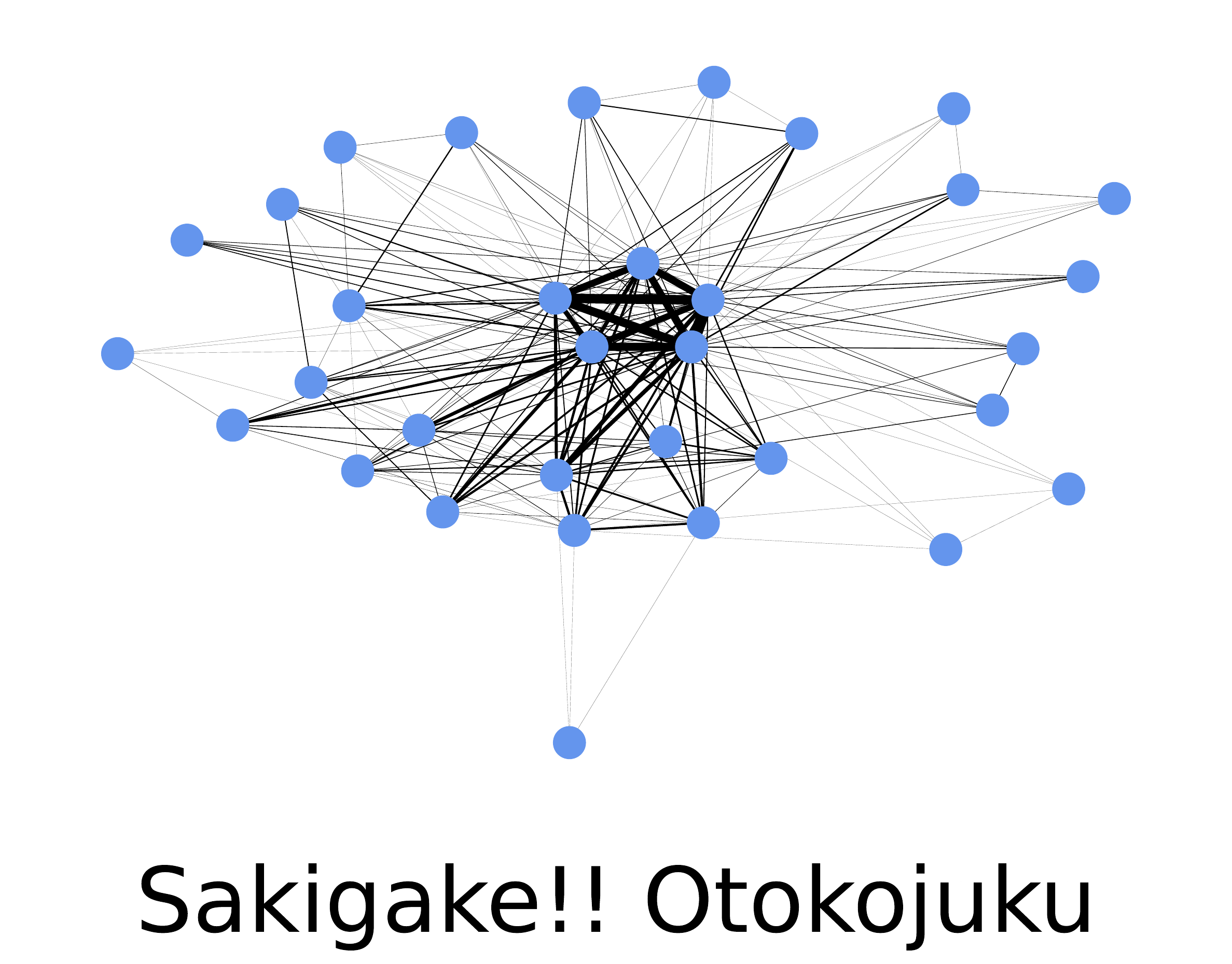}
     \end{subfigure}
     \begin{subfigure}[b]{0.195\textwidth}
         \centering
         \includegraphics[width=\textwidth]{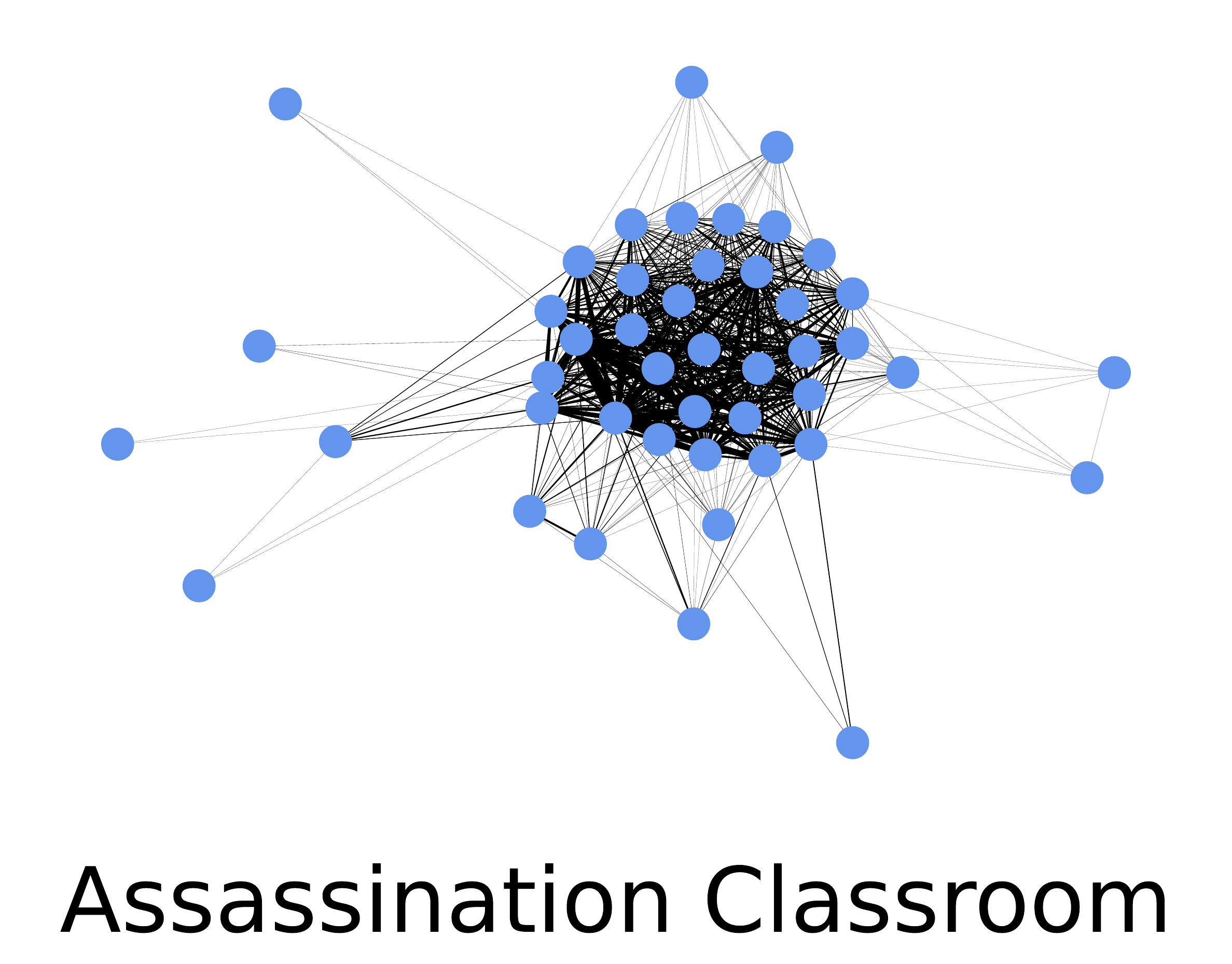}
     \end{subfigure}
     \\
      \begin{subfigure}[b]{0.195\textwidth}
         \centering
         \includegraphics[width=\textwidth]{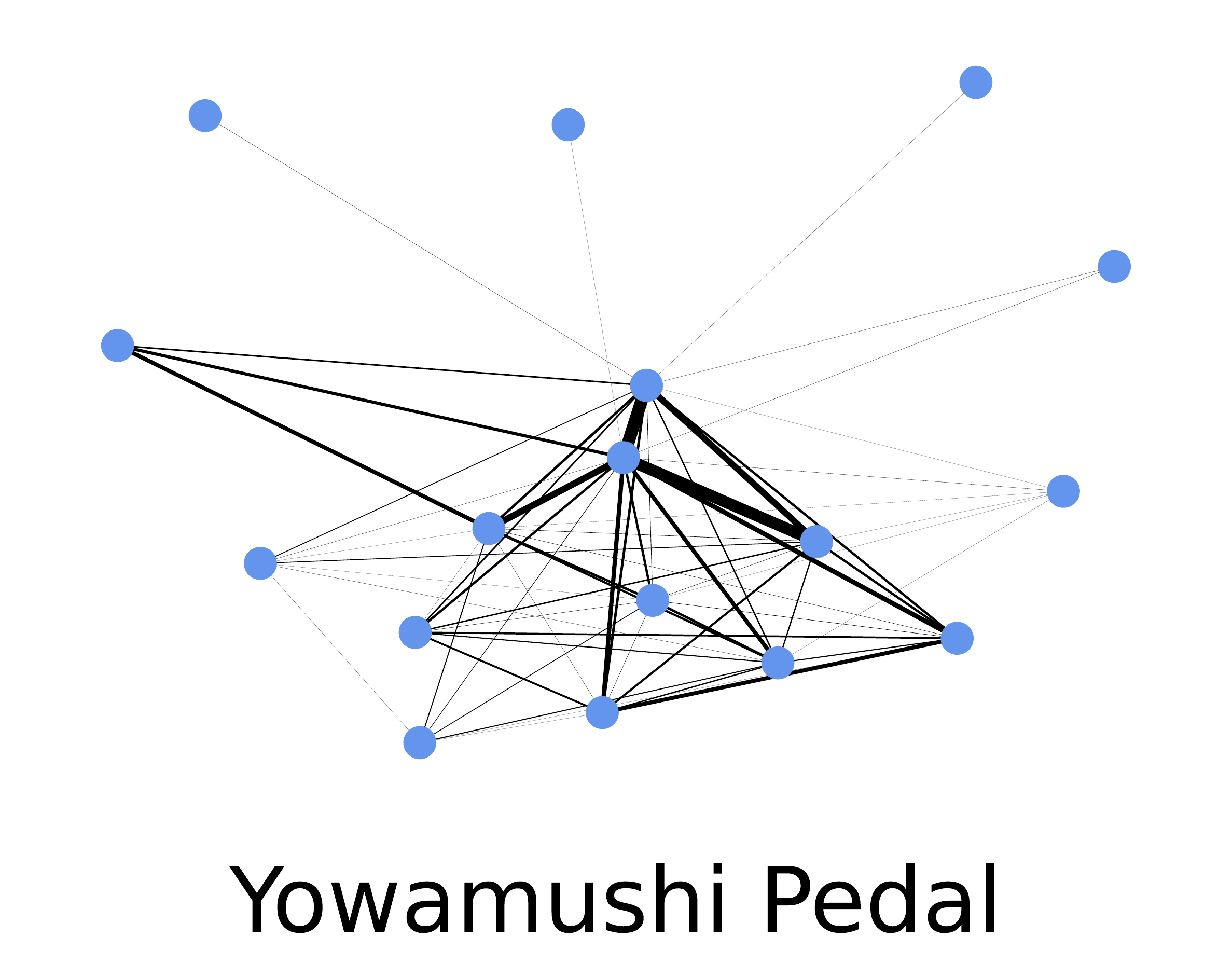}
     \end{subfigure}
     \begin{subfigure}[b]{0.195\textwidth}
         \centering
         \includegraphics[width=\textwidth]{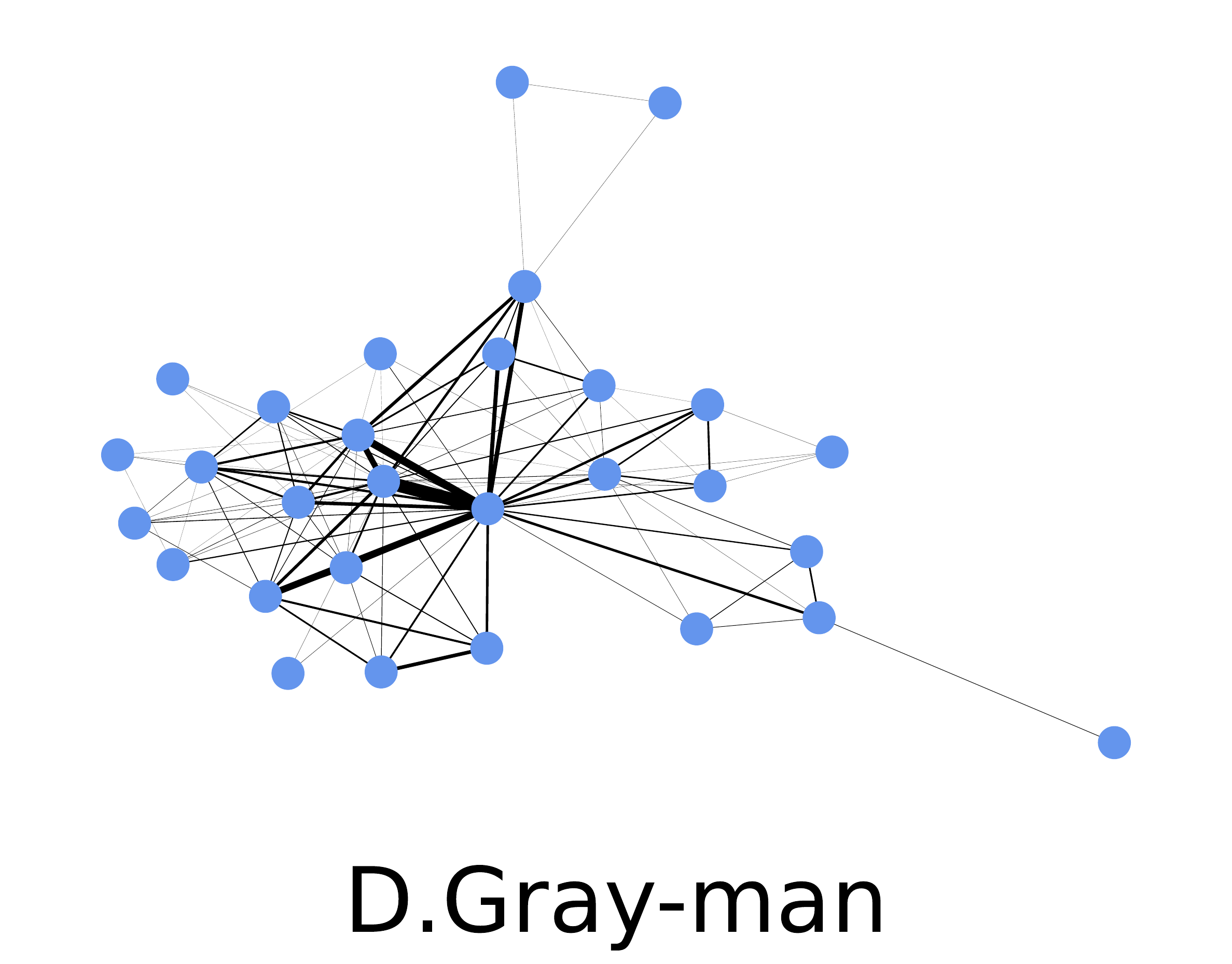}
     \end{subfigure}
     \begin{subfigure}[b]{0.195\textwidth}
         \centering
         \includegraphics[width=\textwidth]{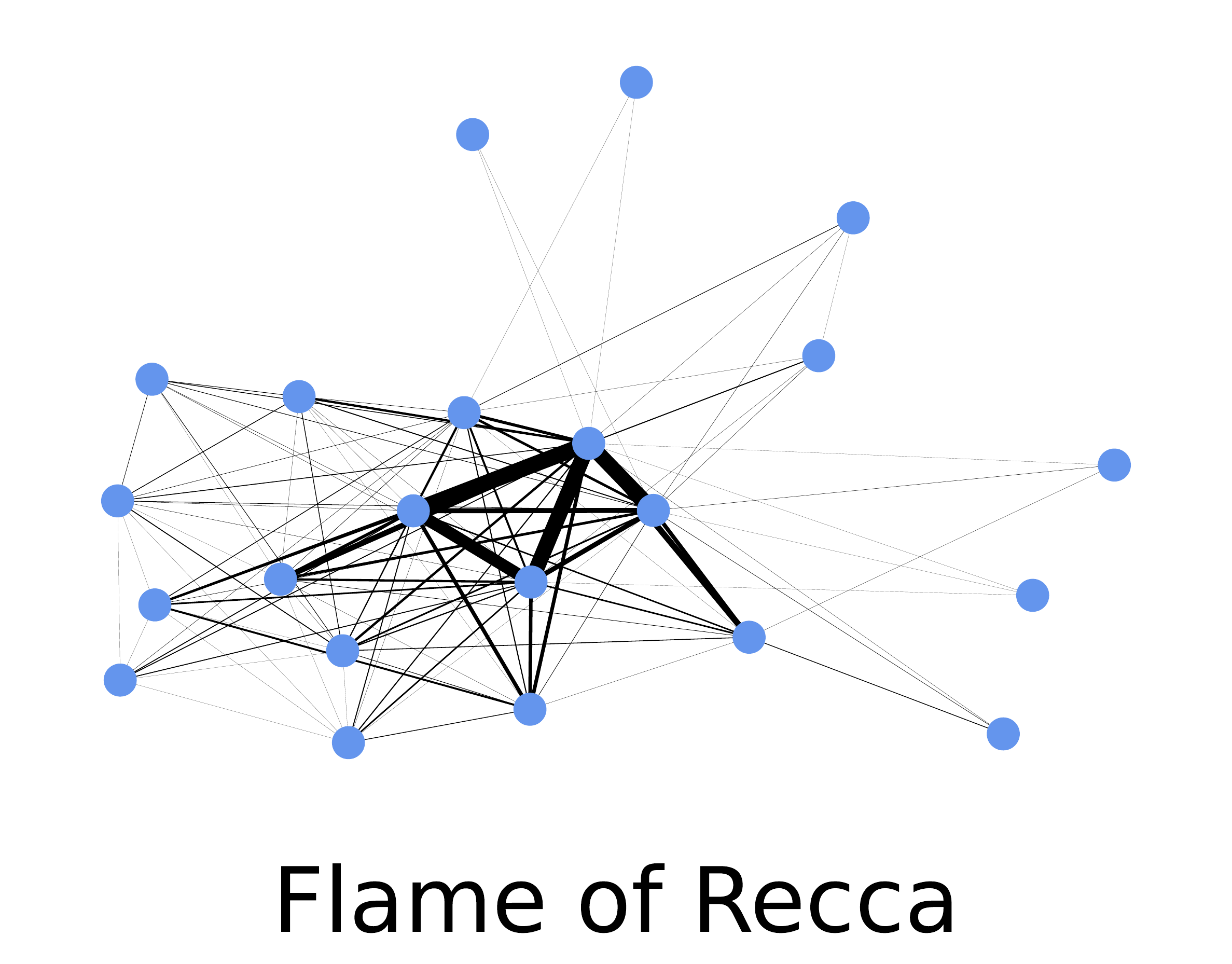}
     \end{subfigure}
     \begin{subfigure}[b]{0.195\textwidth}
         \centering
         \includegraphics[width=\textwidth]{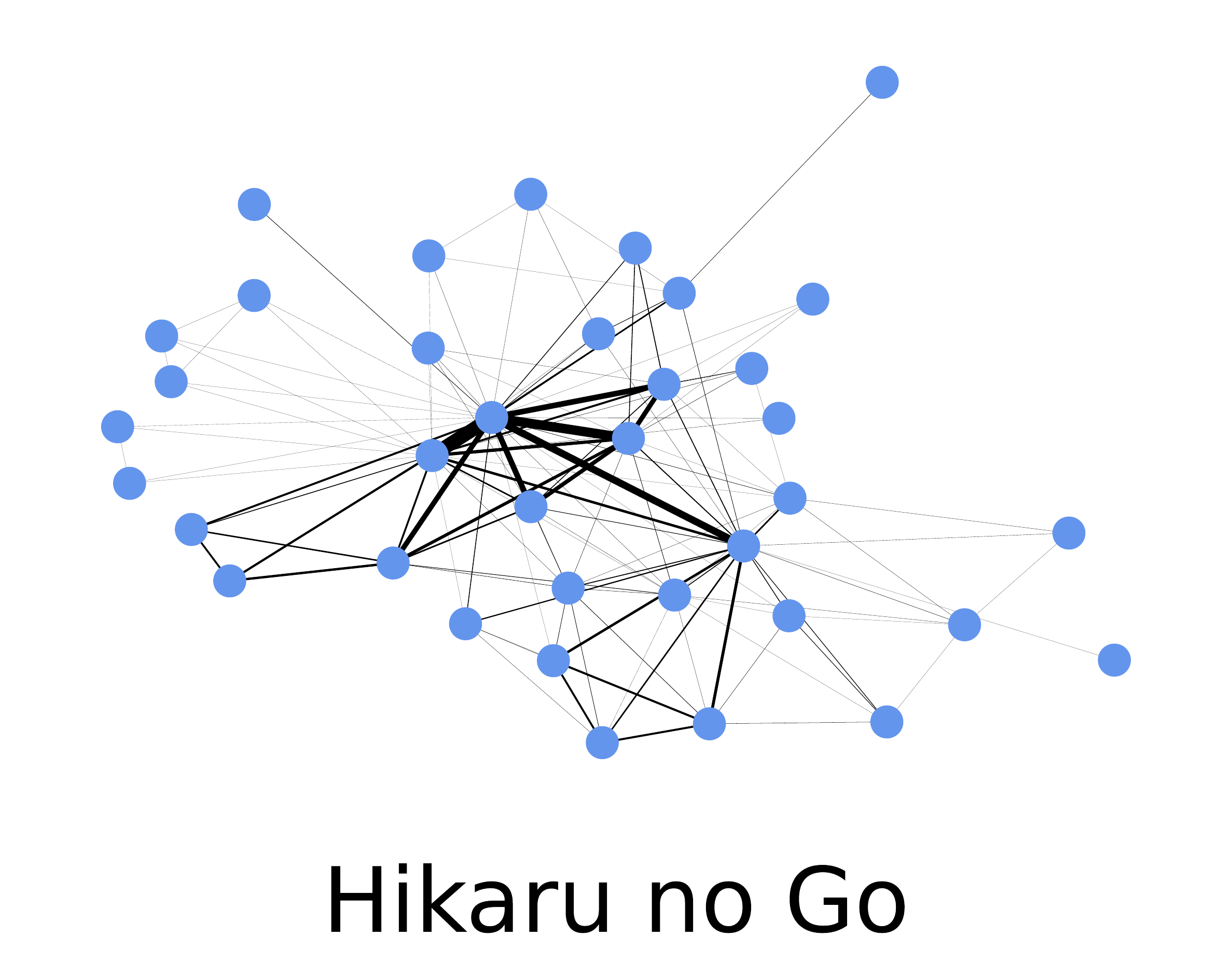}
     \end{subfigure}
     \begin{subfigure}[b]{0.195\textwidth}
         \centering
         \includegraphics[width=\textwidth]{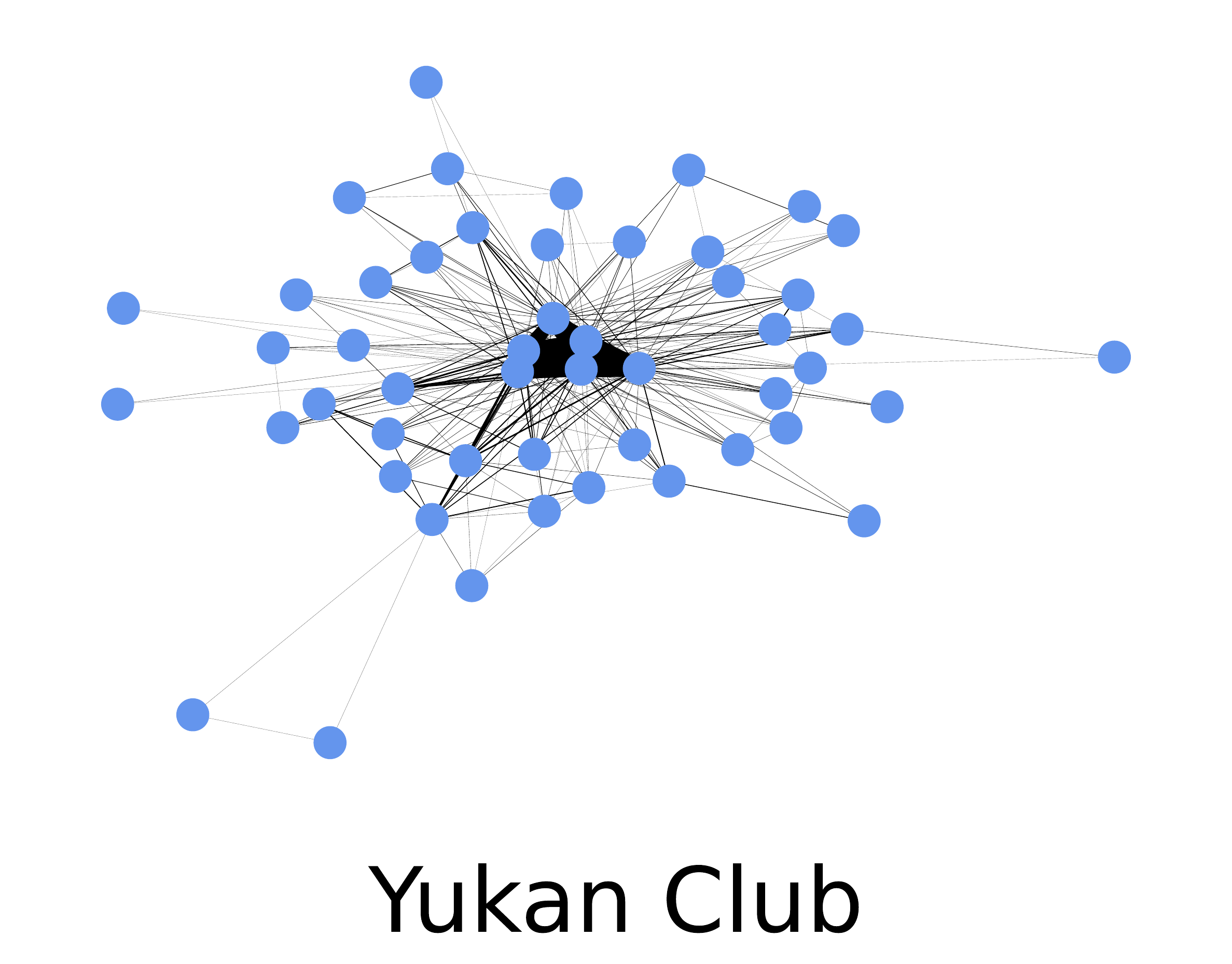}
     \end{subfigure}
     \\
      \begin{subfigure}[b]{0.195\textwidth}
         \centering
         \includegraphics[width=\textwidth]{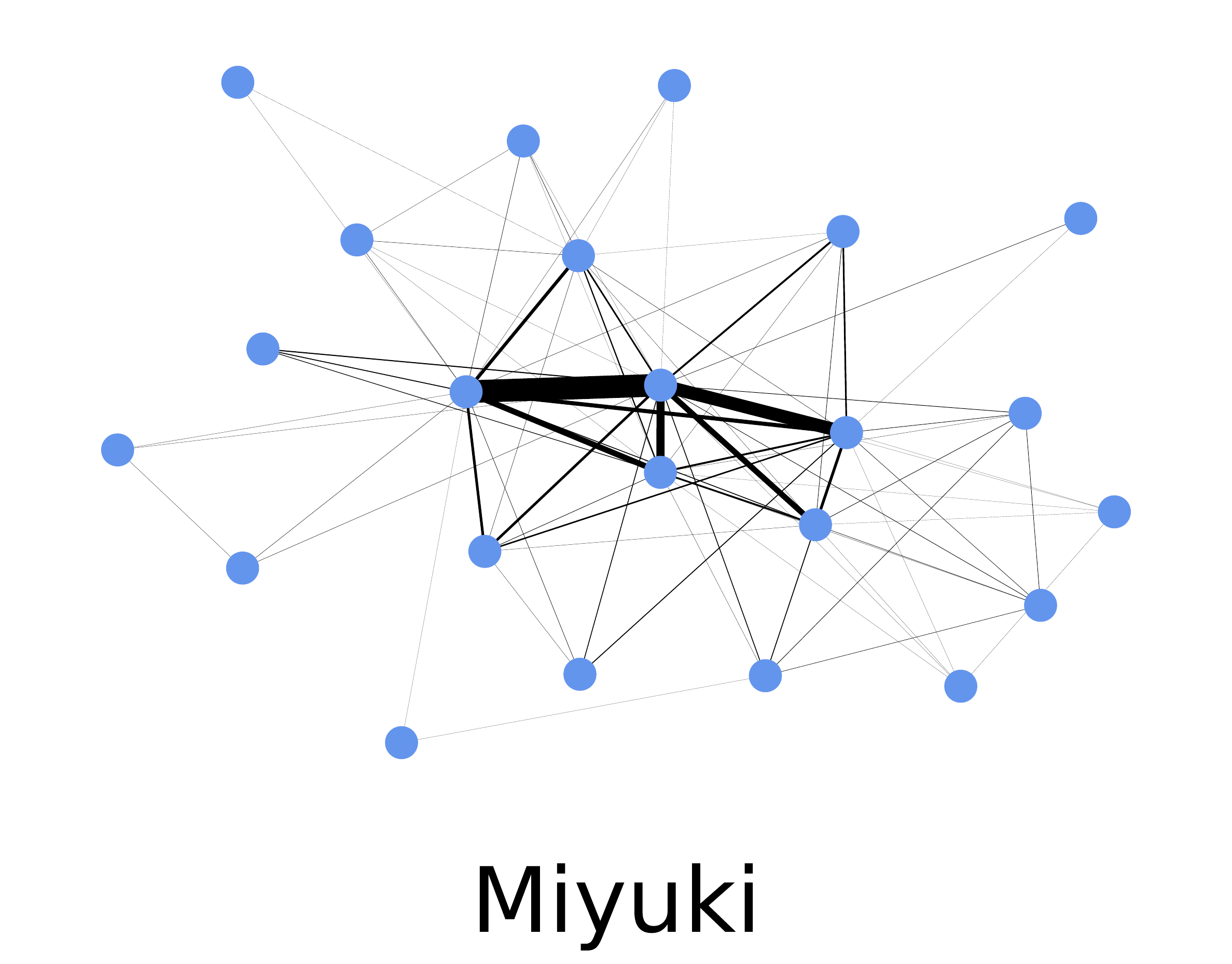}
     \end{subfigure}
     \begin{subfigure}[b]{0.195\textwidth}
         \centering
         \includegraphics[width=\textwidth]{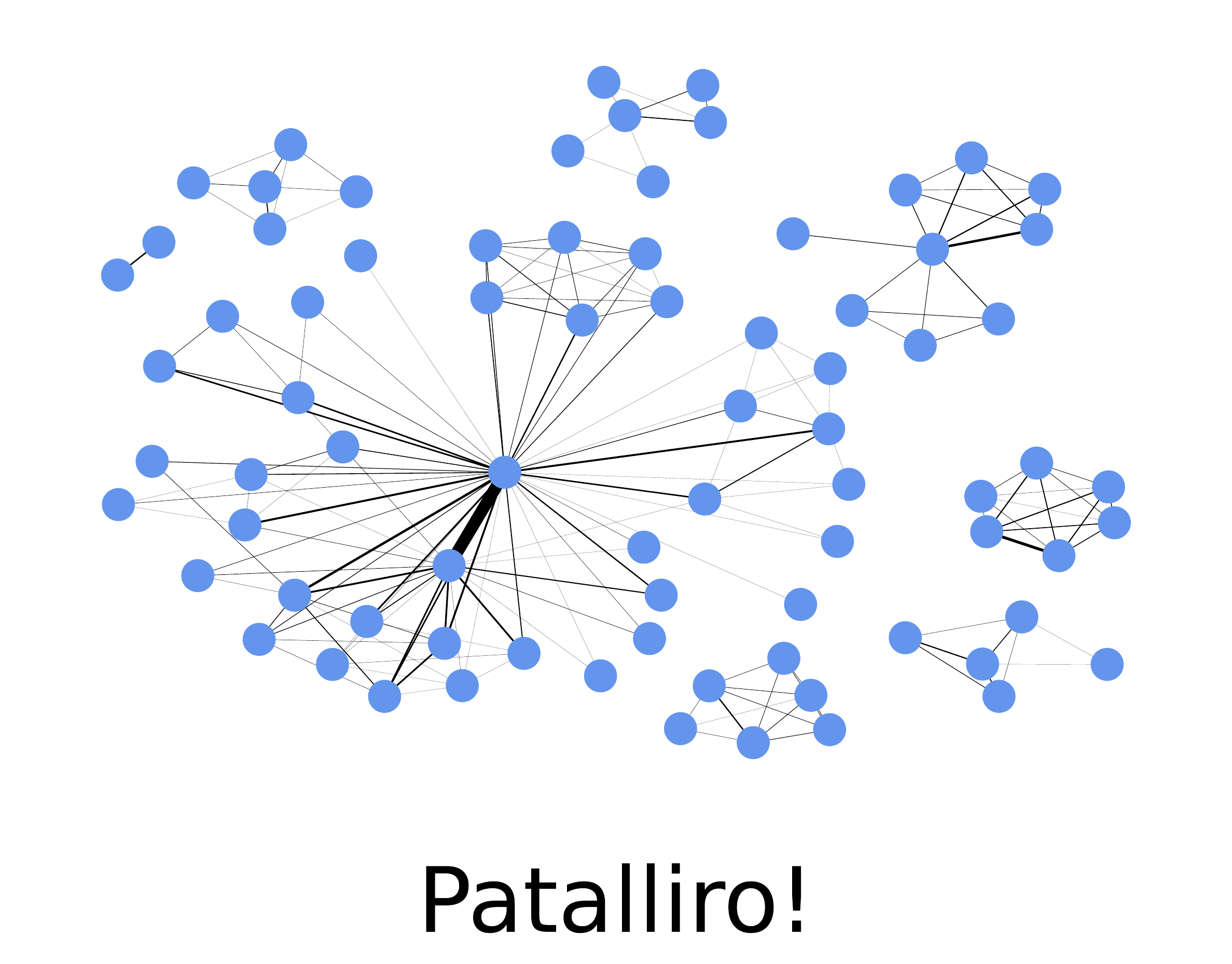}
     \end{subfigure}
     \begin{subfigure}[b]{0.195\textwidth}
         \centering
         \includegraphics[width=\textwidth]{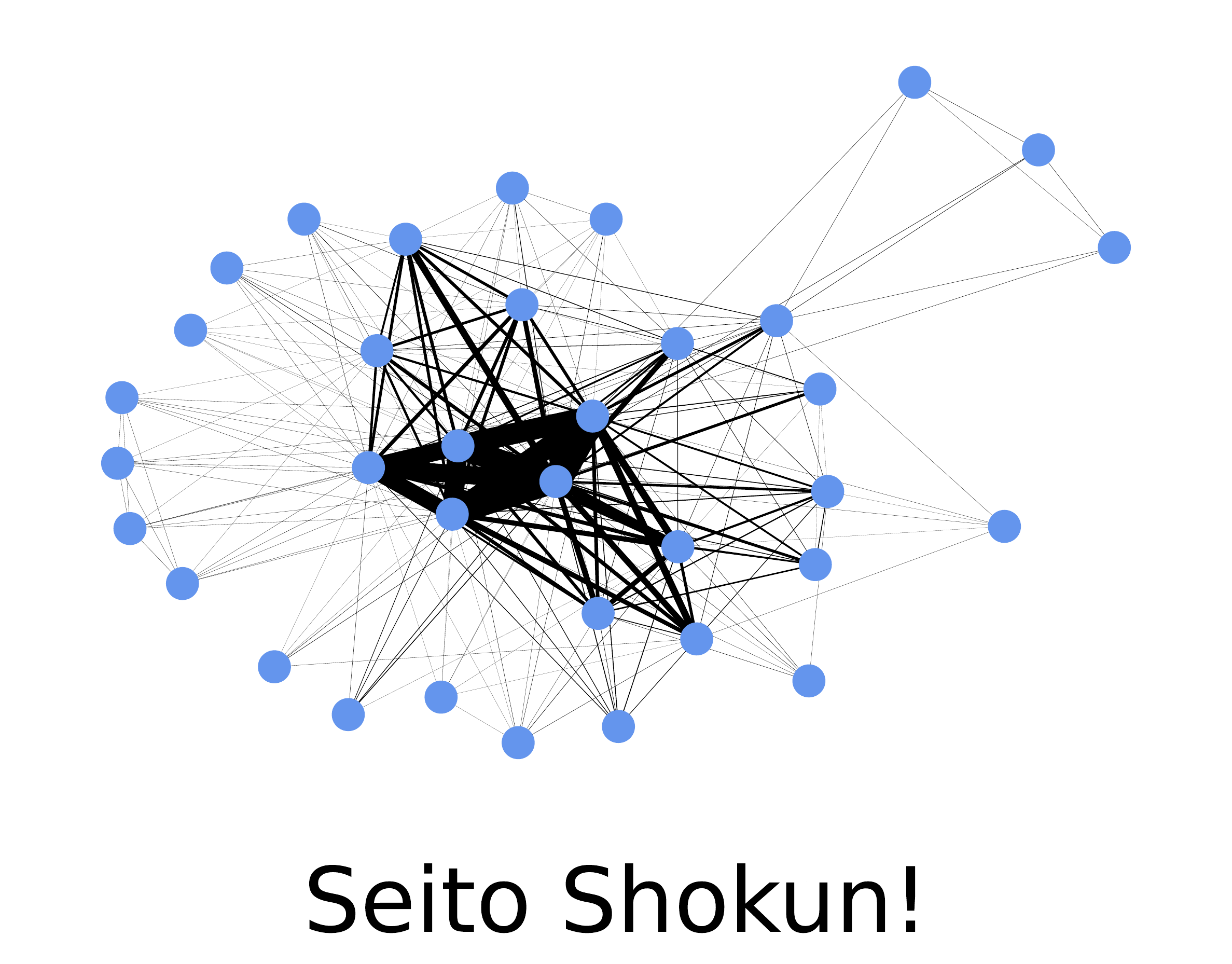}
     \end{subfigure}
     \begin{subfigure}[b]{0.195\textwidth}
         \centering
         \includegraphics[width=\textwidth]{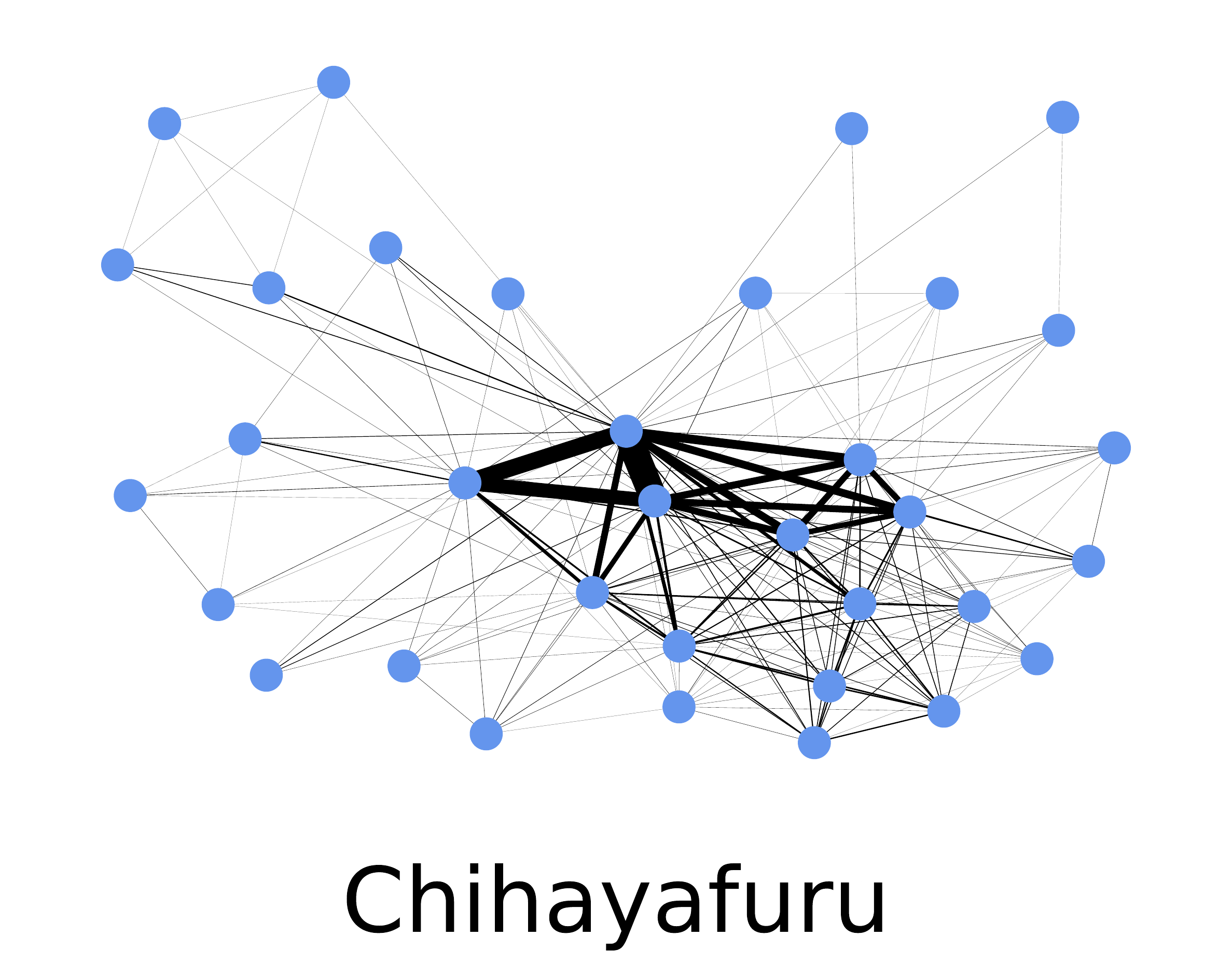}
     \end{subfigure}
     \begin{subfigure}[b]{0.195\textwidth}
         \centering
         \includegraphics[width=\textwidth]{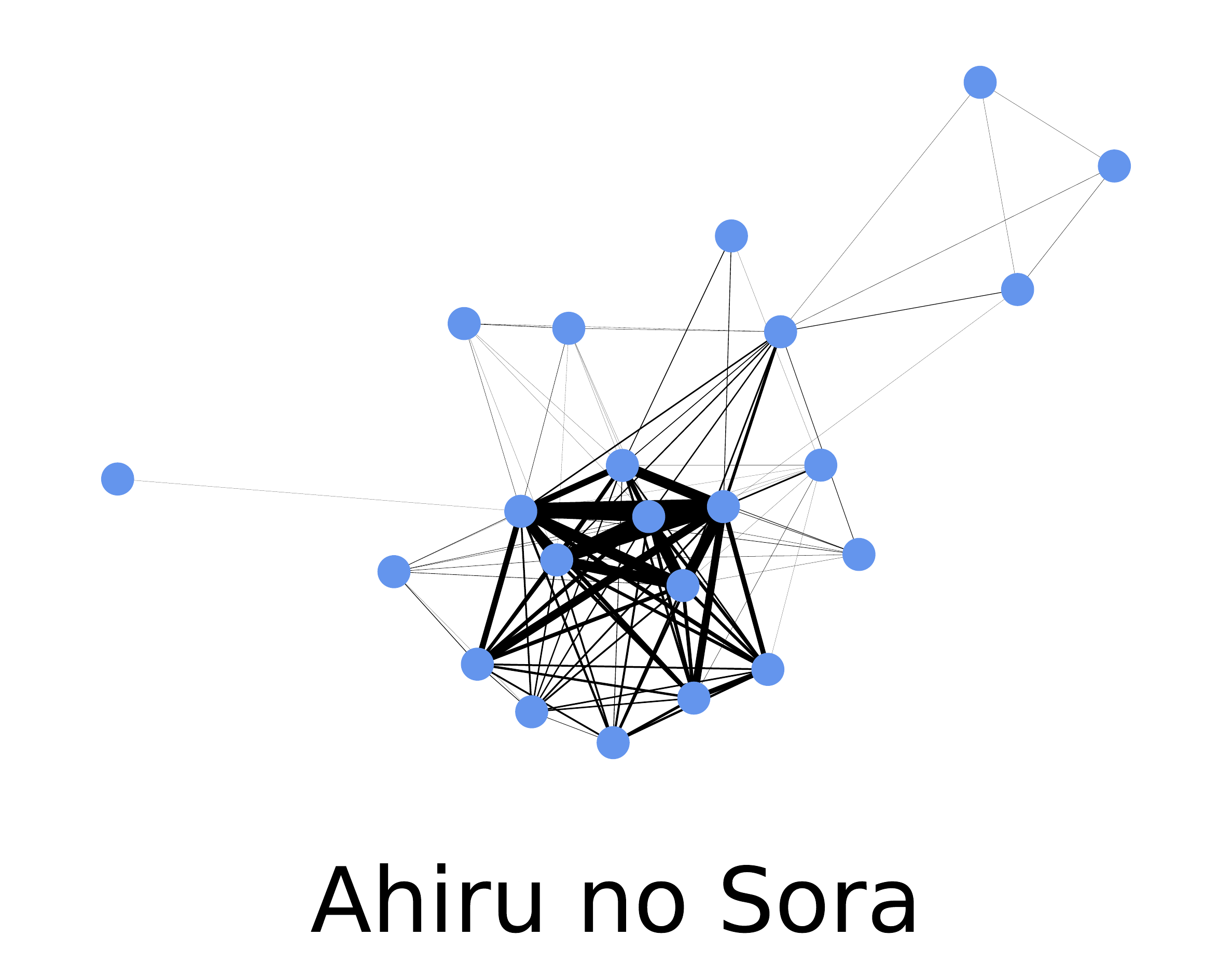}
     \end{subfigure}
     \\
      \begin{subfigure}[b]{0.195\textwidth}
         \centering
         \includegraphics[width=\textwidth]{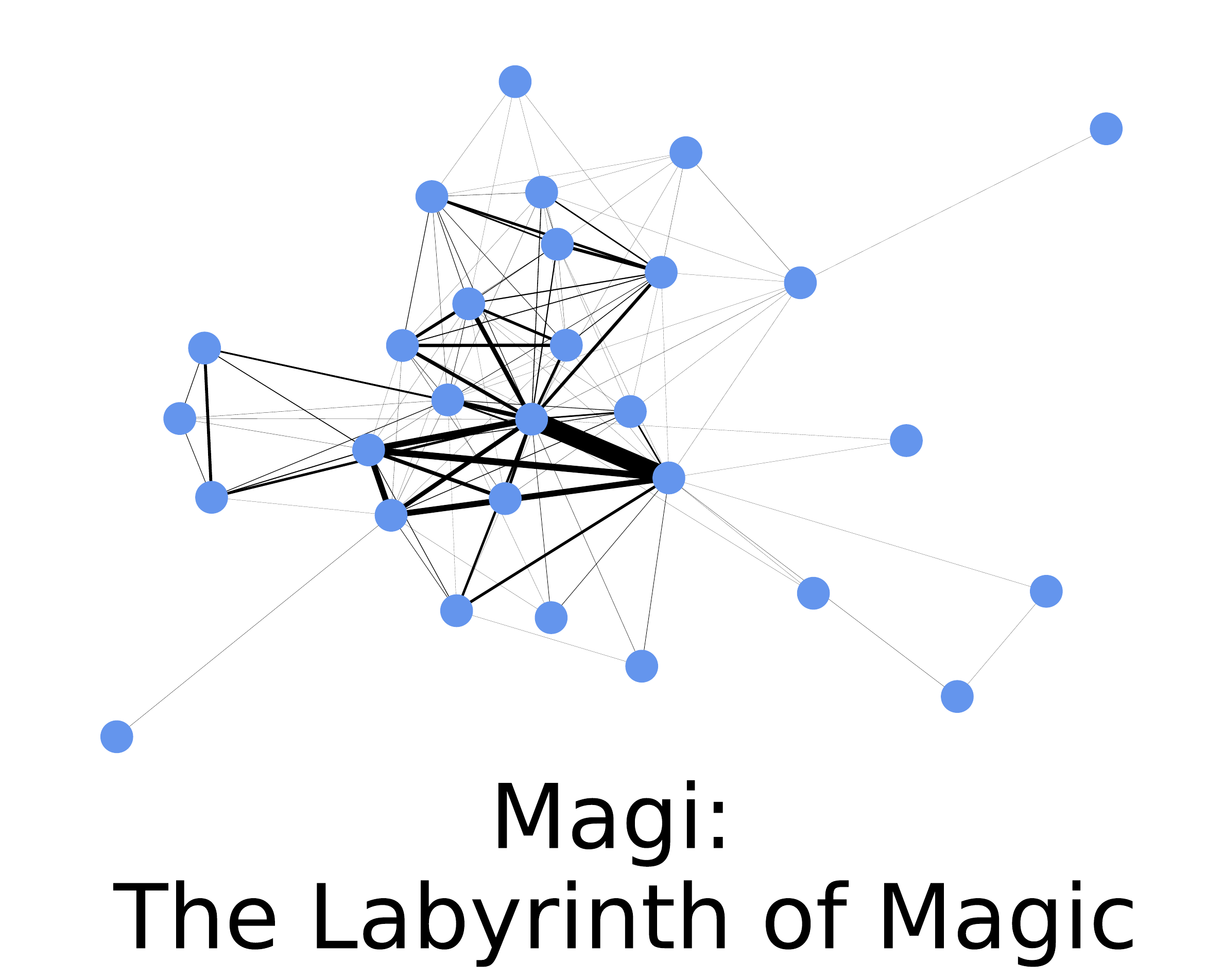}
     \end{subfigure}
     \begin{subfigure}[b]{0.195\textwidth}
         \centering
         \includegraphics[width=\textwidth]{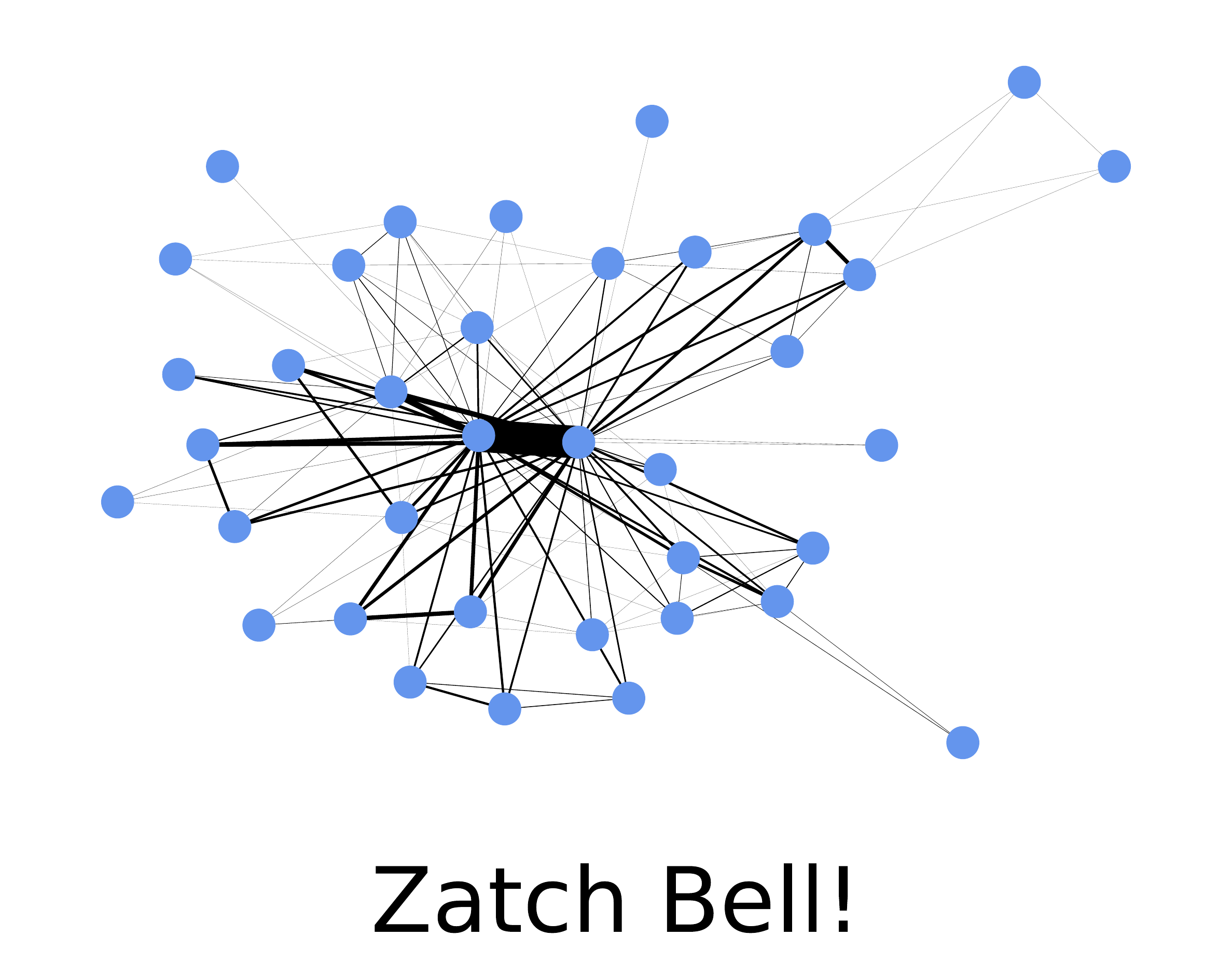}
     \end{subfigure}
     \begin{subfigure}[b]{0.195\textwidth}
         \centering
         \includegraphics[width=\textwidth]{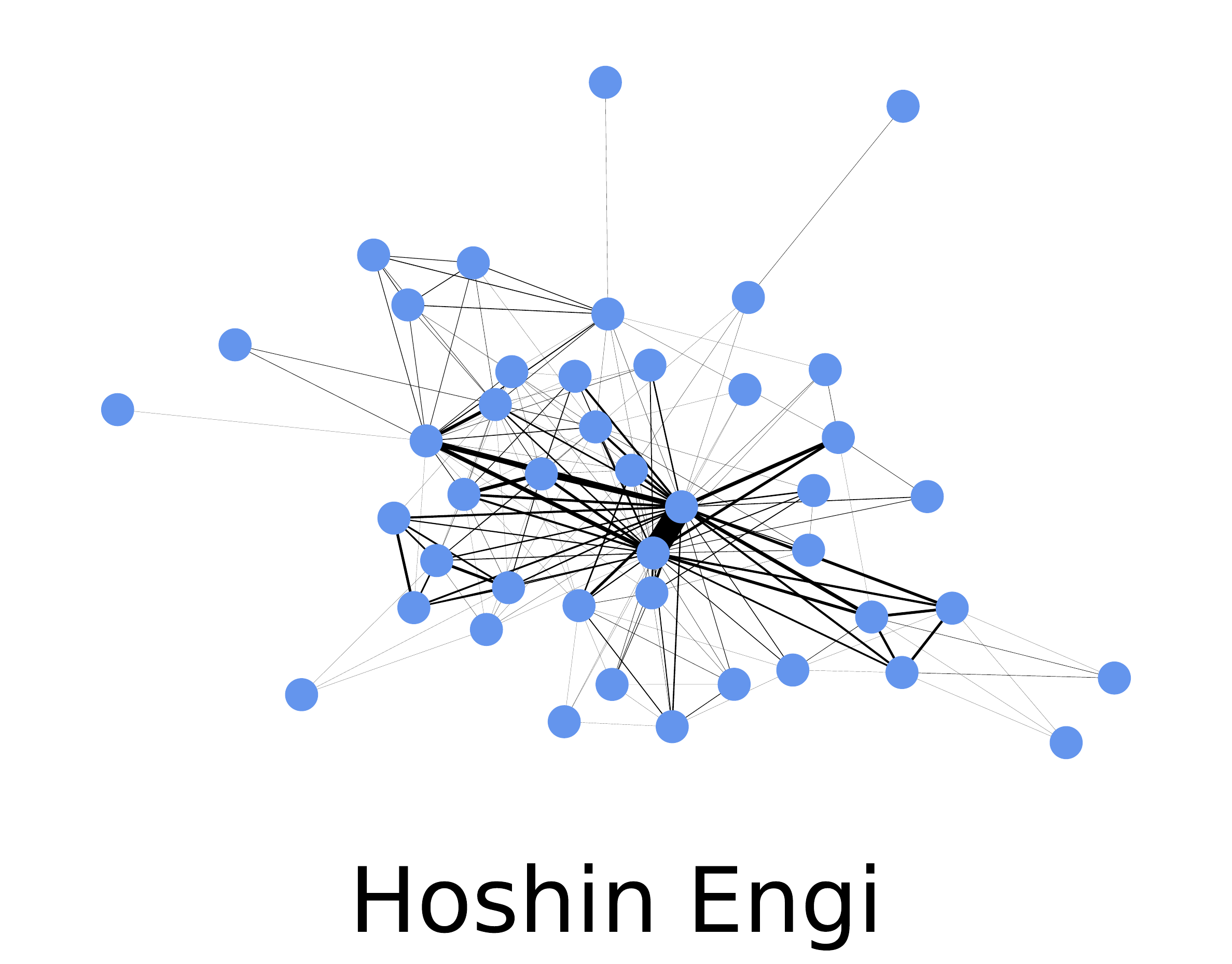}
     \end{subfigure}
     \begin{subfigure}[b]{0.195\textwidth}
         \centering
         \includegraphics[width=\textwidth]{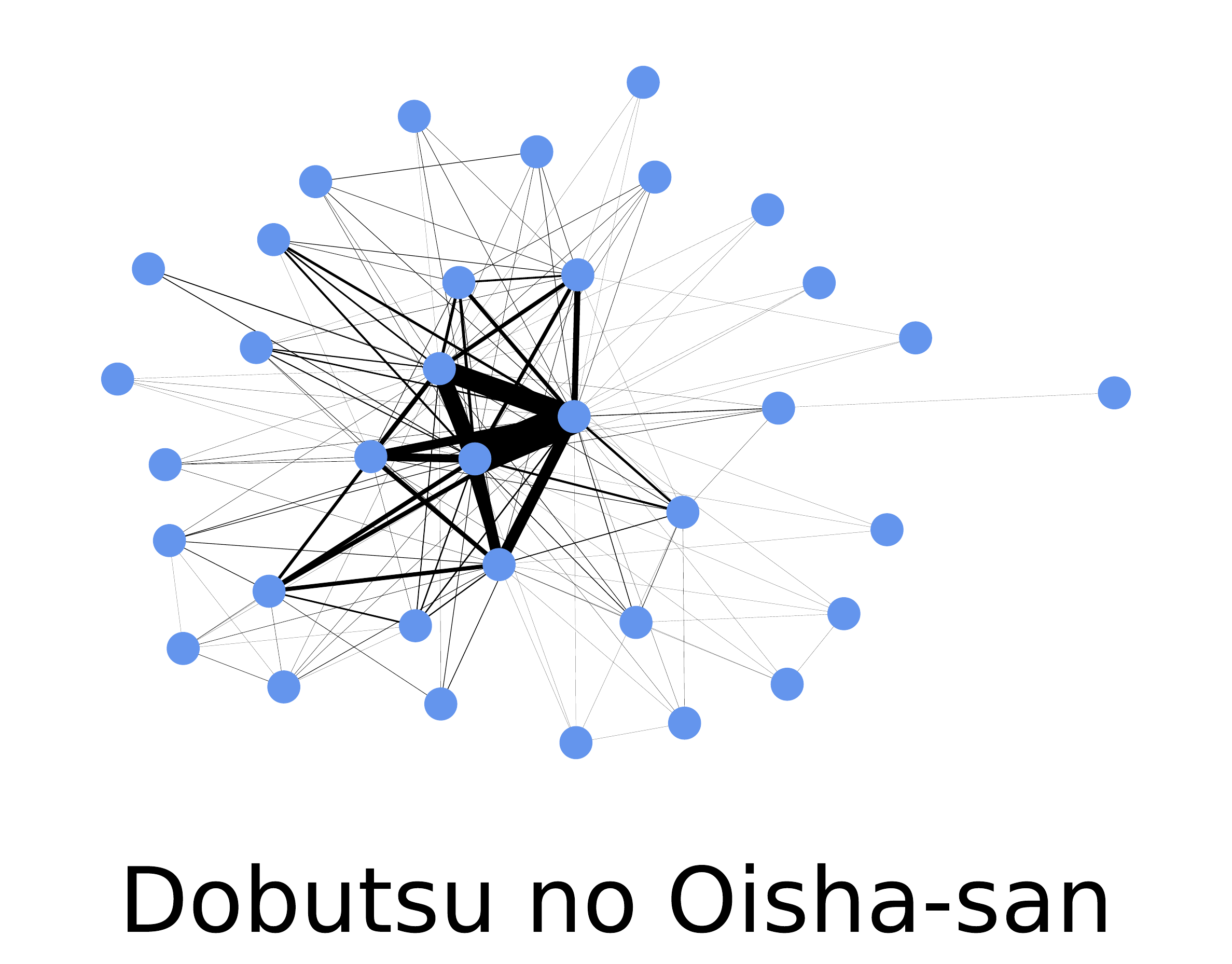}
     \end{subfigure}
     \begin{subfigure}[b]{0.195\textwidth}
         \centering
         \includegraphics[width=\textwidth]{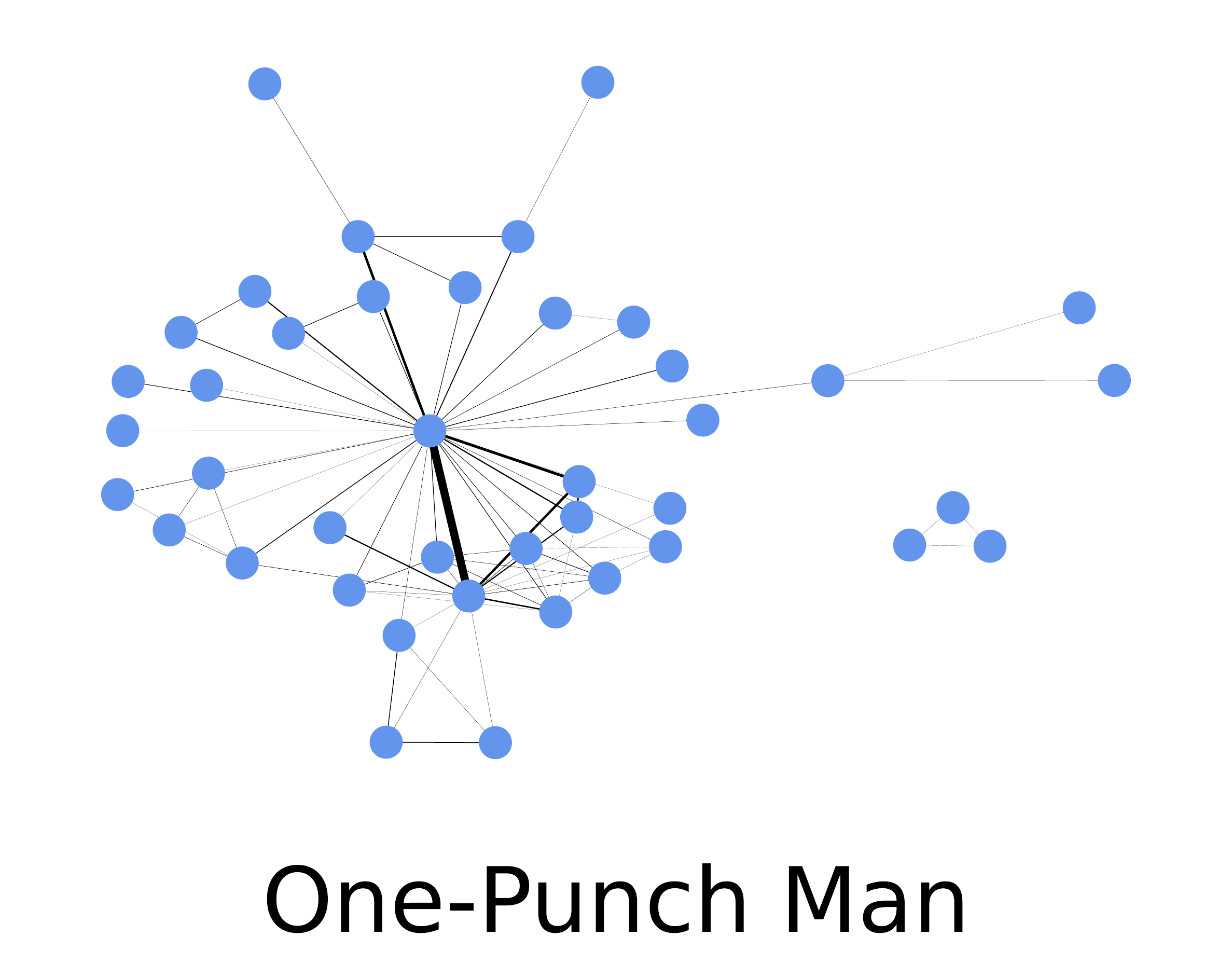}
     \end{subfigure}
     \\
      \begin{subfigure}[b]{0.195\textwidth}
         \centering
         \includegraphics[width=\textwidth]{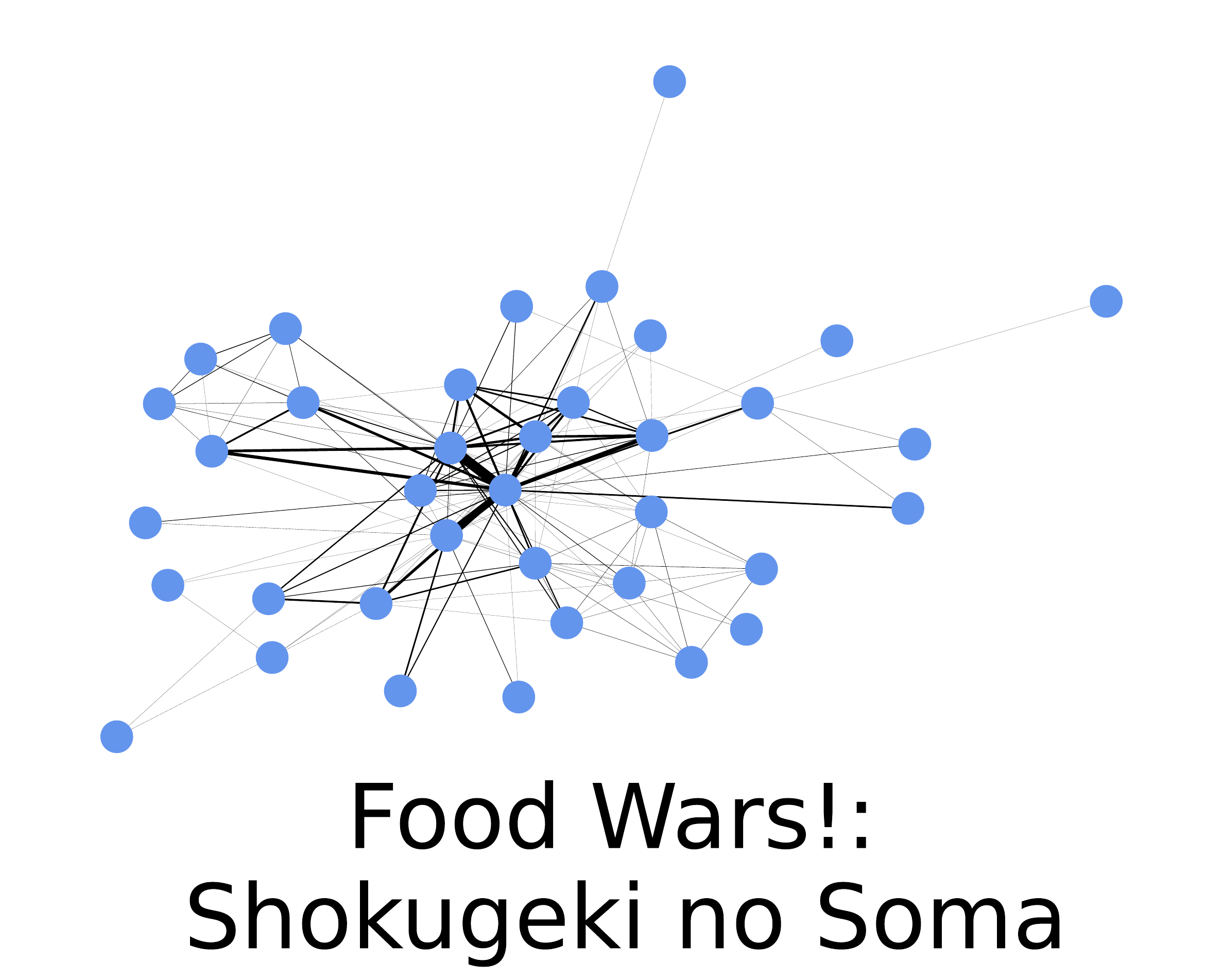}
     \end{subfigure}
     \begin{subfigure}[b]{0.195\textwidth}
         \centering
         \includegraphics[width=\textwidth]{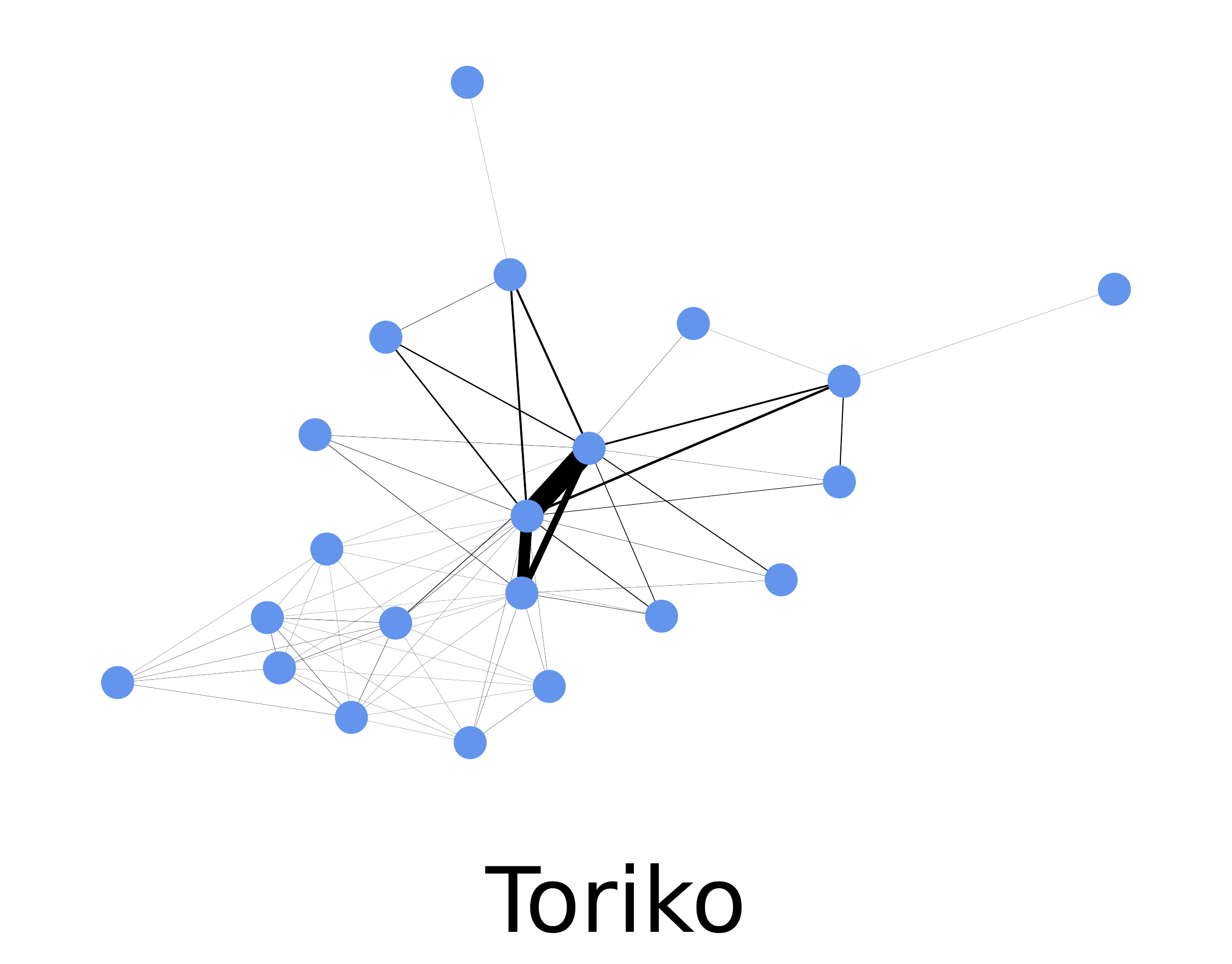}
     \end{subfigure}
     \begin{subfigure}[b]{0.195\textwidth}
         \centering
         \includegraphics[width=\textwidth]{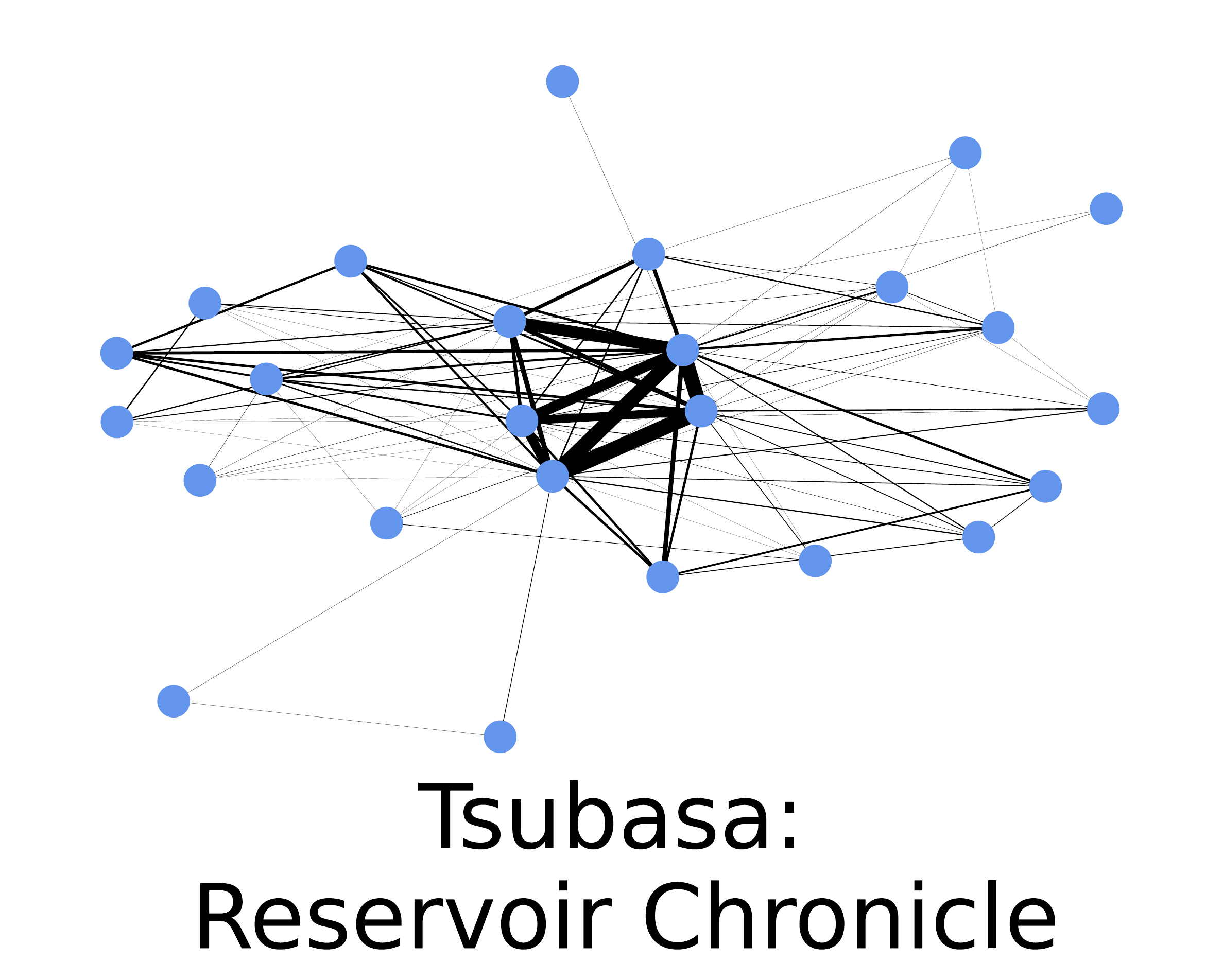}
     \end{subfigure}
     \begin{subfigure}[b]{0.195\textwidth}
         \centering
         \includegraphics[width=\textwidth]{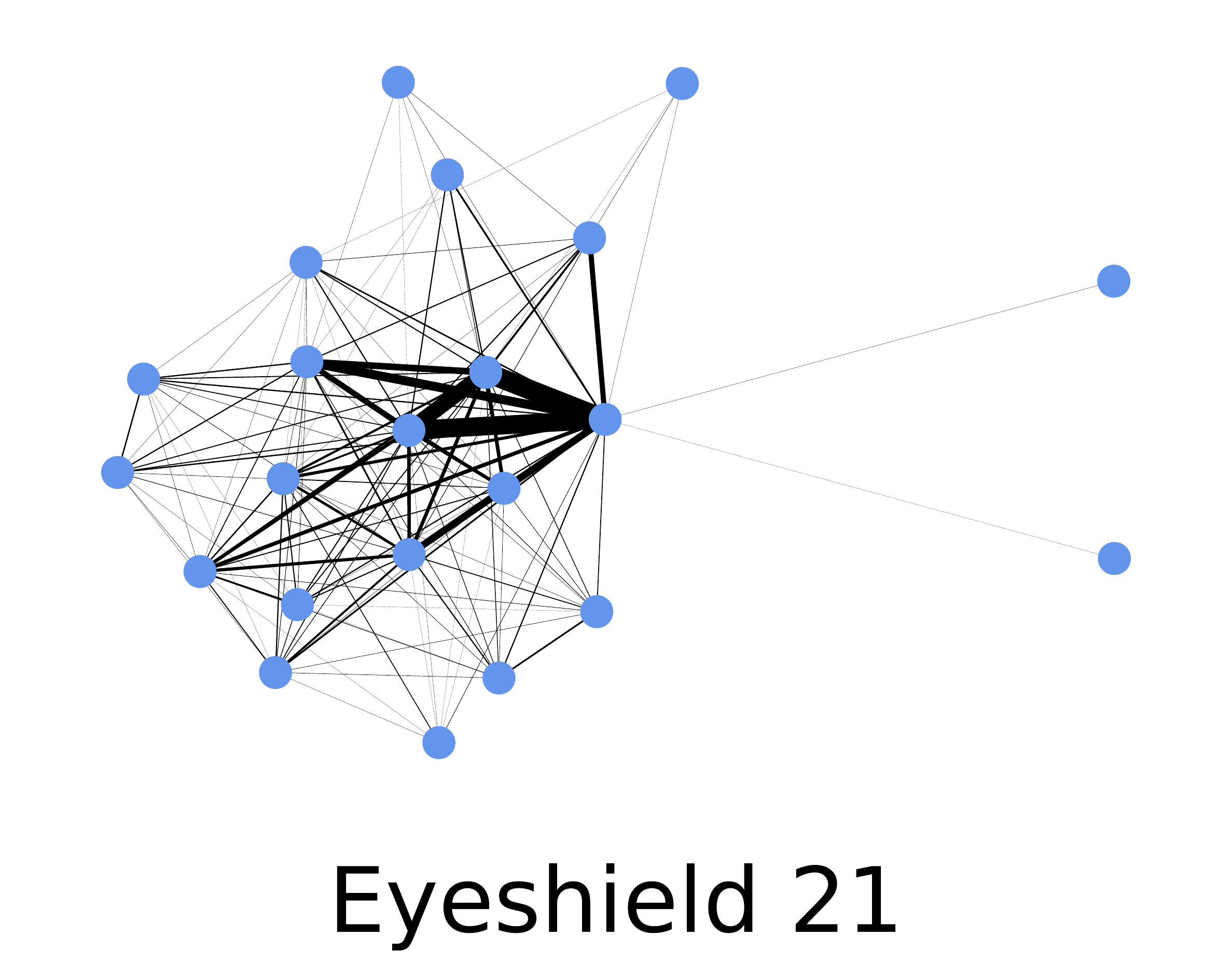}
     \end{subfigure}
     \begin{subfigure}[b]{0.195\textwidth}
         \centering
         \includegraphics[width=\textwidth]{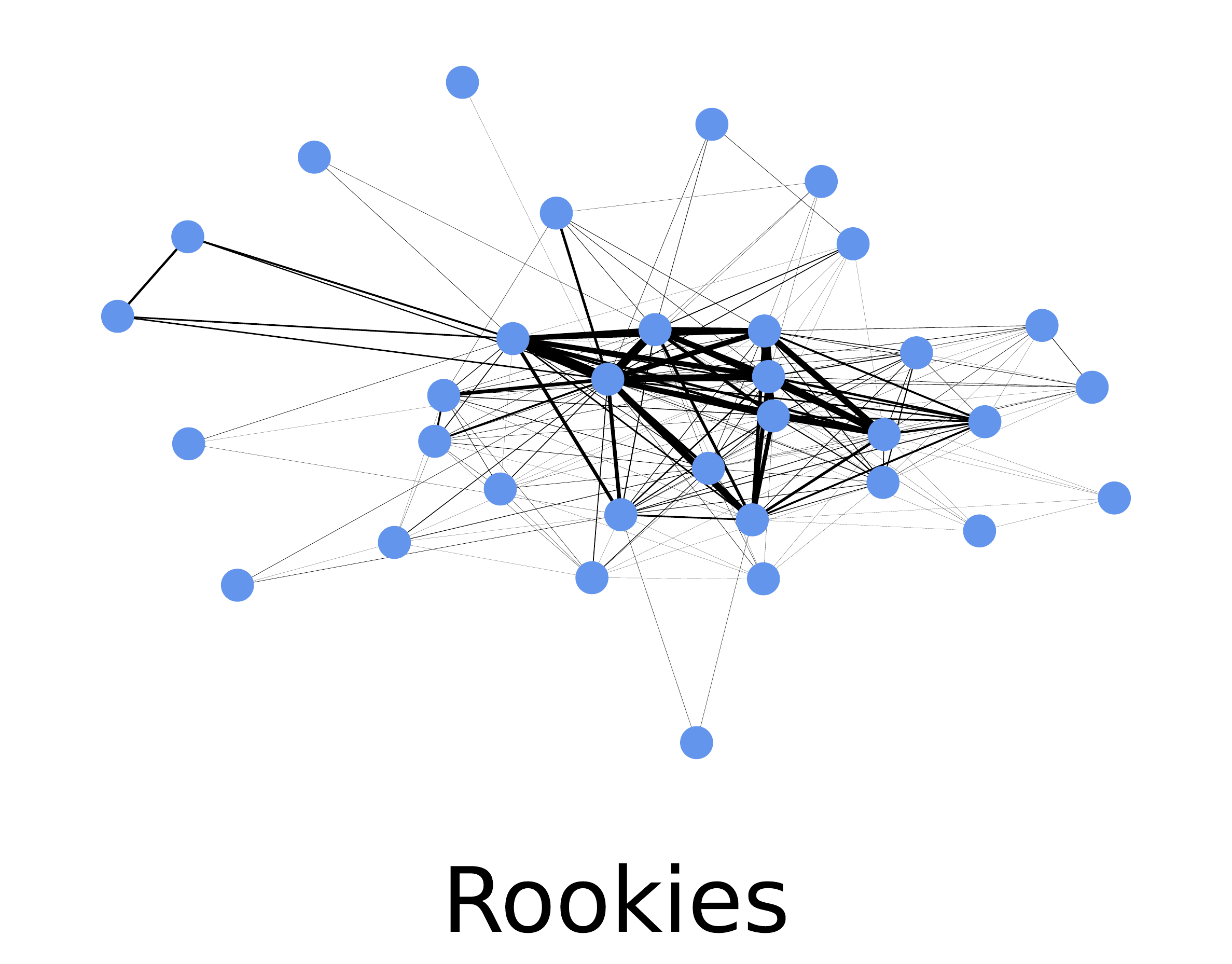}
     \end{subfigure}
     
    \caption{Character networks for 162 manga (continued)}
    
    \label{fig:three graphs}
\end{figure*}

\begin{figure*}\ContinuedFloat
     \centering
      \begin{subfigure}[b]{0.195\textwidth}
         \centering
         \includegraphics[width=\textwidth]{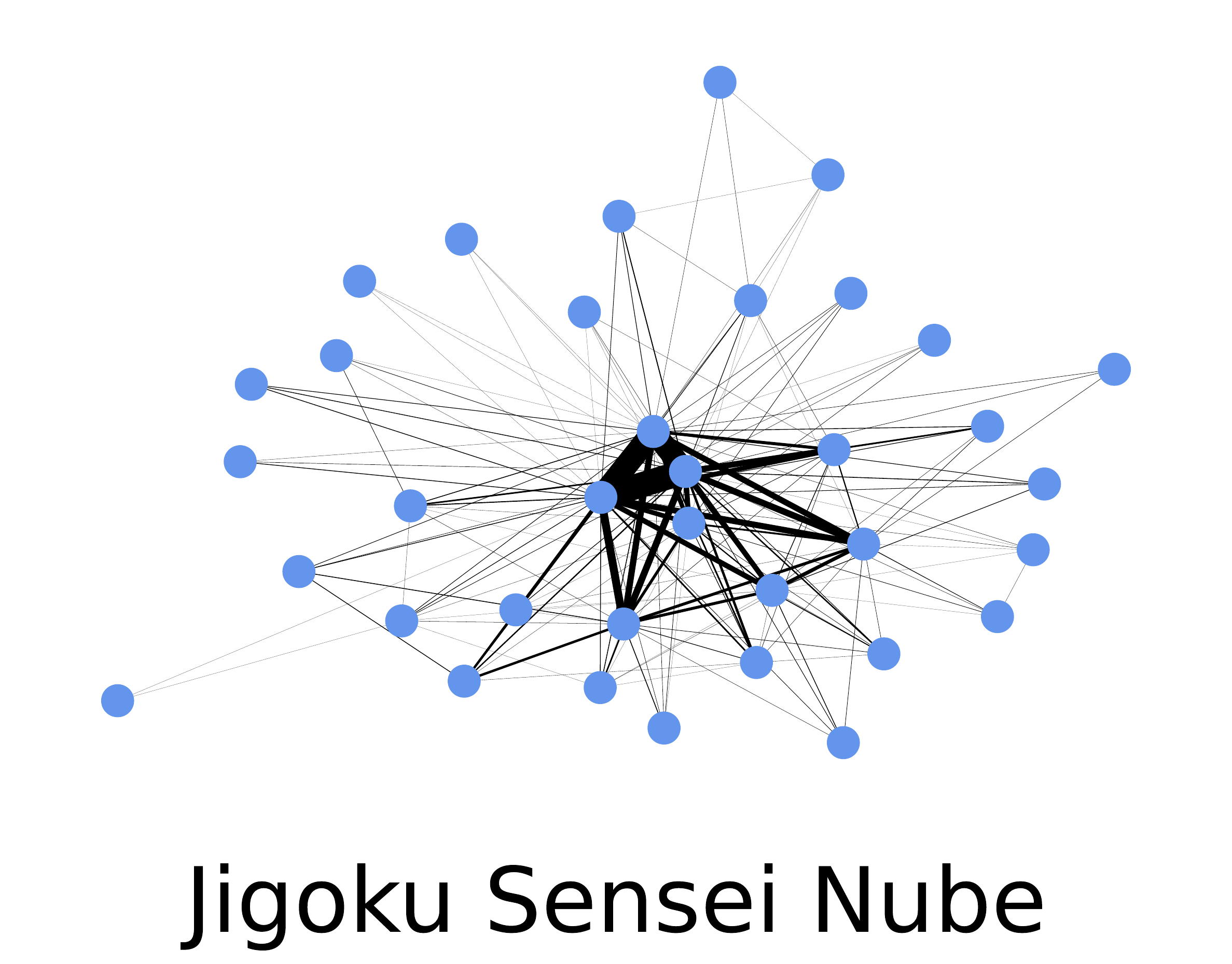}
     \end{subfigure}
     \begin{subfigure}[b]{0.195\textwidth}
         \centering
         \includegraphics[width=\textwidth]{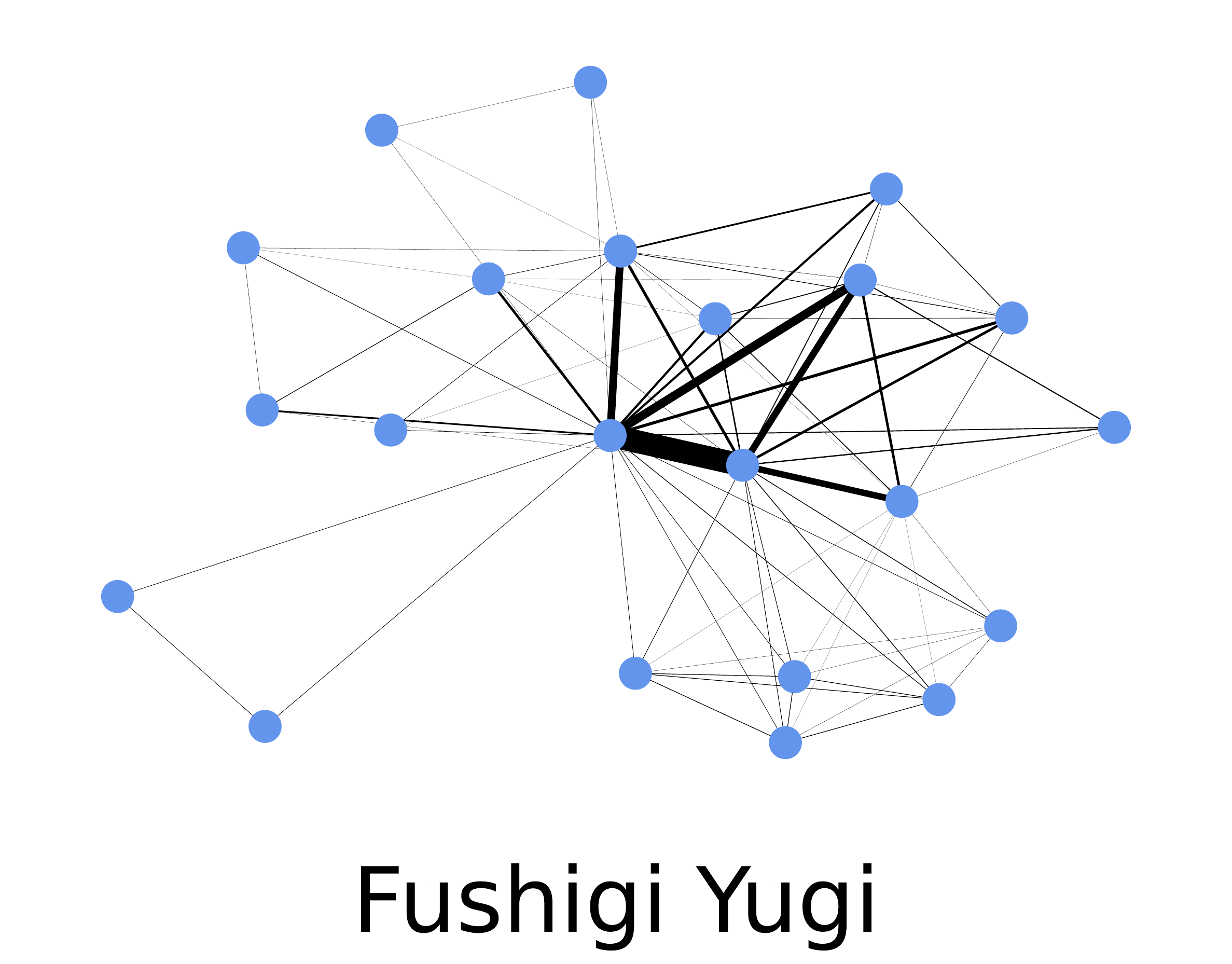}
     \end{subfigure}
     \begin{subfigure}[b]{0.195\textwidth}
         \centering
         \includegraphics[width=\textwidth]{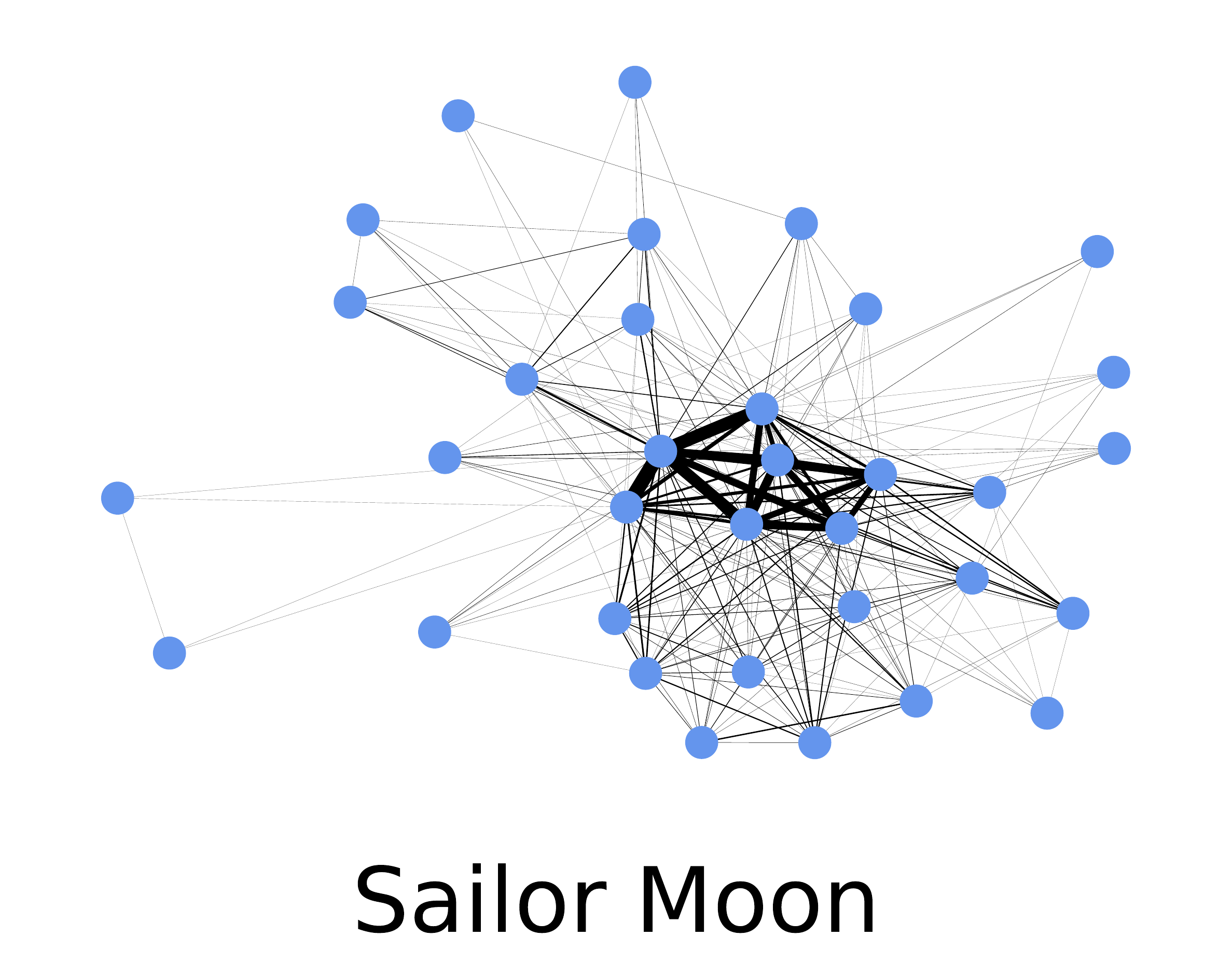}
     \end{subfigure}
     \begin{subfigure}[b]{0.195\textwidth}
         \centering
         \includegraphics[width=\textwidth]{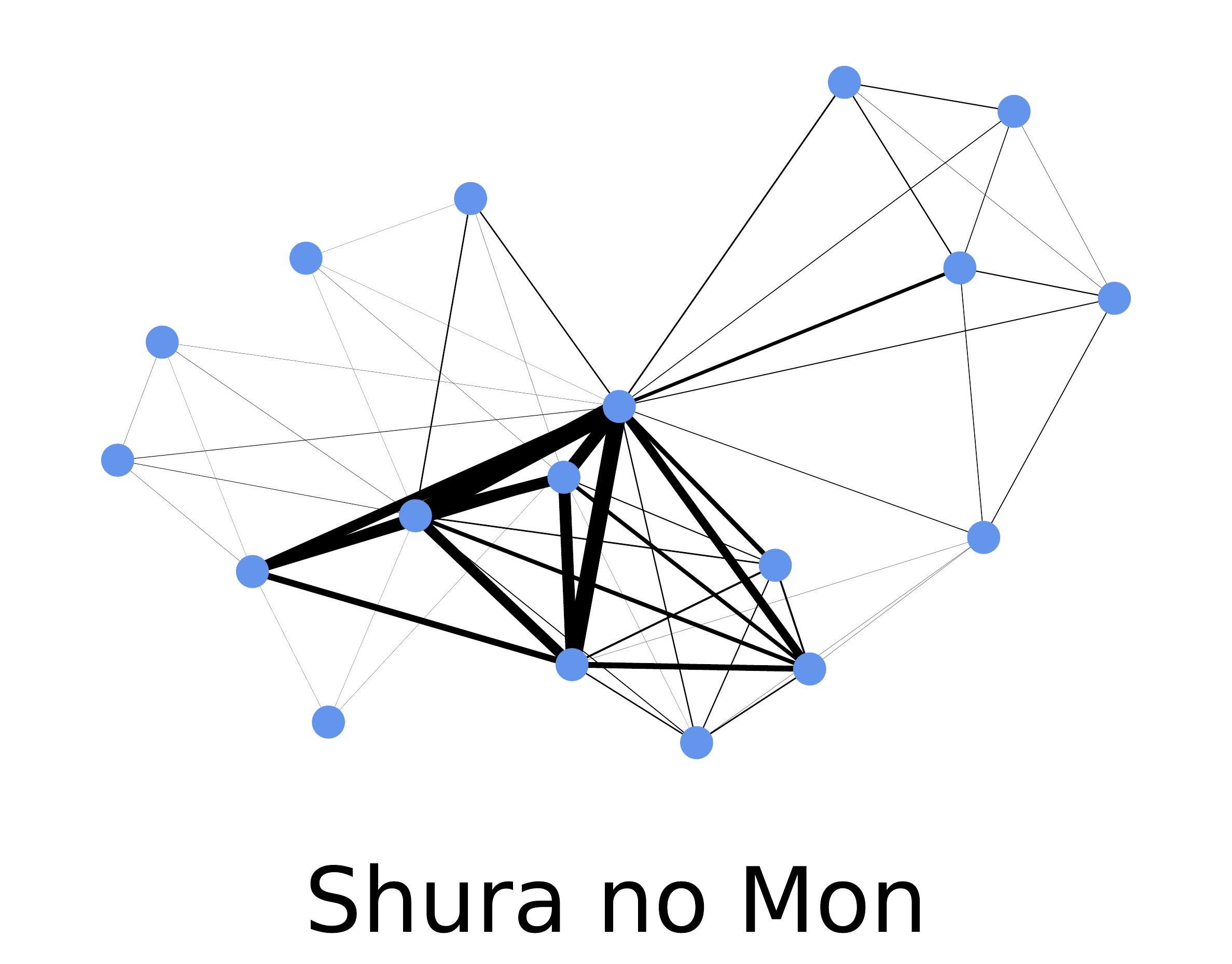}
     \end{subfigure}
     \begin{subfigure}[b]{0.195\textwidth}
         \centering
         \includegraphics[width=\textwidth]{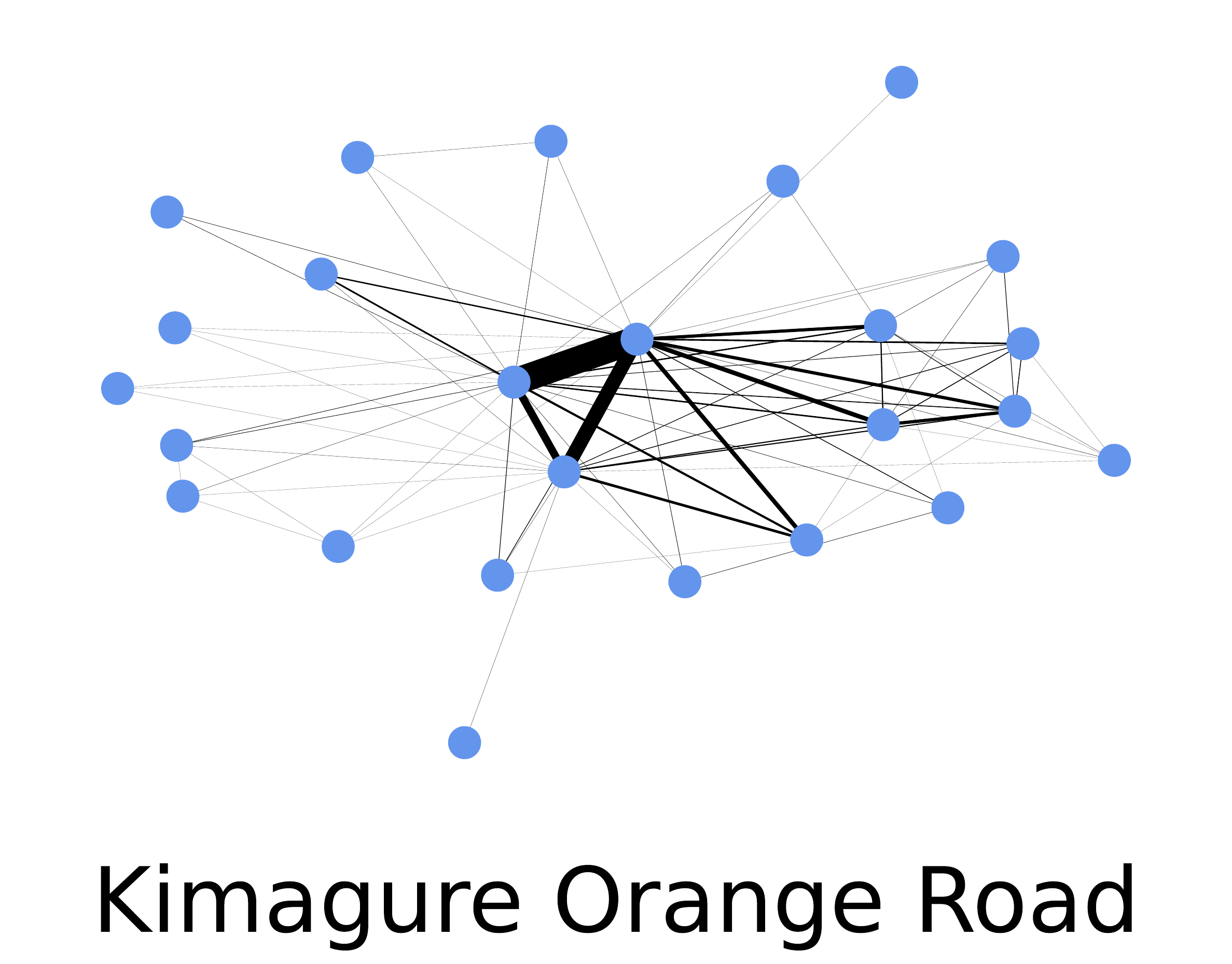}
     \end{subfigure}
     \\
      \begin{subfigure}[b]{0.195\textwidth}
         \centering
         \includegraphics[width=\textwidth]{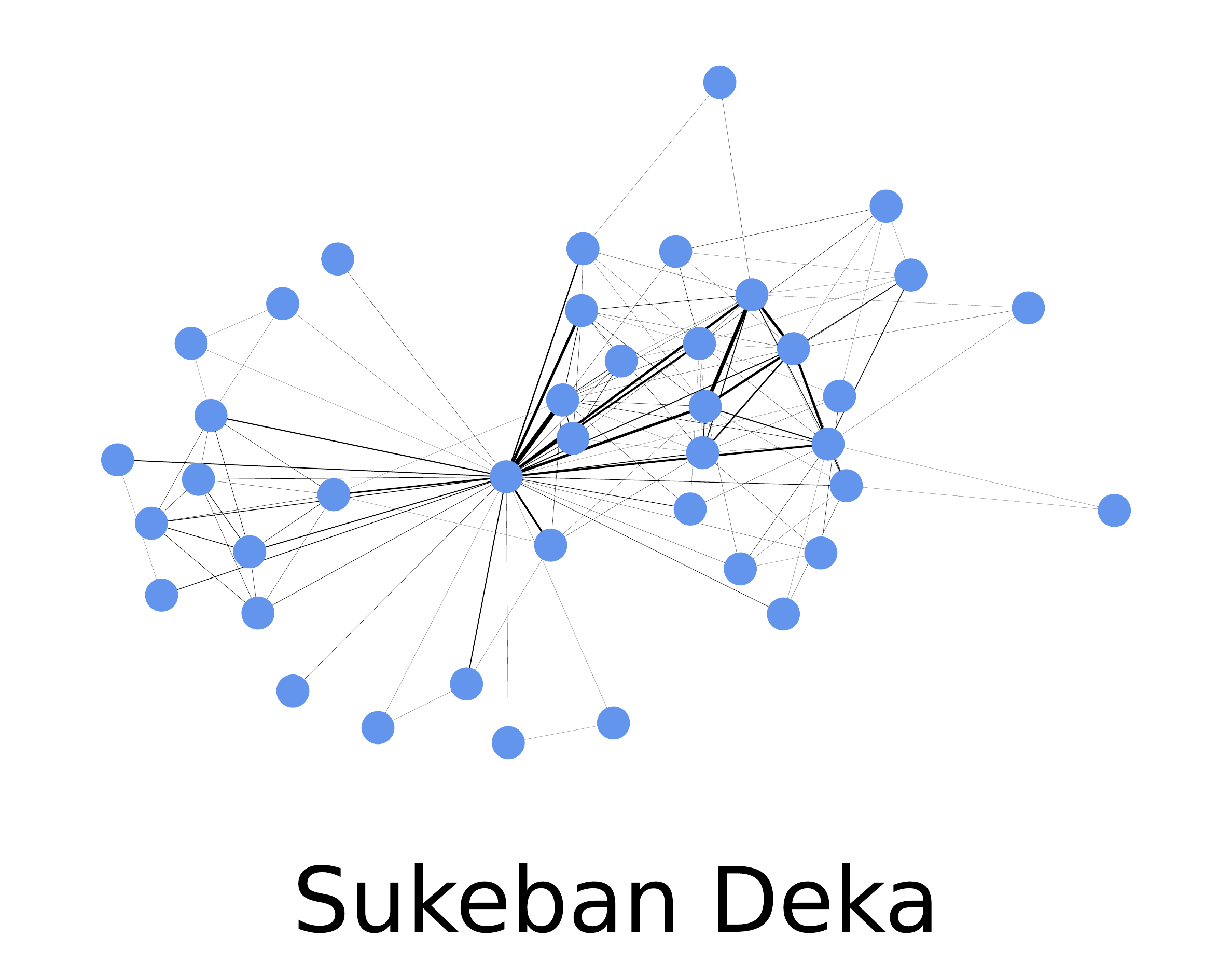}
     \end{subfigure}
     \begin{subfigure}[b]{0.195\textwidth}
         \centering
         \includegraphics[width=\textwidth]{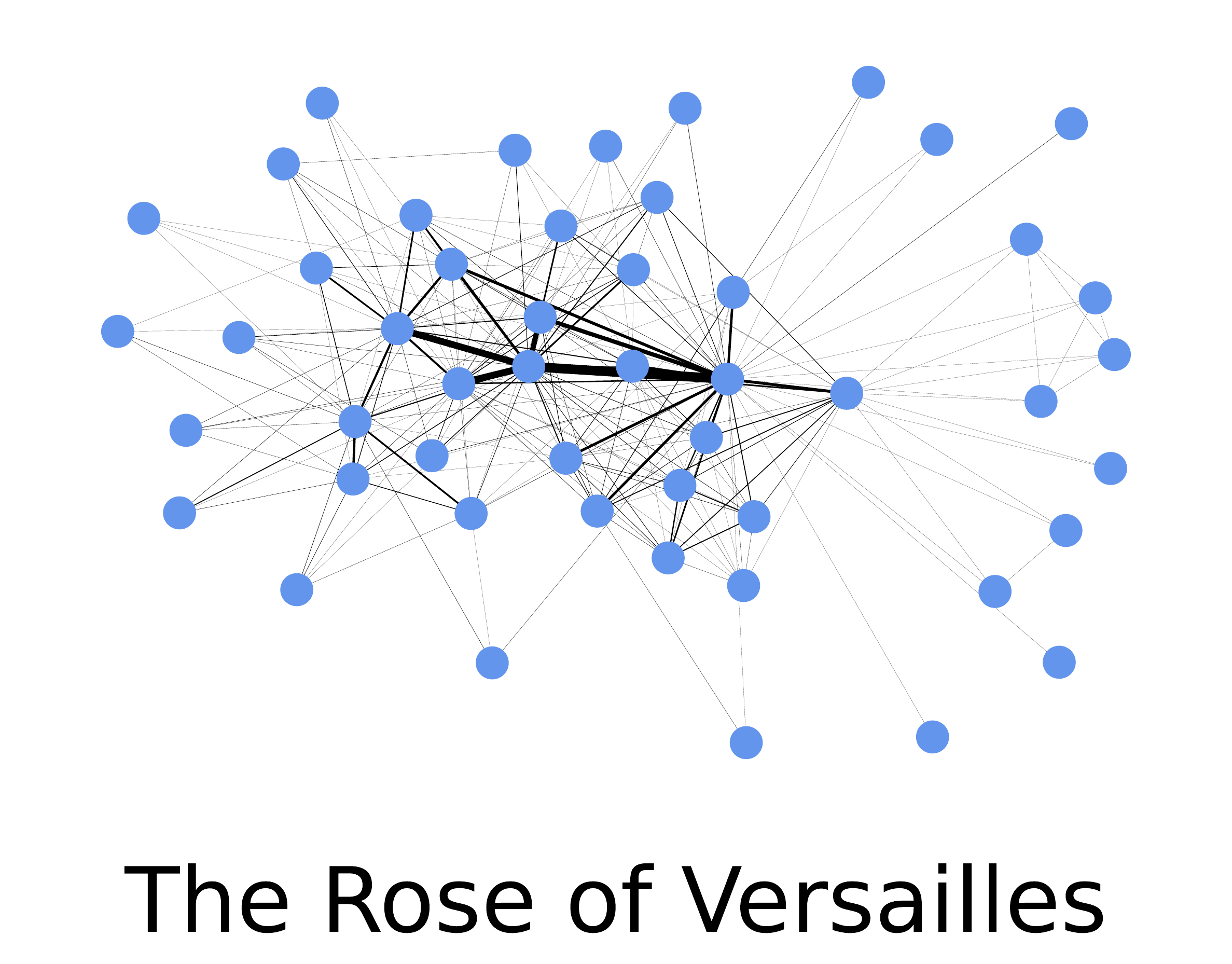}
     \end{subfigure}
     \begin{subfigure}[b]{0.195\textwidth}
         \centering
         \includegraphics[width=\textwidth]{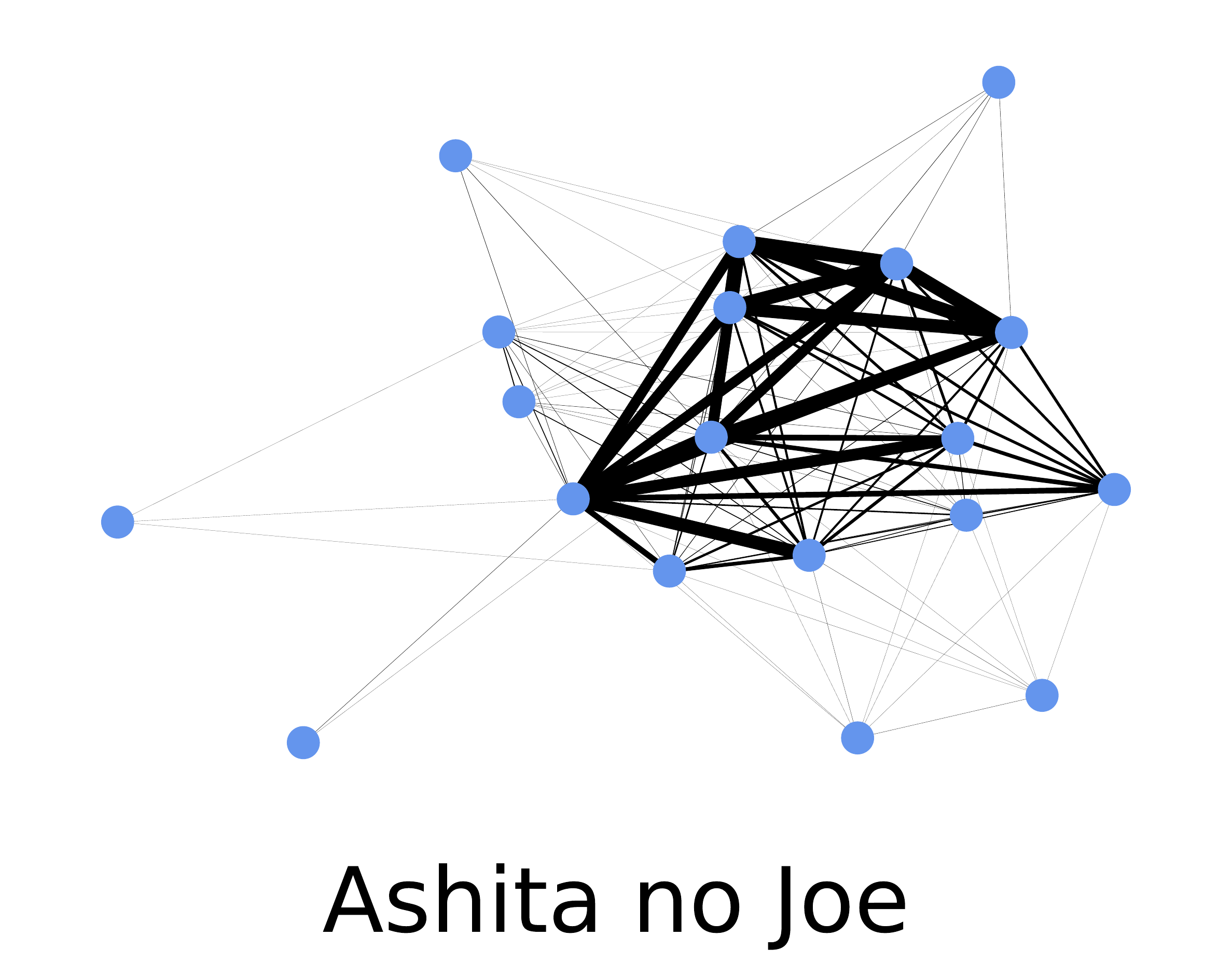}
     \end{subfigure}
     \begin{subfigure}[b]{0.195\textwidth}
         \centering
         \includegraphics[width=\textwidth]{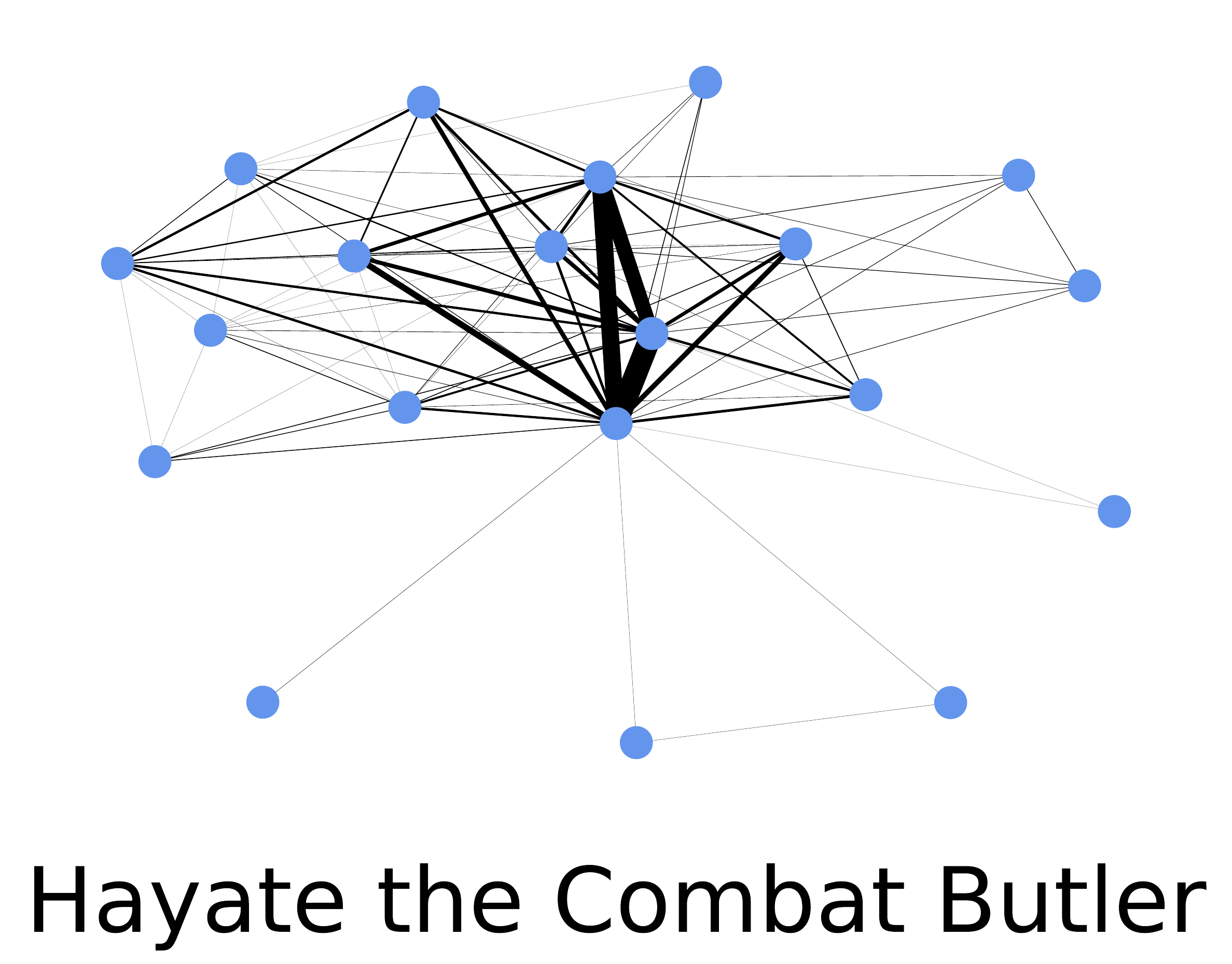}
     \end{subfigure}
     \begin{subfigure}[b]{0.195\textwidth}
         \centering
         \includegraphics[width=\textwidth]{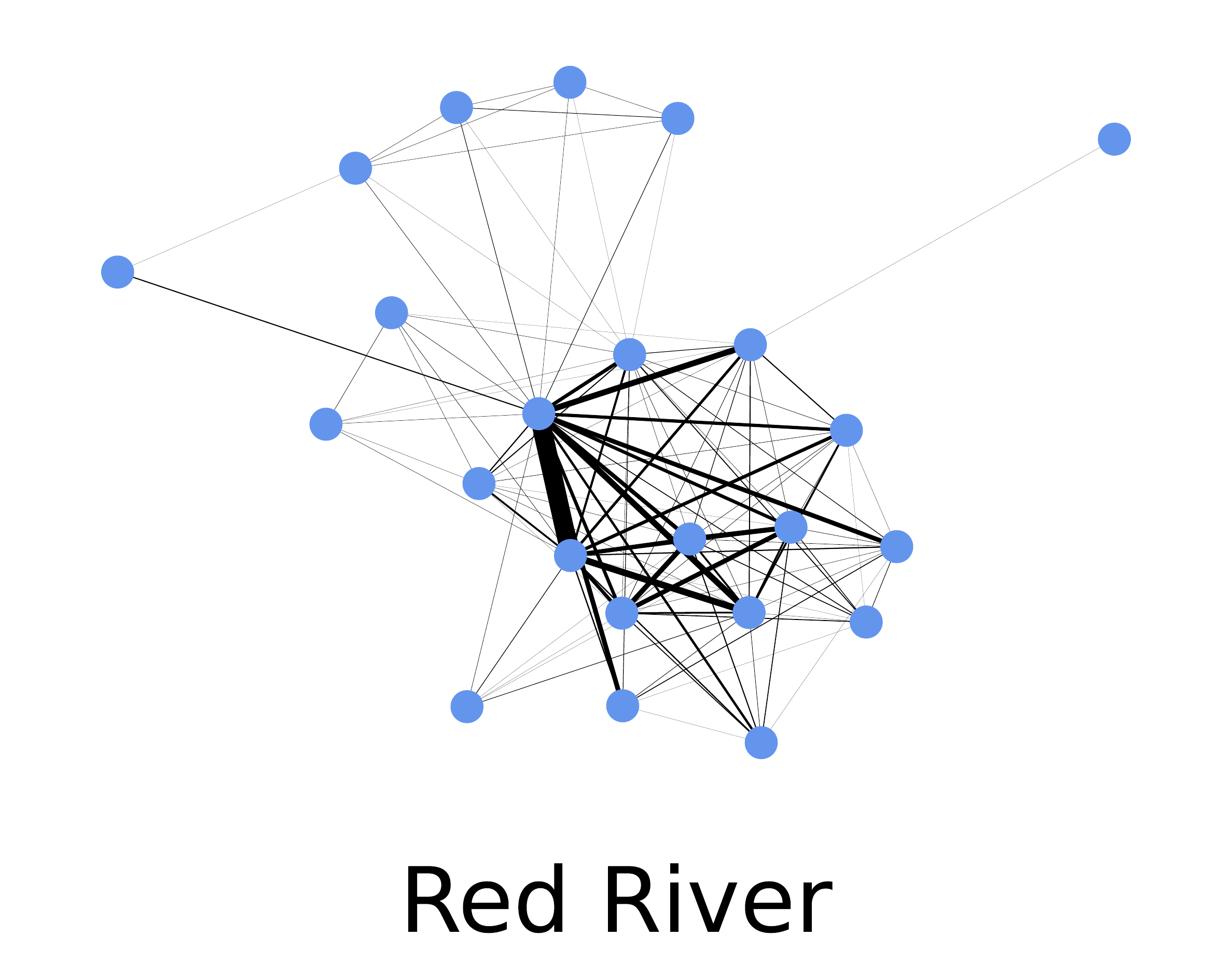}
     \end{subfigure}
    \\
     \begin{subfigure}[b]{0.195\textwidth}
         \centering
         \includegraphics[width=\textwidth]{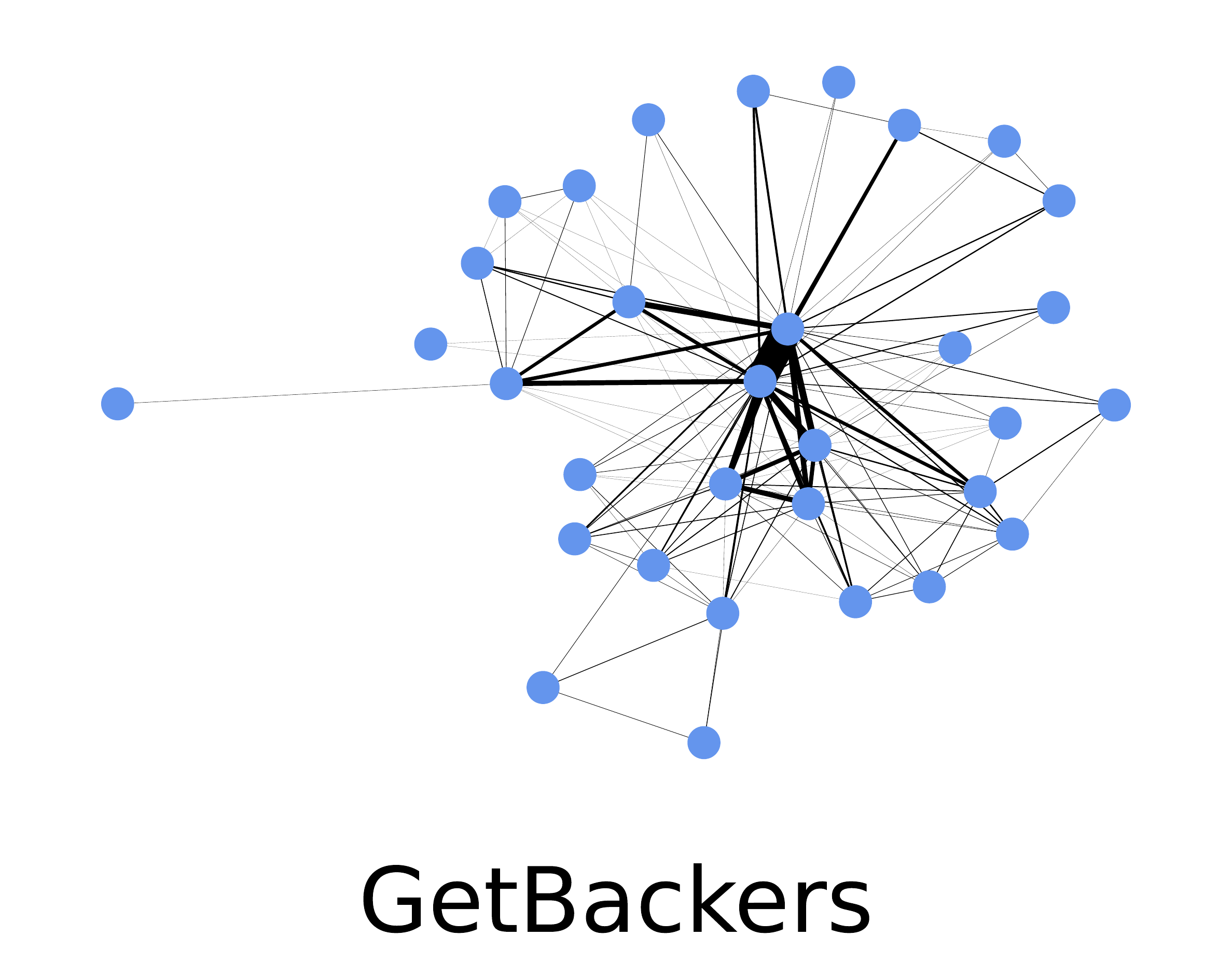}
     \end{subfigure}
     \begin{subfigure}[b]{0.195\textwidth}
         \centering
         \includegraphics[width=\textwidth]{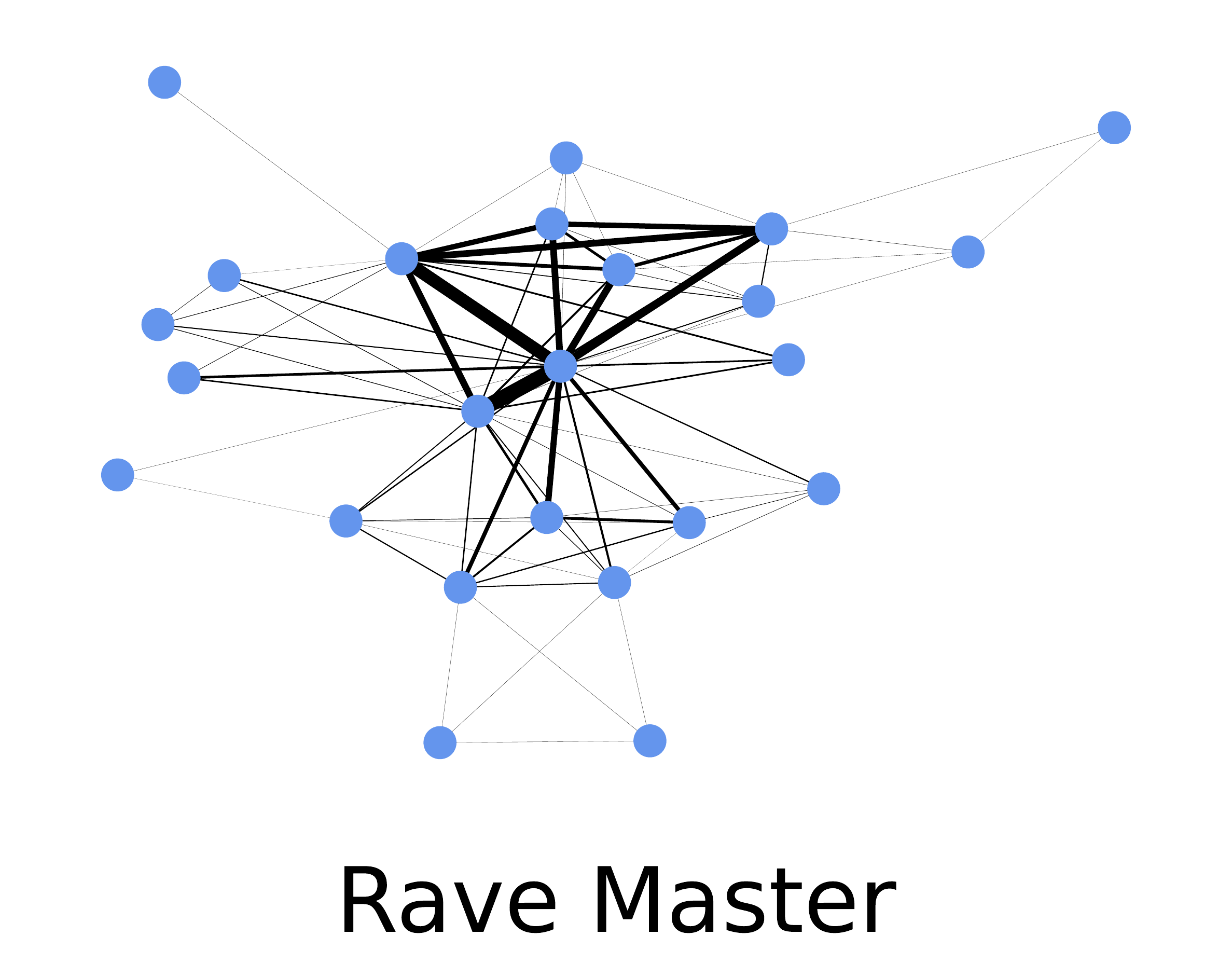}
     \end{subfigure}
     \begin{subfigure}[b]{0.195\textwidth}
         \centering
         \includegraphics[width=\textwidth]{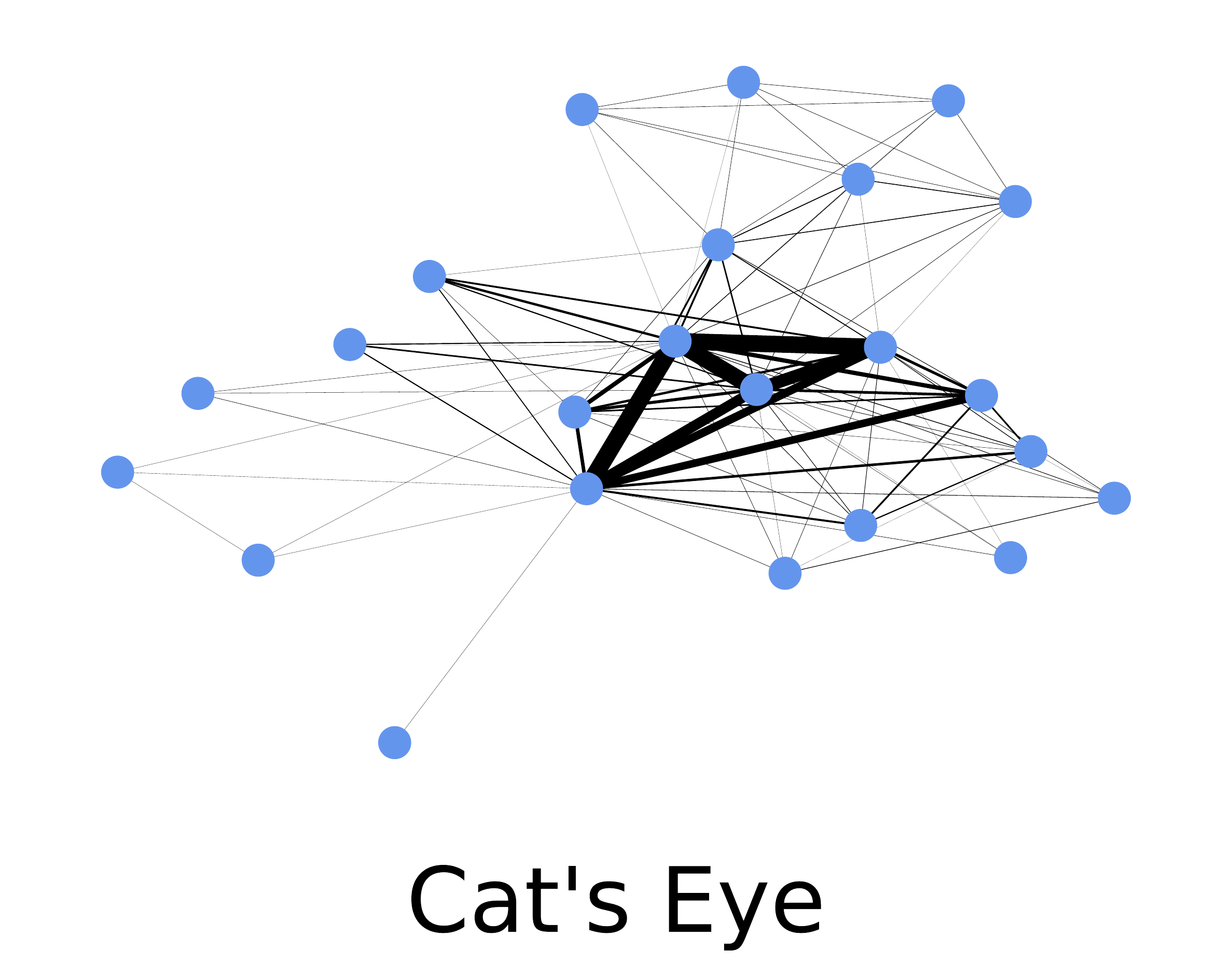}
     \end{subfigure}
     \begin{subfigure}[b]{0.195\textwidth}
         \centering
         \includegraphics[width=\textwidth]{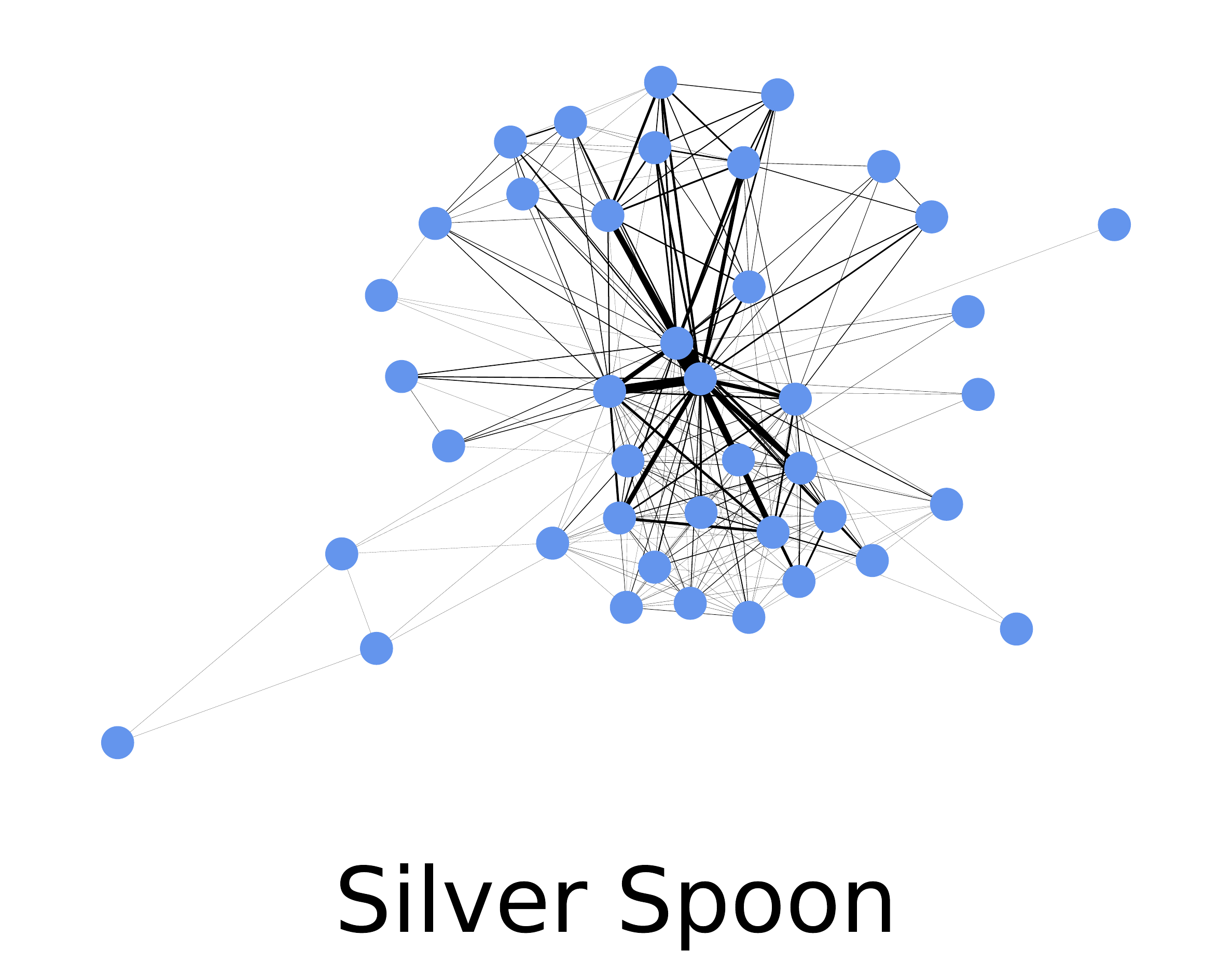}
     \end{subfigure}
     \begin{subfigure}[b]{0.195\textwidth}
         \centering
         \includegraphics[width=\textwidth]{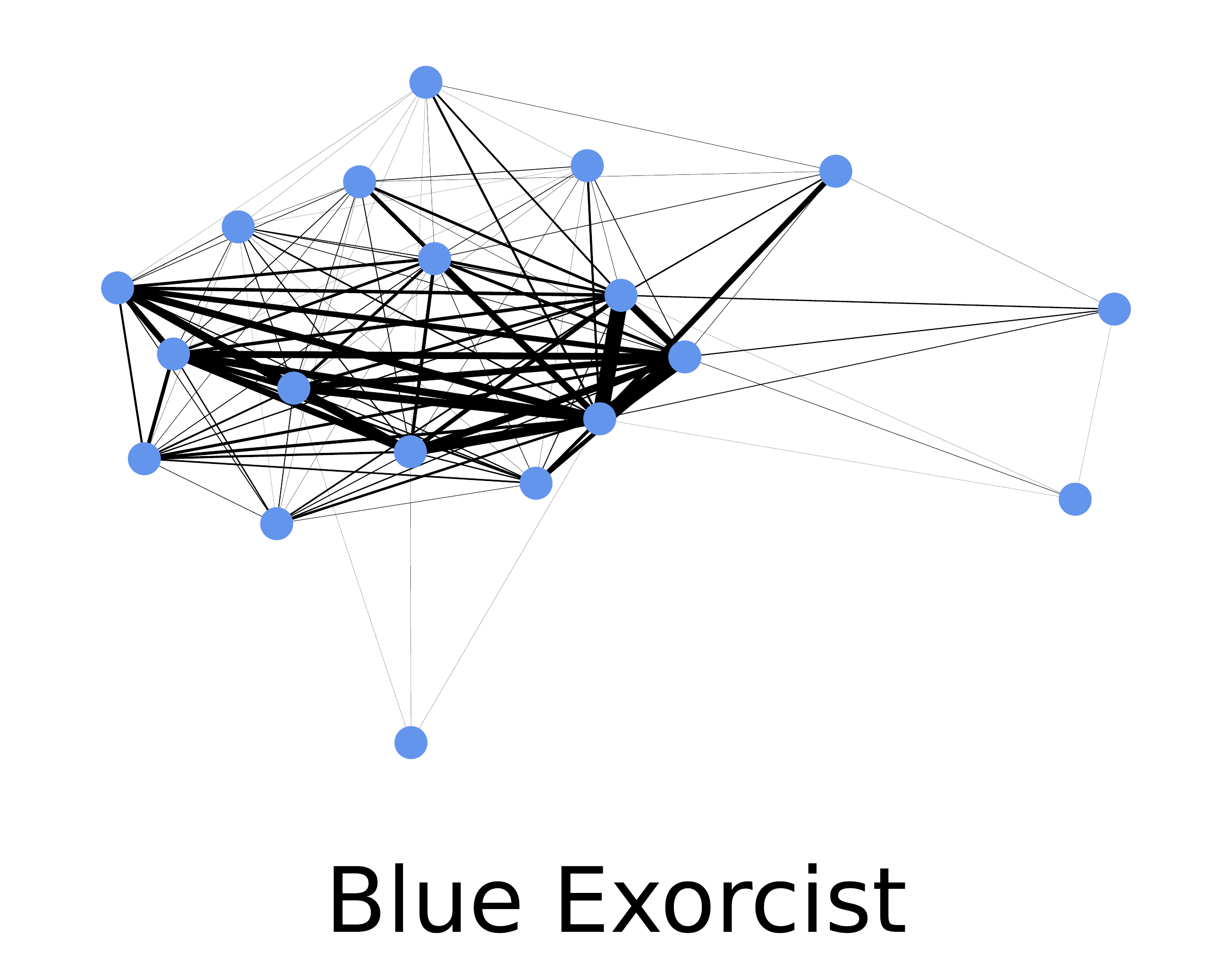}
     \end{subfigure}
     \\
      \begin{subfigure}[b]{0.195\textwidth}
         \centering
         \includegraphics[width=\textwidth]{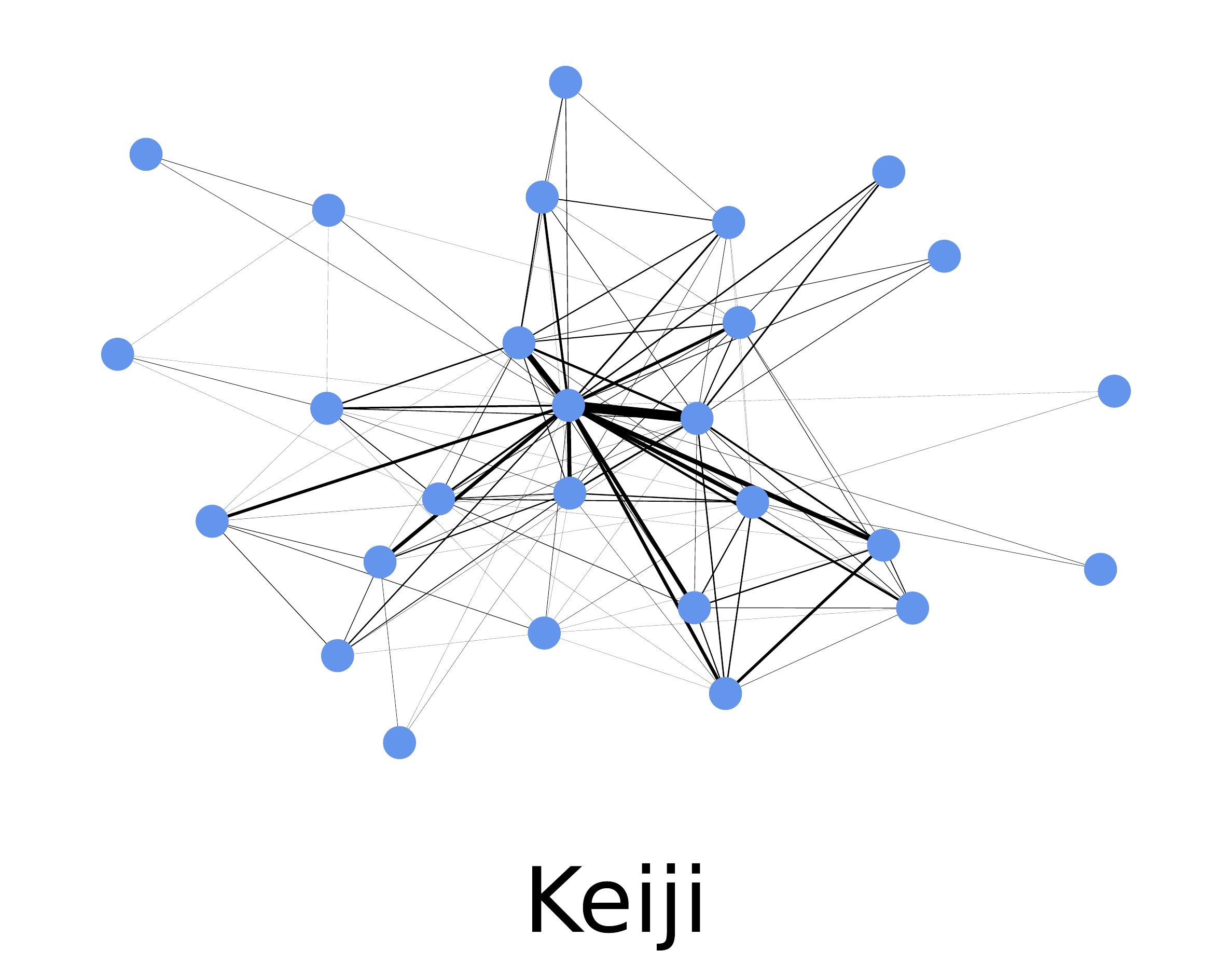}
     \end{subfigure}
     \begin{subfigure}[b]{0.195\textwidth}
         \centering
         \includegraphics[width=\textwidth]{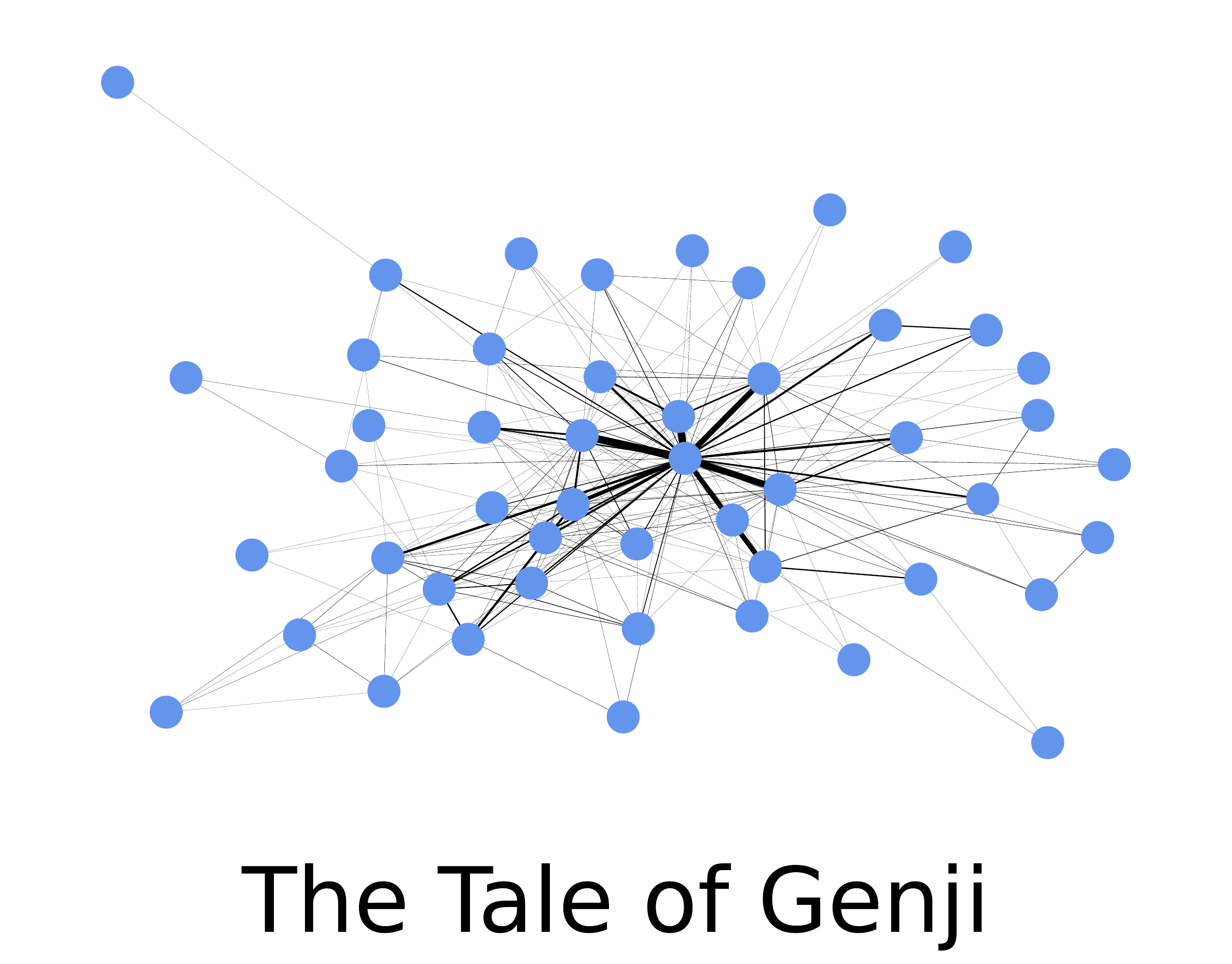}
     \end{subfigure}
     \begin{subfigure}[b]{0.195\textwidth}
         \centering
         \includegraphics[width=\textwidth]{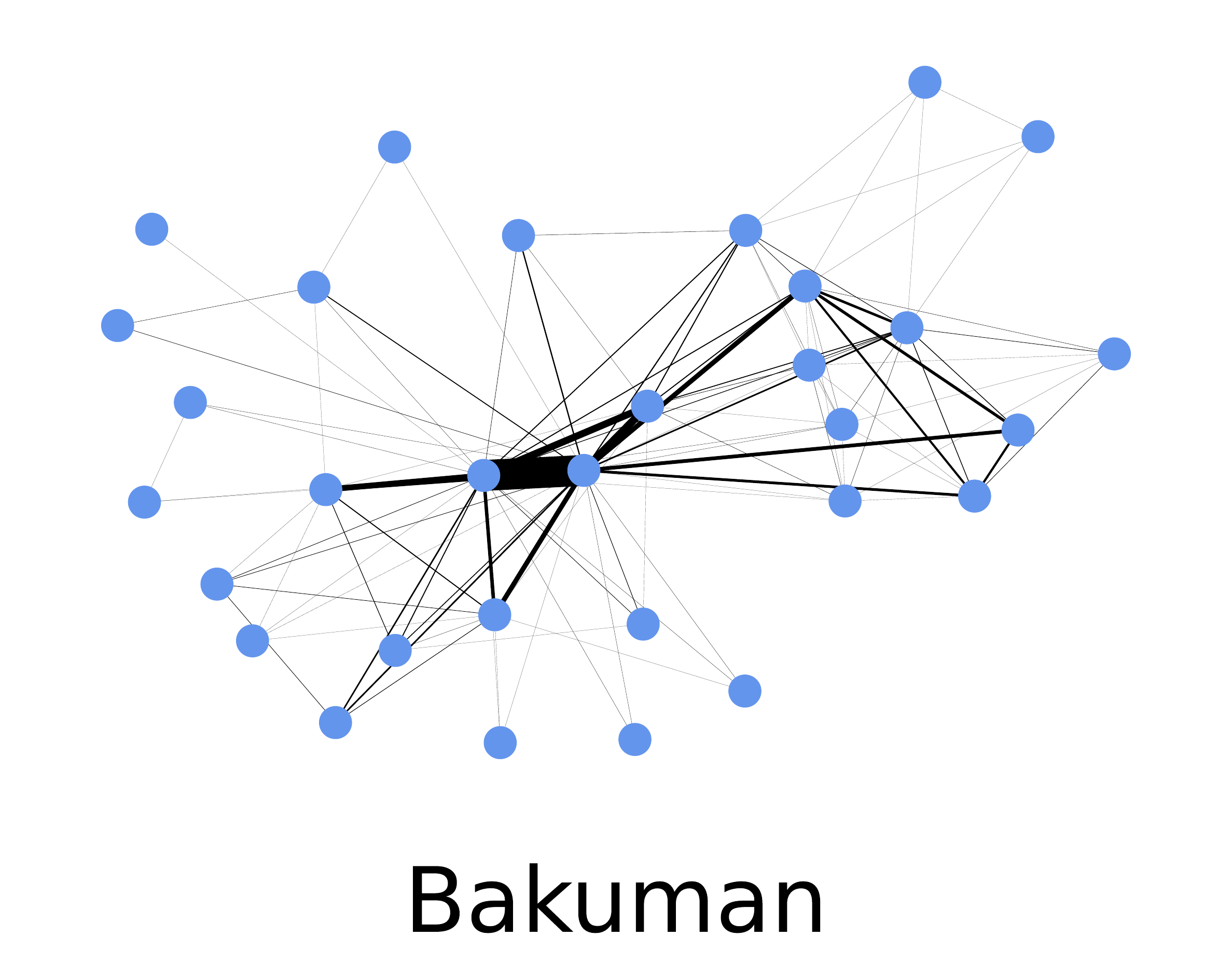}
     \end{subfigure}
     \begin{subfigure}[b]{0.195\textwidth}
         \centering
         \includegraphics[width=\textwidth]{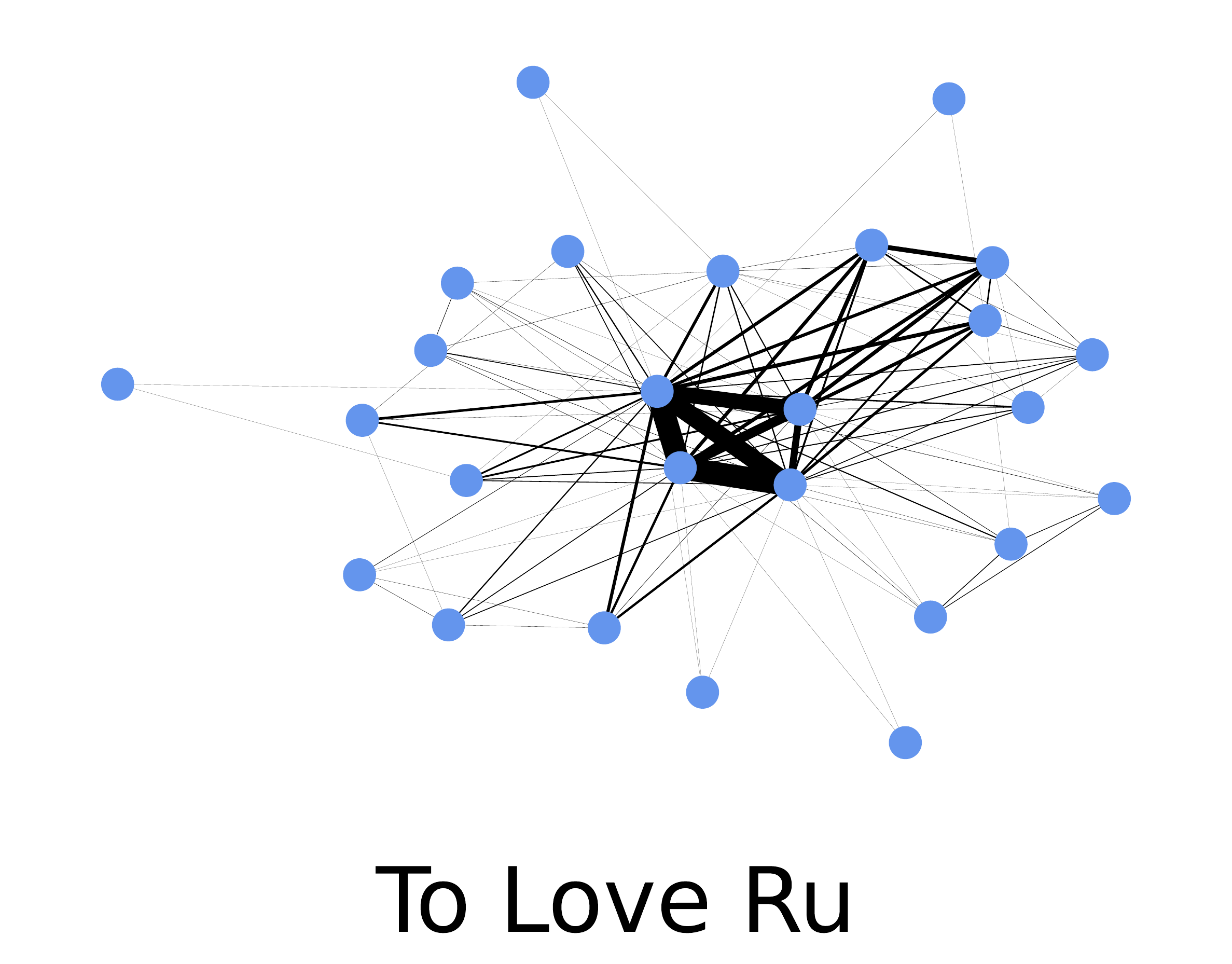}
     \end{subfigure}
     \begin{subfigure}[b]{0.195\textwidth}
         \centering
         \includegraphics[width=\textwidth]{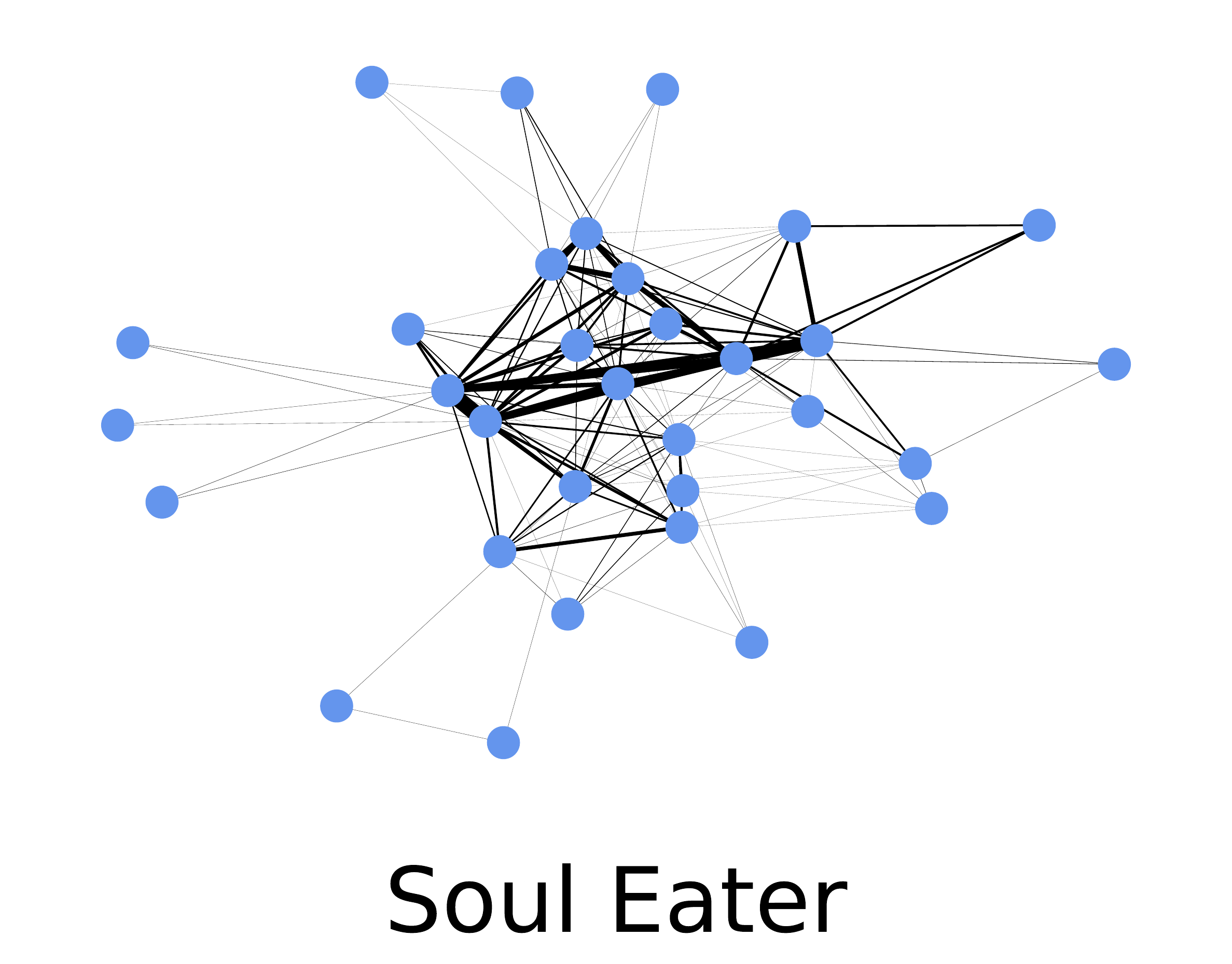}
     \end{subfigure}
     \\
      \begin{subfigure}[b]{0.195\textwidth}
         \centering
         \includegraphics[width=\textwidth]{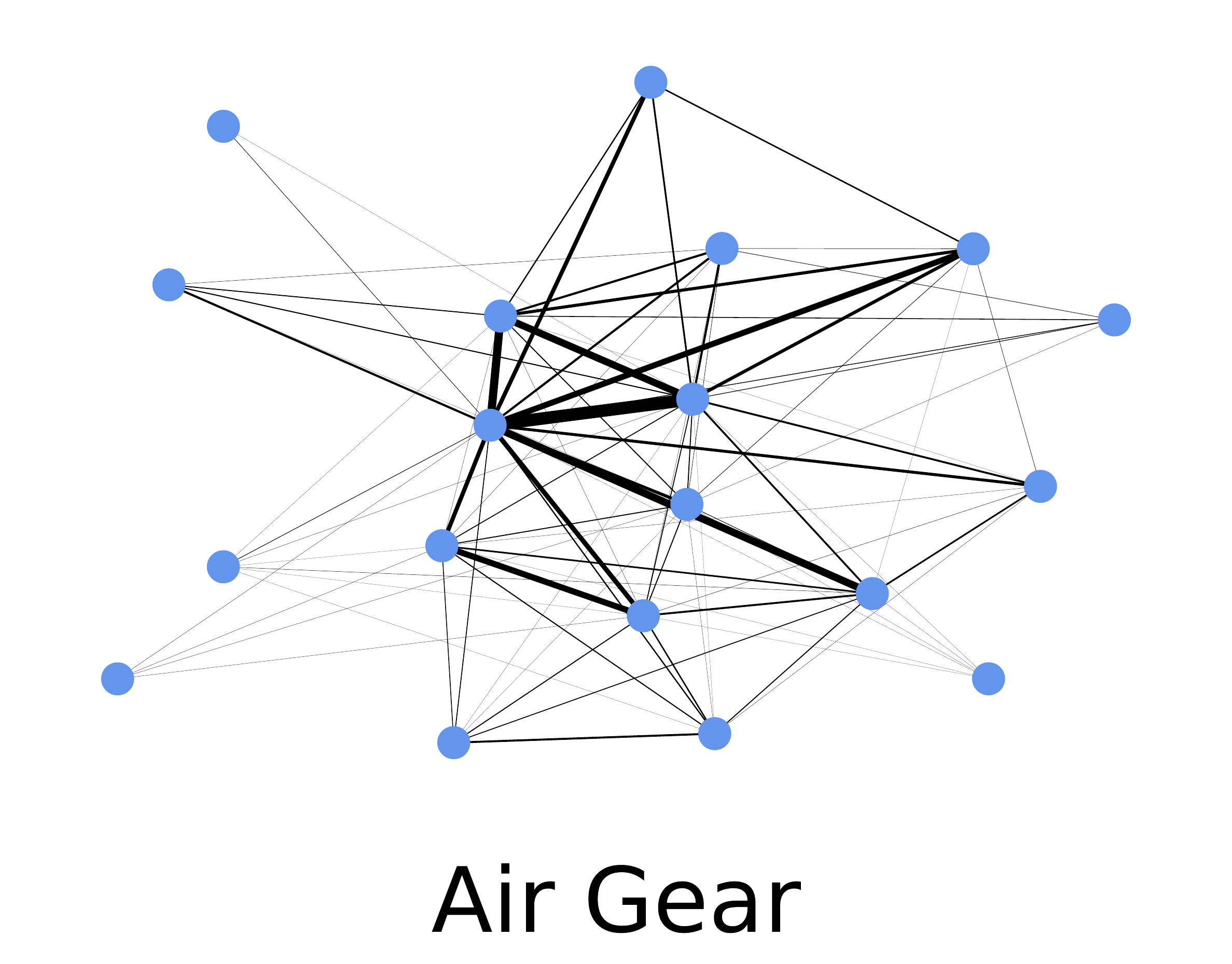}
     \end{subfigure}
     \begin{subfigure}[b]{0.195\textwidth}
         \centering
         \includegraphics[width=\textwidth]{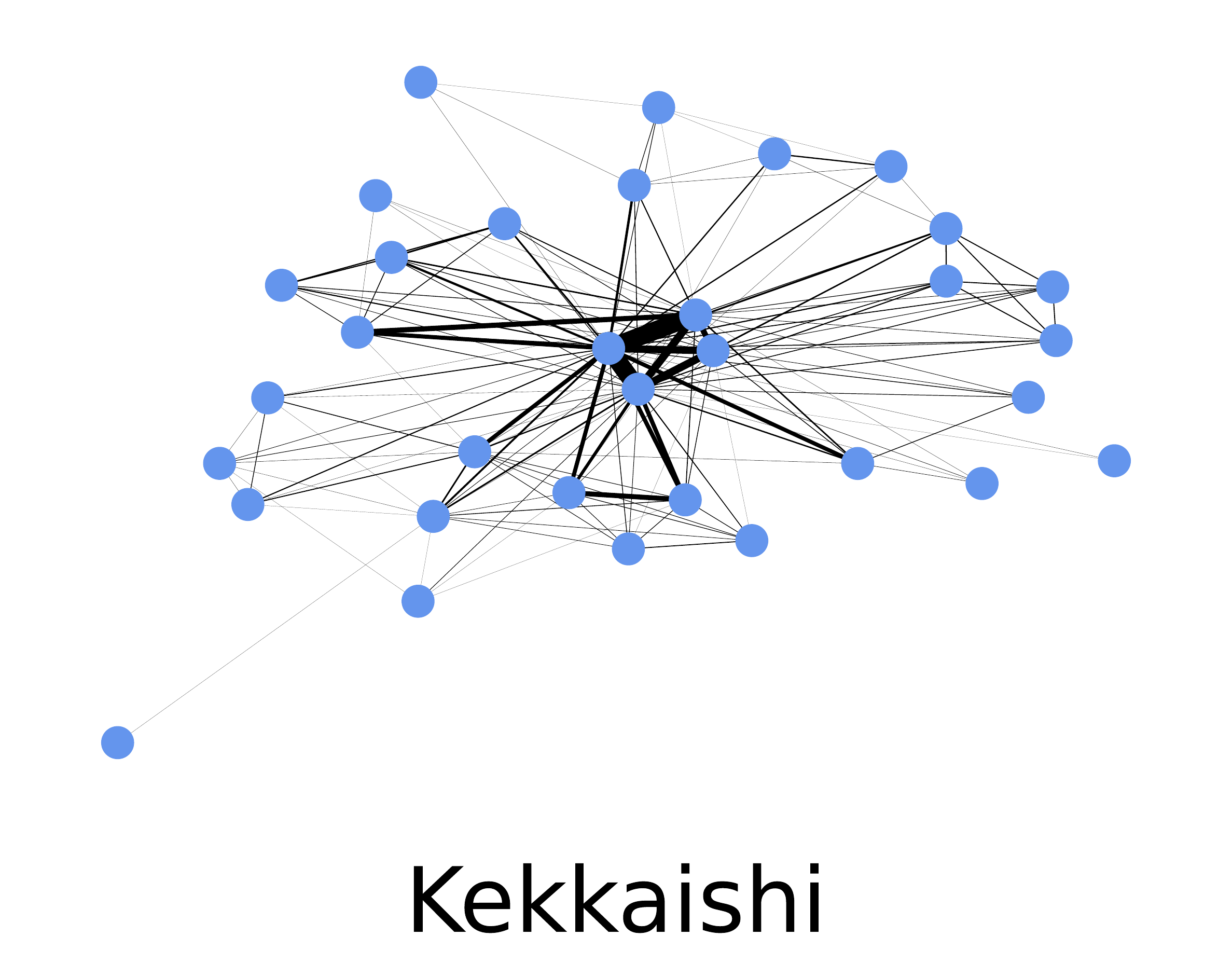}
     \end{subfigure}
     \begin{subfigure}[b]{0.195\textwidth}
         \centering
         \includegraphics[width=\textwidth]{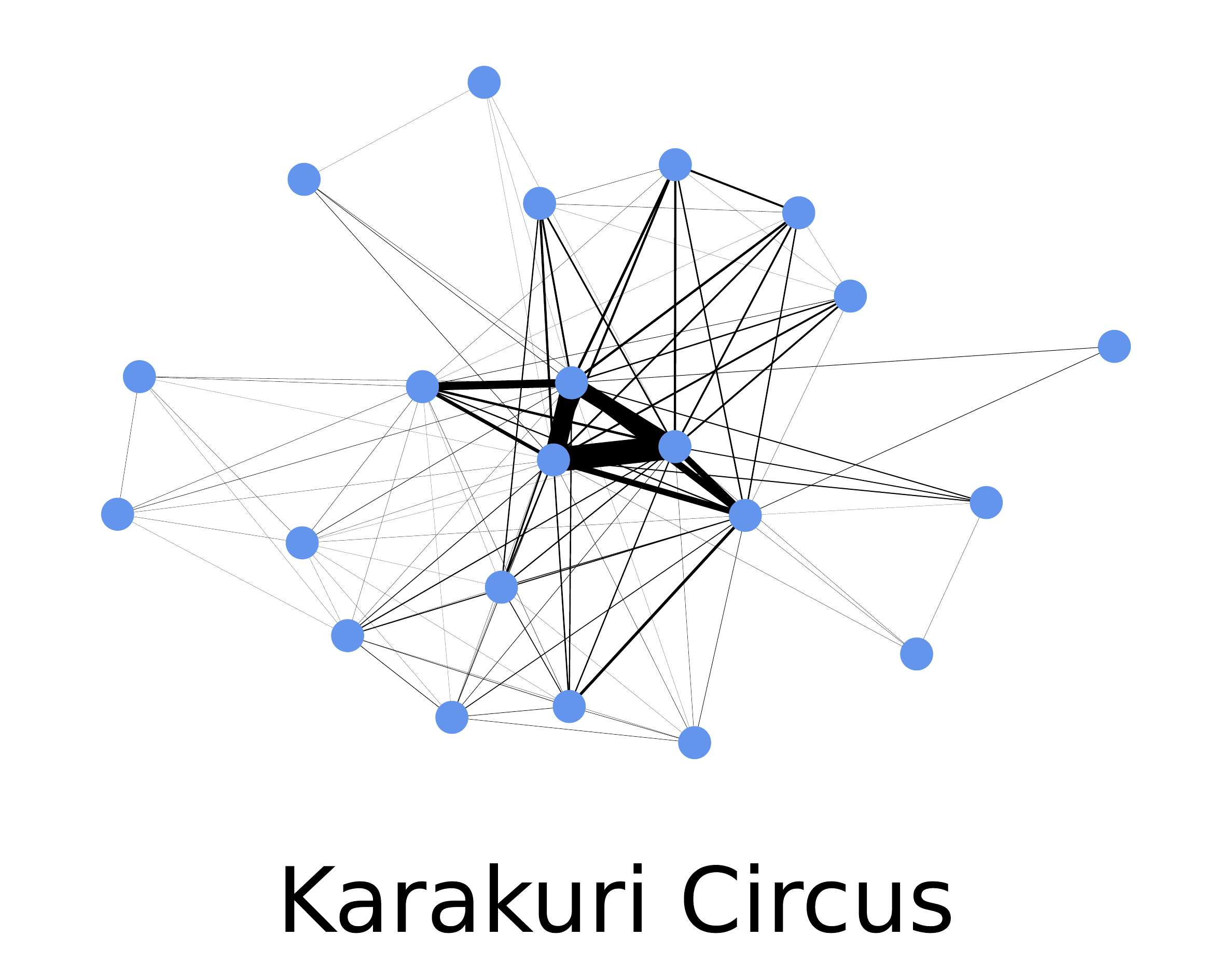}
     \end{subfigure}
     \begin{subfigure}[b]{0.195\textwidth}
         \centering
         \includegraphics[width=\textwidth]{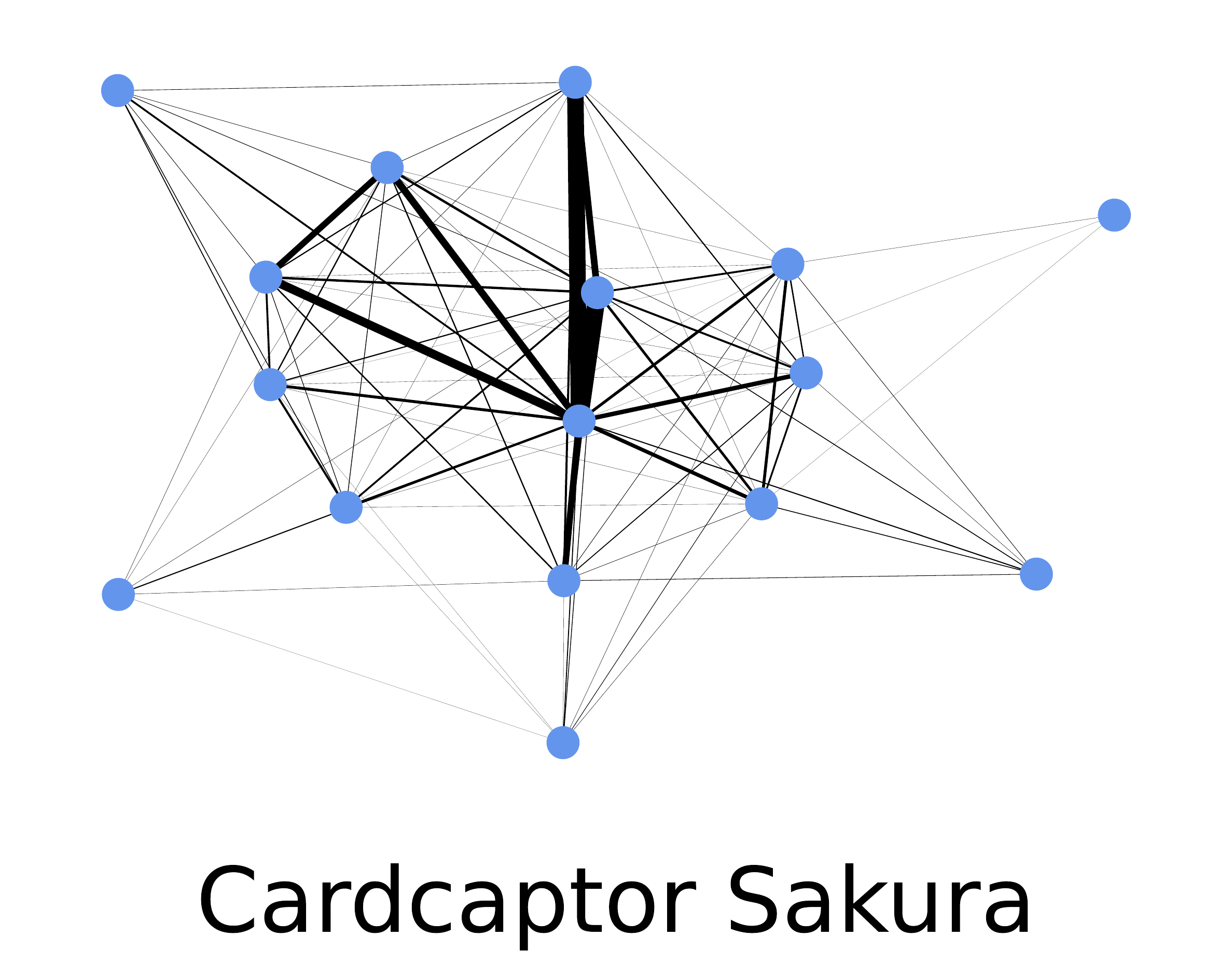}
     \end{subfigure}
     \begin{subfigure}[b]{0.195\textwidth}
         \centering
         \includegraphics[width=\textwidth]{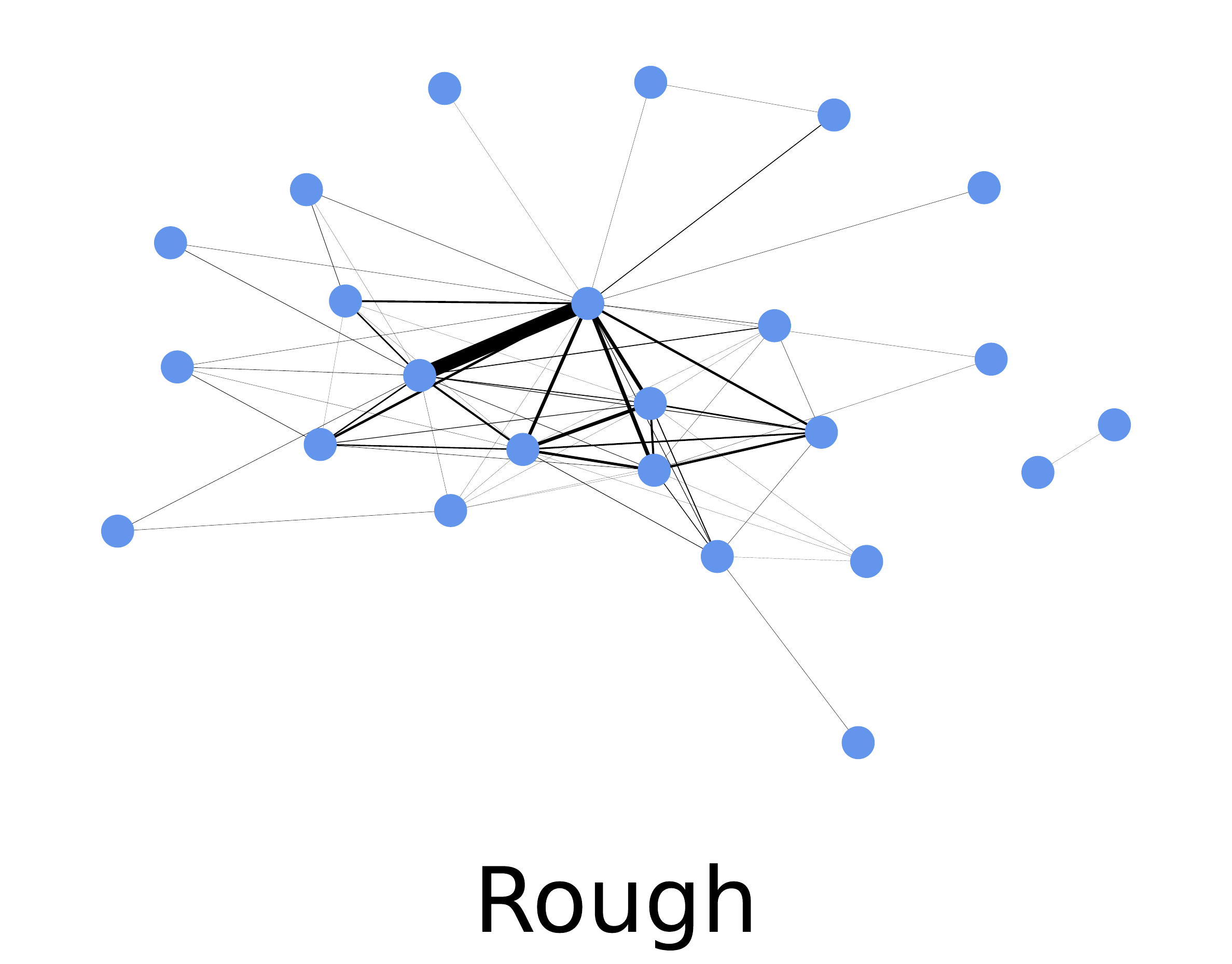}
     \end{subfigure}
     \\
      \begin{subfigure}[b]{0.195\textwidth}
         \centering
         \includegraphics[width=\textwidth]{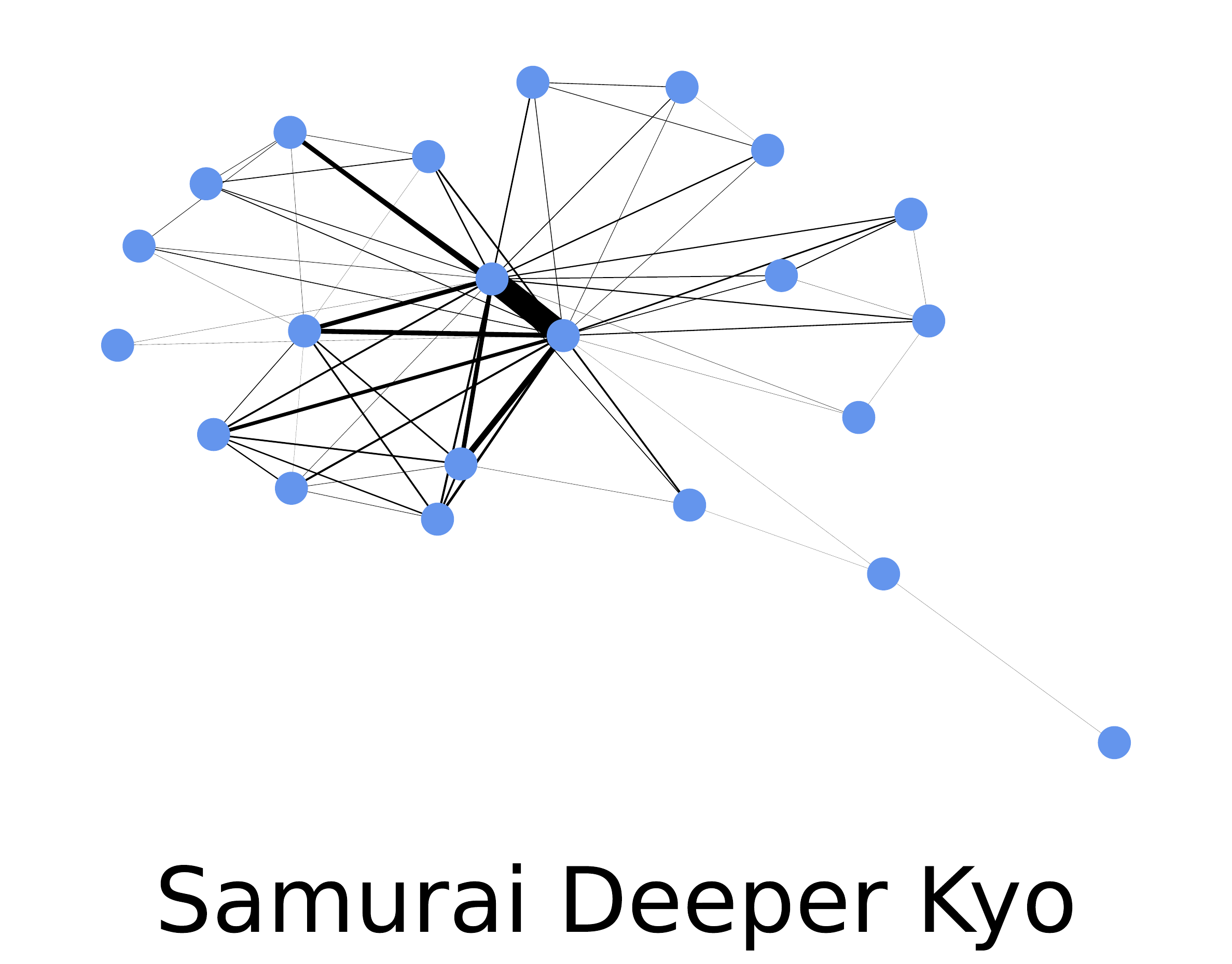}
     \end{subfigure}
     \begin{subfigure}[b]{0.195\textwidth}
         \centering
         \includegraphics[width=\textwidth]{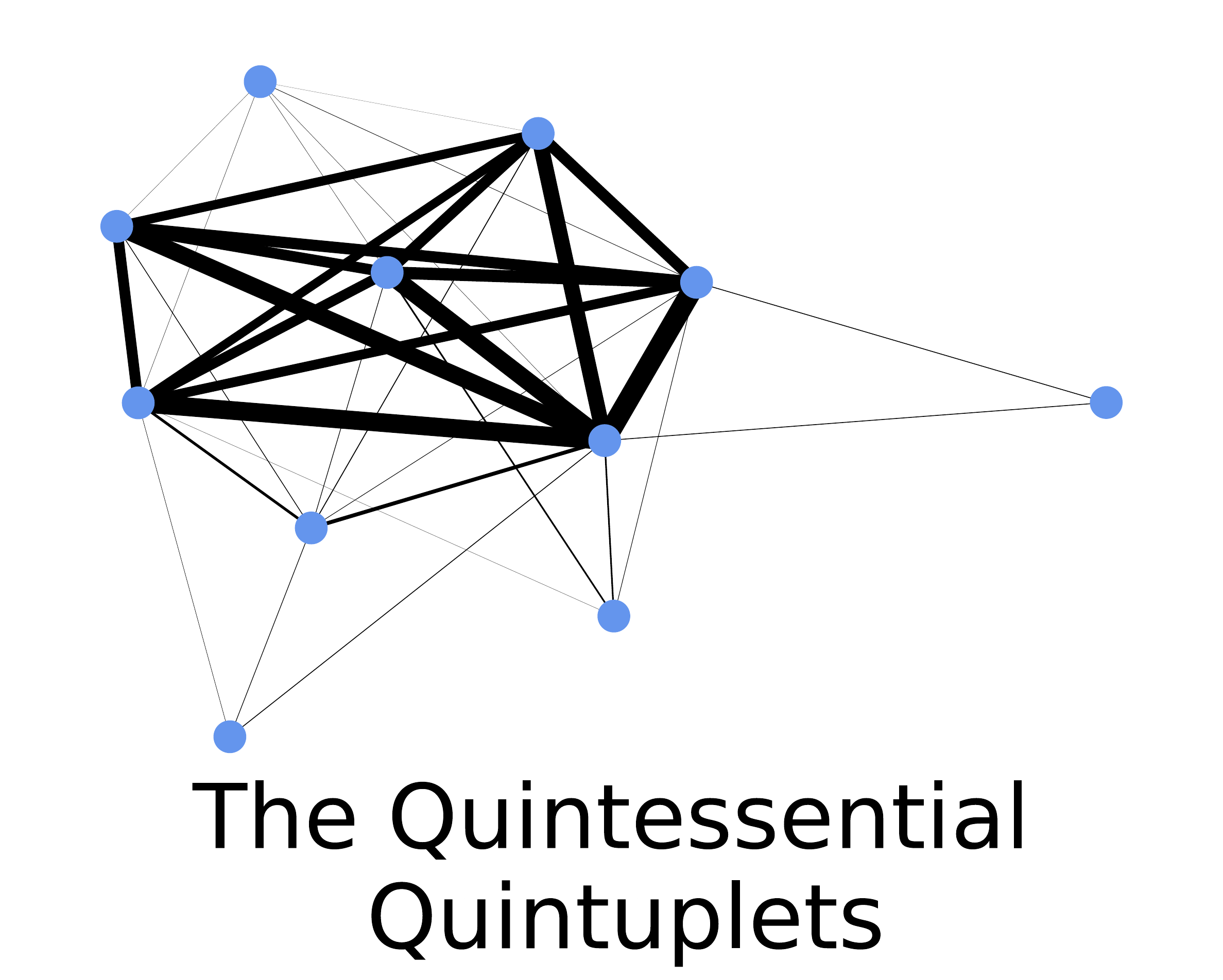}
     \end{subfigure}
     \begin{subfigure}[b]{0.195\textwidth}
         \centering
         \includegraphics[width=\textwidth]{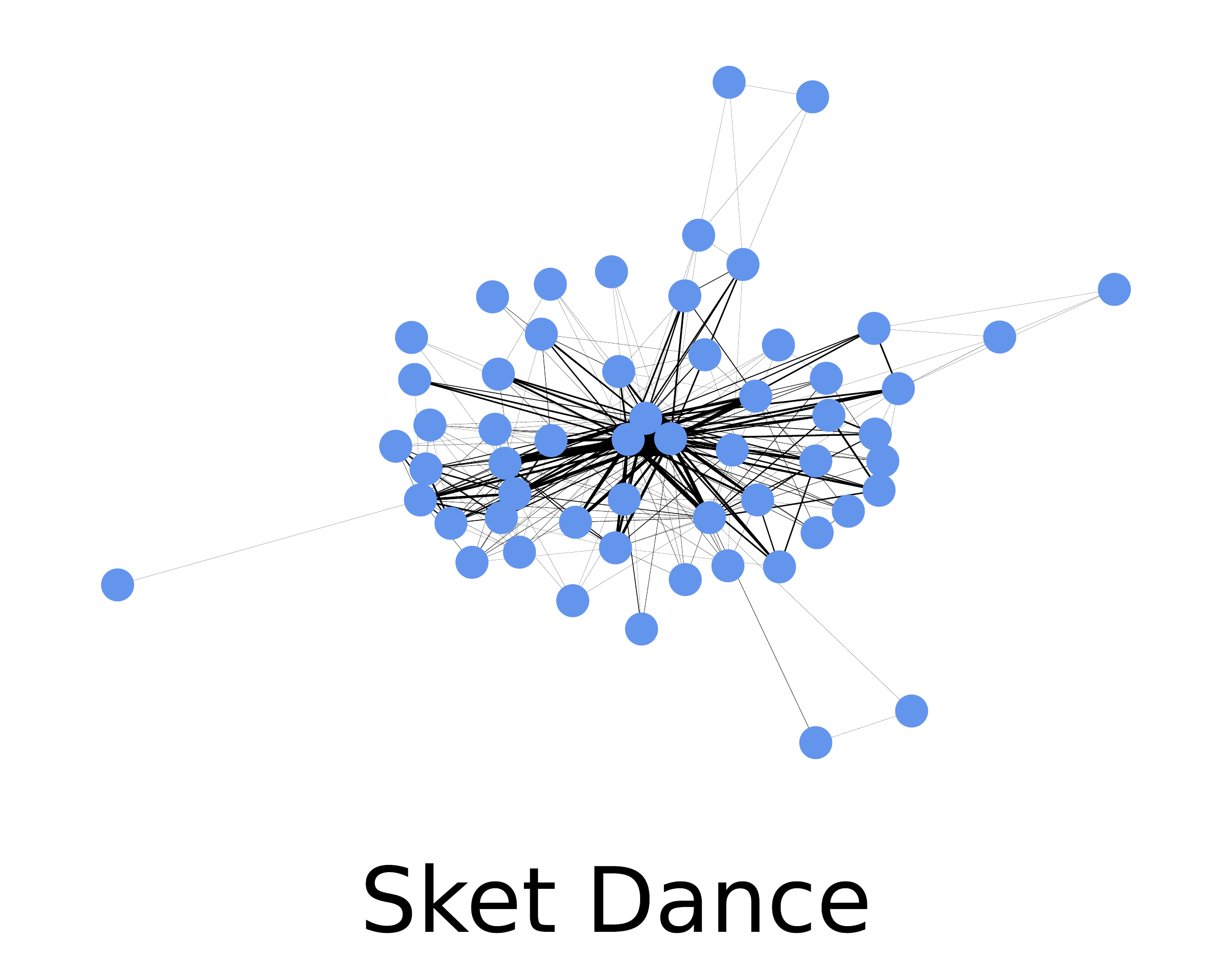}
     \end{subfigure}
     \begin{subfigure}[b]{0.195\textwidth}
         \centering
         \includegraphics[width=\textwidth]{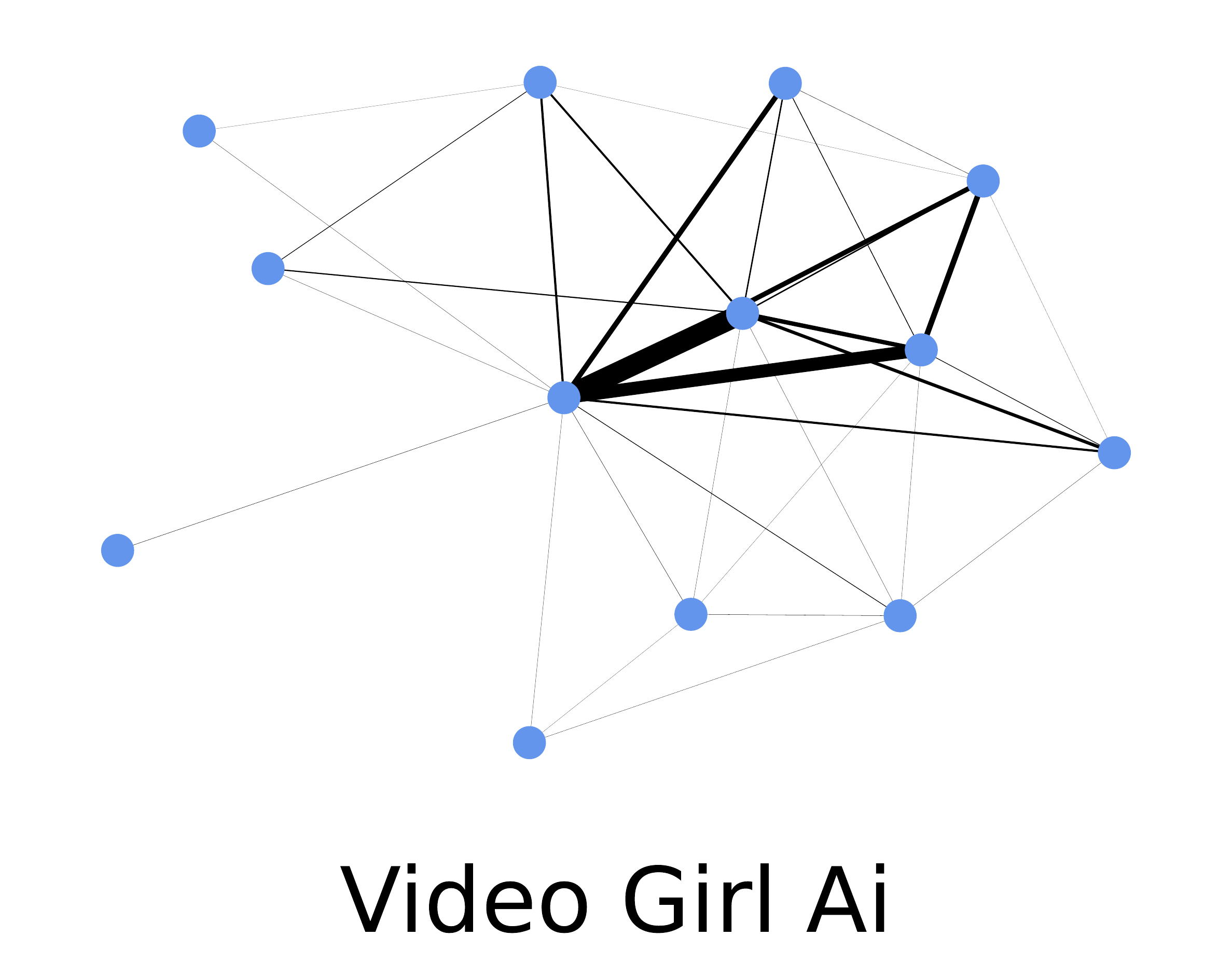}
     \end{subfigure}
     \begin{subfigure}[b]{0.195\textwidth}
         \centering
         \includegraphics[width=\textwidth]{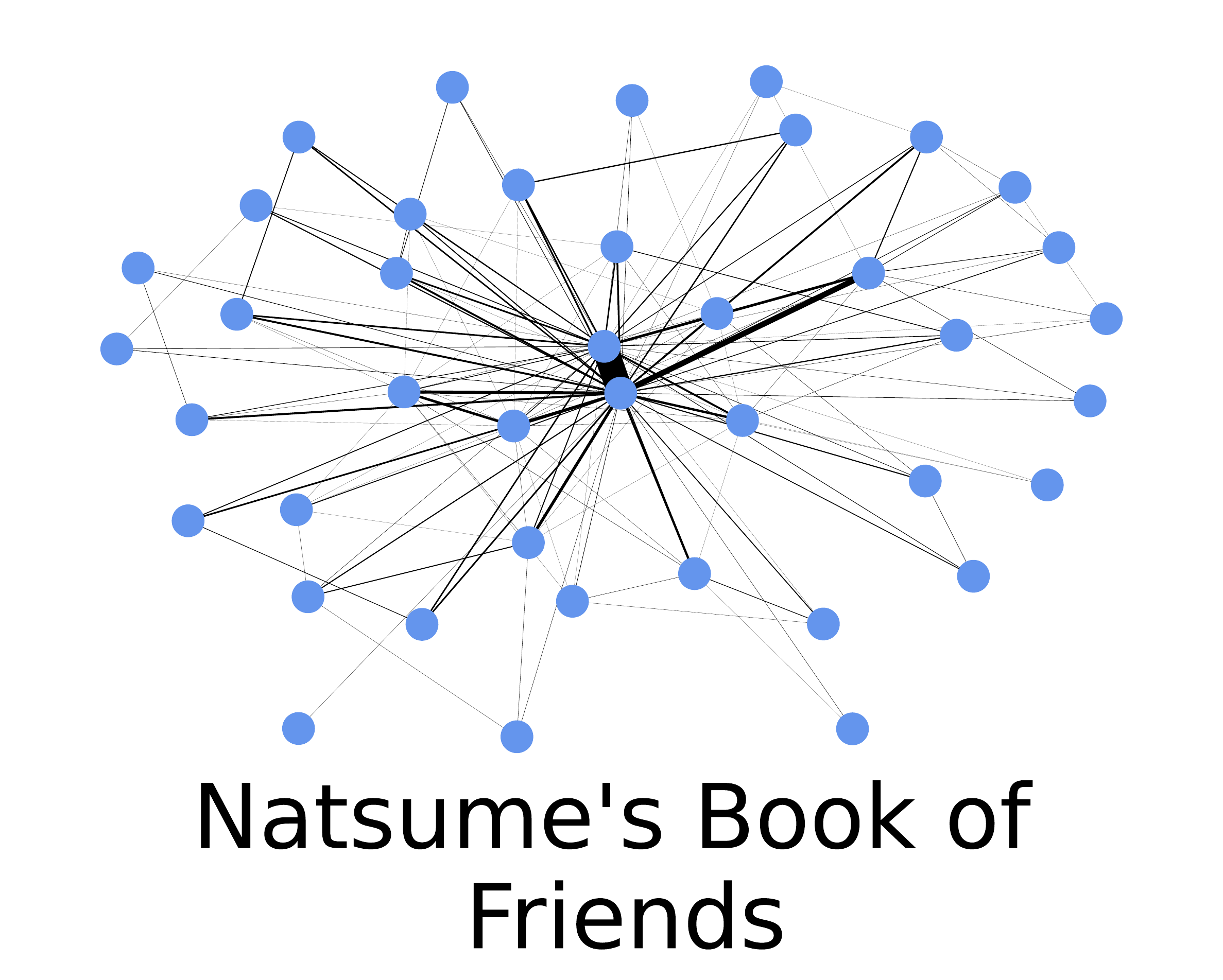}
     \end{subfigure}
     \\
      \begin{subfigure}[b]{0.195\textwidth}
         \centering
         \includegraphics[width=\textwidth]{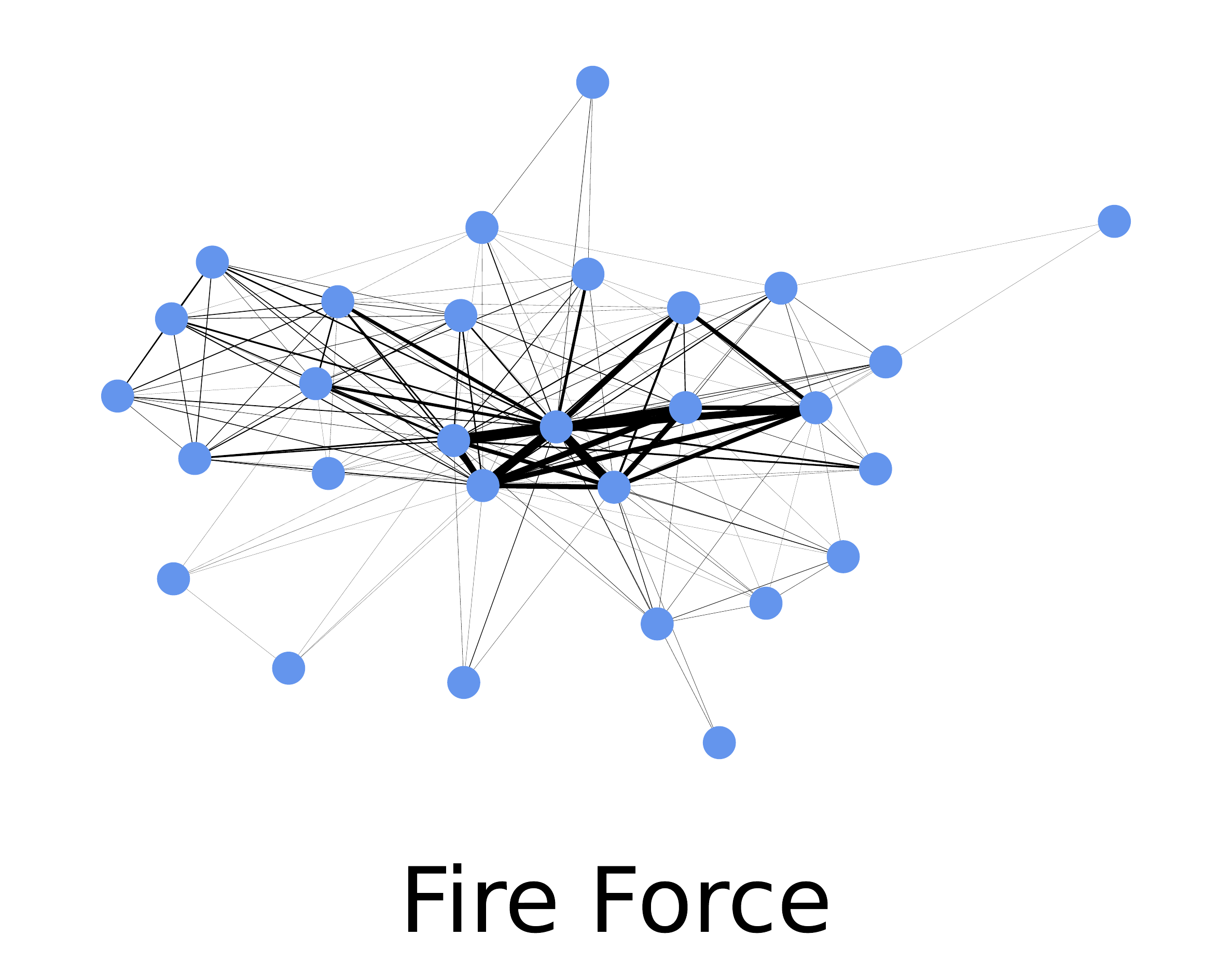}
     \end{subfigure}
     \begin{subfigure}[b]{0.195\textwidth}
         \centering
         \includegraphics[width=\textwidth]{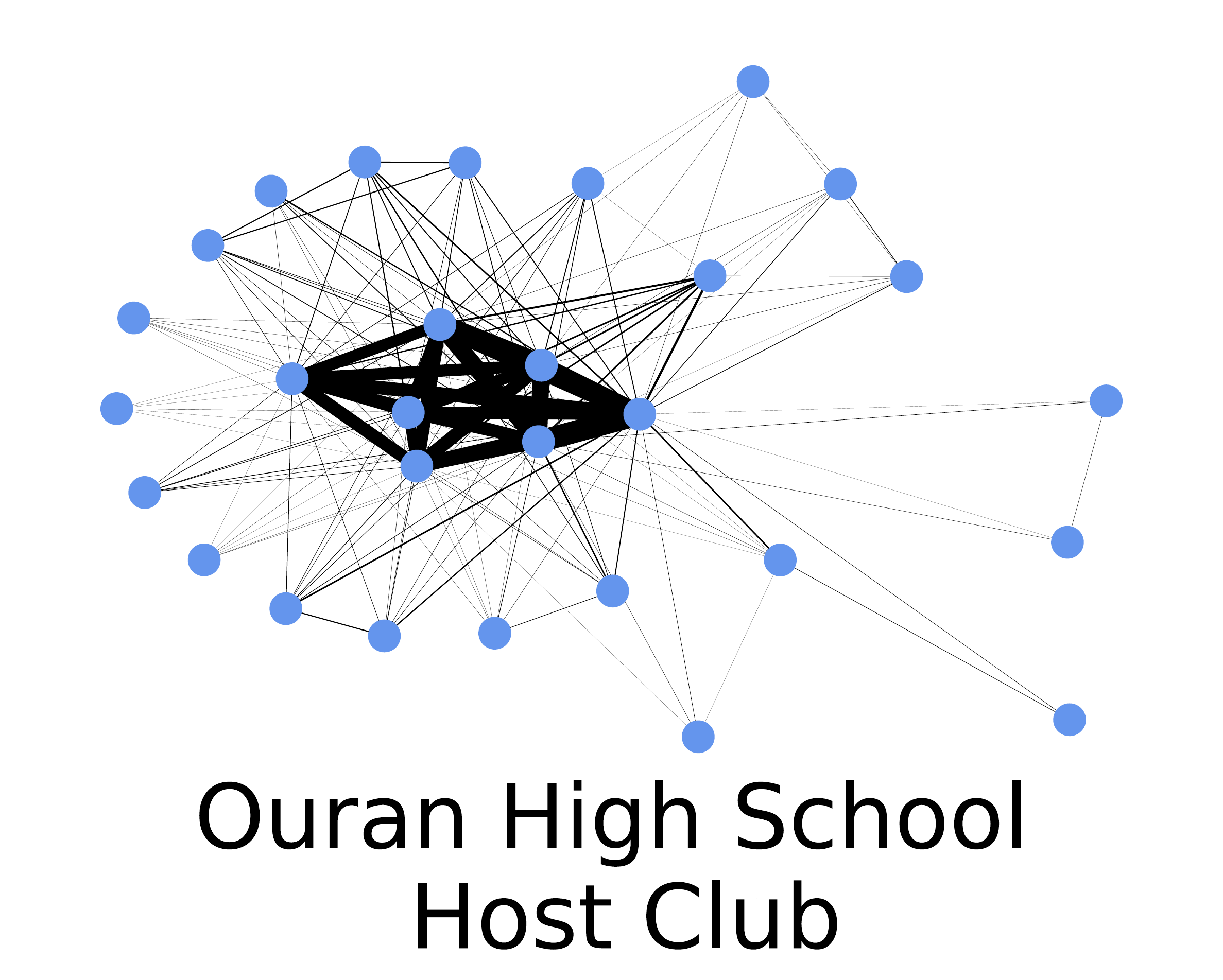}
     \end{subfigure}
     \begin{subfigure}[b]{0.195\textwidth}
         \centering
         \includegraphics[width=\textwidth]{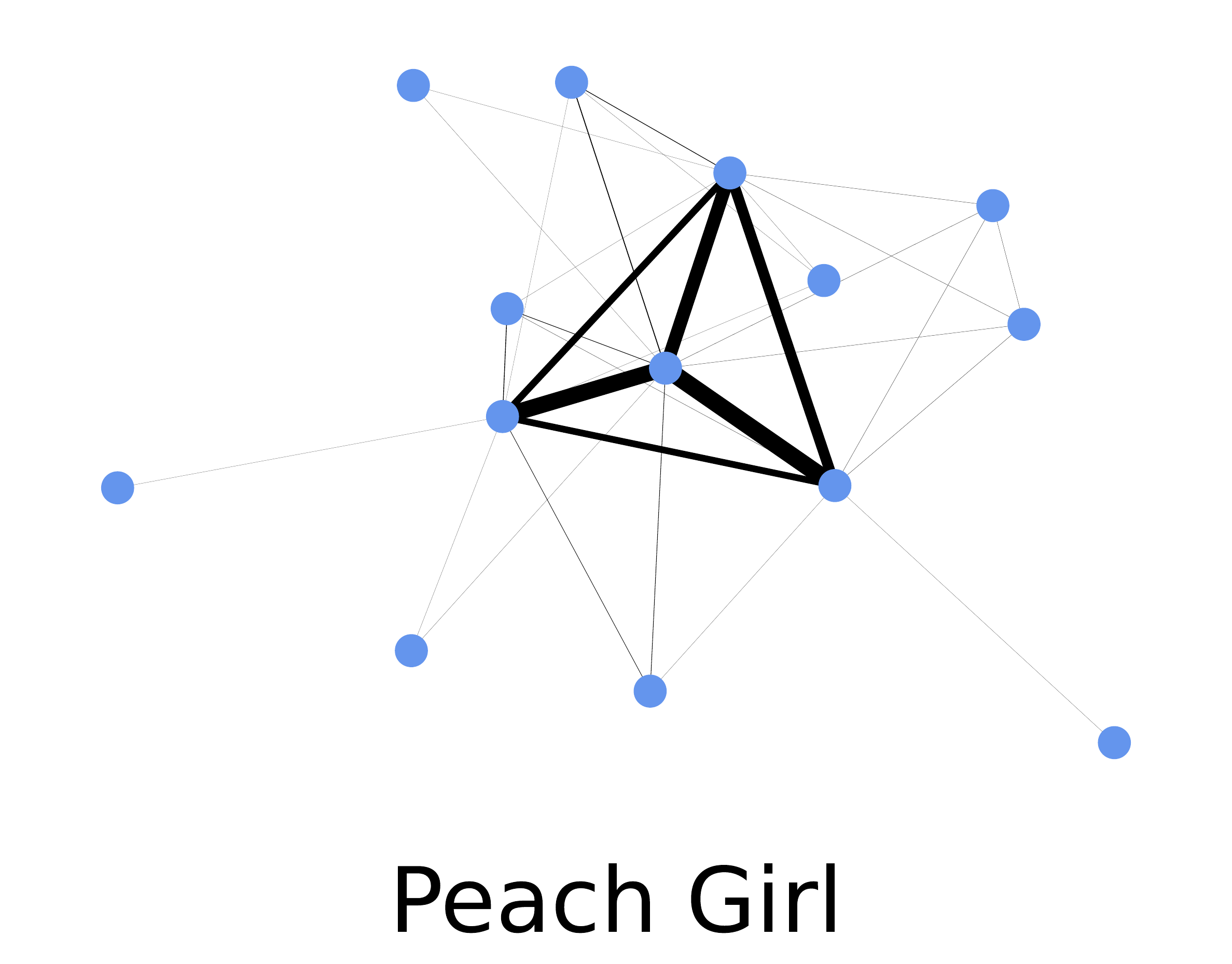}
     \end{subfigure}
     \begin{subfigure}[b]{0.195\textwidth}
         \centering
         \includegraphics[width=\textwidth]{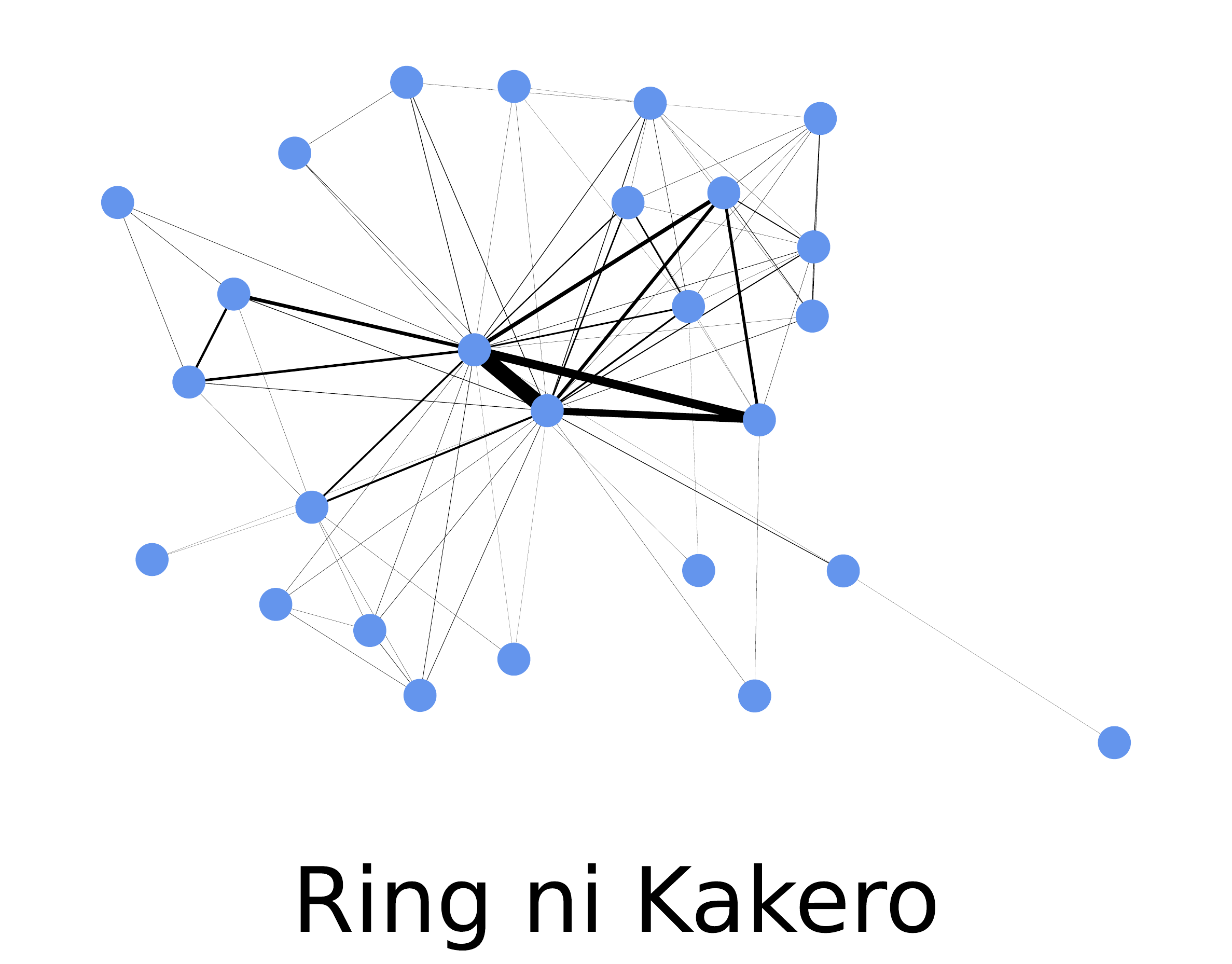}
     \end{subfigure}
     \begin{subfigure}[b]{0.195\textwidth}
         \centering
         \includegraphics[width=\textwidth]{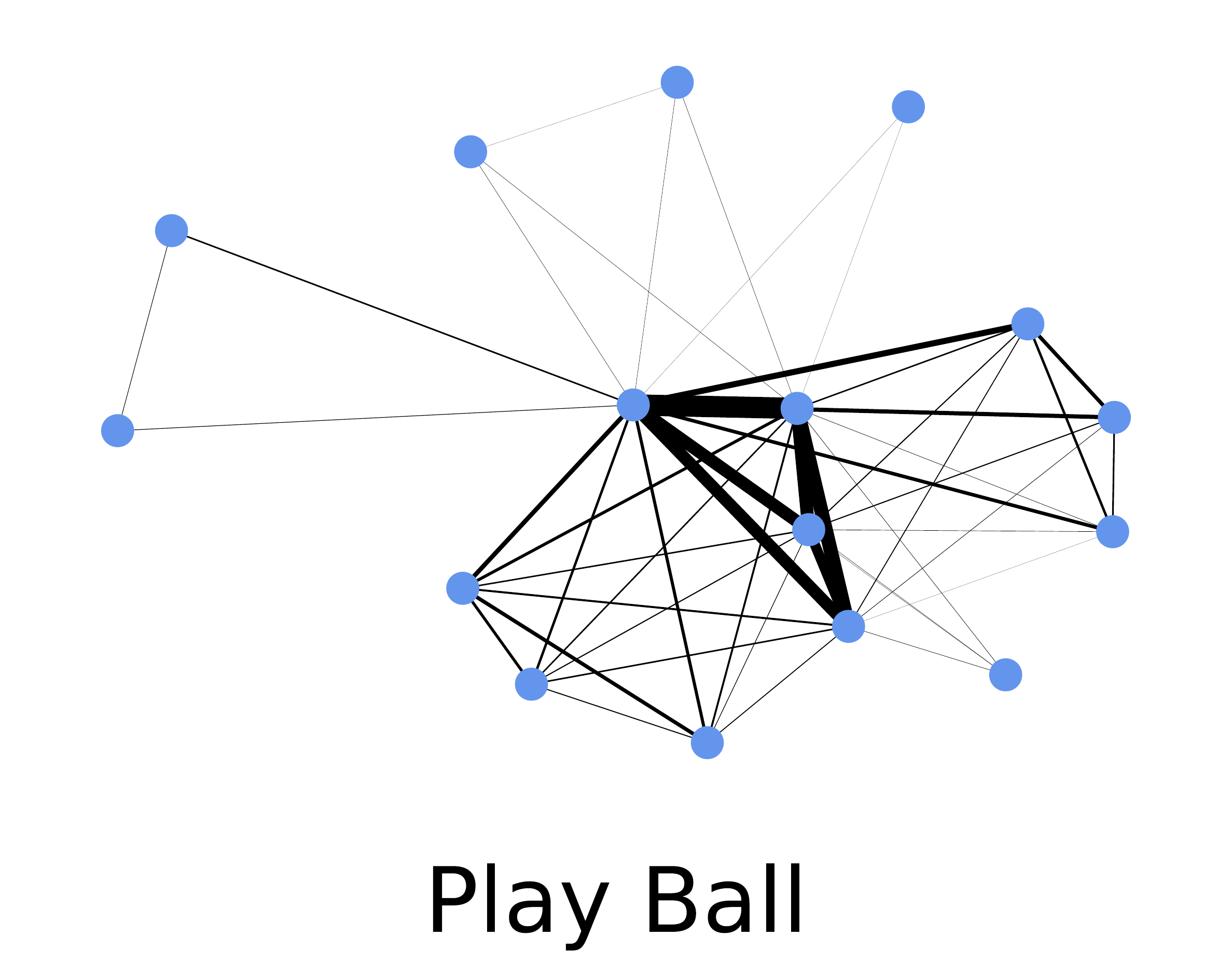}
     \end{subfigure}
     
    \caption{Character networks for 162 manga (continued)}
    
    \label{fig:three graphs}
\end{figure*}

\begin{figure*}\ContinuedFloat
     \centering
      \begin{subfigure}[b]{0.195\textwidth}
         \centering
         \includegraphics[width=\textwidth]{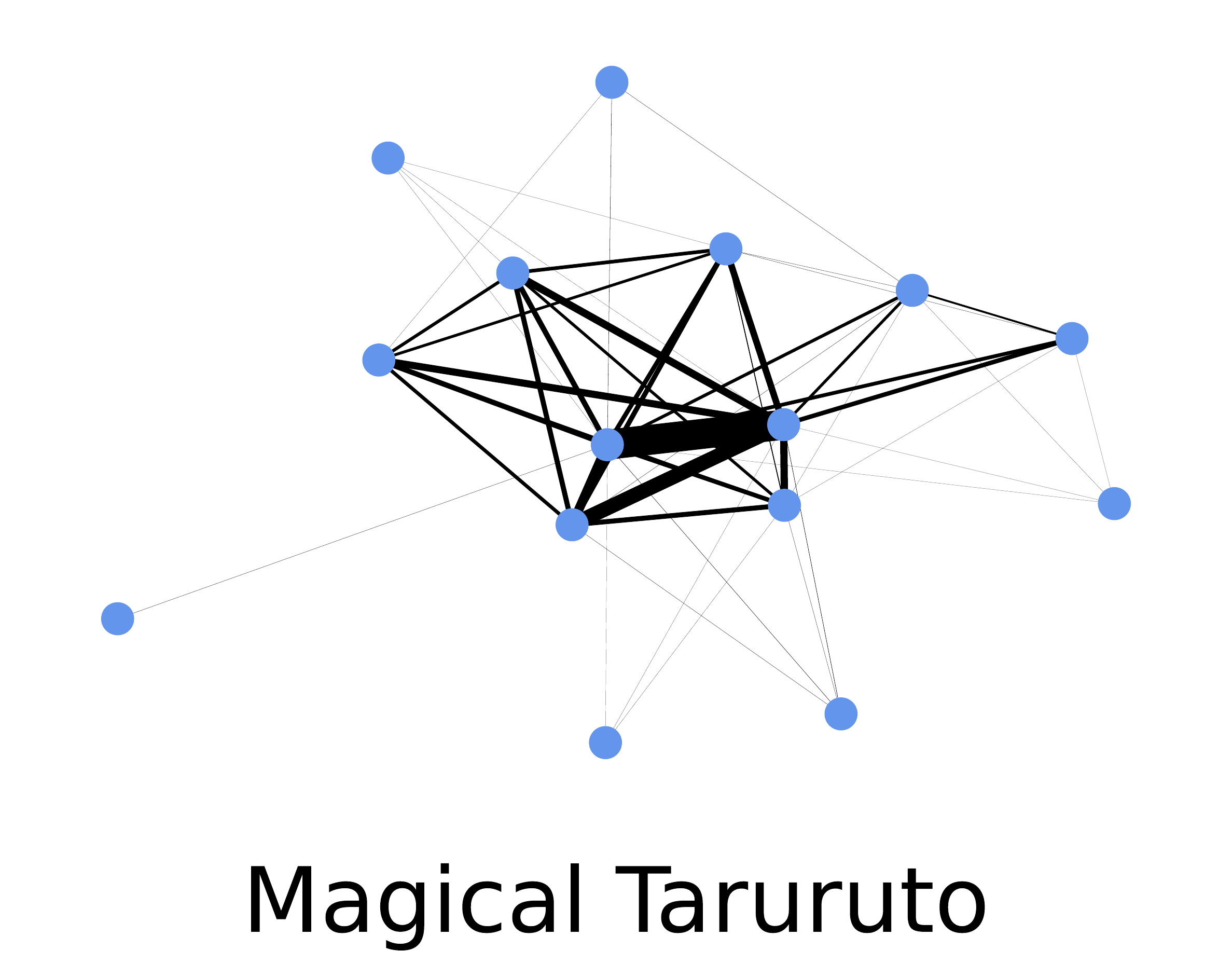}
     \end{subfigure}
     \begin{subfigure}[b]{0.195\textwidth}
         \centering
         \includegraphics[width=\textwidth]{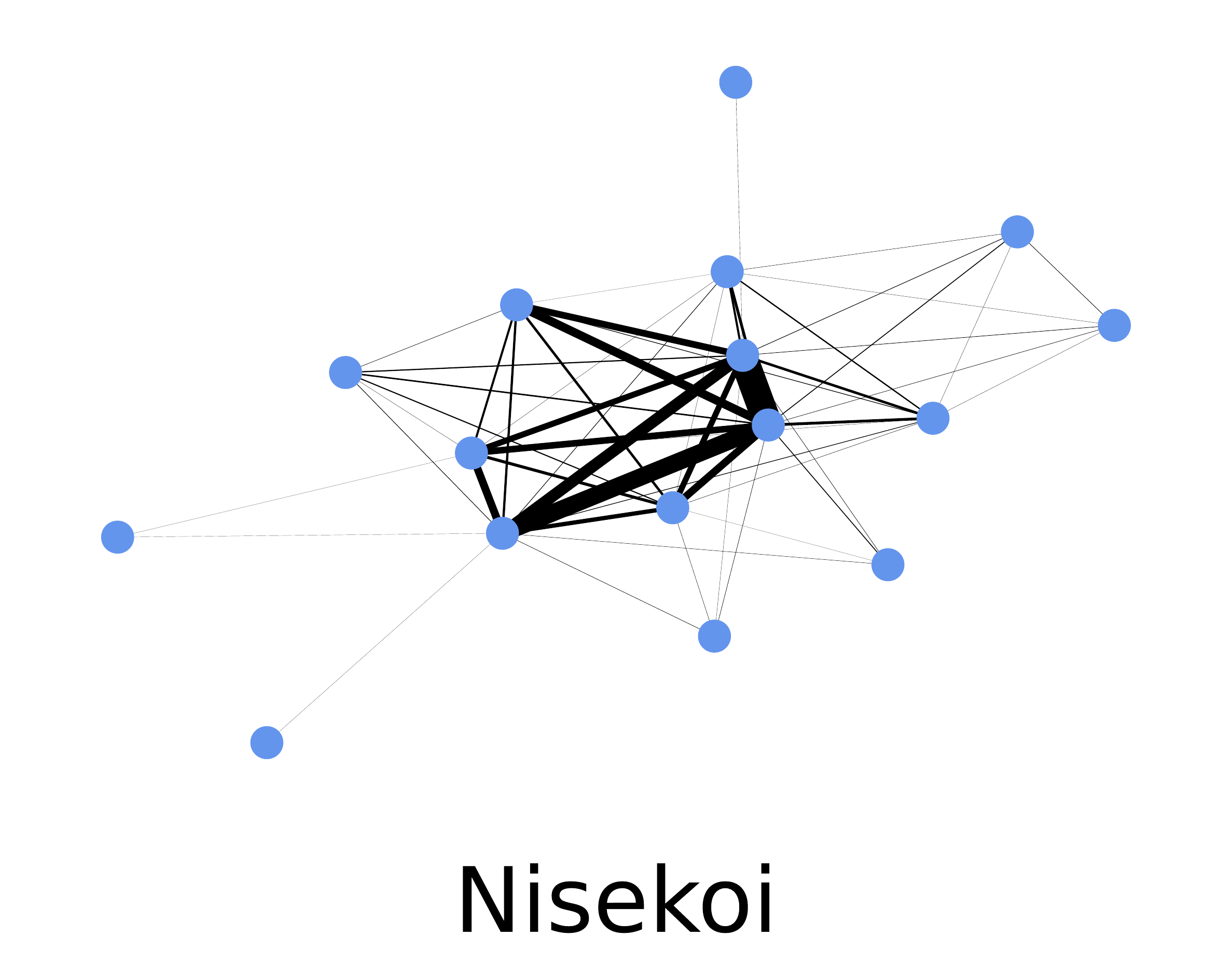}
     \end{subfigure}
     \begin{subfigure}[b]{0.195\textwidth}
         \centering
         \includegraphics[width=\textwidth]{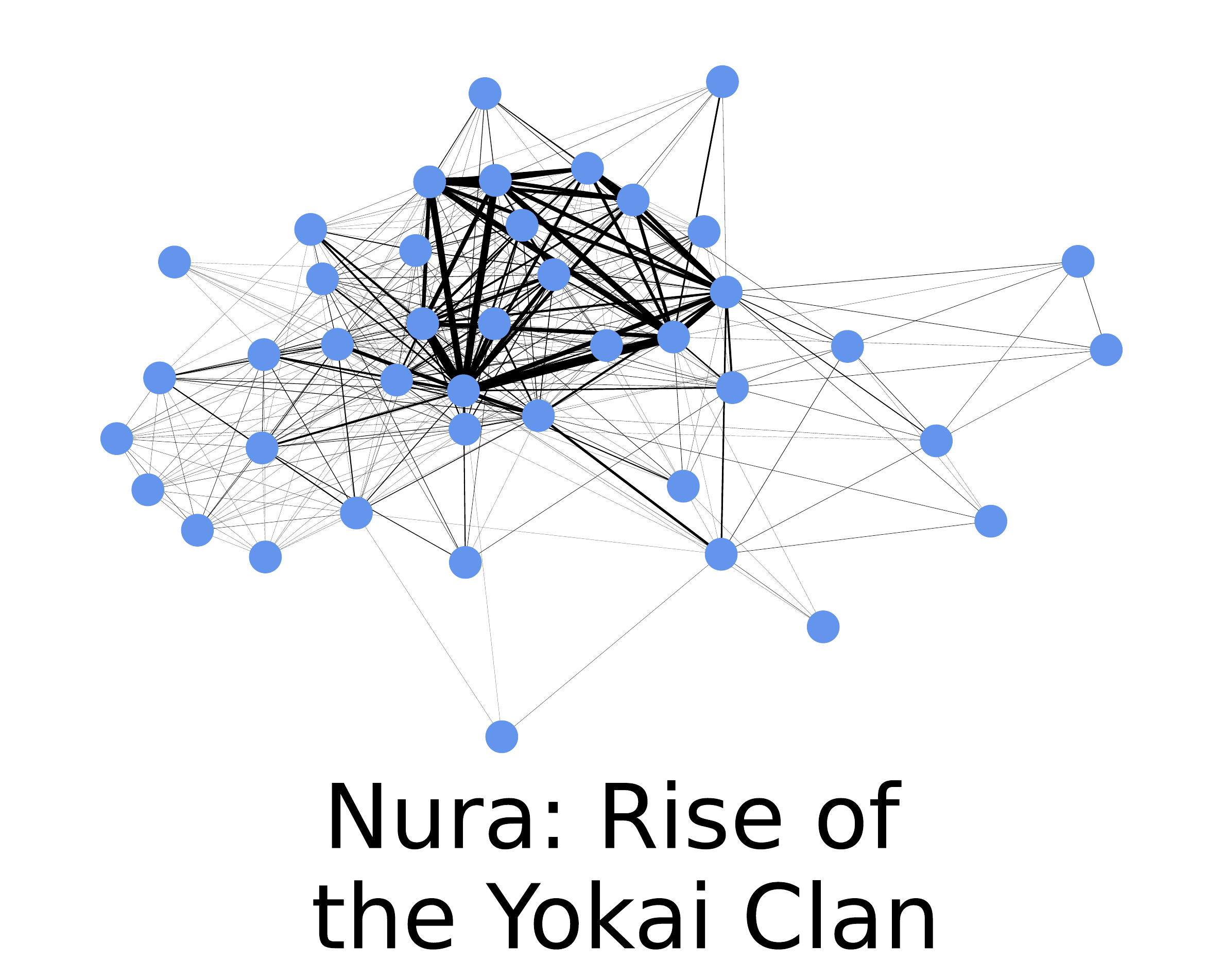}
     \end{subfigure}
     \begin{subfigure}[b]{0.195\textwidth}
         \centering
         \includegraphics[width=\textwidth]{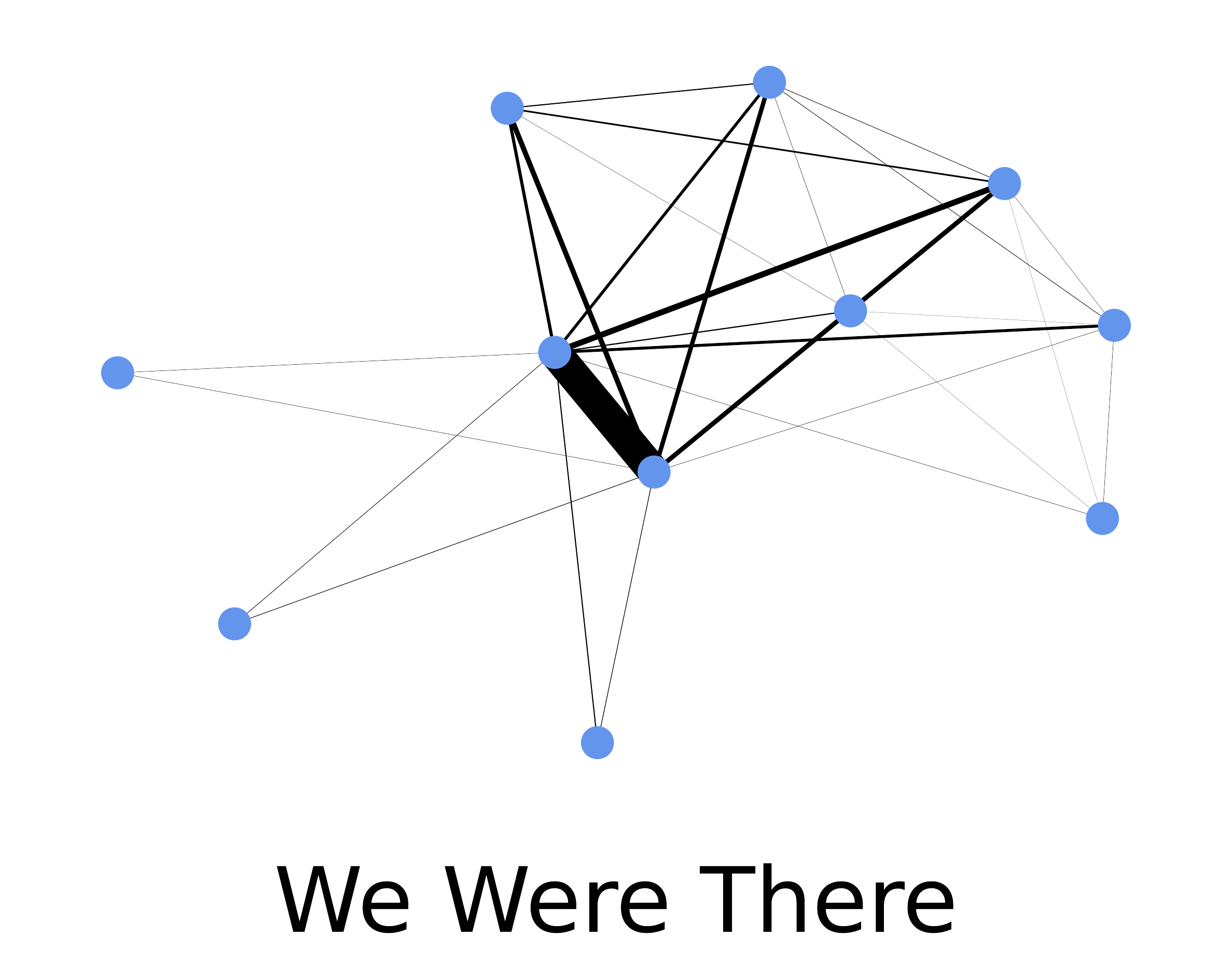}
     \end{subfigure}
     \begin{subfigure}[b]{0.195\textwidth}
         \centering
         \includegraphics[width=\textwidth]{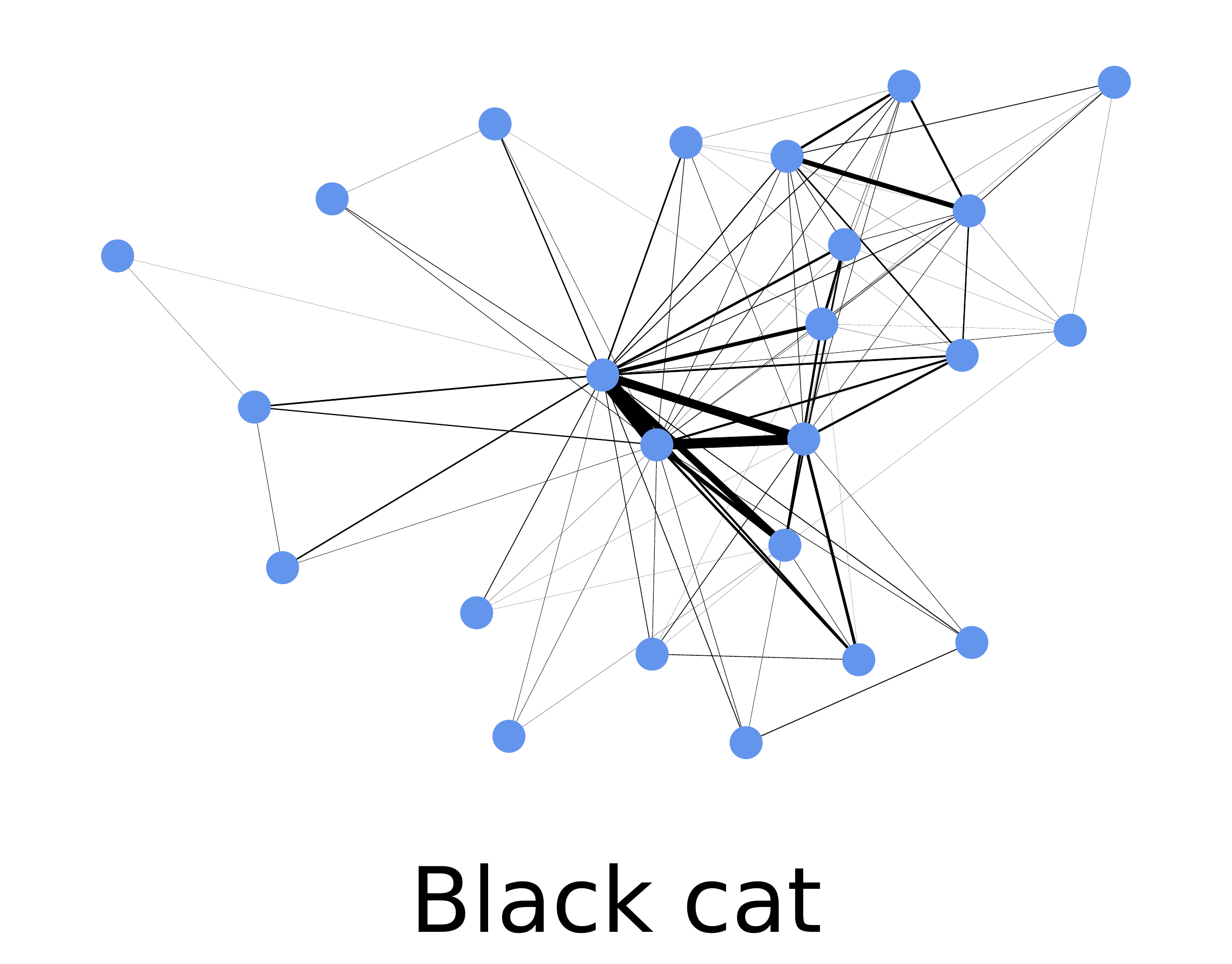}
     \end{subfigure}
     \\
      \begin{subfigure}[b]{0.195\textwidth}
         \centering
         \includegraphics[width=\textwidth]{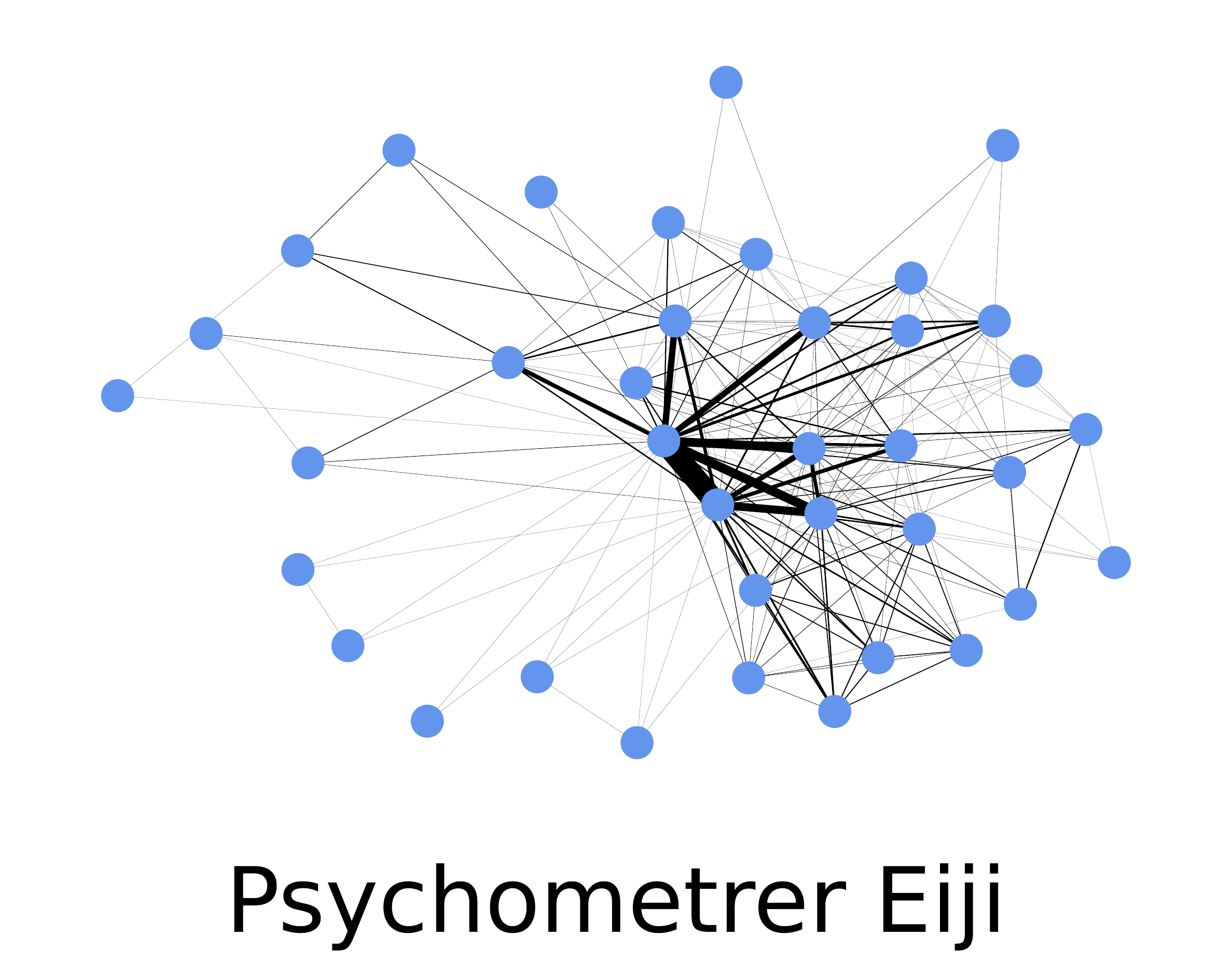}
     \end{subfigure}
     \begin{subfigure}[b]{0.195\textwidth}
         \centering
         \includegraphics[width=\textwidth]{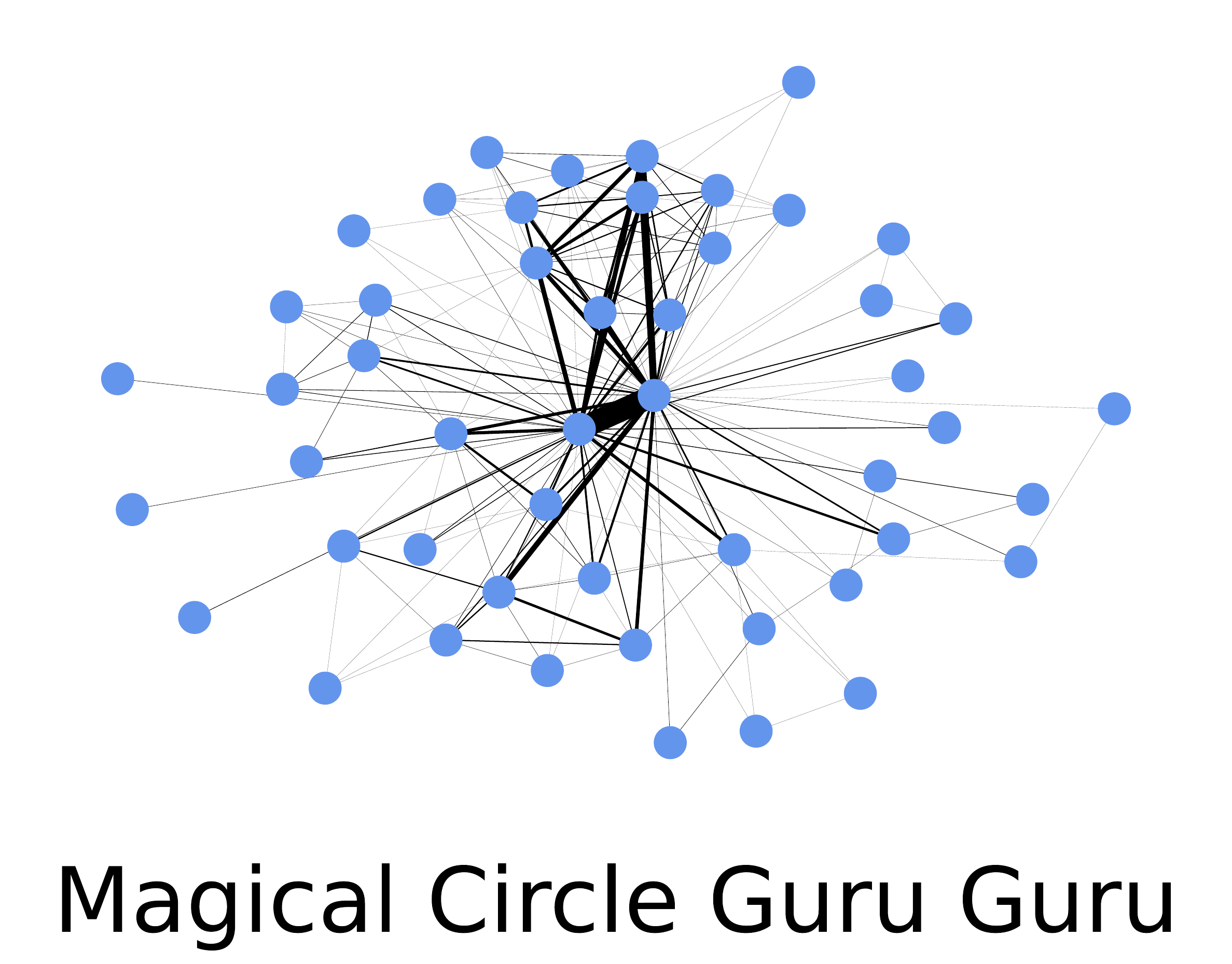}
     \end{subfigure}
     \begin{subfigure}[b]{0.195\textwidth}
         \centering
         \includegraphics[width=\textwidth]{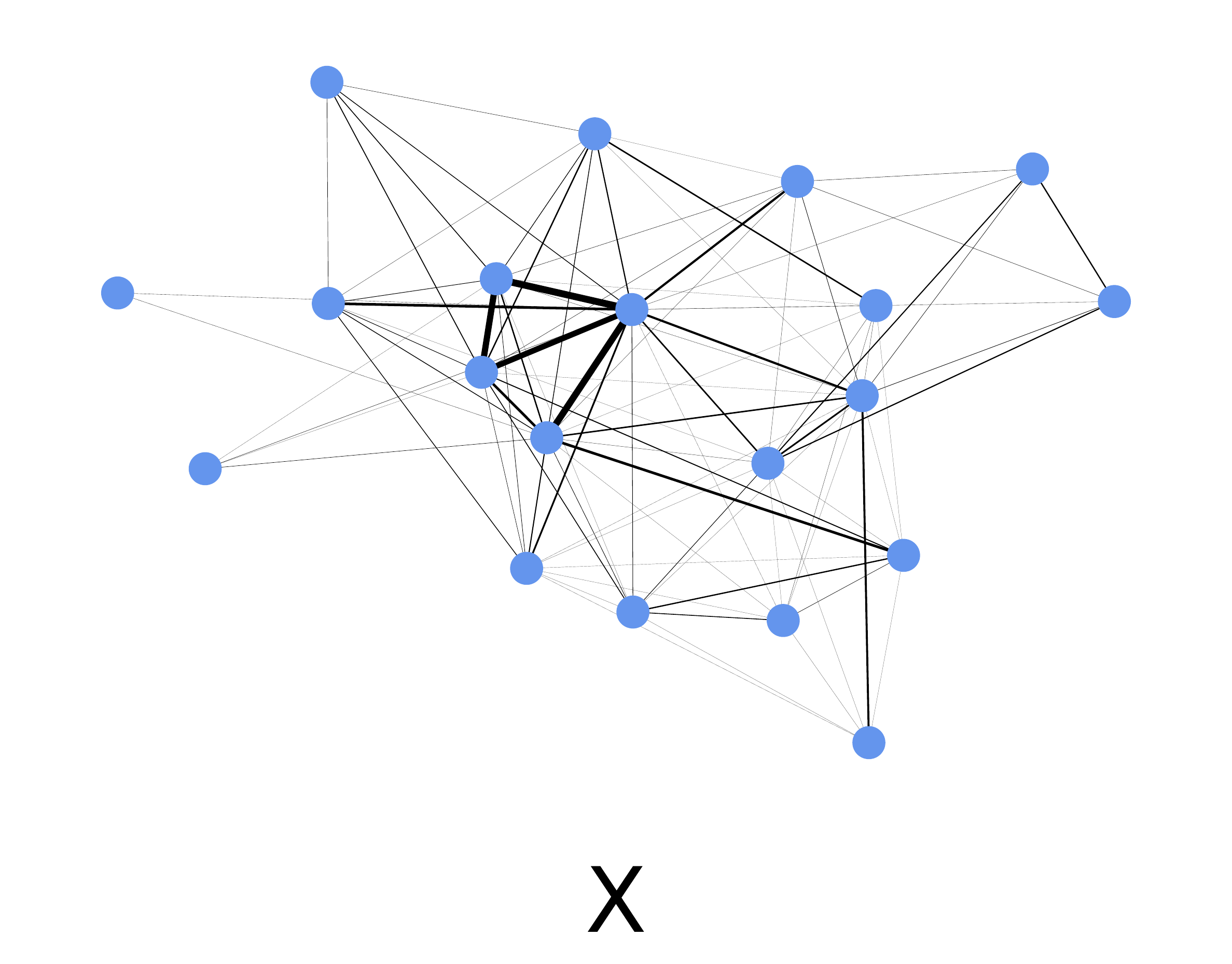}
     \end{subfigure}
     \begin{subfigure}[b]{0.195\textwidth}
         \centering
         \includegraphics[width=\textwidth]{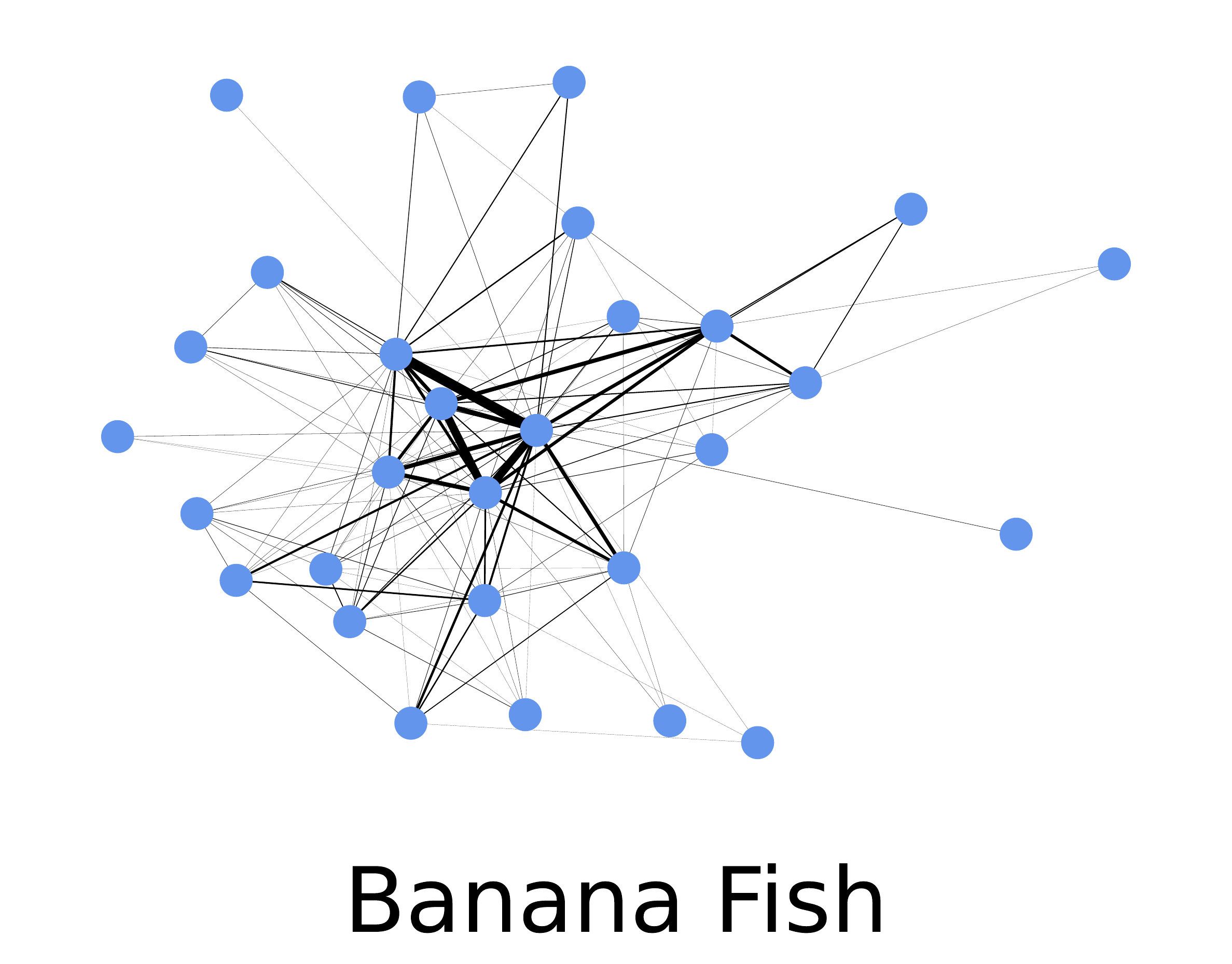}
     \end{subfigure}
     \begin{subfigure}[b]{0.195\textwidth}
         \centering
         \includegraphics[width=\textwidth]{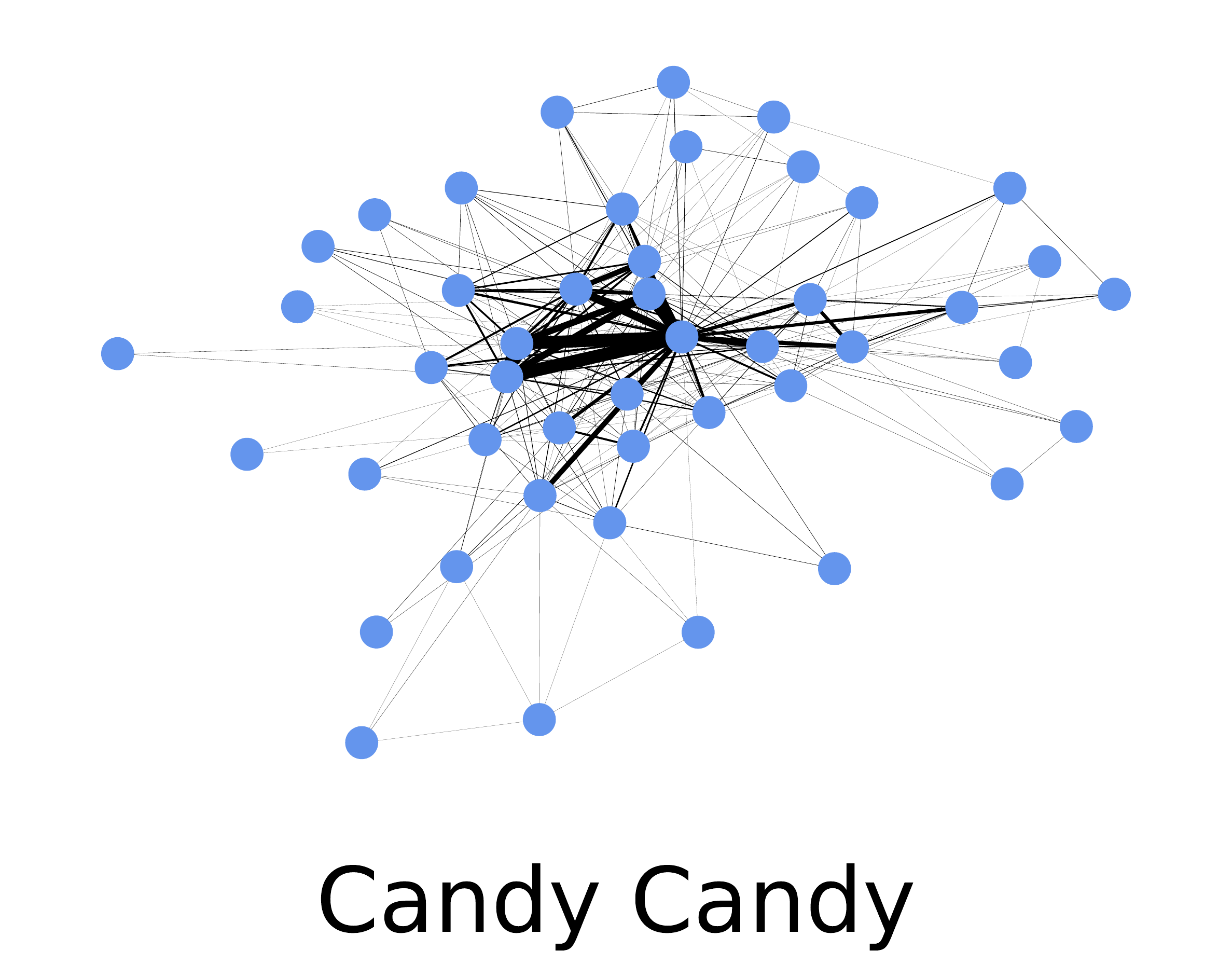}
     \end{subfigure}
     \\
     \begin{subfigure}[b]{0.195\textwidth}
         \centering
         \includegraphics[width=\textwidth]{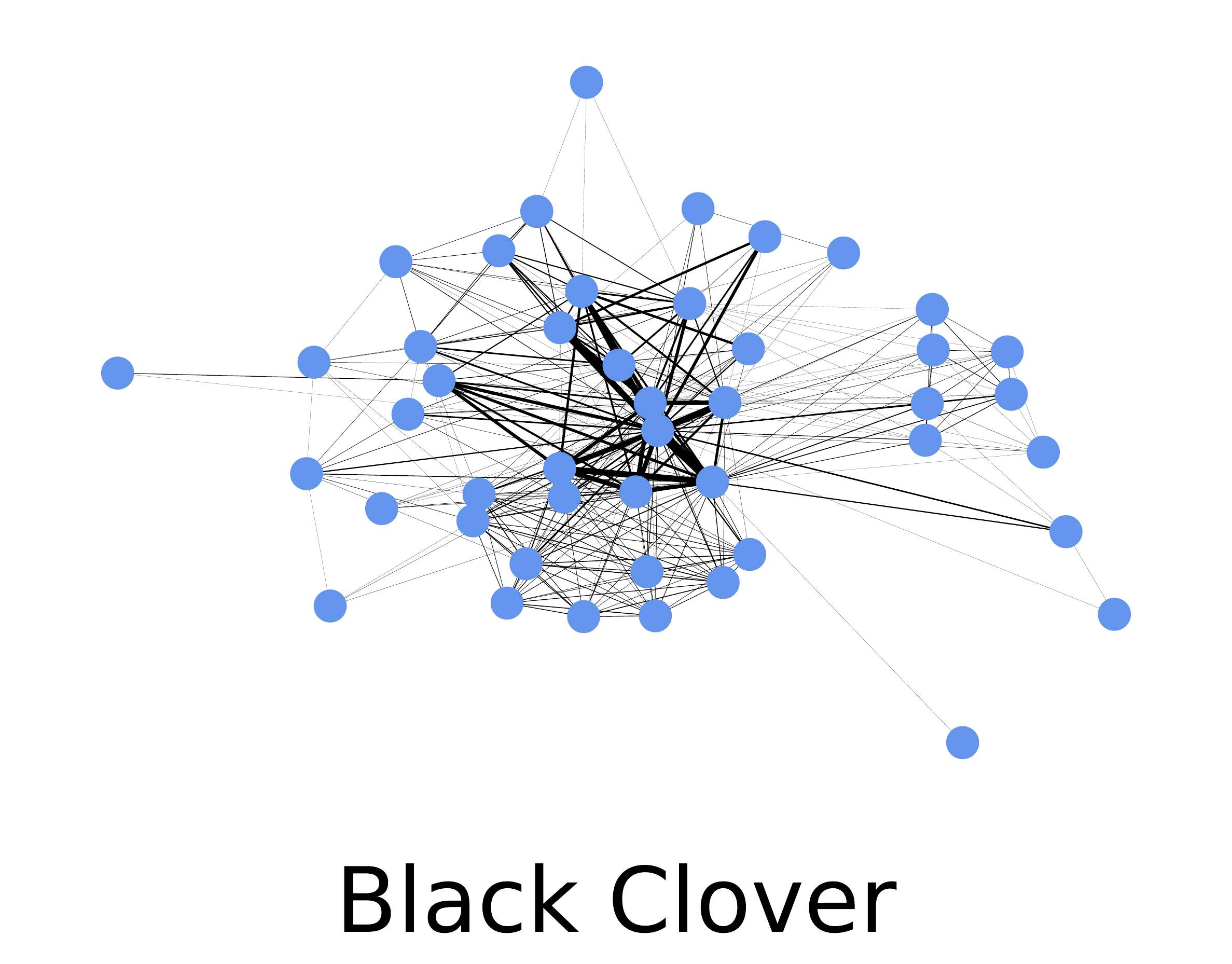}
     \end{subfigure}
     \begin{subfigure}[b]{0.195\textwidth}
         \centering
         \includegraphics[width=\textwidth]{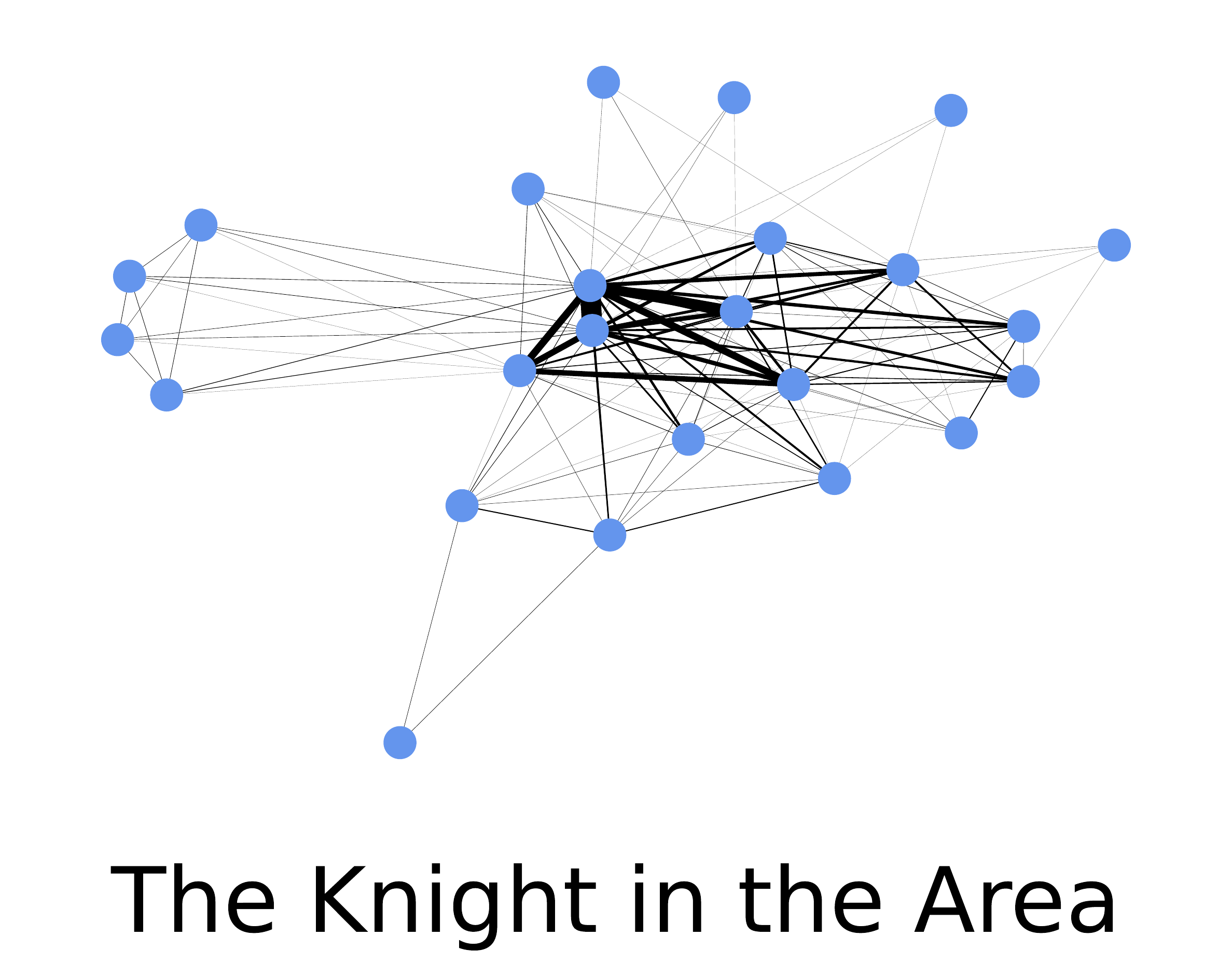}
     \end{subfigure}
     \begin{subfigure}[b]{0.195\textwidth}
         \centering
         \includegraphics[width=\textwidth]{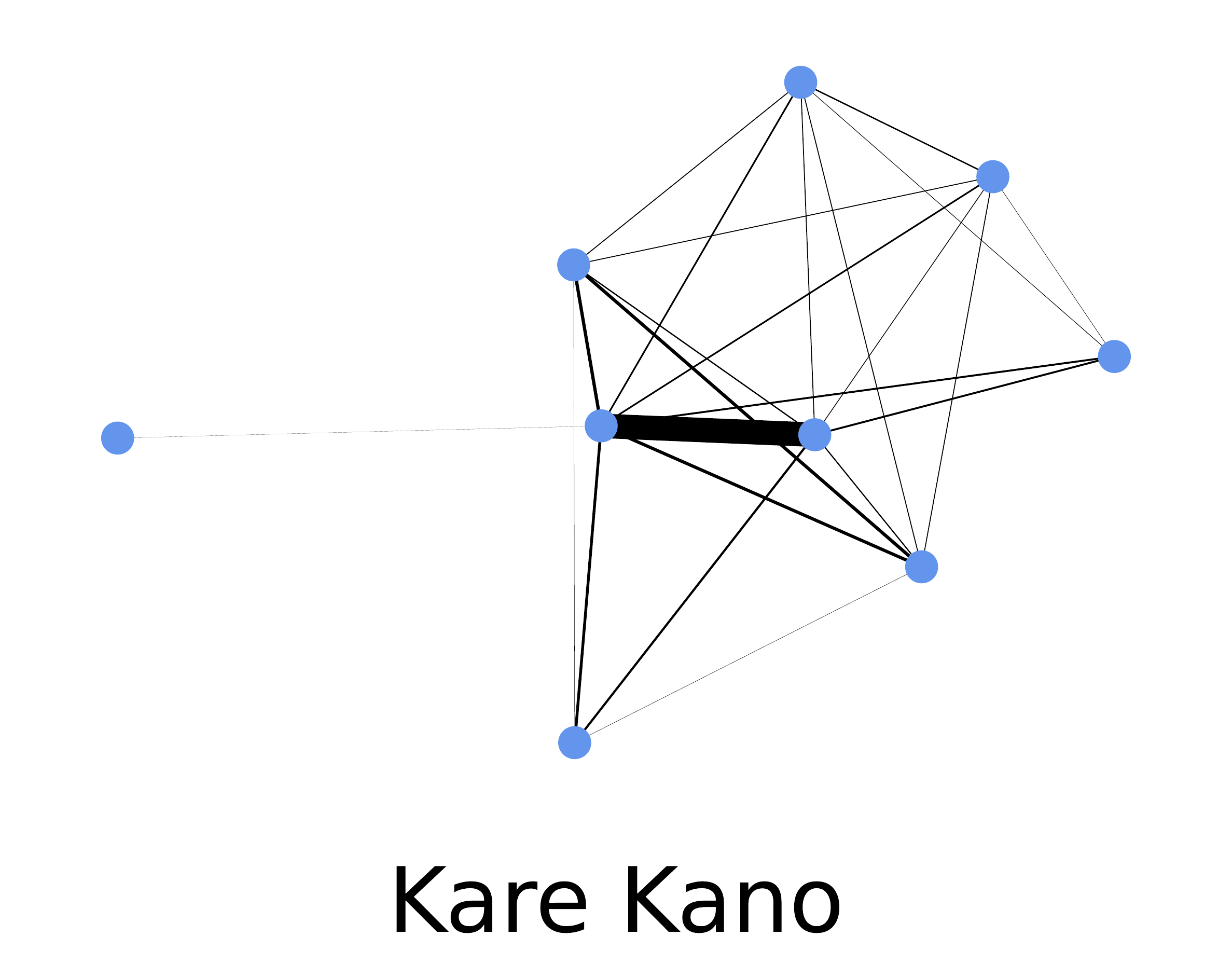}
     \end{subfigure}
     \begin{subfigure}[b]{0.195\textwidth}
         \centering
         \includegraphics[width=\textwidth]{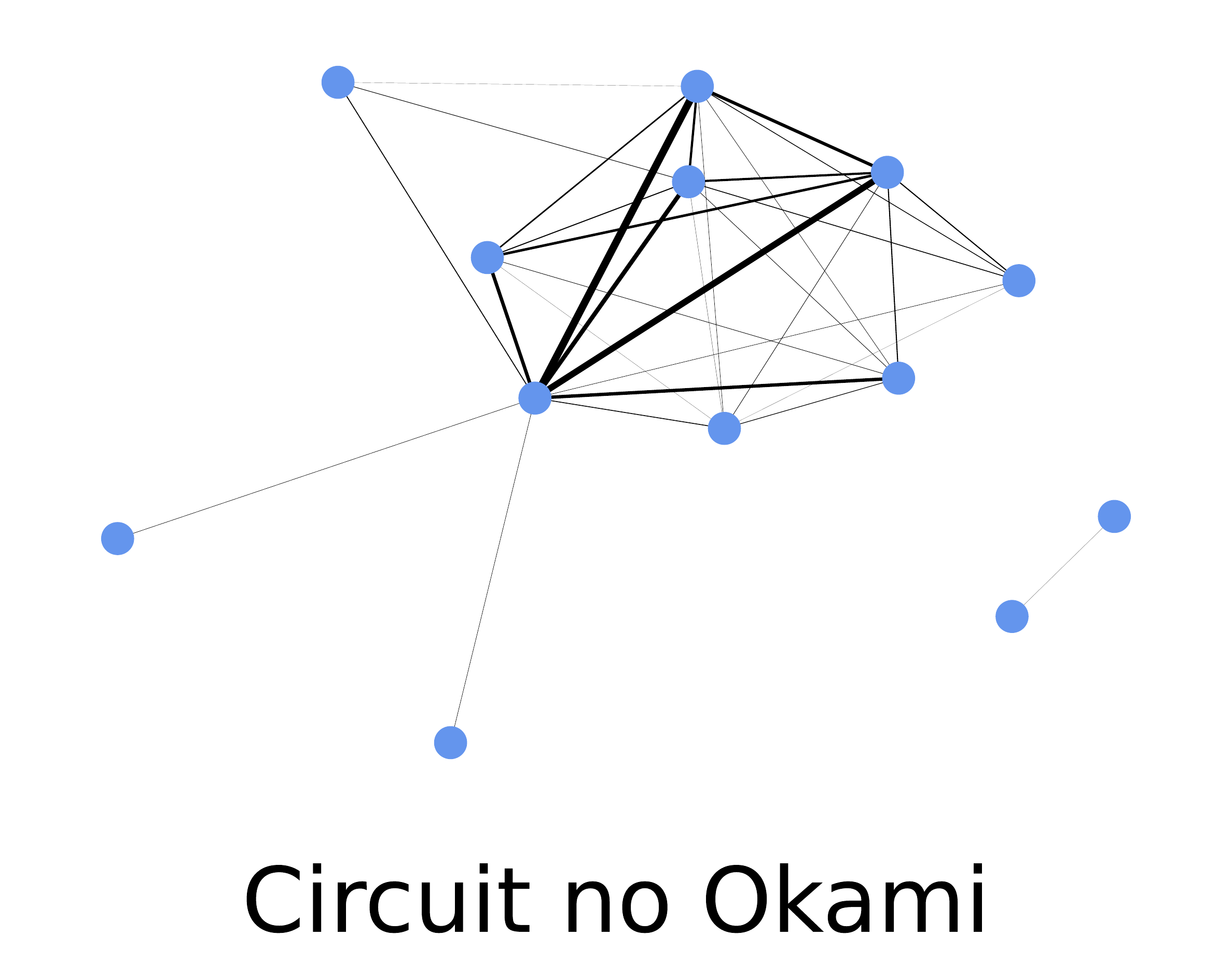}
     \end{subfigure}
     \begin{subfigure}[b]{0.195\textwidth}
         \centering
         \includegraphics[width=\textwidth]{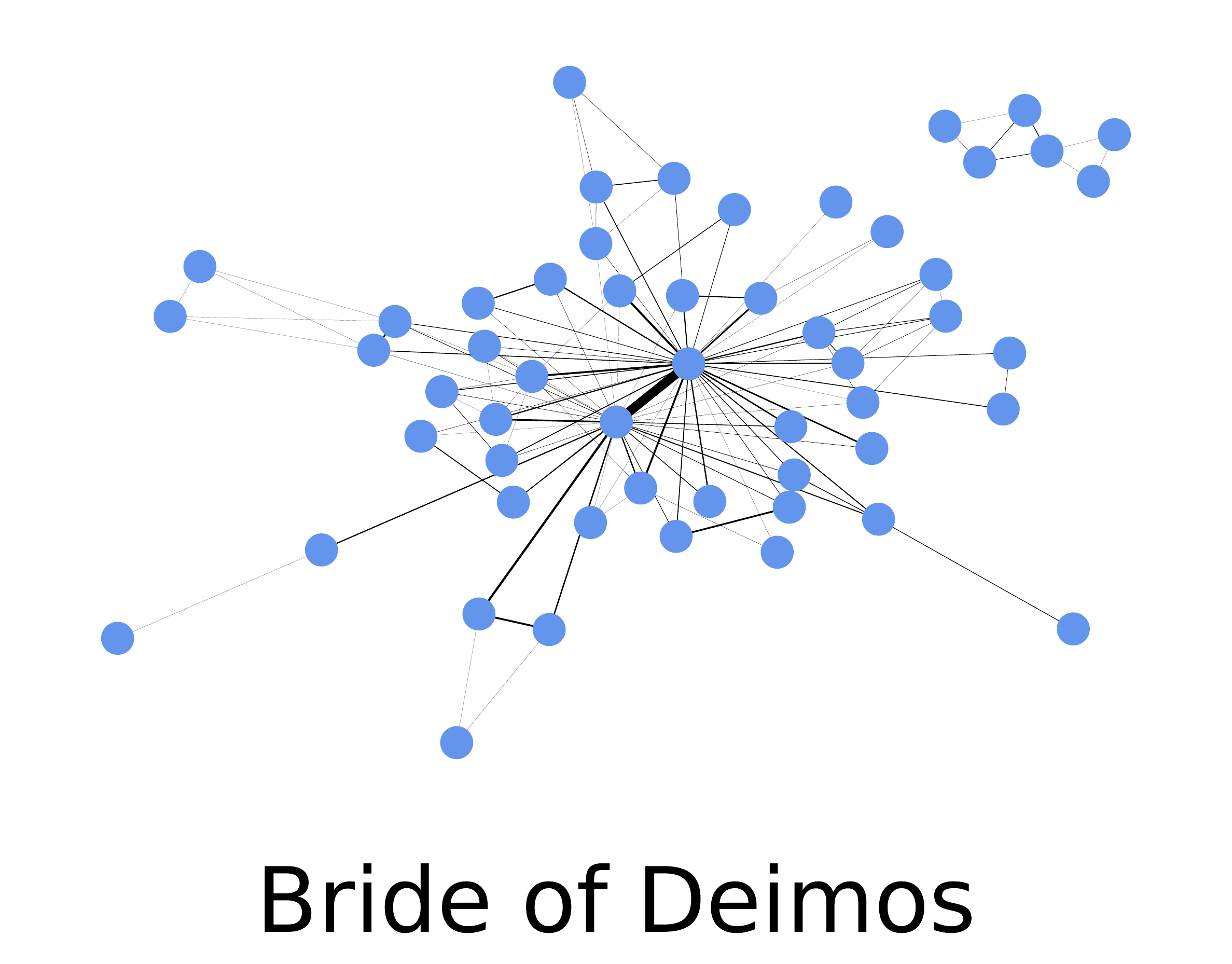}
     \end{subfigure}
     \\
      \begin{subfigure}[b]{0.195\textwidth}
         \centering
         \includegraphics[width=\textwidth]{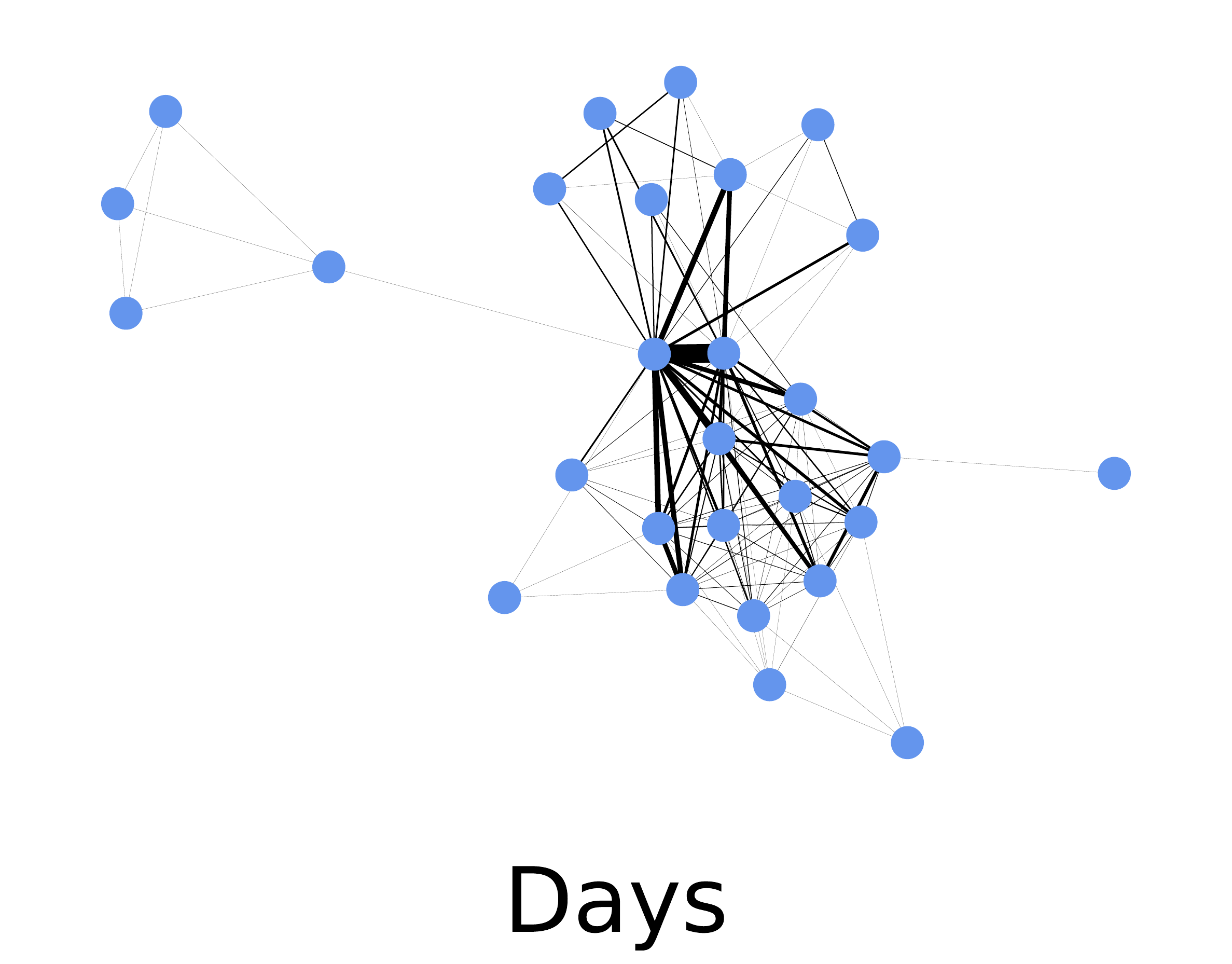}
     \end{subfigure}
     \begin{subfigure}[b]{0.195\textwidth}
         \centering
         \includegraphics[width=\textwidth]{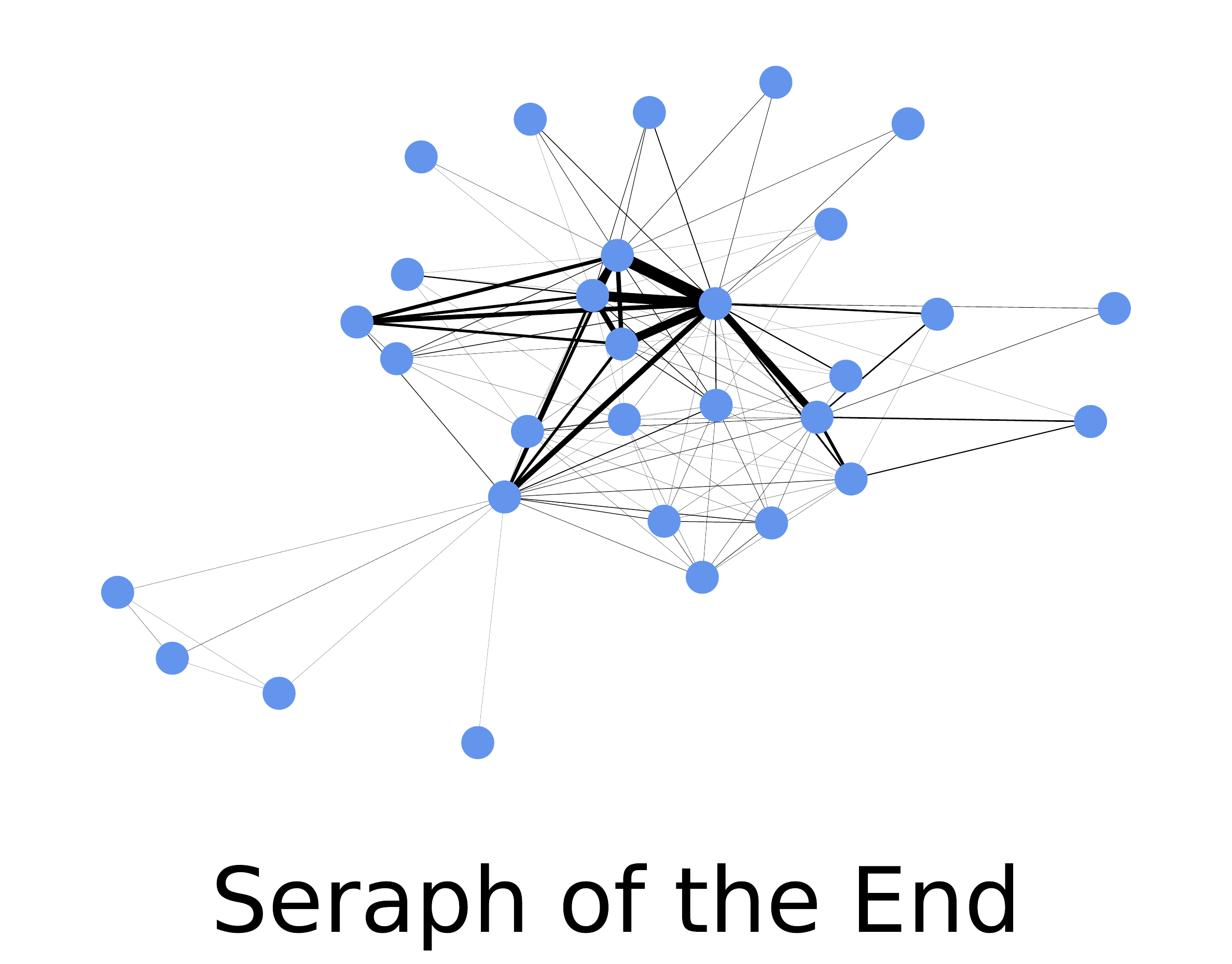}
     \end{subfigure}
     \begin{subfigure}[b]{0.195\textwidth}
         \centering
         \includegraphics[width=\textwidth]{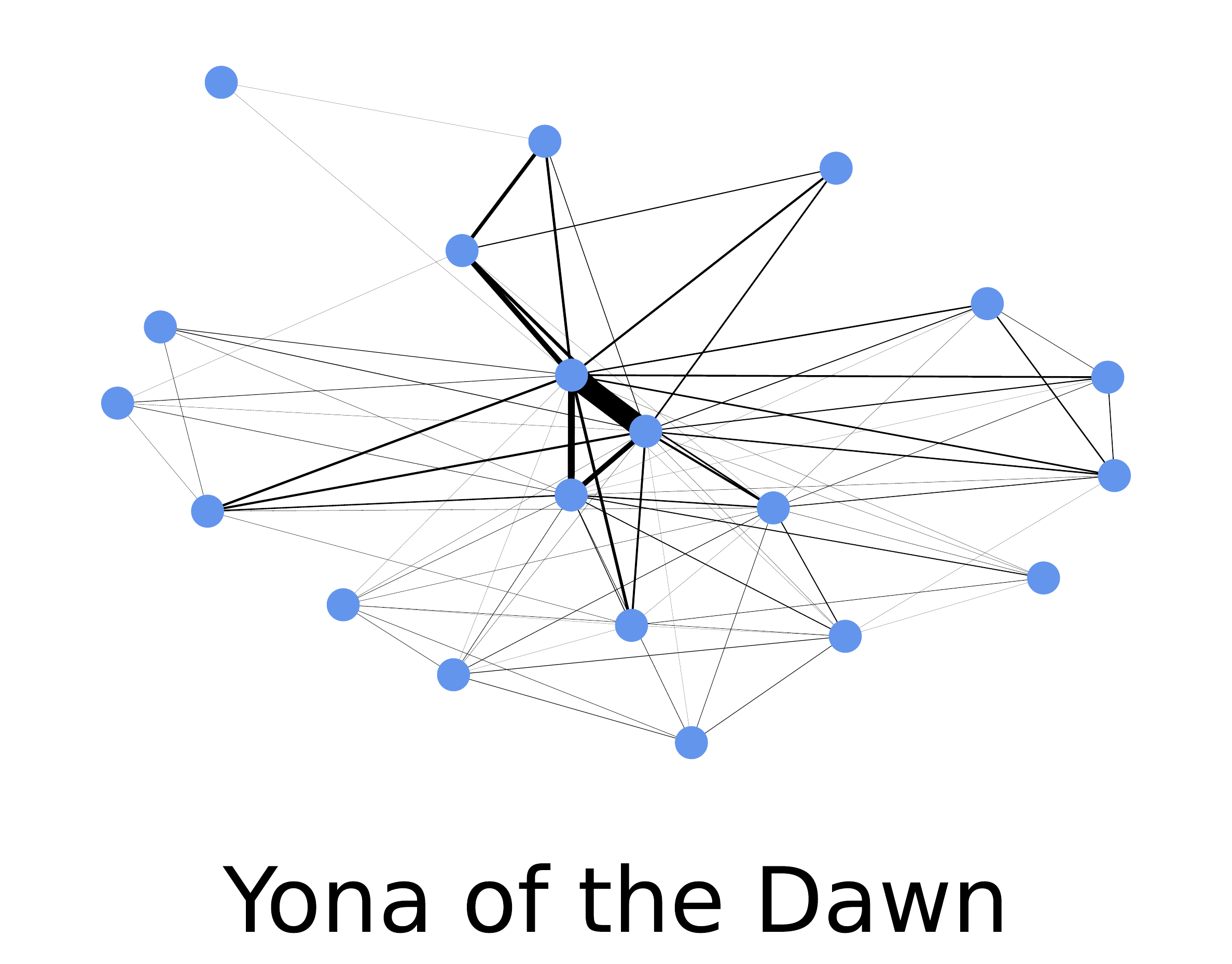}
     \end{subfigure}
     \begin{subfigure}[b]{0.195\textwidth}
         \centering
         \includegraphics[width=\textwidth]{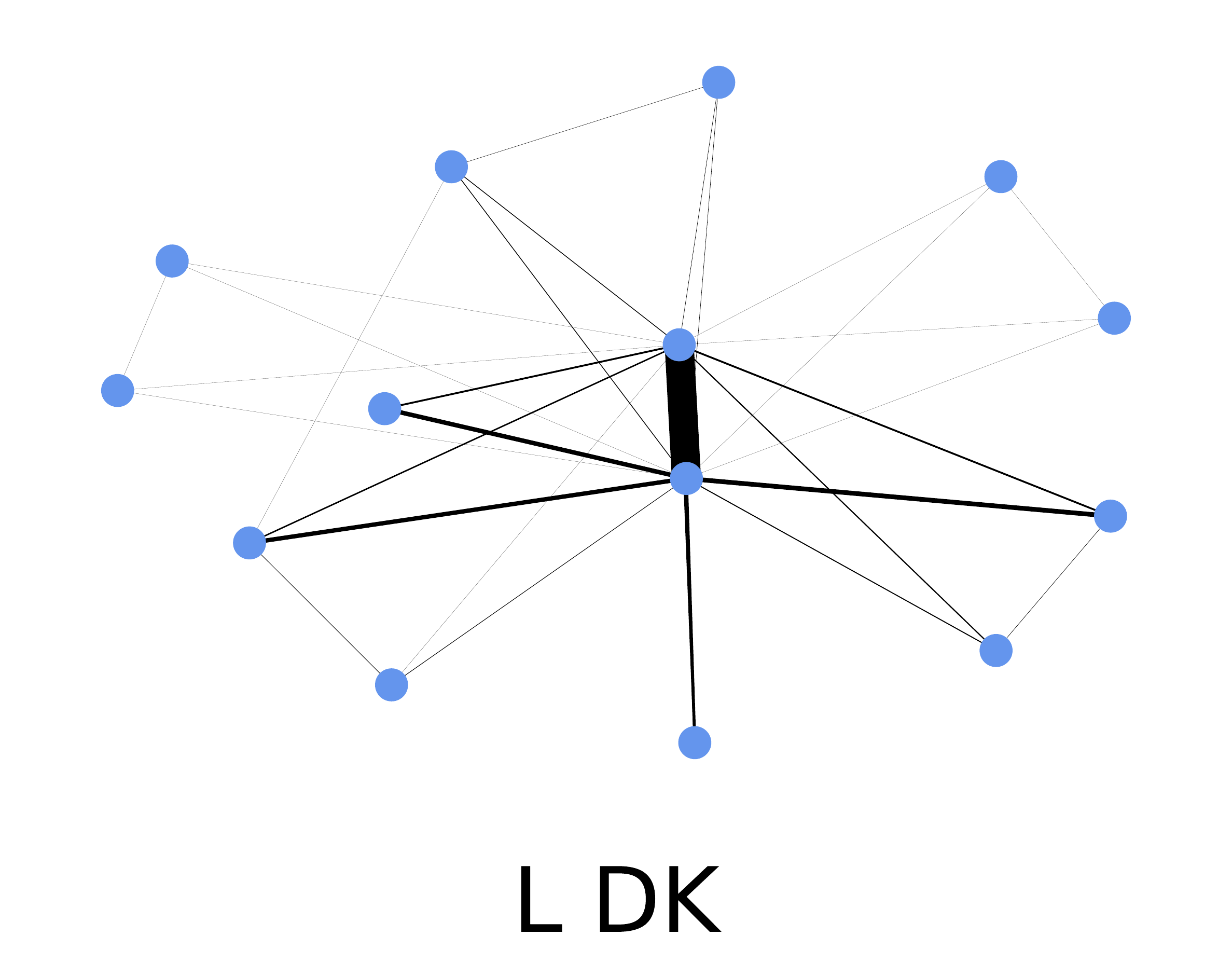}
     \end{subfigure}
     \begin{subfigure}[b]{0.195\textwidth}
         \centering
         \includegraphics[width=\textwidth]{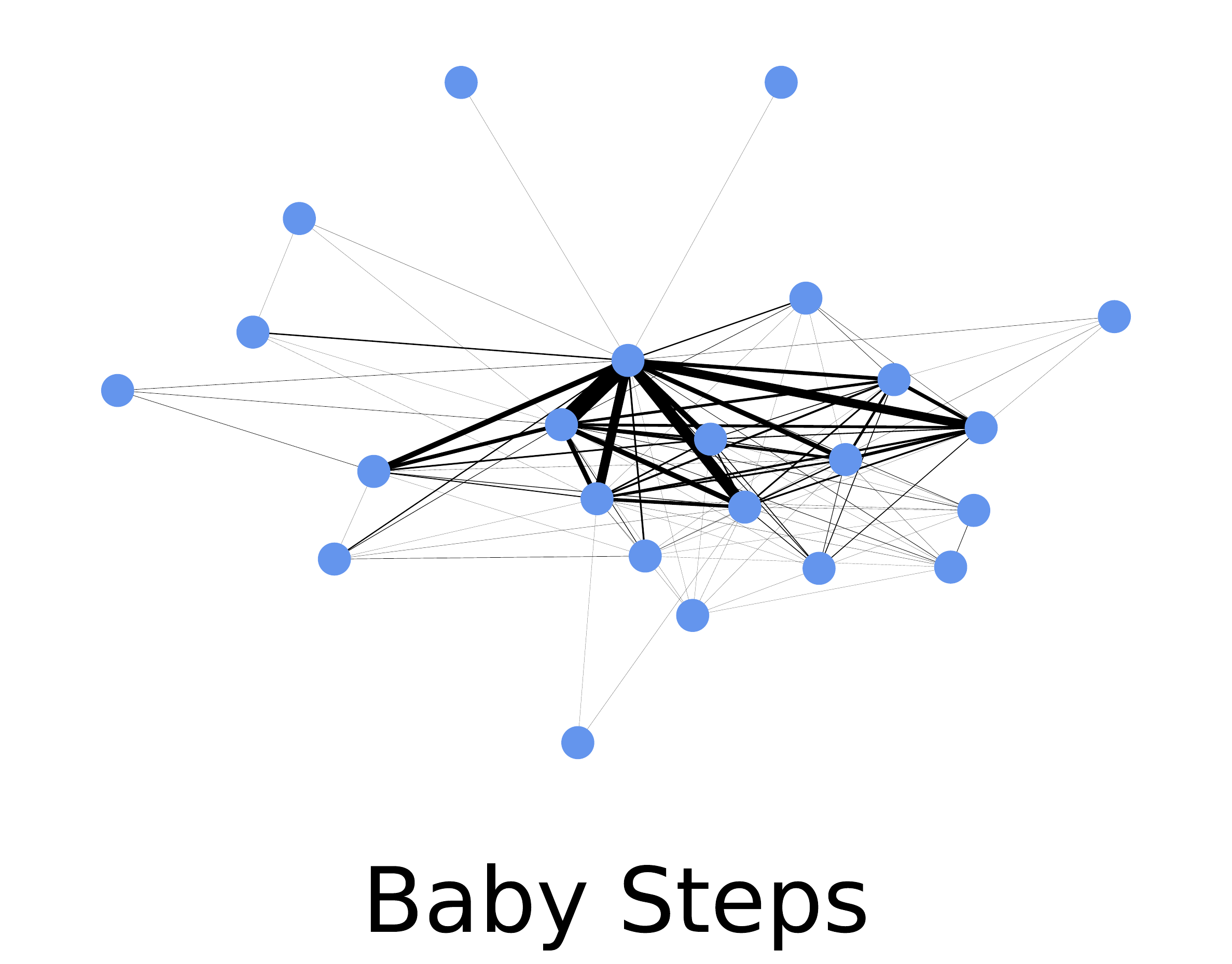}
     \end{subfigure}
     \\
      \begin{subfigure}[b]{0.195\textwidth}
         \centering
         \includegraphics[width=\textwidth]{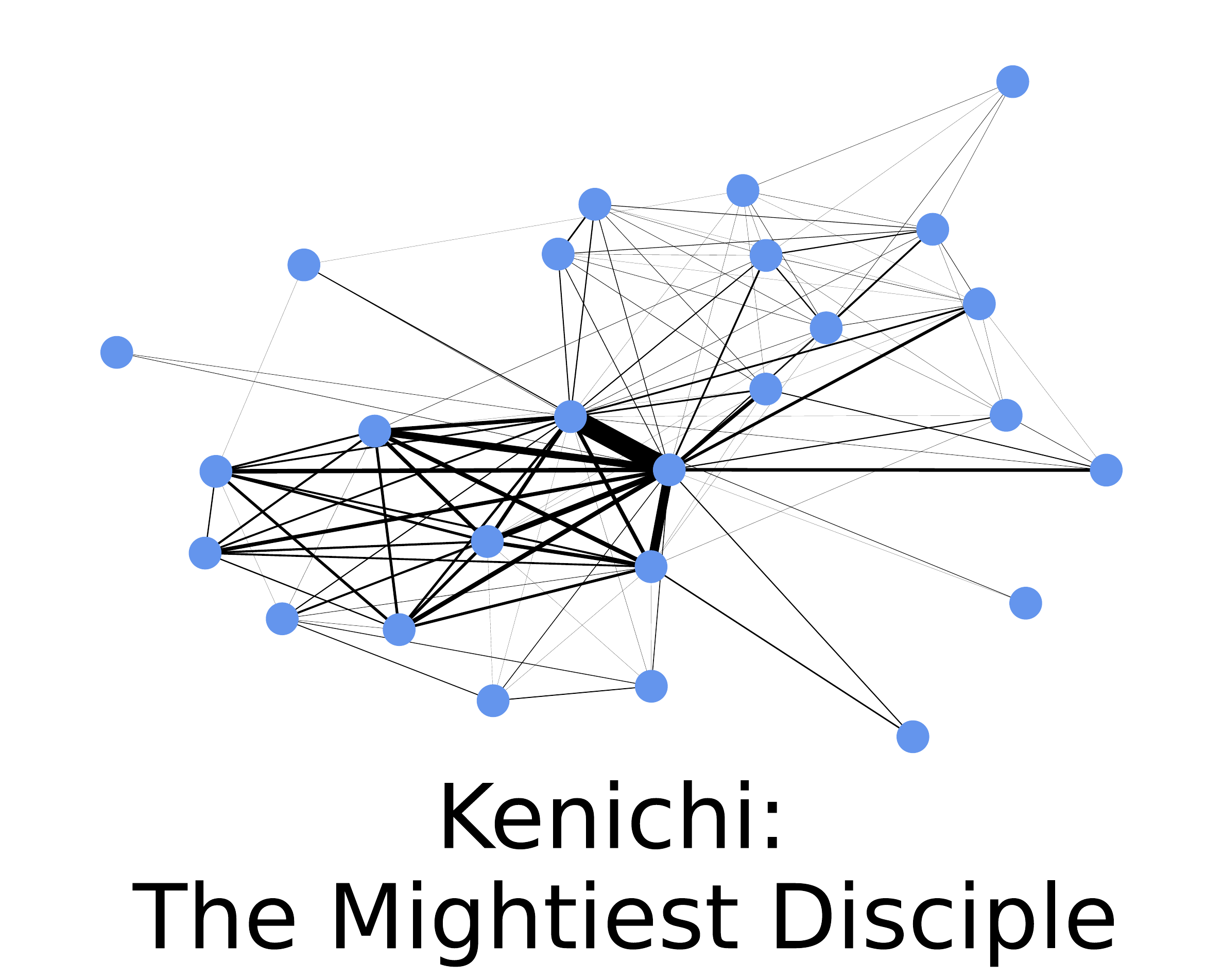}
     \end{subfigure}
     \begin{subfigure}[b]{0.195\textwidth}
         \centering
         \includegraphics[width=\textwidth]{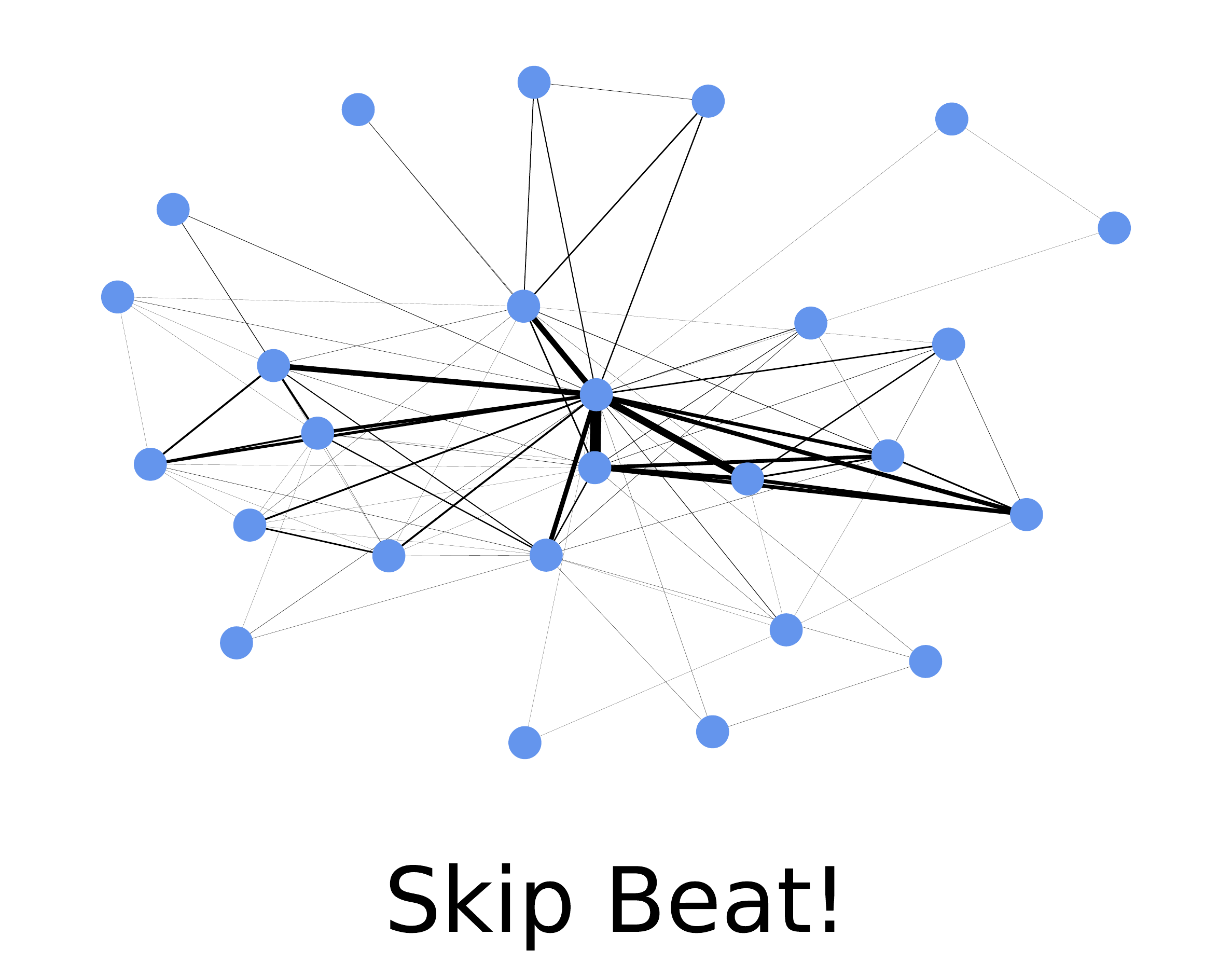}
     \end{subfigure}
     \begin{subfigure}[b]{0.195\textwidth}
         \centering
         \includegraphics[width=\textwidth]{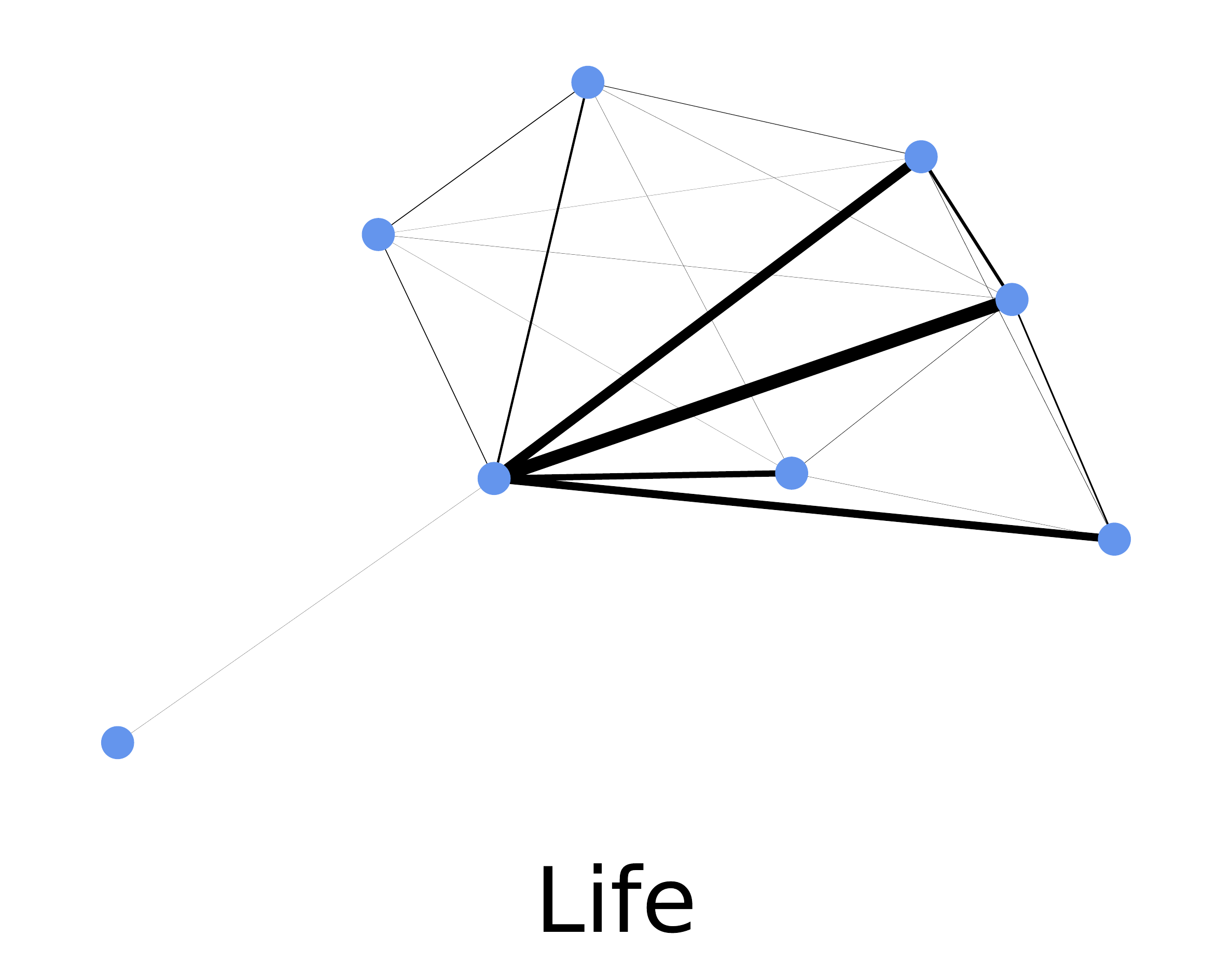}
     \end{subfigure}
     \begin{subfigure}[b]{0.195\textwidth}
         \centering
         \includegraphics[width=\textwidth]{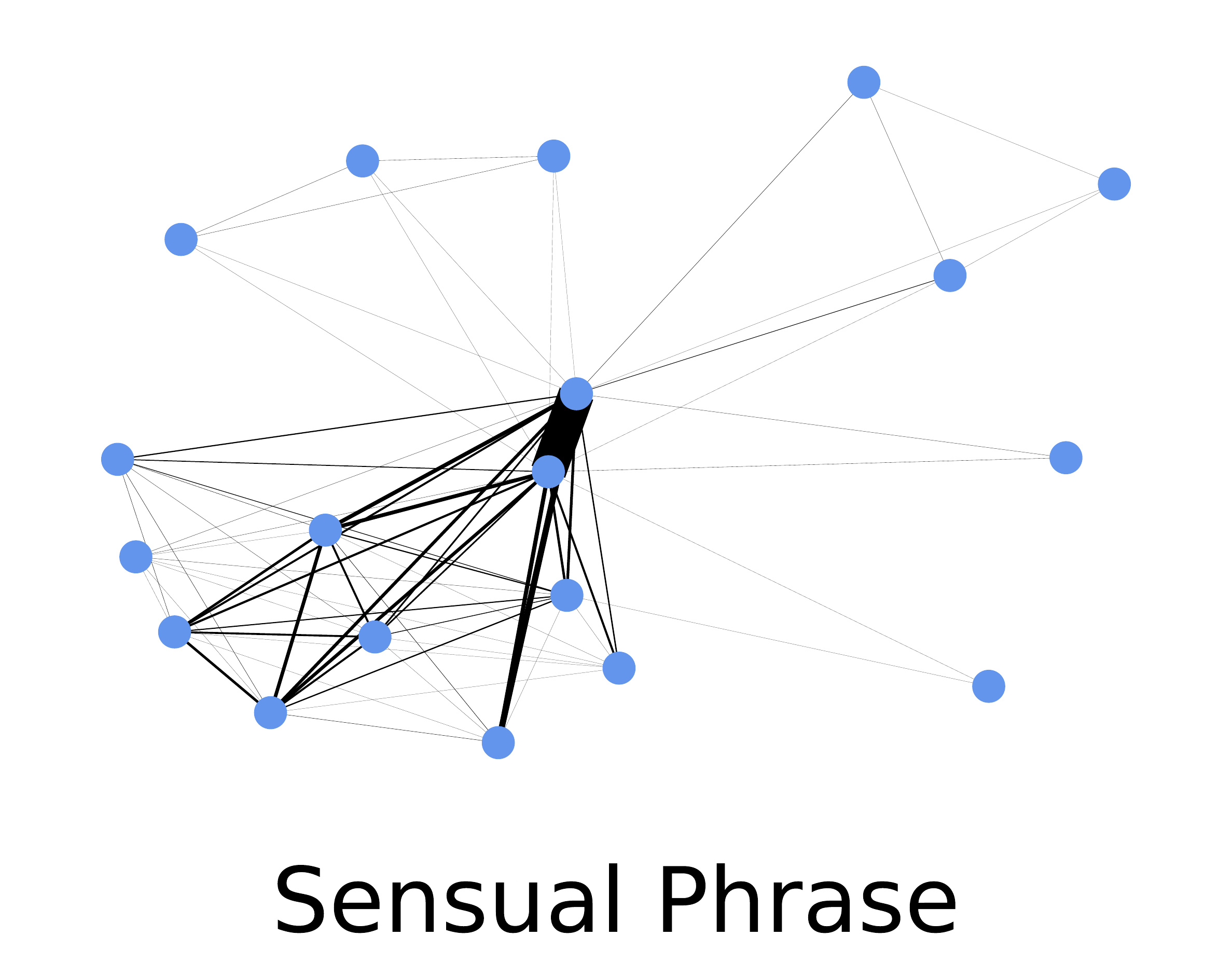}
     \end{subfigure}
     \begin{subfigure}[b]{0.195\textwidth}
         \centering
         \includegraphics[width=\textwidth]{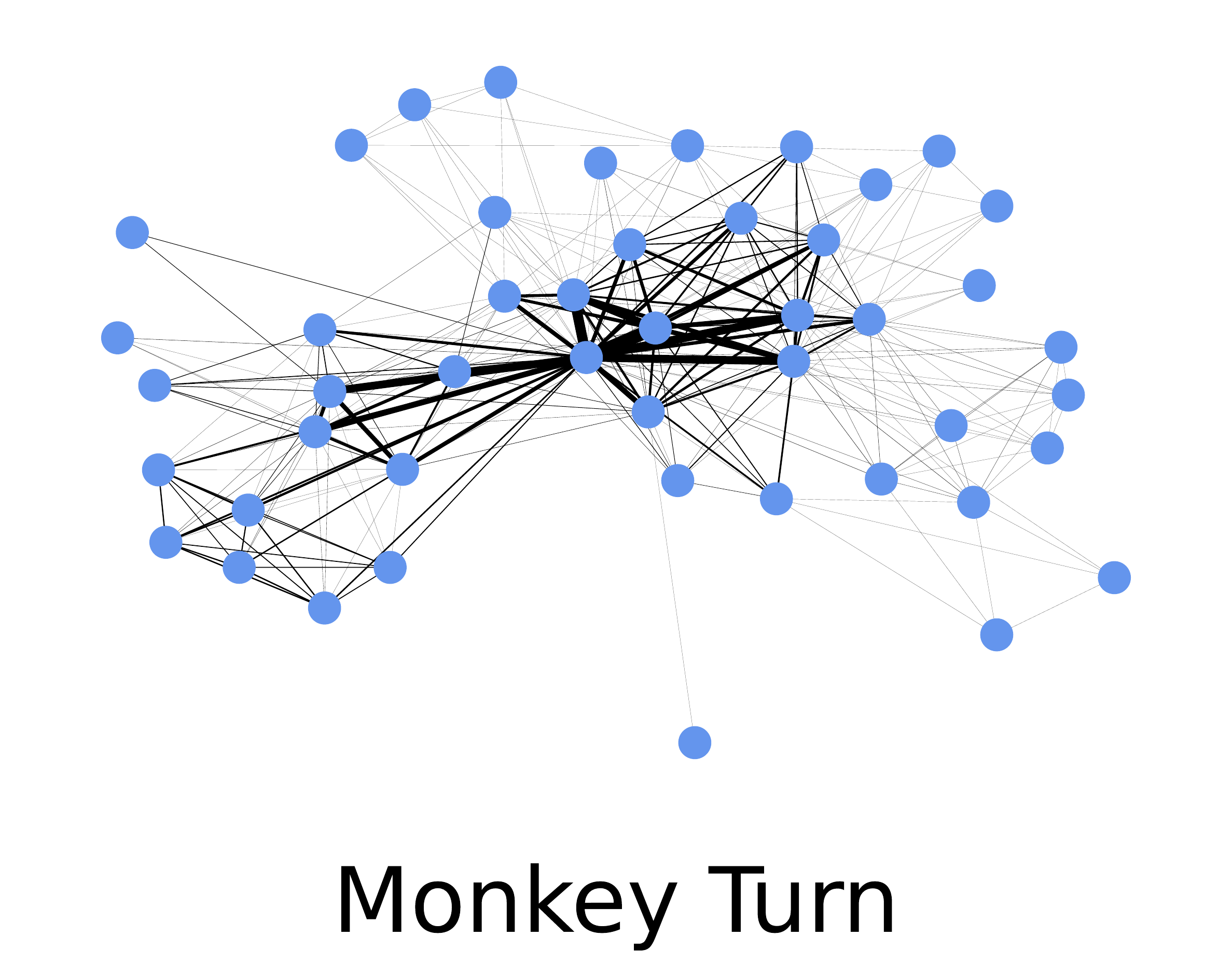}
     \end{subfigure}
     \\
      \begin{subfigure}[b]{0.195\textwidth}
         \centering
         \includegraphics[width=\textwidth]{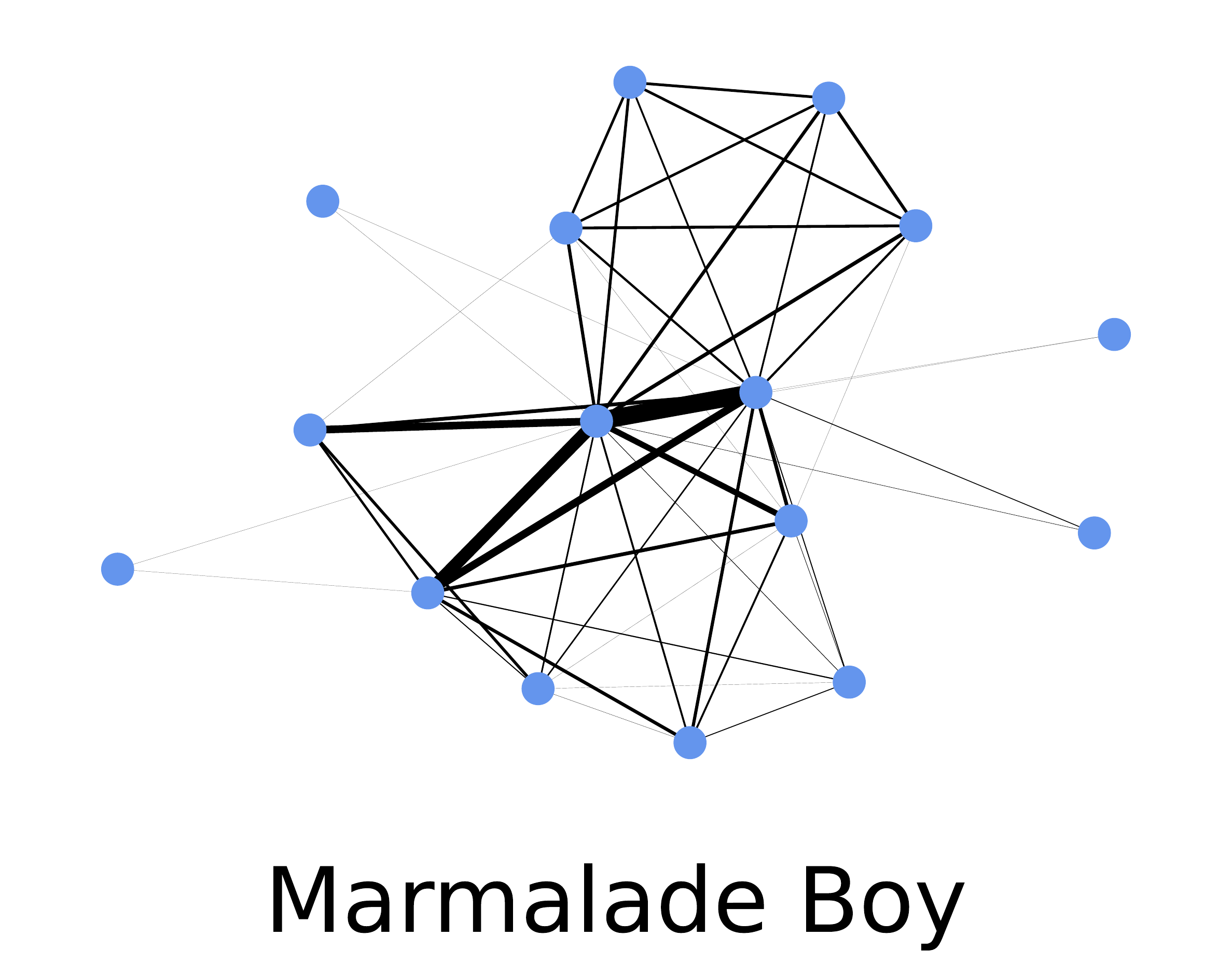}
     \end{subfigure}
     \begin{subfigure}[b]{0.195\textwidth}
         \centering
         \includegraphics[width=\textwidth]{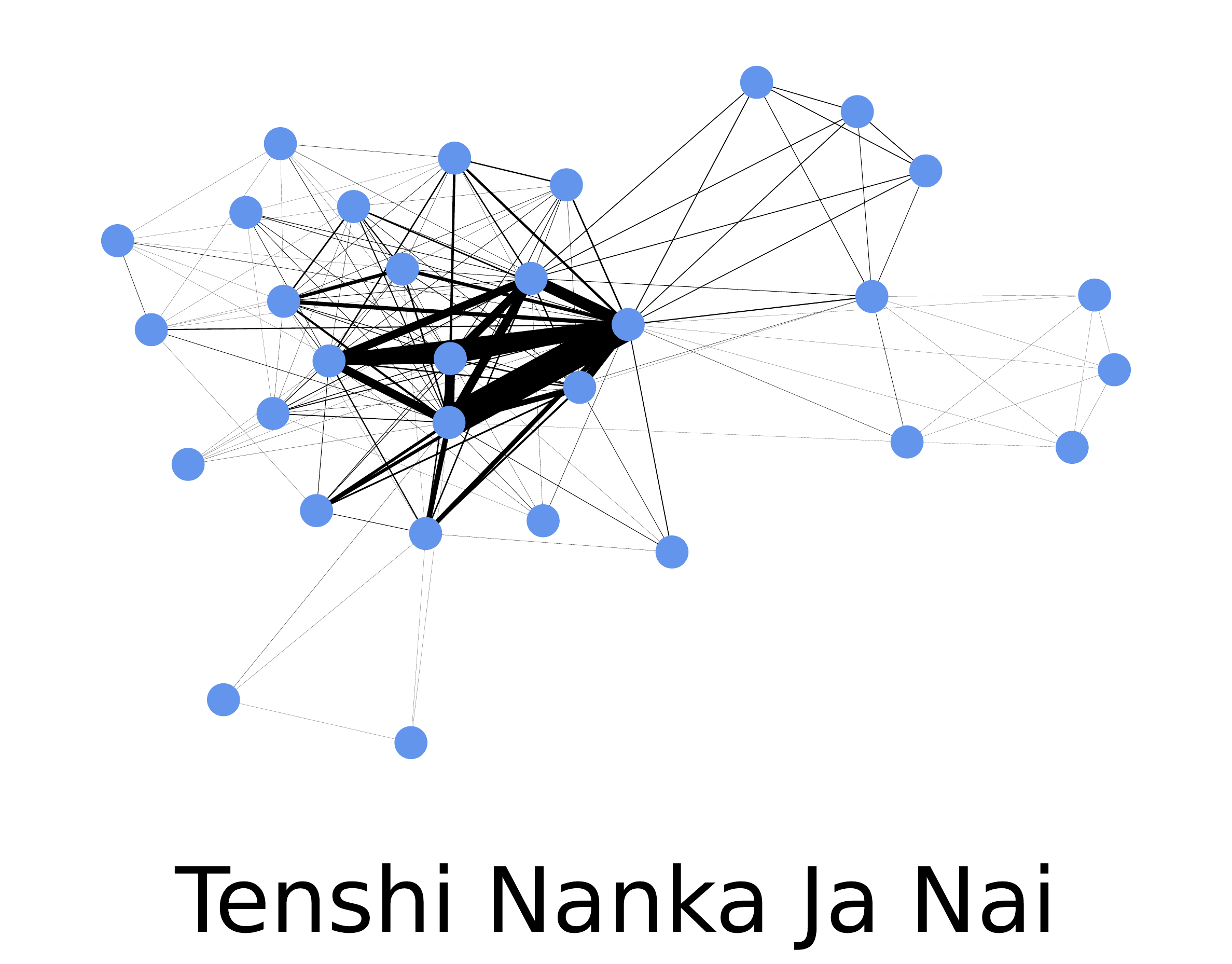}
     \end{subfigure}
     \begin{subfigure}[b]{0.195\textwidth}
         \centering
         \includegraphics[width=\textwidth]{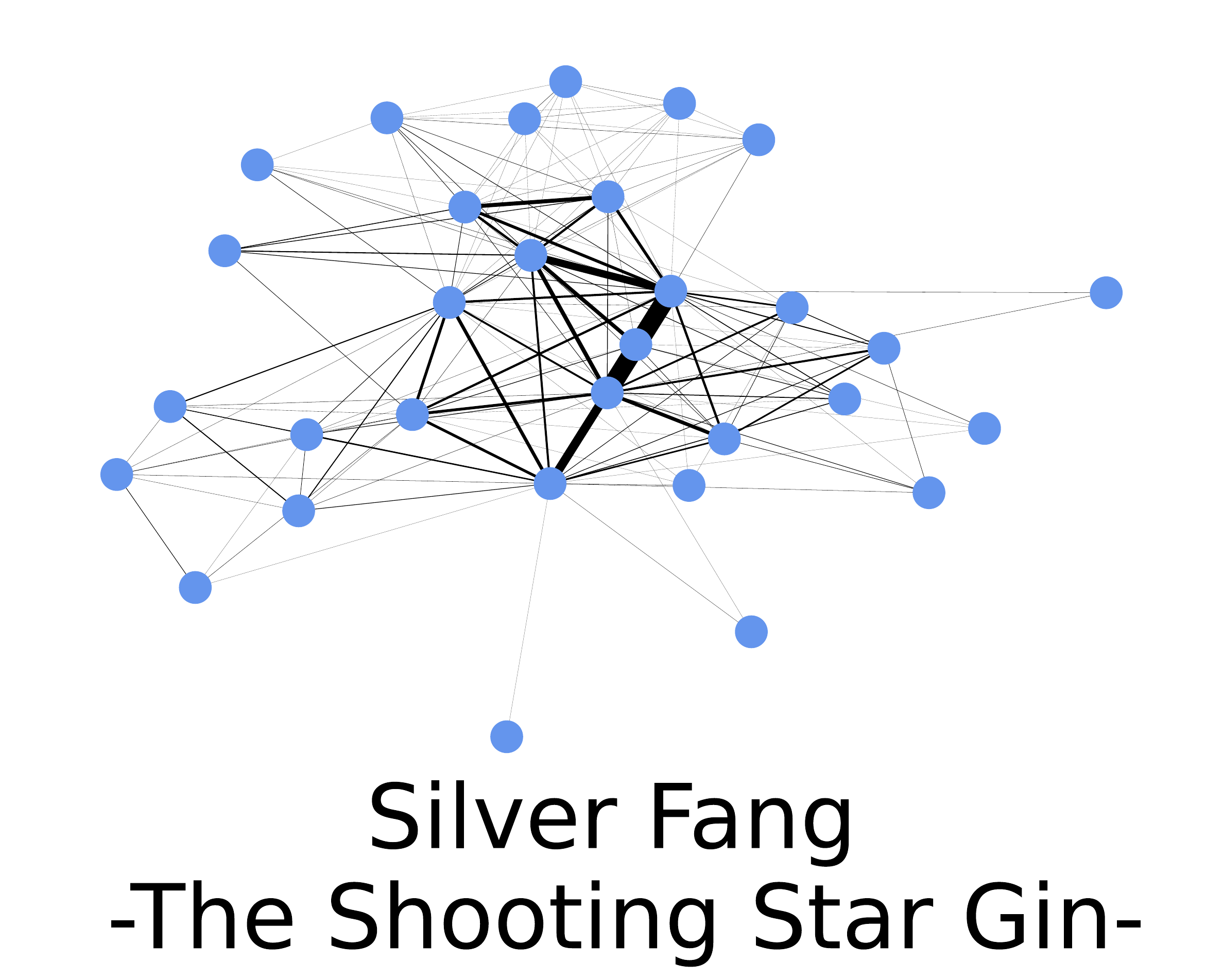}
     \end{subfigure}
     \begin{subfigure}[b]{0.195\textwidth}
         \centering
         \includegraphics[width=\textwidth]{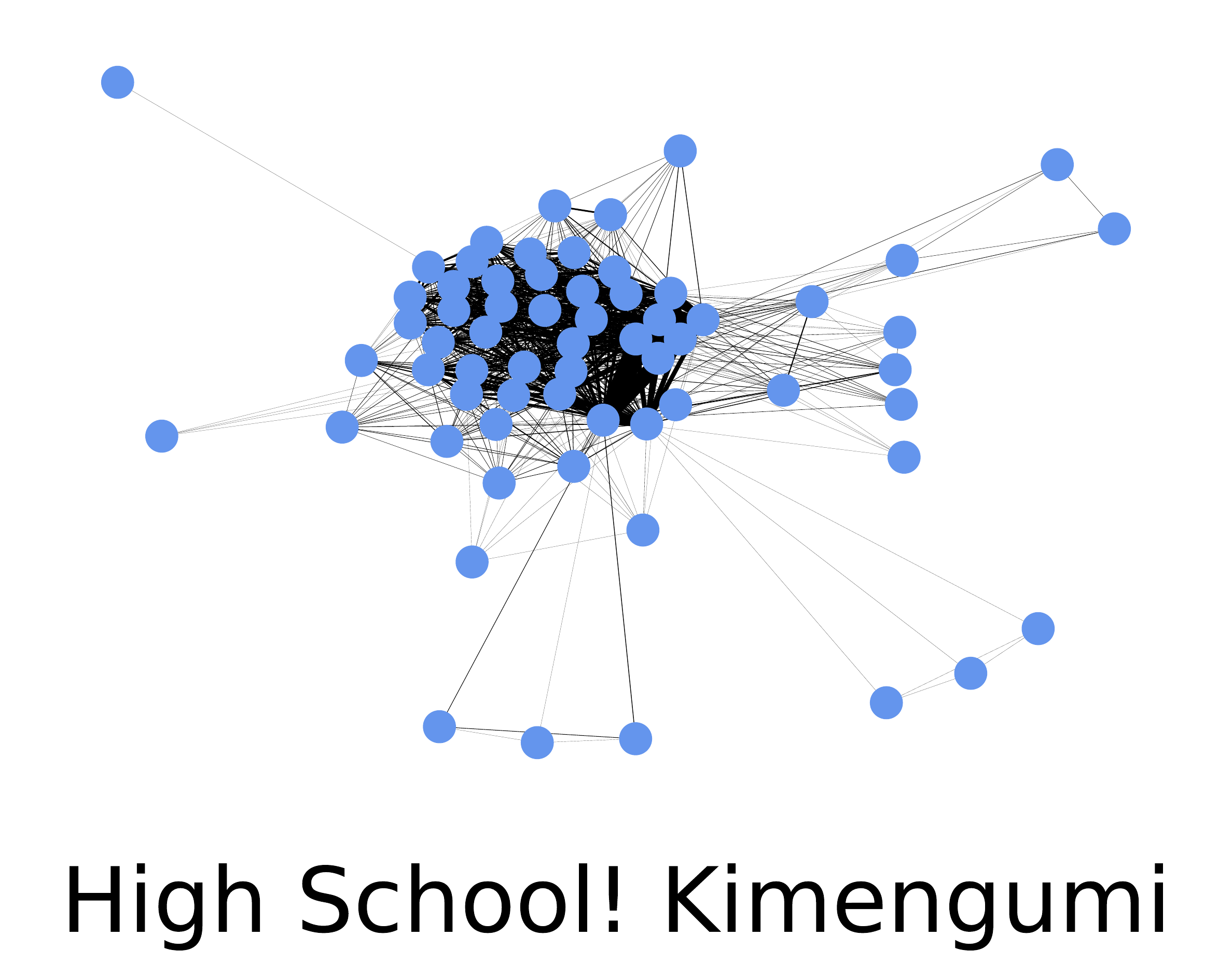}
     \end{subfigure}
     \begin{subfigure}[b]{0.195\textwidth}
         \centering
         \includegraphics[width=\textwidth]{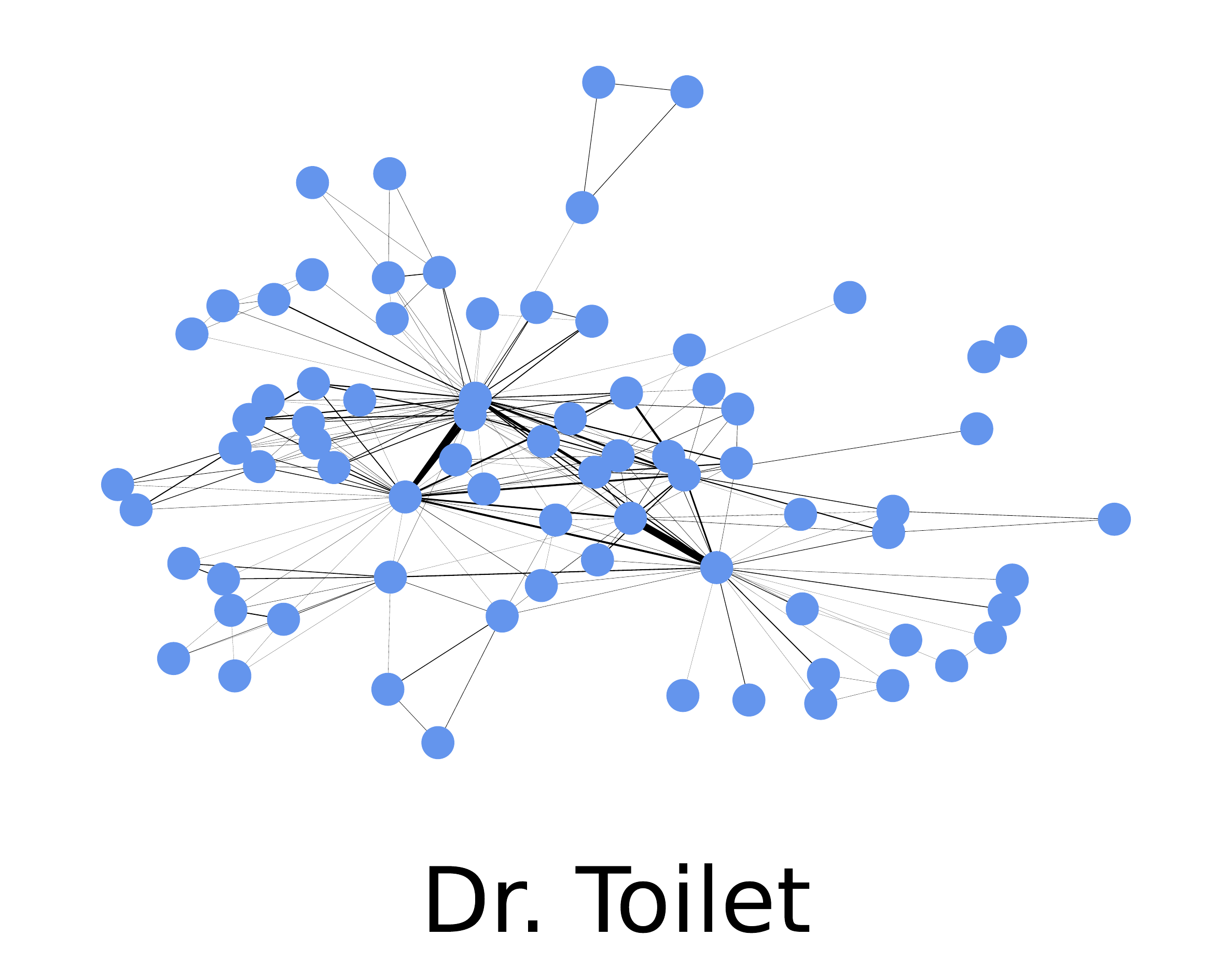}
     \end{subfigure}
     \\
      \begin{subfigure}[b]{0.195\textwidth}
         \centering
         \includegraphics[width=\textwidth]{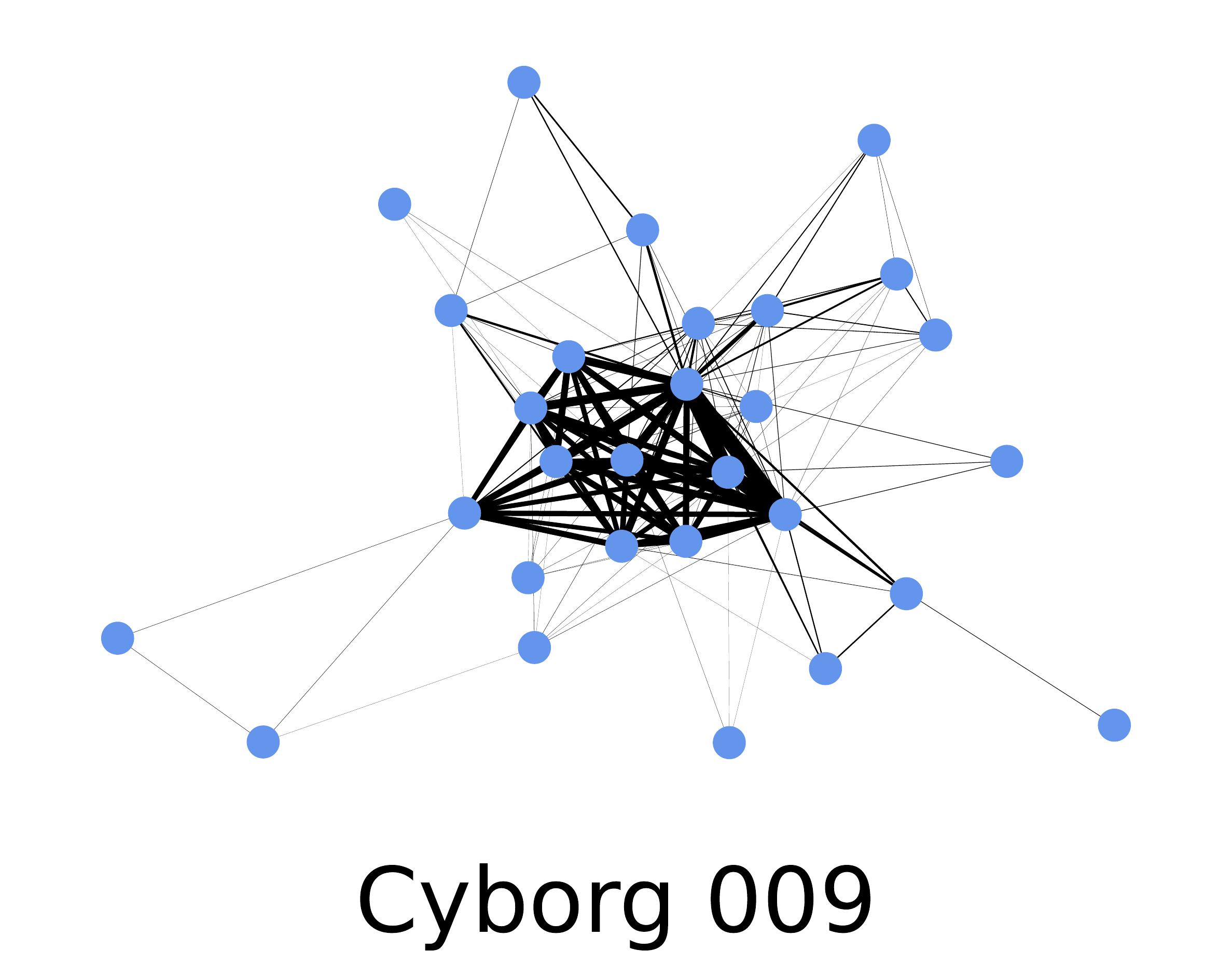}
     \end{subfigure}
     \begin{subfigure}[b]{0.195\textwidth}
         \centering
         \includegraphics[width=\textwidth]{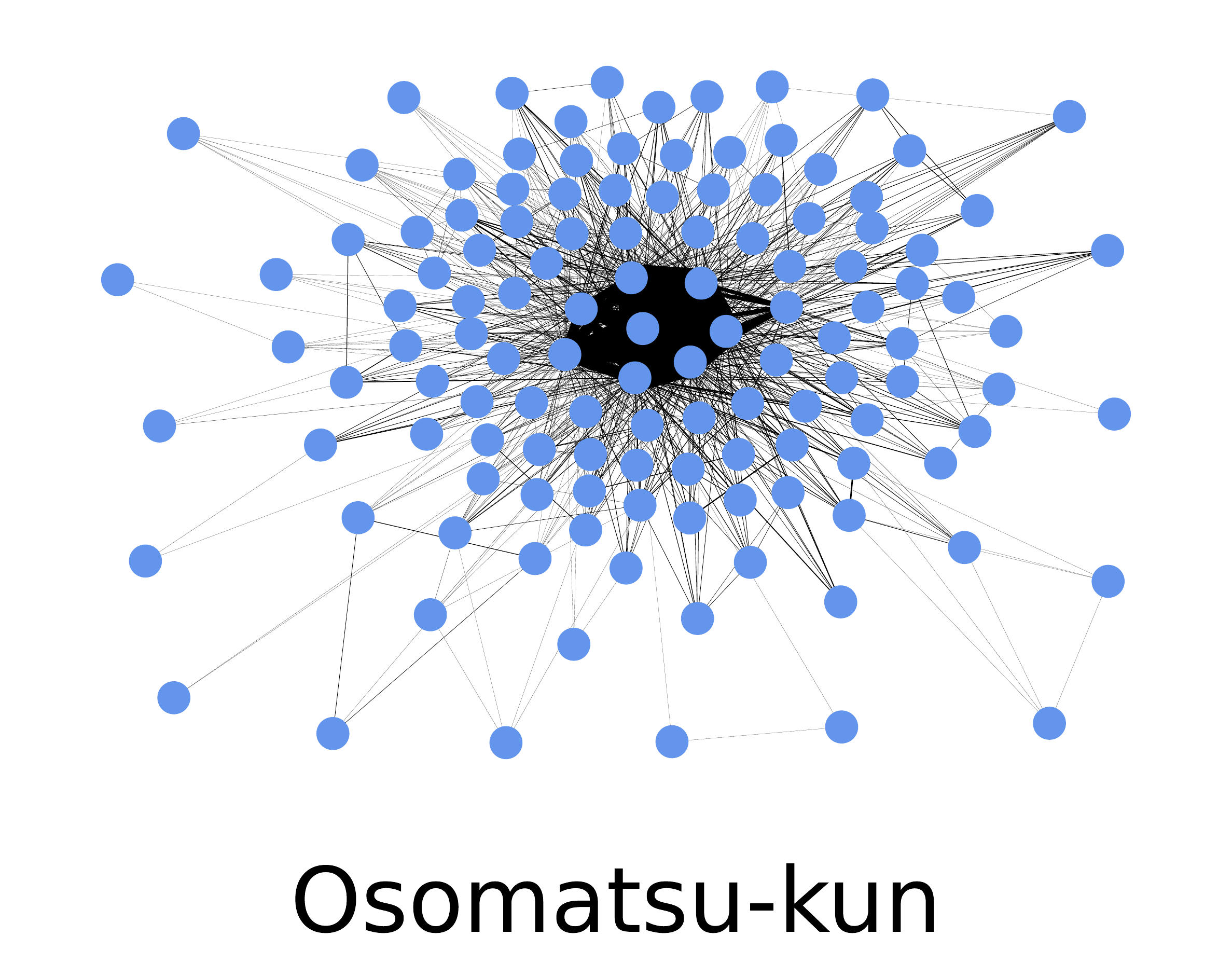}
     \end{subfigure}

    \caption{Character networks for 162 manga (continued)}
    
    \label{fig:three graphs}
\end{figure*}




\end{document}